\documentclass[12pt]{article}
\usepackage[sectionbib]{natbib}
\usepackage{array,epsfig,rotating}
\usepackage[]{hyperref}
\usepackage{sectsty, secdot}
\usepackage{mathtools,subcaption,mathrsfs,graphicx}
\usepackage{geometry}
\usepackage{amssymb}
\usepackage{amsfonts}
\usepackage{multirow}
\usepackage{amsthm}
\usepackage{placeins}

\newtheorem{theorem}{Theorem}
\newtheorem{algorithm}{Algorithm}
\newtheorem{lemma}{Lemma}
\newtheorem{corollary}{Corollary}

\theoremstyle{definition}
\newtheorem{definition}{Definition}

\newtheorem{remark}{Remark}

\DeclareMathOperator{\margmin}{\mathrm{argmin}}
\newcommand*{\addheight}[2][.5ex]{%
  \raisebox{0pt}[\dimexpr\height+(#1)\relax]{#2}%
}
\DeclarePairedDelimiter{\ceil}{\lceil}{\rceil}
\newcommand*{\QED}{\hfill\ensuremath{\square}}

\begin{document}

\title{Principal Sub-manifolds}
\author{Zhigang Yao\footnote{Department of Statistics and Applied Probability, University of Singapore, Singapore}, Benjamin Eltzner\footnote{Max-Planck-Institute for Multidisciplinary Sciences, Goettingen, Germany}, Tung Pham\footnote{School of Mathematics and Statistics, University of Melbourne, Australia}}
\maketitle

\begin{abstract}
  We propose a novel method of finding principal components in multivariate data sets that lie on an embedded nonlinear Riemannian manifold within a higher-dimensional space. Our aim is to extend the geometric interpretation of PCA, while being able to capture non-geodesic modes of variation in the data. We introduce the concept of a principal sub-manifold, a manifold passing through a reference point, and at any point on the manifold extending in the direction of highest variation in the space spanned by the eigenvectors of the local tangent space PCA. Compared to recent work for the case where the sub-manifold is of dimension one \citet{Panaretos2014}--essentially a curve lying on the manifold attempting to capture one-dimensional variation--the current setting is much more general. The principal sub-manifold is therefore an extension of the principal flow, accommodating to capture higher dimensional variation in the data. We show the principal sub-manifold yields the ball spanned by the usual principal components in Euclidean space. By means of examples, we illustrate how to find, use and interpret a principal sub-manifold and we present an application in shape analysis.
\end{abstract}

\vspace{9pt}
\noindent {\it Key words and phrases:}
manifold, principal component analysis, tangent space, dimension reduction, shape analysis.

\def\thefigure{\arabic{figure}}
\def\thetable{\arabic{table}}

\renewcommand{\theequation}{\thesection.\arabic{equation}}

\section{Introduction}
\label{sec:Intro}

Many quantities of interest are best described as points in a non-Euclidean space, not as vectors in a vector space. The most well-known example are directional data represented on a circle or sphere in \emph{directional statistics}, which has been discussed as early as \citet{Fisher1953}. Higher dimensional manifold data spaces arise in the description of shapes in terms of landmarks, e.g. by \cite{Kendall1989}. In many cases, data lie close to a low dimensional sub-manifold of the data space. Approaches to restrict consideration to such a sub-manifold broadly fall into two categories. There are approaches to represent data \emph{explicitly} on a \textit{known sub-manifold} embedded in the data space \citep{Kendall1999,Patrangenaru2015}. Alternative approaches represent the data on an \emph{unknown sub-manifold} in the sense that it is not embedded in the original data space, which is determined by \emph{manifold learning} \citep{Roweis2000,Donoho2003,zhang2004,Guhaniyogi2016,yao-pnas2023}. In this paper, we discuss a method which provides an explicit, embedded sub-manifold of a manifold data space. The setting of a manifold data space in which a data sub-manifold is embedded becomes increasingly important, as many procedures in medical imaging \citep{Gerber2010,Souvenir2007} and computer vision produce high-dimensional manifold data \citep{Pennec2006,Pennec1997}. Such methods are so far underdeveloped because conventional statistical methodology for vector spaces cannot be easily adapted to manifold spaces. The simplest case is that existence and uniqueness of the commonly used notion of sample mean is not guaranteed anymore on a manifold \citep{karcher1977,Kendall1989}. To quantify statistical variation on more complex features such as curves and surface a strategy of developing statistical tools in parallel with their Euclidean counterparts is highly relevant. 

Previous approaches to determine an explicit data sub-manifold in a manifold data space are typically framed as efforts to generalize principal component analysis (PCA) to manifold data spaces and broadly fall into two categories. In the \emph{forward approach}, the sub-manifold is built up by increasing dimension stepwise. Tangent space PCA \citep{Fletcher2007} attempts to project the manifold data by simply lifting them to the relevant tangent space and approximating the data distribution locally at the lifting point on the manifold with the induced Euclidean distribution. This approach only works well if the data are fairly concentrated. An alternative line of work seeks instead to directly use geodesics, which generalize the Euclidean straight lines to manifolds. Most notable are principal geodesics \citep{Fletcher2004} and a sequence of improvements \citep{Huckemann2006,Huckemann2010,Kenobi2010,Jung2012,Sommer2013,Pennec2015,Eltzner2017}. In shape space, many approaches use the pre-shape space of oriented shapes and the carefully deal with the quotient space structure. Using spline functions on manifolds, \citet{Jupp1987} and \citet{Kumi2007} develop smooth curves by unrolling and unwrapping the shape space. The \emph{backward approach} carries out the procedure in reverse order from higher to lower dimension \citep{backward2010}, discarding the direction of lowest variation at each step. The most well known approach of this type is principal nested spheres by \citet{Jung2012}, which fits a a sequence of nested sub-spheres with decreasing dimension, by minimizing the residuals of the projected data in each step.

A recent approach, which retains the classical PCA interpretation at each point of the curve, is the principal flow \citep{Panaretos2014}. The flow attempts to follow the main direction of the data cloud locally and offers a trade-off between data fidelity and curve regularity. Differing from the principal flow that starts from a given reference point, \cite{Yao2019Fixed} further develops a fixed boundary flow with fixed starting and ending point for multivariate datasets lying on an embedded non-linear Riemannian manifold. More recently, inspired by finding an optimal boundary between the two classes of data lying on manifolds, \cite{Yao2019principal} invent a novel approach – the principal boundary. From the perspective
of classification, the principal boundary is defined as an optimal curve that moves in between the
principal flows traced out from two classes of data, and at any point on the boundary, it maximizes the margin between the two classes.
In the present paper we tackle the challenging higher dimensional generalization of principal flows. The idea is to generalize the flow to a surface or more generally a sub-manifold. To find a suitable sub-manifold, we start from any point of interest on the manifold, preferably close to a large number of data points, just like we do for the principal flow; but unlike the principal flow that moves only along the maximum direction of variation of the data, we let the sub-manifold expand in all directions along multiple dimensions simultaneously. Instead of following a given shape template in every direction, the sub-manifold expands guided by the eigenvectors of the local covariance matrix.

During the preparation of this paper, a preprint was published, which further elaborates on the geometric theory underpinning principal submanifolds, see \cite{Akhoj2023}.

\begin{figure}[ht!] 
  \centering
  \begin{subfigure}[b]{0.4\textwidth}
    \centering
    \includegraphics[width=0.8\linewidth]{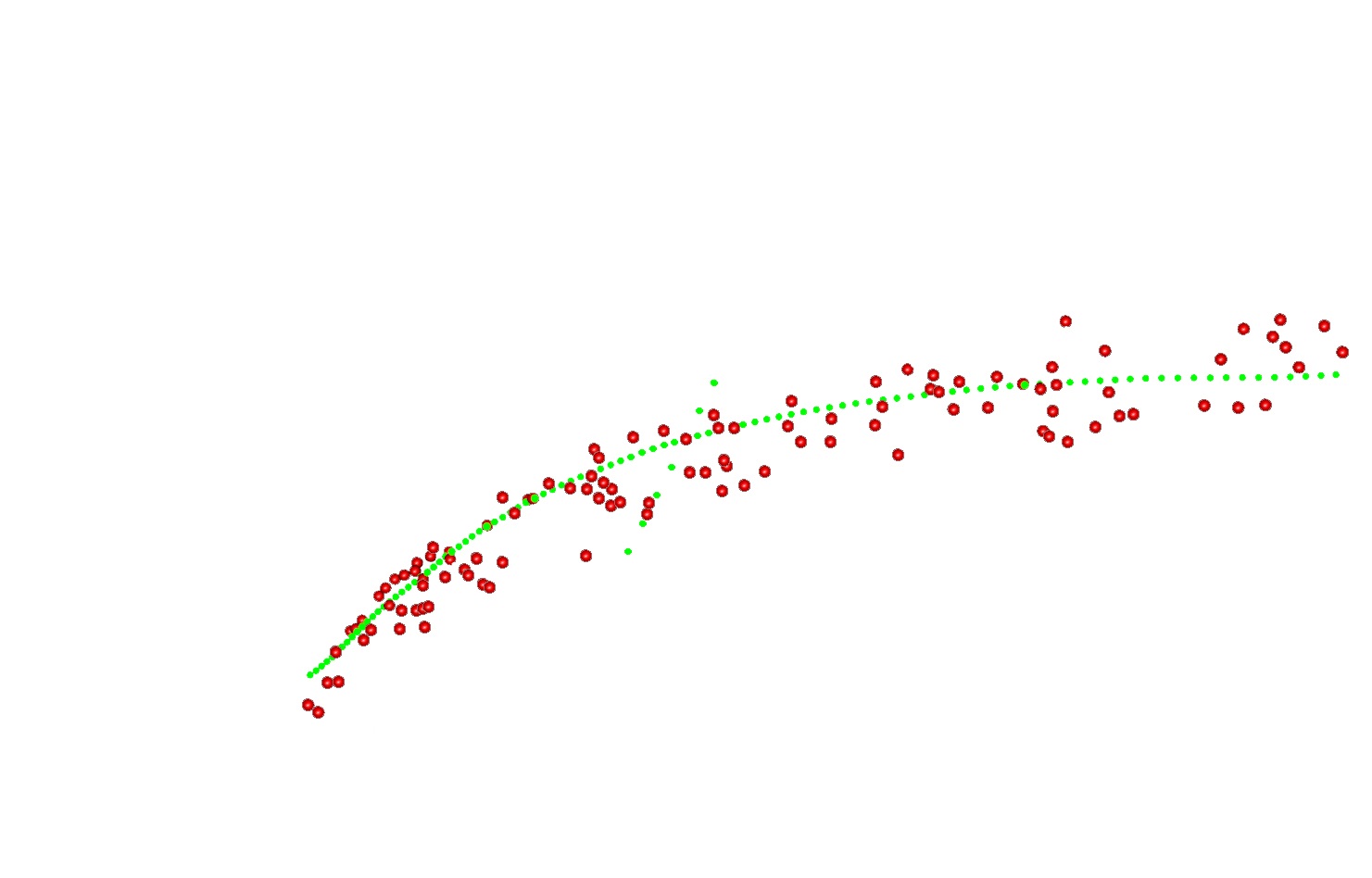}
    \caption{}
  \end{subfigure}%
  %\hspace{0.5 in}
  \begin{subfigure}[b]{0.4\textwidth}
    \centering
    \includegraphics[width=0.8\linewidth]{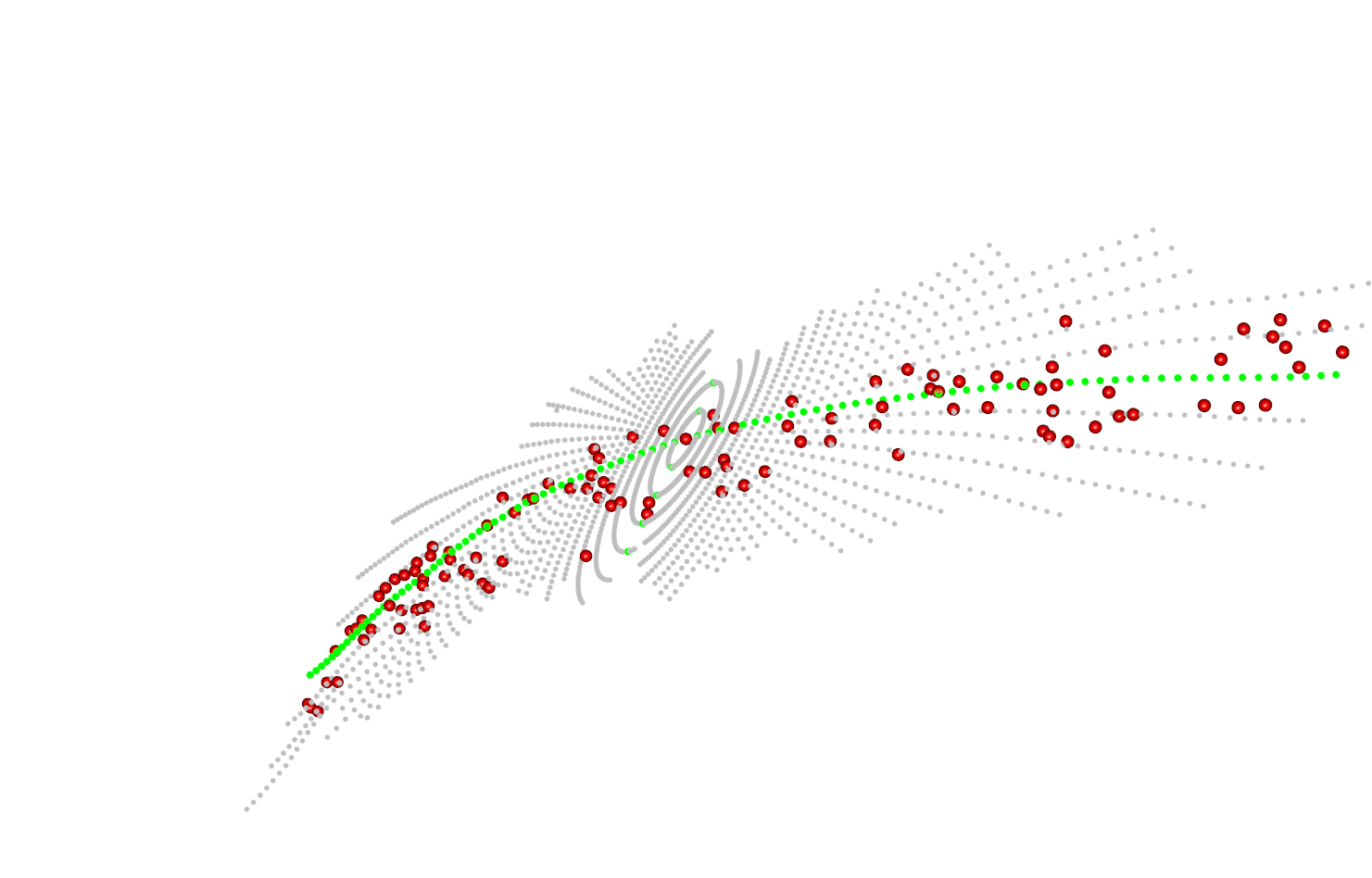}
    \caption{}
  \end{subfigure}% 
  \caption{Visualization of the projected two-dimensional sub-manifold for data on $S^3$. (a) Principal flow; (b) Principal sub-manifold. The data points are labeled in red, with the first and second principal flows (in green) going through the starting point. The sub-manifold (in gray) are the estimated principal sub-manifold.  For visualization purpose, the sub-manifold, the first and second principal direction and the data points have been projected to the first three eigenvectors of the covariance matrix at the starting point.}
  \label{illustrate-flow-sub-manifold}
\end{figure}

In Appendix \ref{app:1-illustration} we present a simulation of a data set which is close to a two-dimensional sub-manifold of $S^3 \subset \mathbb{R}^4$. Figure \ref{illustrate-flow-sub-manifold} shows the data, the superimposed principal flow and the estimated two-dimensional principal sub-manifold. Since the surface extends in two dimensions it can for more variance of the data points.

The optimization to determine the principal sub-manifold subject to smoothness constraints is a challenging problem. The same problem has appeared in finding the principal flow, but for higher dimensional surfaces the problem of parametrization is much more involved. We introduce two points of view on principal sub-manifolds; the first, conceptual point of view is parametrization invariant, while the second, more concrete point of view uses a specific parametrization of the surface. Since the latter point of view is more amenable to an explicit construction, our algorithm of finding the principal sub-manifold is based thereon.

We formally define the principal sub-manifold (Section 2.4) as a sub-manifold in which at any point of the sub-manifold, the tangent space of the sub-manifold attempts to be close to that of the data manifold; intuitively, this definition is analogous to the definition of the principal flow. We show that in case of a flat space, the principal sub-manifold reduces to a ball spanned by the usual principal components, in which the dimension of the sub-manifold corresponds to the number of principal components. The principal sub-manifold also provides a complementary notation to that of a principal surface by \citet{hastie1989}, as a self-consistent surface defined in Euclidean space. 

\section{Principal Sub-manifolds} 

\subsection{Preliminaries}
Suppose that $\left\{x_1,\ldots,x_n\right\}$ are $n$ data points on a complete Riemannian manifold $(\mathcal{M},g)$ of dimension $m$, isometrically embedded in a linear space $\mathbb{R}^d$, where $m <d$. The manifold $(\mathcal{M},g)$ is considered known throughout all of the following. The principal sub-manifolds which are introduced here are understood to be submanifolds of this manifold $(\mathcal{M},g)$.

Let $U \subset \mathbb{R}^{d}$ be an open set, which satisfies that there is an $\epsilon > 0$ such that $\{x \in \mathbb{R}^{d} : \exists y \in \mathcal{M} \text{ such that } |x-y| < \epsilon\} \subseteq U$. Throughout the paper, we assume that there exists a differentiable function $F: U \rightarrow \mathbb{R}^{d-m}$ such that
\begin{align*}
  \mathcal{M} := \left\{x \in \mathbb{R}^{d}: F(x)= 0  \right\}.
\end{align*} 
For each $x\in \mathcal{M}$, the tangent space at $x$, denoted by $T_x\mathcal{M}$ is characterized by the equation
\begin{align*}
  T_x\mathcal{M} = \left\{y \in \mathbb{R}^d: y^T \nabla F(x) = 0 \right\}.
\end{align*} 
Here, $\nabla F(x)$ is the $d \times (d-m)$ derivative matrix of $F$
evaluated at $x \in \mathcal{M}$, assumed to be of full rank everywhere on
$\mathcal{M}$. This full rank assumption implies that the components of $F$ are functionally independent in a suitable sense. This tangent space $T_{x}\mathcal{M}$ provides a local vector space approximation of the manifold $\mathcal{M}$ analogous to the derivative of a real-valued function that provides a local approximation of the function. Let $g$ be a smooth family of inner products associated with the manifold $\mathcal{M}$:
\begin{align*}
  g_x : T_x \mathcal{M} \times T_x\mathcal{M} \rightarrow \mathbb{R}
\end{align*}
Then for any $v\in T_x\mathcal{M}$, the norm of $v$ is defined by
\begin{align*}
  \|v \| = \sqrt{g_x(v,v)}.
\end{align*}

\begin{definition} \label{geo}
  An arc length parametrized curve $\gamma: [0,\delta] \rightarrow \mathcal{M}$ is a geodesic if and only if its tangent vector $\dot{\gamma}(t) \in T_{\gamma(t)}\mathcal{M} \subset \mathbb{R}^d$ satisfies
  \begin{align*}
    \frac{d (\dot{\gamma})}{dt}=0, \quad t \in [0, \delta].
  \end{align*}
  This means that $\frac{d (\dot{\gamma})}{dt}$ considered as a vector in $\mathbb{R}^d$ is normal to $T_{\gamma(t)}\mathcal{M}$ at any time $t$.
\end{definition}

We define the manifold exponential map
\begin{equation} \label{exp}
  \mbox{{\bf exp}}_{x}: T_x \mathcal{M} \rightarrow \mathcal{M}%\quad \mbox{by } \mbox{{\bf exp}}_{x}(v)=\gamma_v(1)
\end{equation}
for $\|v\| \leq \delta$ by $\mbox{{\bf exp}}_{x}(v)=  \gamma(\|v\|)$ where $\gamma$ is a geodesic starting from $\gamma(0) = x$ with initial velocity $\dot{\gamma}(0) = v/\|v\|$. For a suitable neighborhood $U_x \subset \mathcal{M}$ of $x$, which excludes the cut locus of $x$, the logarithm map
\begin{equation} \label{log}
  \mbox{{\bf log}}_x : U_x \rightarrow T_x \mathcal{M}.
\end{equation}
is the inverse of the exponential map.

\begin{figure}[ht!]
  \begin{center} 
    \includegraphics[width=1.8in]{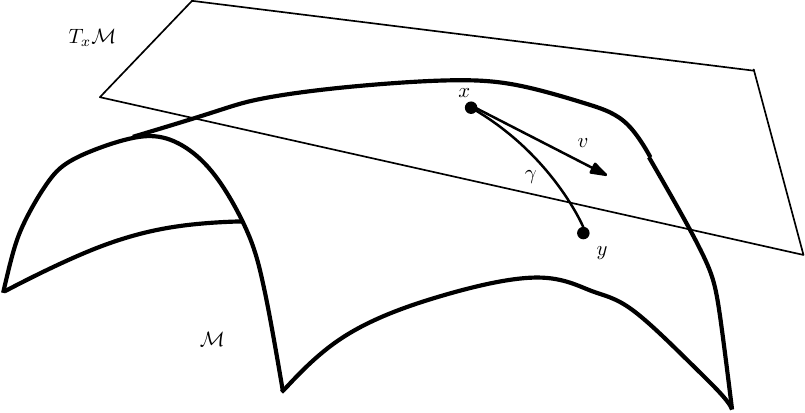}
  \end{center}
  \caption{The vector $v$ on the tangent subspace $T_x \mathcal{M}$ at $x$. The endpoint of vector $v$ is the image of $y=\mbox{{\bf exp}}_x(v)$  under the mapping defined in \eqref{log}.}
  \label{tangent}
\end{figure}
Let $x, y \in \mathcal{M}$. Denote the set of all (piecewise) smooth curves $\gamma(t): [0,1] \rightarrow \mathcal{M}$ with endpoints such that $\gamma(0)=x$ and $\gamma(1)=y$ by $\Gamma_{x,y}$. The {\it geodesic distance} from $x$ to $y$ is defined as 
\begin{equation}\label{lencurve}
  d_{\mathcal{M}}(x,y)= \inf_{\gamma \in \Gamma_{x,y}} \ell(\gamma)
\end{equation}
where $\ell(\gamma)=\int_{[0,1]} \left\|\dot{\gamma}(t)\right\|dt= \int_{[0,1]} g_{\gamma(t)}\left( \dot{\gamma}(t), \dot{\gamma}(t) \right) ^{\frac{1}{2}} dt$. Minimizing \eqref{lencurve} yields geodesics as in Definition \ref{geo}, providing the shortest distance between two points $x$ and $y$ in $\mathcal{M}$.

\subsection{Principal sub-manifolds}

The concept of a principal sub-manifold is strongly inspired by the principal flow, see \citet{Panaretos2014}, we review this concept in Appendix \ref{app:2-princ-flows}.

In the following, we will go beyond previous work by introducing both population and sample principal submanifolds and discussing statistical properties. Consider a probability measure $\mathbb{P}$ on the manifold $\mathcal{M}$ as well as data $\left\{x_1, \cdots, x_n\right\} \subset \mathcal{M}$. We will give the definition of a multi-dimensional sub-manifold $\mathcal{N} \subset \mathcal{M}$, based on a reference point $x \in \mathcal{M}$. For any point $x$ in $\mathcal{M}$, following Equation \eqref{covar} in Appendix \ref{app:2-princ-flows}, define the population and sample local tangent covariance matrices on $\mathcal{M}$
\begin{align}
  \Sigma_{x} &= \frac{1}{\int_\mathcal{M}\kappa_h(y,x) d\mathbb{P}(y)}  \int_\mathcal{M} {\bf log}_{x}(y) \otimes {\bf log}_{x}(y) \kappa_h(y,x) d\mathbb{P}(y), \label{eq:sigma-pop}\\
  \widehat{\Sigma}_{x,n} &= \frac{1}{\sum_{i}\kappa_h(x_i,x)}  \sum_{i=1}^n{\bf log}_{x}(x_i) \otimes {\bf log}_{x}(x_i) \kappa_h(x_i,x). \label{eq:sigma-sample}
\end{align}
Let $\big\{\lambda_1(x),\ldots,\lambda_k(x)\big\}$ and $\big\{e_{1}(x), \ldots, e_{k}(x)\big\}$ be the first $k$ eigenvalues and eigenvectors of $\Sigma_{x}$ and let $\big\{\widehat{\lambda}_{n,1}(x),\ldots,\widehat{\lambda}_{n,k}(x)\big\}$ and $\big\{\widehat{e}_{n,1}(x),\ldots,\widehat{e}_{n,k}(x)\big\}$ be the first $k$ eigenvalues and eigenvectors of $\widehat{\Sigma}_{x,n}$. Denote the linear subspace spanned by $\big\{e_1(x),\ldots, e_k(x) \big\}$ as $W(x)$ and the linear subspace spanned by $\big\{\widehat{e}_{n,1}(x),\ldots,\widehat{e}_{n,k}(x)\big\}$ as $\widehat{W}_n(x)$. For both cases, assume that each of the $k$ principal components is smooth. Then $W$ and $\widehat{W}_n(x)$ are smooth distributions as defined by Definition \ref{def:distribution}.

\begin{definition} \label{def:distribution}
  Let $\pi: T \mathcal{M} \to \mathcal{M}$ be the canonical projection, such that for every subset $U \subset \mathcal{M}$ the set $TU := \pi^{-1}(U) \subset T \mathcal{M}$ is well defined and contains the tangent spaces $T_p \mathcal{M}$ for all $p \in U$. Assume a subset of the tangent bundle $\mathcal{D} \subseteq T \mathcal{M}$ with the property that for every $p \in \mathcal{M}$ there is some open set $p \in U_p \subset{\mathcal{M}}$ such that there is a set of continuous vector fields $\mathfrak{X} = \{X_1, \dots, X_k\}$, defined on $U_p$ which gives rise to a homeomorphism $\mathcal{D} \cap TU_p \leftrightarrow U_p \times \mathbb{R}^k$.
  \begin{itemize}
    \item[(i)] Then the bundle $\mathcal{D}$ is called a \emph{distribution} of dimension $k$.
    \item[(ii)] If the vector fields used in the definition are $C^r$, $\mathcal{D}$ is called a \emph{$C^r$-distribution} of dimension $k$.
    \item[(iii)] If for all local vector fields $X_i, X_j \in \mathfrak{X}$ defined on the same $U_p$ we have $[X_i,X_j] \in \mathcal{D} \cap TU_p$, the distribution is called \emph{involutive}.
  \end{itemize}
\end{definition}

In the following, we will denote components in $\mathbb{R}^d$ by latin letters from the middle of the alphabet and components in $\mathcal{N}$ and $W$ ranging from $1$ to $k$ by greek letter from the beginning of the alphabet.

In Appendix \ref{app:6-lagrangian}, we propose a description of principal sub-manifolds in terms of a Lagrangian problem, analogous to \citet{Panaretos2014}. However, since the solution technique used there is not immediately applicable here, we instead propose an operational definition of principal sub-manifolds and provide a simpler ``greedy'' algorithm in Section \ref{sec:greedy-algo} to approximate them.

Since principal sub-manifolds will be defined locally, we introduce the notion of a local submanifold.

\begin{definition}[Local Submanifold]\label{def:local-submf}
  Assume a sub-manifold, described by the image of an injective smooth function
  \begin{align} \label{eq:princ-submf-map}
    N: \mathbb{R}^k \supset U \to \mathcal{N} \subset \mathcal{M} \subset \mathbb{R}^d\, .
  \end{align}
  In this expression, the image $\mathcal{N} := \{N(t)\}$ is the principal sub-manifold. We denote the space of local $k$-dimensional sub-manifolds containing some point $A \in \mathcal{M}$, i.e. $A \in \mathcal{N}$ by this assumption, and satisfying $\forall N \in \mathcal{N} \, : \, d_{\mathcal{N}}(N,A) < L$ by $\textnormal{SubM}(A, L, k, \mathcal{M})$. Here $d_{\mathcal{N}}$ is the metric on $\mathcal{N}$ induced by the metric on $\mathcal{M}$.
\end{definition}

Next, we define an integral sub-manifold of the distributions $W(x)$ or $\widehat{W}_n(x)$. If such integral sub-manifolds exist, the principal sub-manifolds are defined to be these integral sub-manifolds.

\begin{definition}[Integral Submanifold] \label{def:int-submf}
  A sub-manifold $\mathcal{N} \subset \mathcal{M}$ is called an \emph{integral sub-manifold} of the distribution $W$, if for every point $q \in \mathcal{N}$ the tangent space is spanned by the distribution vector fields $T_q\mathcal{N} = W(q) := \mathop{span} \{X_1(q), \dots, X_k(q)\}$.
\end{definition}

The following is a theorem from differential geometry on the existence of integral sub-manifolds.

\begin{theorem}
  For any point $p \in \mathcal{M}$ a distribution $W$ can give rise to at most one integral sub-manifold containing $p$. A distribution $W$ defines a unique $C^2$ integral sub-manifold for each point $p\in \mathcal{M}$ if and only if $W$ is involutive.
\end{theorem}

\begin{remark}
  Involutiveness is a strong property that is not generically satisfied for the spans of eigenvectors of local covariance matrices which we consider. As an example for a simple non-involutive distribution that could arise in our setting, consider the two everywhere orthonormal vector fields $X = \cos(y) \partial_x + \sin(y) \partial_z$ and $Y = \partial_y$ on $\mathbb{R}^3$. These define a non-involutive distribution since $[X,Y] = \sin(y) \partial_x - \cos(y) \partial_z \notin \mathrm{span}(X,Y)$. The defining construction for principal sub-manifolds as presented in Section~\ref{sec:greedy-algo} always yields an interpretable sub-manifold, however it is not in general an integral sub-manifold.
\end{remark}

In order to show the connection between population and sample principal sub-manifolds, we discuss asymptotics results in Appendix \ref{app:3-proofs-thms-3+4}.

\subsection{Asymptotic Theory for Principal Submanifolds}

For the asymptotic theory discussed here, we assume that integral submanifolds of $W(x)$ and $\widehat{W}_n(x)$ exist and the principal submanifolds are defined as these. The difference between population and sample principal sub-manifolds rests entirely on the distinction whether the distribution $W(x)$ or $\widehat{W}_n(x)$ are used, since principal sub-manifolds are defined as integral manifolds of these distributions. In Theorems \ref{thm:consistency-covar} and \ref{thm:consistency-submf} we also assume a fixed kernel with bandwidth $h$ used to define both $W(x)$ and $\widehat{W}_n(x)$. We do not place specific restrictions onto the kernel or the bandwidth except for the high-level requirement that the resultant distributions $W(x)$ and $\widehat{W}_n(x)$ be involutive.

The first step towards establishing consistency and asymptotics of sample principal sub-manifolds with respect to population principal sub-manifolds requires establishing these properties for the distributions $W(x)$ and $\widehat{W}_n(x)$. Note that here and in the following we denote by $\angle(v,w)$ the angle between two vectors $v$ and $w$.

\begin{theorem}[Consistency of Local Covariance] \label{thm:consistency-covar}
  If for every $x \in B_{\epsilon}(A)$ we have $\lambda_k(x) > \lambda_{k+1}(x)$, then we have for every $ \delta > 0$ and for every sequence $a_n \to 0$
  \begin{align*}
    \lim_{n\to \infty} \sup_{x \in B_{\epsilon}(A)} \mathbb{P} \left(a_n n^{1/2} \angle \left( \widehat{W}_n(x), W(x)\right) > \delta \right) = 0 \, .
  \end{align*} 
\end{theorem}

\begin{proof}
  Using the CLT for principal components by \cite{Anderson63} the result follows immediately.
\end{proof}

This result does not immediately yield a consistency result for principal sub-manifolds. In fact, since $\widehat{W}_n(x)$ will in general deviate from $W(x)$ even at the reference point $A$, the two sub-manifolds may diverge proportionately to the distance $L$ from the reference point $A$.

\begin{theorem}[Consistency of Local Principal Submanifolds] \label{thm:consistency-submf}
  Assume that for every $x \in B_{\epsilon}(A)$ we have $\lambda_k(x) > \lambda_{k+1}(x)$. Furthermore, assume a sequence $\{L_n \in \mathbb{R}^+\}_{n \in \mathbb{N}}$ which satisfies $n^{1/4} L_n \to 0$, and consider a sequence of $\{A_n \in \mathbb{R}^k\}_{n \in \mathbb{N}}$ satisfying $n^{1/2} d_{\mathcal{M}} (A_n, A) \to 0$ for some point $A \in \mathcal{N}$ on the population principal submanifold. Define local sample principal submanifolds $\widehat{\mathcal{N}}_n \in \textnormal{SubM}(A_n, L_n, k, \mathcal{M})$, then we have for every $ \delta > 0$
  \begin{align*}
    \lim_{n\to \infty} \mathbb{P} \left( n^{1/2} \max_{x \in \widehat{\mathcal{N}}_n} \min_{y \in \mathcal{N}} d_{\mathcal{M}} \left( x, y\right) > \delta \right) = 0 \, .
  \end{align*} 
\end{theorem}

\begin{proof}
  The proof can be found in Appendix \ref{app:3-proofs-thms-3+4}.
\end{proof}

The asymptotic results given above assume a fixed kernel $\kappa$ and a fixed bandwidth $h$. This is due to the fact that the population principal submanifold is defined as an integral manifold of a geometric distribution defined by an optimization criterion. This might not be immediately intuitive and one might rather have the picture in mind of the population manifold being a true smooth object and the sample to be a noisy discrete representation thereof drawn via a generative model. We will therefore discuss a simple a generative model and show a consistency result for it.

\begin{theorem}
  Consider a submanifold $\mathcal{N}_0 \subset \mathcal{M} = \mathbb{R}^d$ and the multivariate normal probability density $\phi(x;\mu, \Sigma)$ on $\mathbb{R}^d$. Then define a one-parameter family of probability densities
  \begin{align*}
    \phi_{n,\mathcal{N}_0}(x) = \int_{\mathcal{N}_0} \phi(x; y, \sigma_n \cdot \textnormal{Id}) dy \, .
  \end{align*}
  Consider a sequence of bandwidths $\{ h_n \in \mathbb{R}^+\}_{n\in\mathbb{N}}$ with $h_n \to 0$, a sequence of noise levels $\{ \sigma_n \in \mathbb{R}^+\}_{n\in\mathbb{N}}$ with $n^{1/4} \sigma_n / h_n \to 0$, and a sequence $\{L_n \in \mathbb{R}^+\}_{n \in \mathbb{N}}$ with $n^{1/4} L_n \to 0$.
  Then, a sequence of local population principal submanifold $\mathcal{N}_n \in \textnormal{SubM}(A, L_n, k, \mathcal{M})$ with reference point $A \in \mathcal{N}_0$ defined by the sequence of local covariance fields
  \begin{align*}
    \Sigma_{n,x} &= \frac{1}{\int_\mathcal{M}\kappa_{h_n}(y,x) d\mathbb{P}(y)}    \int_\mathcal{M} {\bf log}_{x}(y) \otimes {\bf log}_{x}(y) \kappa_{h_n}(y,x) d\mathbb{P}(y),
  \end{align*}
  leading to a sequence of distributions $W_n$ satisfies for every $ \delta > 0$
  \begin{align*}
    \lim_{n\to \infty} n^{1/2} \max_{x \in \mathcal{N}_n} \min_{y \in \mathcal{N}_0} d_{\mathcal{M}} \left( x, y\right) = 0 \, .
  \end{align*} 
\end{theorem}

\begin{proof}
  The proof can be found in Appendix \ref{app:3-proofs-thms-3+4}.
\end{proof}

\section{Determination of Principal sub-manifold}
\subsection{An algorithm for principal sub-manifold} \label{sec:greedy-algo}

Recall that the principal flow is the solution of an optimization problem in equations \eqref{flow+} and \eqref{flow-}. Finding such a solution requires an extensive search for a critical point of a Euler-Lagrange problem that involves integrating the vector field along the curve. Because it is a one dimensional curve, standard numerical methods can be applied as shown in \citet{Panaretos2014}, reducing it to a problem of determining the solution of a system of ordinary differential equations (ODEs). As seen in equation \eqref{eq:lagrange1} in Appendix \ref{app:6-lagrangian}, when it comes to a sub-manifold, things turn out to be quite different. The corresponding optimization problem for sub-manifolds is much more complex and it is not clear how to approach the problem numerically. The main reason is that the Lagrangian theory leads to a partial differential equation for which the method used in \citet{Panaretos2014} is not applicable, whereas in the case of principal flow one has a simple ordinary differential equation.

%As guiding principle, consider the notion of an integral sub-manifold of a distribution.
%
%\begin{definition} \label{def:int-submf}
%  A sub-manifold $\mathcal{N} \subset \mathcal{M}$ is called an \emph{integral sub-manifold} of the distribution $W$, if for every point $q \in \mathcal{N}$ the tangent space is spanned by the distribution vector fields $T_q\mathcal{N} = W(q) := \mathop{span} \{X_1(q), \dots, X_k(q)\}$.
%\end{definition}
%
%The following is a theorem from differential geometry on the existence of integral sub-manifolds.
%
%\begin{theorem}
%  For any point $p \in \mathcal{M}$ a distribution $W$ can give rise to at most one integral sub-manifold containing $p$. A distribution $W$ defines a unique $C^2$ integral sub-manifold for each point $p\in \mathcal{M}$ if and only if $W$ is involutive.
%\end{theorem}

With this in mind, it is clear that the algorithm we provide should approximate curves in an integral sub-manifold, whenever the distribution is involutive.

Some complications arise when working with a sub-manifold of higher dimension than two. One problem is that computational complexity increases exponentially with dimension, which can be easily understood since the number of points in a simple rectangular grid depends exponentially on dimension. For the algorithm we present here there is an additional complication for higher dimensions, which we briefly mention below. We will discuss how to determine a two-dimensional principal sub-manifold as a special case and present an algorithm for the rest of the paper.

Principal sub-manifolds are always constructed starting from some initial point $A \in \mathcal{M}$. A possible initial point, which is used in some of the applications below, is the Fr\'echet mean defined below.

\begin{definition}
  The Fr\'{e}chet sample mean, $\bar{x} \in \mathcal{M}$, for a sample of data points $\left\{x_1, \cdots, x_n\right\} \in \mathcal{M}$ is a minimizer of the {\it Fr\'{e}chet sample variance}, if the minimizer is unique:
  \begin{align*}
    \bar{x}= \margmin_{x \in \mathcal{M}} \limits \frac{1}{n} \sum_{i=1}^{n} d^2_{\mathcal{M}}(x, x_i).
  \end{align*}
\end{definition}

Consider $\mathcal{N} \in \textnormal{SubM}(A, \epsilon, k, \mathcal{M})$ where $k=2$. We will not perform an analytical optimization and rather present an approximating algorithm. To this end, we will work with a natural parametrization of $\mathcal{N}$ induced by the vector space structure of ${\bf log}_A(\mathcal{N})$. Thus, recalling the defining map from equation \eqref{eq:princ-submf-map}, we use $N := {\bf exp}_A$ and $U := B_\epsilon(0)$ in order to parameterize any sub-manifold $\mathcal{N} \in \textnormal{SubM}(A, \epsilon, 2, \mathcal{M})$.

We now define an equidistant pattern of directions $P := \{ 2 j \pi / L  \, |\, j = 1, \dots L \} \subset S^{1} \subset \mathbb{R}^2$ and use it to define a set of points in $B_\epsilon(0) \subset \mathbb{R}^2$ as
\begin{equation*}
  \forall l \in P \subset \mathbb{R}^2 \, : \quad P_l := \left\{ t l \big| t \in \{t_0 = 0, t_1, \dots , T_{N(l)}\} \subset [0, \epsilon] \right\} \subset \mathbb{R}^2 \, ,
\end{equation*}
where $N(l)$ is the number of levels for the $l$th direction. These are mapped by ${\bf exp}_A$ into point sets, which we call \textit{rays},
\begin{align}
  \forall l \in P \, : \quad \mathcal{A}_{l} :=&~ \left\{ A_{l,0} = A, A_{l,1} = {\bf exp}_A(1 \cdot l), \ldots, A_{l,N(l)} = {\bf exp}_A(N(l) \cdot l)\right\}\nonumber\\
  &~ \subset \mathcal{M} \, , \label{A_L}
\end{align}
where we choose $L=180$.

For dimension $k > 2$ one can also define patterns $P \subset S^{k-1}$ of finitely many directions, which are close to evenly and uniformly distributed on $S^{k-1}$. However, one cannot achieve the same regularity as for an equidistant pattern in $S^1$. This is an additional complication which arises when trying to use the algorithm presented here in higher dimension.

In words, the key is to represent $\mathcal{N}$ discretely by a collection of ordered rays, each representing a certain amount of data variation and all of them spanning the sub-manifold of maximal variation. The rays are expected to grow and expand along all directions. While the principal flow tries to match its tangent vector to the first eigenvector at a certain point, the principal sub-manifold tries to find the best direction that belongs to the plane spanned by the first few eigenvectors, as represented by the Lagrangian $\mathscr{L}_2$. In this sense, the directions in which the sub-manifolds expands provide an extra dimension to build up the target sub-manifold of maximal variation. A set of such rays representing an approximation to the sub-manifold $\mathcal{N}$ at $A$---in every possible direction of variation---remain to be found.

\begin{figure}[ht!]
  \centering
  \includegraphics[clip, trim=0 2in  0 0, scale=0.7]{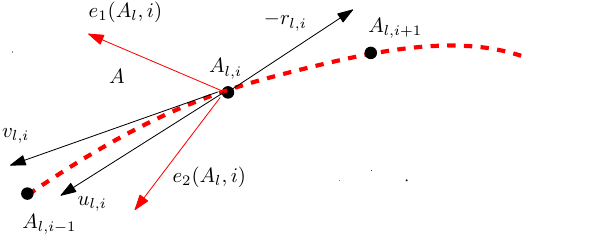}
  \caption{Illustration of the Algorithm. The goal is to determine $A_{l,i+1}$, given the current point $A_{l,i-1}$ and the previous point $A_{l,i}$:  $v_{l,i}$ is the tangent vector of the curve $A_{l,i}A_{l, i-1}$, $u_{l, i}$ is the projection of $v_{l,i}$ onto the tagent plane spanned by $e_1(A_{l,i})$ and $e_2(A_{l,i})$. The point $A_{l,i+1}$ is found by mapping a small move $\epsilon'$ from $A_{l,i}$ along the direction of $r_{l,i}=-u_{l,i}/ \big\|u_{l,i} \big\|$.}
  \label{algorithm_illus}
\end{figure}

We call all the rays for all directions a \textit{principal sub-manifold} $\mathcal{N}$. 
A complete algorithm (Algorithm \ref{algo1}) can be found in Appendix \ref{app:7-greedy-algorithm}. Here, we elaborate the core of the algorithm (see Figure~\ref{algorithm_illus}): given direction $l$, we are at the $i$th level, $A_{l,i}$, there are three steps to go through to find the $(i+1)$th level, $A_{l,i+1}$
\begin{itemize}
  \item[(1)] {\it Reorientation}:  identify the current tangent vector of the curve $A_{l, i}A_{l,i-1}$ and determine the direction for the next move
  
  \item[(2)] {\it Projection}: expand the rays from the points $A_{l,i}$ along the direction $r_{l,i}$ by a step of $\epsilon'$ and arrive at point $A_{l,i+1}$
  
  \item[(3)] {\it Updating}: project the data points $x_j$'s$(1 \leq j \leq n)$ onto the point $A_{l,i+1}$, and re-calculate the tangent plane at $A_{l,i+1}$  
\end{itemize}
In Step (1), given the current point $A_{l,i}$ and the previous point $A_{l,i-1}$, we obtain the tangent vector of the curve 
$A_{l,i}A_{l,i-1}$ by backward projection
\begin{align*}
  v_{l,i}={\bf log}_{A_{l,i}}\big(A_{l,i-1}\big) \, .
\end{align*}
The best knowledge we have about the ray at $A_{l,i}$ is $v_{l,i}$. Let $u_{l,i}$ be the direction for the next move and define the projected vector
\begin{equation*}
  \tilde{v}_{l,i} := W(A_{l,i})^T W(A_{l,i}) v_{l,i} = \Big\langle v_{l,i}, e_1\big(A_{l,i}\big) \Big\rangle   e_1\big(A_{l,i}\big)  +\Big\langle v_{l,i}, e_2\big(A_{l,i}\big) \Big\rangle  e_2\big(A_{l,i}\big)
\end{equation*}
where $e_1\big(A_{l,i}\big)$ and $ e_2\big(A_{l,i}\big)$ are the first and second eigenvector of $\Sigma_{A_{l,i}}$. We discuss two alternatives to determine $u$. Let $W_{l,i}$ denote the plane spanned by $e_1\big(A_{l,i}\big)$ and $ e_2\big(A_{l,i}\big)$.

\begin{itemize}
  \item [(a)] The straight forward choice $u_{l,i} := \tilde{v}_{l,i}$ amounts to \textit{projection} to $W_{l,i}$.
  \item [(b)] Choosing $u_{l,i} := 2 \tilde{v}_{l,i} - v_{l,i}$ amounts to \textit{reflection} at $W_{l,i}$.
\end{itemize}
While the reflection is a less obvious choice, it can achieve better data fidelity for large, but slowly varying curvature. We illustrate this point in Appendix \ref{app:9-reflection}.

In Step (2), we move $A_{l,i}$ on the tangent plane by a step of $\epsilon'$ along $r_{l,i}$,  where
\begin{equation*}
  r_{l,i} = -\epsilon' \times  \frac{u_{l,i}}{\big\|u_{l,i} \big\|}, 
\end{equation*}
and then map it back to the manifold $\mathcal{M}$ 
\begin{equation*}
  A_{l,i+1} = {\bf exp}_{A_{l,i}}(r_{l,i}).
\end{equation*}
Note that $u_{l,i}$ is not of unit length, and the negative sign appears as $u_{l,i}$ is obtained from $v_{l,i}$.

In Step (3), updating the covariance matrix at $A_{l, i+1}$ is necessary when local data points change significantly, where the covariance matrix is updated by replacing $\Sigma_{A_{l,i}}$ with $\Sigma_{A_{l,i+1}}$.

It is crucial to make sure that the rays always move forward and do not return to points already explored by the ray itself or another ray. Additionally, a stop condition is necessary such hat the principal sub-manifold does not extend far beyond the data domain. In accordance with the stopping rule used in \citet{Panaretos2014}, we can terminate the process when the {\it length} of the $l$th ray, i.e., 
\[
\ell_{\mathcal{A}_l} =\sum_{i=1}^{N(l)-1}d(A_{l,i}, A_{l,i+1}),
\]
exceeds 1. %where $\ell_{\mathcal{A}_l}$ is the corresponding length of $\mathcal{A}_l$. 
The length of $l$th ray does not necessarily have to be equal. There may exist other stopping rules that one can use. Among them, we should also consider that for all $j$,
\[
\big\|{\bf log}_{A_{l,i+1}}(x_j) \big\|> \delta \mbox{ or } \Big \langle {\bf log}_{A_{l,i+1}}(A_{l,i}) , {\bf log}_{A_{l,i+1}}(x_j) \Big \rangle \geq 0,
\] 
which implies that either there are not enough data points in the neighborhood or $A_{l,i+1}$ is already outside the convex hull of the $x_j$'s under the logarithm map.\\

\begin{remark}
  Both $\epsilon'$ and $\delta$ are pre-defined parameters. We suggest to choose $\epsilon'$ preferably with small values to ensure the stability of the local move on the tangent plane, while the choice of $\delta$ depends more on the data dispersion and configuration, which might vary from case to case.
\end{remark}

For $h \to \infty$ and a flat manifold the principal component distribution is constant over the whole space and therefore it is clear that the greedy algorithm leads to straight lines spanning the linear subspace spanned by the $k$ largest principal components. In this sense, the limiting case of standard PCA is trivial. In Appendix \ref{app:8-greedy-convergence} we investigate the convergence behavior of the greedy algorithm in the limit $\epsilon' \to 0$ if the length of all curves is fixed in advance. In general, it is very difficult to show that solution curves of the greedy algorithm approximate solution curves to either Lagrangian. Instead, we show that the curves converge to the integral sub-manifold if the distribution is involutive, as expected. This convergence result is very important, because it shows that the greedy algorithm leads to meaningful results in all cases where a unique ``true'' geometric solution in terms of an integral sub-manifold exists.

\subsection{Visualization of the principal sub-manifold}
The principal sub-manifold in general cannot be fully visualized when its dimension exceeds one. Consider a simple case where the data lies in $S^3 \subset \mathbb{R}^{4}$; the principal sub-manifold is then a subset of $S^3$; that is, it is equivalent to visualizing a two-dimensional manifold in a four-dimensional space. However, a meaningful representation of the sub-manifold is still quite relevant for understanding the shape of the manifold, at least partially. We propose two ways of visualizing the principal sub-manifold. The first one is to represent the sub-manifold in terms of principal direction rays. The second one is to visualize the sub-manifold in the projected manifold space. 
%this is the same space where the manifold data is collected. 

\begin{itemize}
  \item parameterize the sub-manifold in polar coordinates and represent it by the shapes of principal direction rays  
  \item project the sub-manifold by multiplying a projection matrix in which the basis is formed by eigenvectors from the covariance matrix at the starting point
\end{itemize}

\begin{figure}[ht!]
  \centering
  \begin{subfigure}[b]{0.35\textwidth}
    \includegraphics[width=1.1\linewidth]{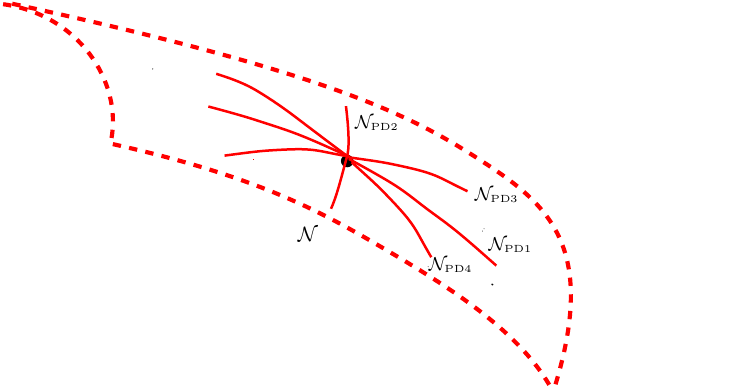}
    \caption{}
  \end{subfigure}%
  \hspace{.25 in}
  \begin{subfigure}[b]{0.35\textwidth}
    \includegraphics[width=1.1\linewidth]{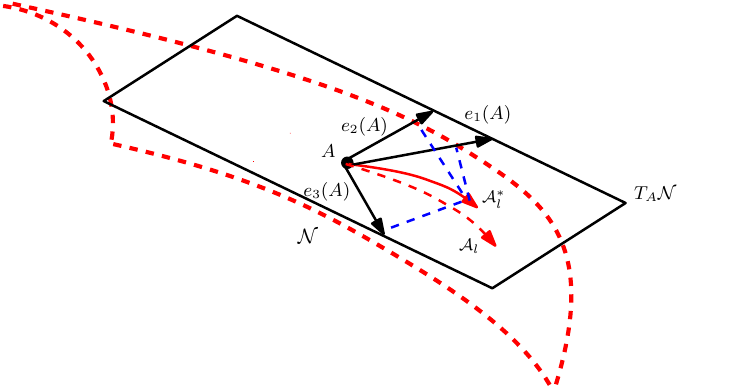}
    \caption{}
  \end{subfigure}%
  \caption{Visualization of a principal sub-manifold. (a) Visualize the sub-manifold by eight principal direction rays. (b) Visualize the sub-manifold by projecting to the three largest eigenvectors of covariance matrix at ${A}$.}
  \label{manifold_visual}
\end{figure}

{\it Visualization in principal direction rays}: Choose a number of directions from the starting point of the algorithm and visualize the sub-manifold by using the corresponding rays. Recall that the entire sub-manifold can be expressed as follows
\begin{equation*}
  \mathcal{N}=   
  \tiny{
    \big[\mathcal{A}_1, \mathcal{A}_2,\cdots, \mathcal{A}_{L}\big]^{\tiny{\mbox{T}}}}.
\end{equation*}
Although we denote $\mathcal{N}$ as a ``matrix'', the actual length of each row (i.e., $\mathcal{A}_l, 1 \leq l \leq L$) may vary. To visualize the sub-manifold, we select a candidate set $\mathcal{L}_s \subset \mathcal{L}$ and map the corresponding rows of $\mathcal{N}$ into the corresponding rays in shape coordinates. Thus, the principal direction rays of the sub-manifold shall be represented by    
\begin{equation*}\label{shape-map} 
  x_{l, i}= f^{-1}(A_{l, i}),  \quad \forall l \in \mathcal{L}_s, 1 \leq i \leq N(l),
\end{equation*}
where $f$ is the embedding function. In the case of Kendall shape space, the resultant $x_{l, 1}, \cdots, x_{l, N(l)}$ is a collection of $N(l)$ $k$-ads.

Among all $l$'s, the two {\it principal directions} of the sub-manifold are defined as follows. Recall polar coordinates on the image $\theta=2l \pi /L$, where $L=180$. The first principal direction, denoted as  $\mathcal{N}_{\tiny{\mbox{PD1}}}$, is the curve corresponding to $\theta= \pi \mbox{ and }\theta=  2 \pi$ in polar coordinates on the image; this is equivalent to $l$ equals $90$ and $180$ such that
\[
\mathcal{N}_{\tiny{\mbox{PD1}}}=\mathcal{A}_{90} \cup \mathcal{A}_{180}.
\]
The second principal direction, denoted as $\mathcal{N}_{\tiny{\mbox{PD2}}}$, corresponds to the curve with $\theta= \pi/2 \mbox{ and }\theta=  3\pi/2$ in polar coordinates on the same image; this is equivalent to $l$ equals $45$ and $135$ such that
\[
\mathcal{N}_{{\tiny \mbox{PD2}}}=\mathcal{A}_{45} \cup \mathcal{A}_{135}.
\]

\noindent In addition, it is suggested to also include the curves, $\mathcal{N}_{\tiny{\mbox{PD3}}}$, corresponding to $\theta= \pi/4 \mbox{ and }\theta=  5 \pi/4$ as well as the ones, $\mathcal{N}_{\tiny{\mbox{PD4}}}$, corresponding to $\theta=3 \pi/4 \mbox{ and }\theta= 7 \pi/4$. Adding two extra directions gives additional details about the sub-manifold. 

We remark here that although we have used $\mathcal{N}_{{\tiny \mbox{PD1}}}-\mathcal{N}_{{\tiny \mbox{PD4}}}$ as the principal directions, they are by no means the simple extension of the usual principal components or any variants thereof. Figure \ref{subman-digit3-mean} gives an example of such a configuration of shapes. The entire image contains 9 by 9 small shapes. The central figure is the mean shape. Row 5 represents the shapes of $\mathcal{N}_{\tiny{\mbox{PD1}}}$. Column 5 is the shapes of $\mathcal{N}_{\tiny{\mbox{PD2}}}$. The main diagonal contains the shapes of $\mathcal{N}_{\tiny{\mbox{PD3}}}$. The other diagonal contains the shapes $\mathcal{N}_{\tiny{\mbox{PD4}}}$.\\

{\it Visualization in projected space}:  Alternatively, one may wish to represent the sub-manifold using a projected sub-manifold rather than itself. The latter serves as a much simplified version of the original one and it is more interpretable, provided that the majority of variation of the principal sub-manifold can be explained by a reduced one. Compared to the previous representation, this visualization preserves the resolution of the sub-manifold. 

To fix representation, we center the $\mathcal{N}$ row-wise by $A$ and obtain the centered sub-manifold 
\begin{equation*}
  \mathcal{N}^{*} = \left(\tiny{\mathcal{N}}^{*}_{l,i} \right)_{1 \leq l \leq L, 1 \leq i \leq N(l)}
\end{equation*}
where $\mathcal{N}^*_{l,i}=A_{l,i}-A$ where $1 \leq l \leq L, 1 \leq i \leq N(l)$. Clearly, for $i=1$, $\mathcal{N}^*_{l,i}=\mathbf{0}$. Let the projection matrix for $\mathcal{N}^{*}$ be
\begin{equation*}
  \Psi = \left({\psi}_{l,i} \right)_{1 \leq l \leq L, 1 \leq i \leq N(l)},
\end{equation*} 
where ${\psi}_{l,i}$ is the projection matrix for $A_{l,i}$. Usually, we choose ${\psi}_{l,i}=E_3$ where $E_3=\left[e_1(A), e_2(A), e_3(A) \right]^{\tiny{\mbox{T}}}$ of $\Sigma_{A}$.
The process is carried out by multiplying $\mathcal{N}^{*}$ element-wise by the projection $\Psi$, so that 
\begin{equation*}
  \mathcal{N}^{\tiny{\mbox{pro}}} = \Psi  \odot \mathcal{N}^{*}
\end{equation*}
where $ \odot$ is the element-wise product such that $\mathcal{N}^{\tiny{\mbox{pro}}}_{l,i}={\psi}_{l,i} \tiny{\mathcal{N}}^{*}_{l,i}$. Figure \ref{manifold_visual}(b) illustrates the main idea: the red dashed arrow starting from $A$ denotes the $l$th ray (or a vector of $(A_{l,1},\ldots, A_{l, N(l)})$) of the principal sub-manifold $\mathcal{N}$; the red solid arrow denotes the projected $l$th ray of the principal sub-manifold, $\mathcal{A}^*_l$. The point $A$ is now regarded as the new origin under the new coordinate system, correspondingly. Moreover, the data points $x_j$'s are projected in the same way by
\[ 
x^*_j =E_3 (x_j-A),  \quad 1 \leq j \leq n.
\] 
In general, the projected points $x^*_j$'s are expected to lie closely to the sub-manifold $\mathcal{N}^{\tiny{\mbox{pro}}}$, provided that the projection matrix has accounted for most of the variability. 

\section{Applications}
This section contains an illustration of principal sub-manifolds on a data set of handwritten digits. Additional simulations are presented in Appendix \ref{app:10-simulations-s3} and \ref{app:11-simulations-s2} and two more applications can be found in Appendix \ref{app:12-digits-princ-var} and \ref{app:13-leaves}. Since we are concerned with landmark shapes, we provide a brief introduction to that topic in Appendix \ref{app:14-landmark-shapes}.

\begin{figure}[ht!]%[ht]
  \centering
  \includegraphics[width=0.5in]{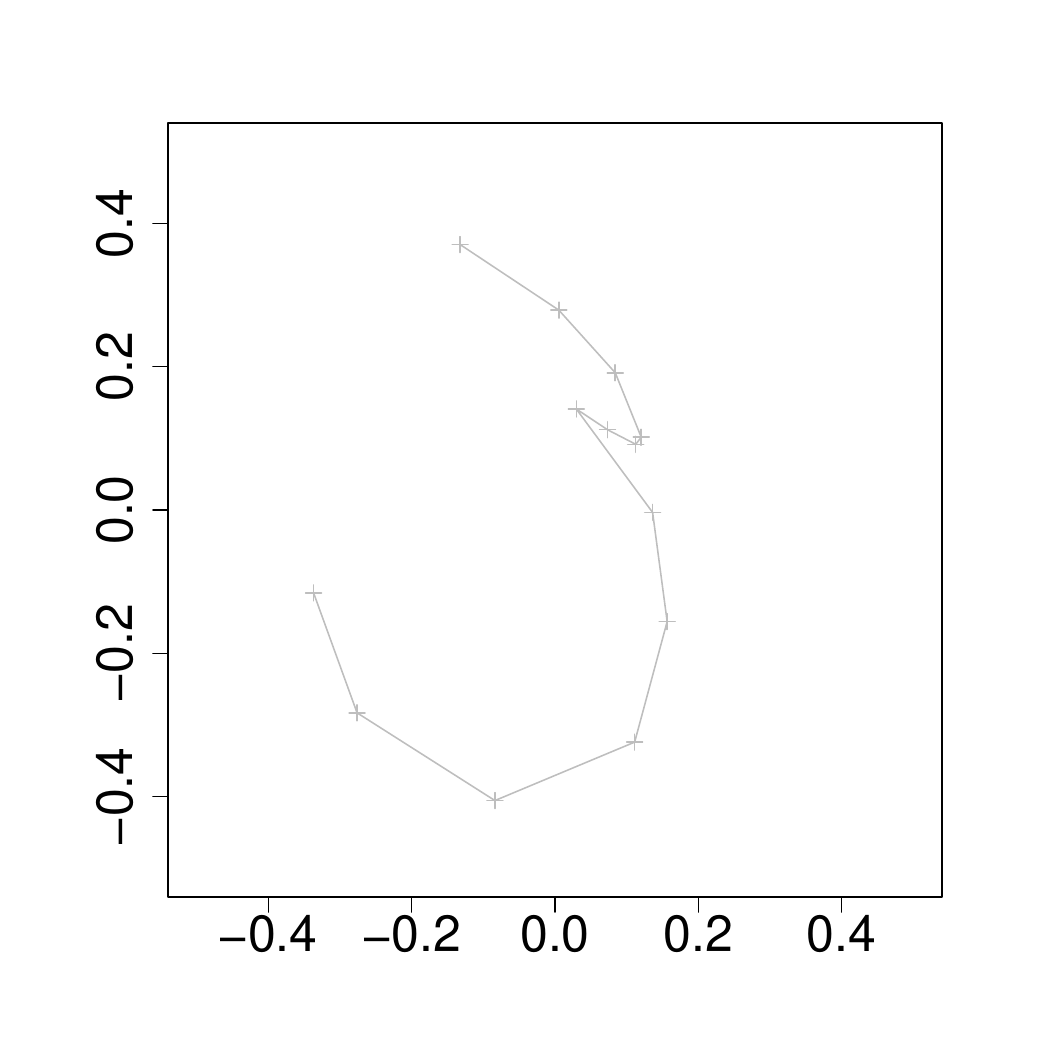}
  \includegraphics[width=0.5in]{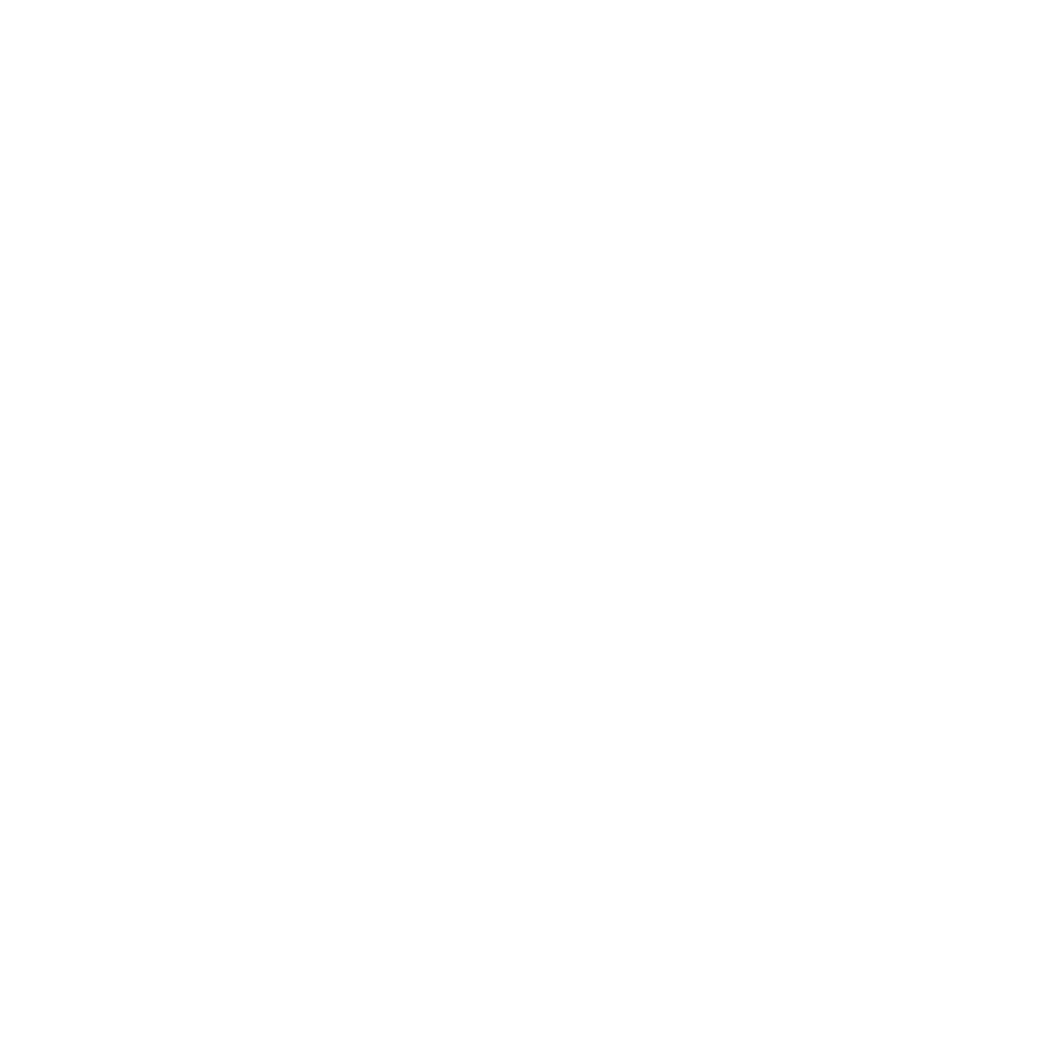}
  \includegraphics[width=0.5in]{PDF/grid-digit3-mean/empty}
  \includegraphics[width=0.5in]{PDF/grid-digit3-mean/empty}
  \includegraphics[width=0.5in]{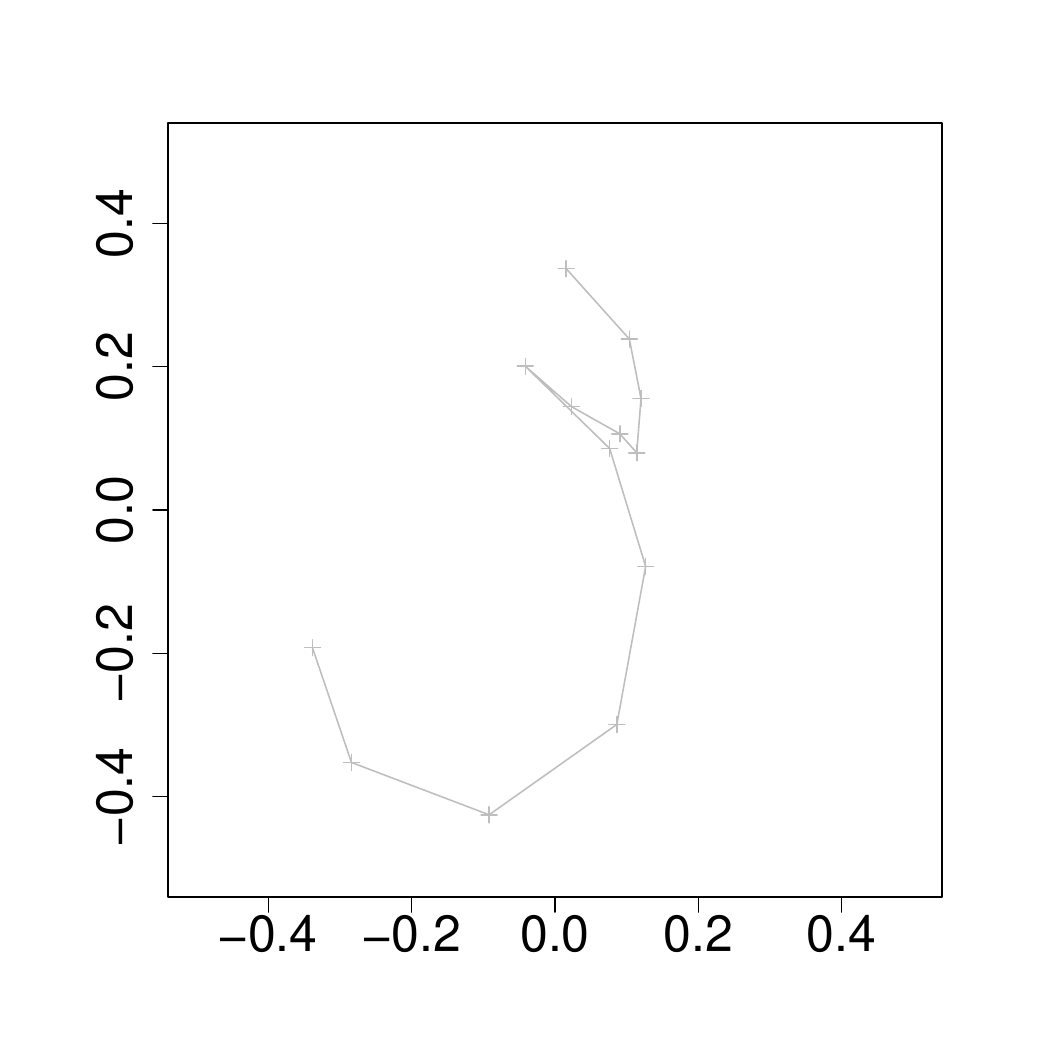}
  \includegraphics[width=0.5in]{PDF/grid-digit3-mean/empty}
  \includegraphics[width=0.5in]{PDF/grid-digit3-mean/empty}
  \includegraphics[width=0.5in]{PDF/grid-digit3-mean/empty}
  \includegraphics[width=0.5in]{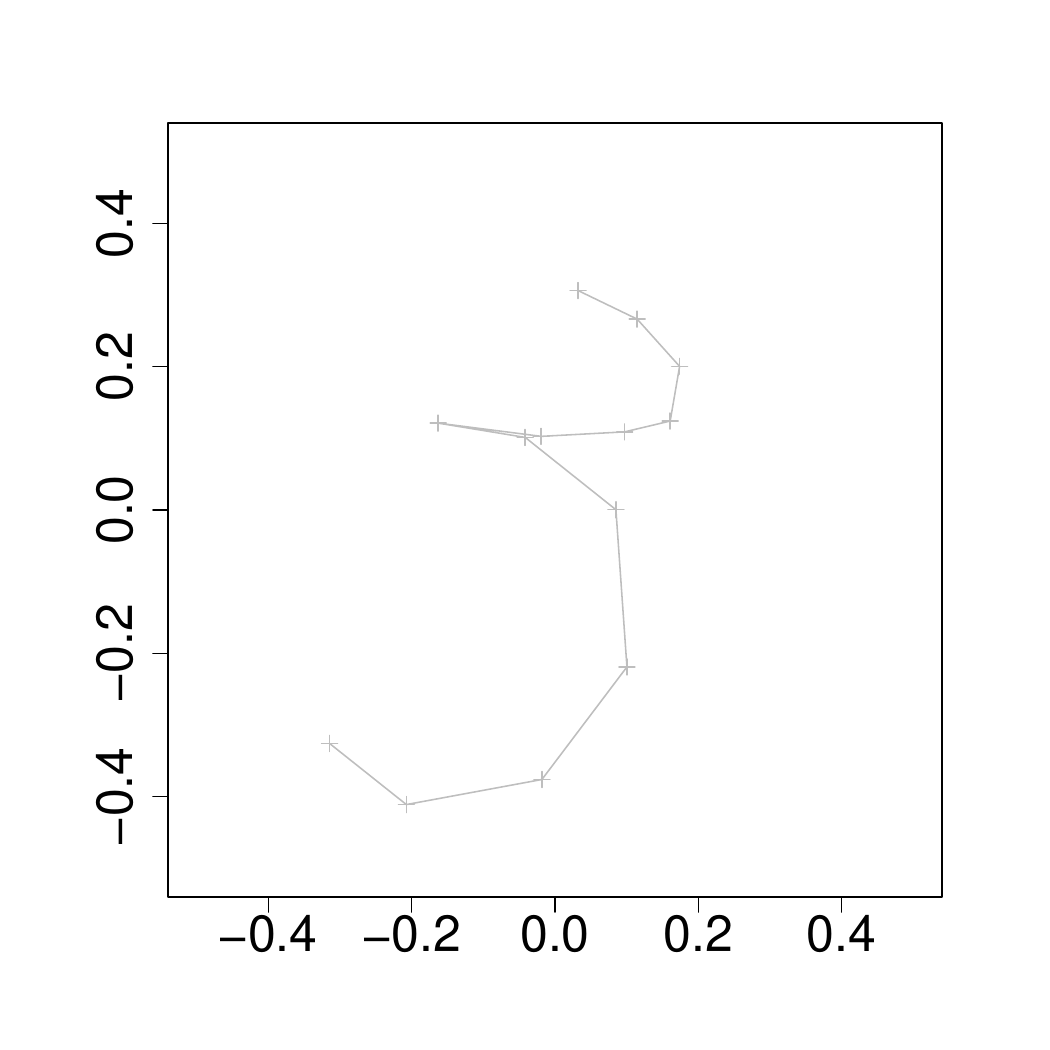}\\
  \includegraphics[width=0.5in]{PDF/grid-digit3-mean/empty}
  \includegraphics[width=0.5in]{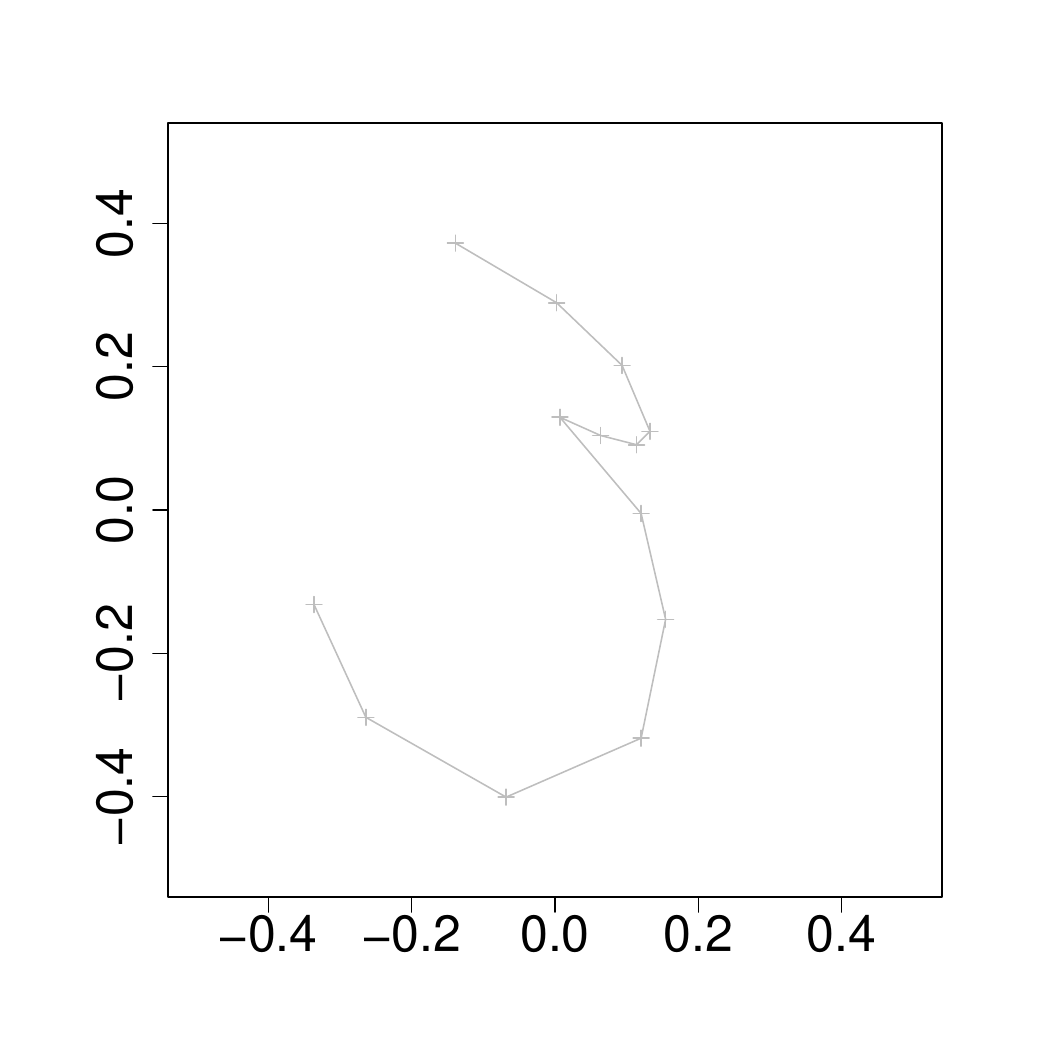}
  \includegraphics[width=0.5in]{PDF/grid-digit3-mean/empty}
  \includegraphics[width=0.5in]{PDF/grid-digit3-mean/empty}
  \includegraphics[width=0.5in]{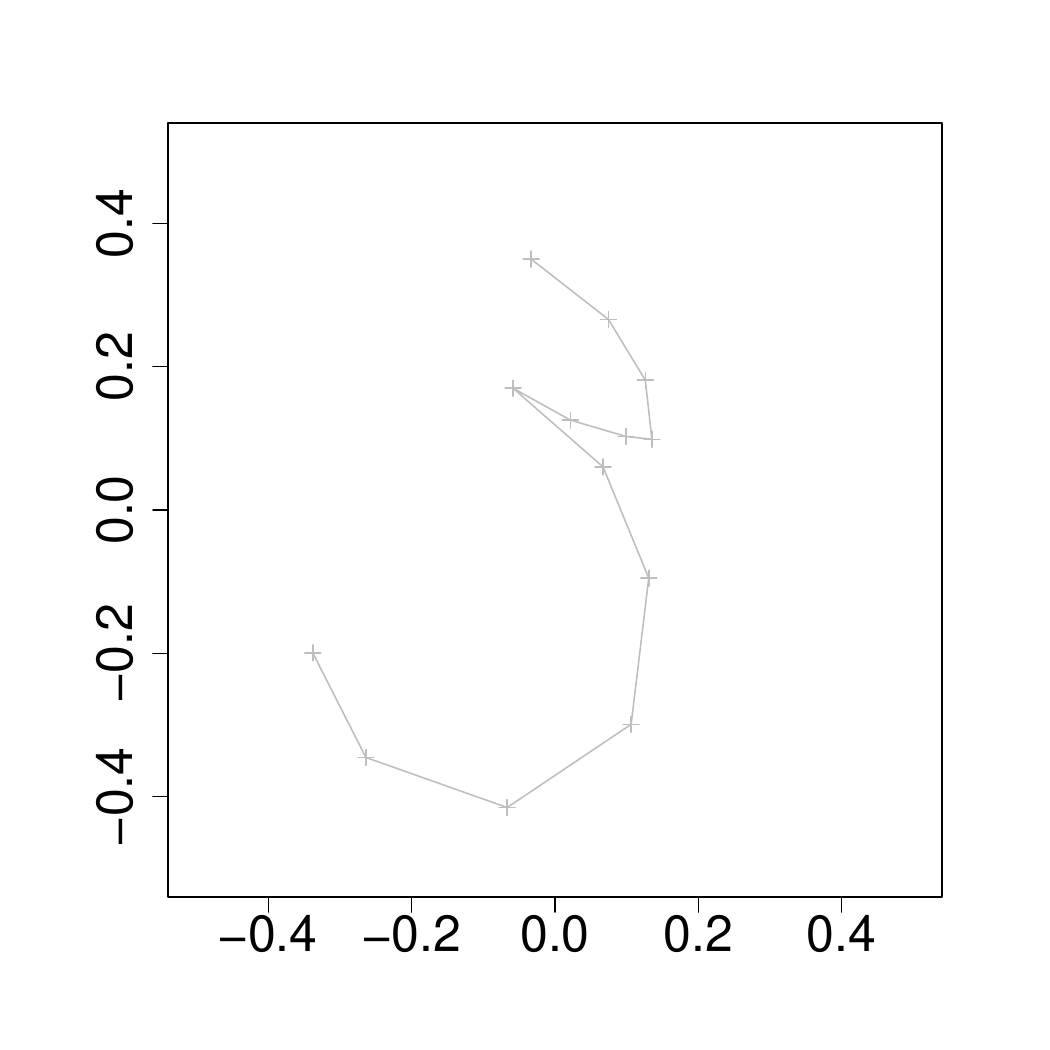}
  \includegraphics[width=0.5in]{PDF/grid-digit3-mean/empty}
  \includegraphics[width=0.5in]{PDF/grid-digit3-mean/empty}
  \includegraphics[width=0.5in]{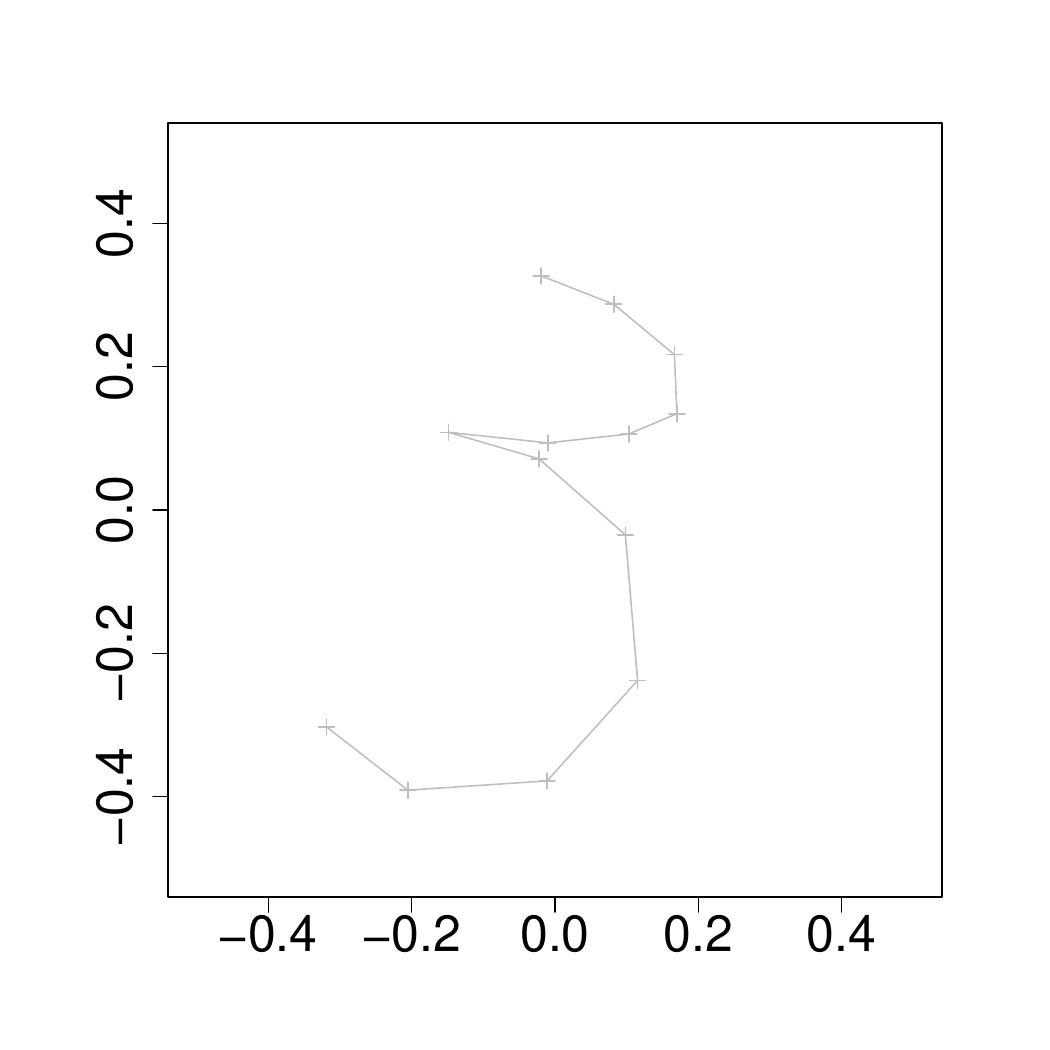}
  \includegraphics[width=0.5in]{PDF/grid-digit3-mean/empty}\\
  \includegraphics[width=0.5in]{PDF/grid-digit3-mean/empty}
  \includegraphics[width=0.5in]{PDF/grid-digit3-mean/empty}
  \includegraphics[width=0.5in]{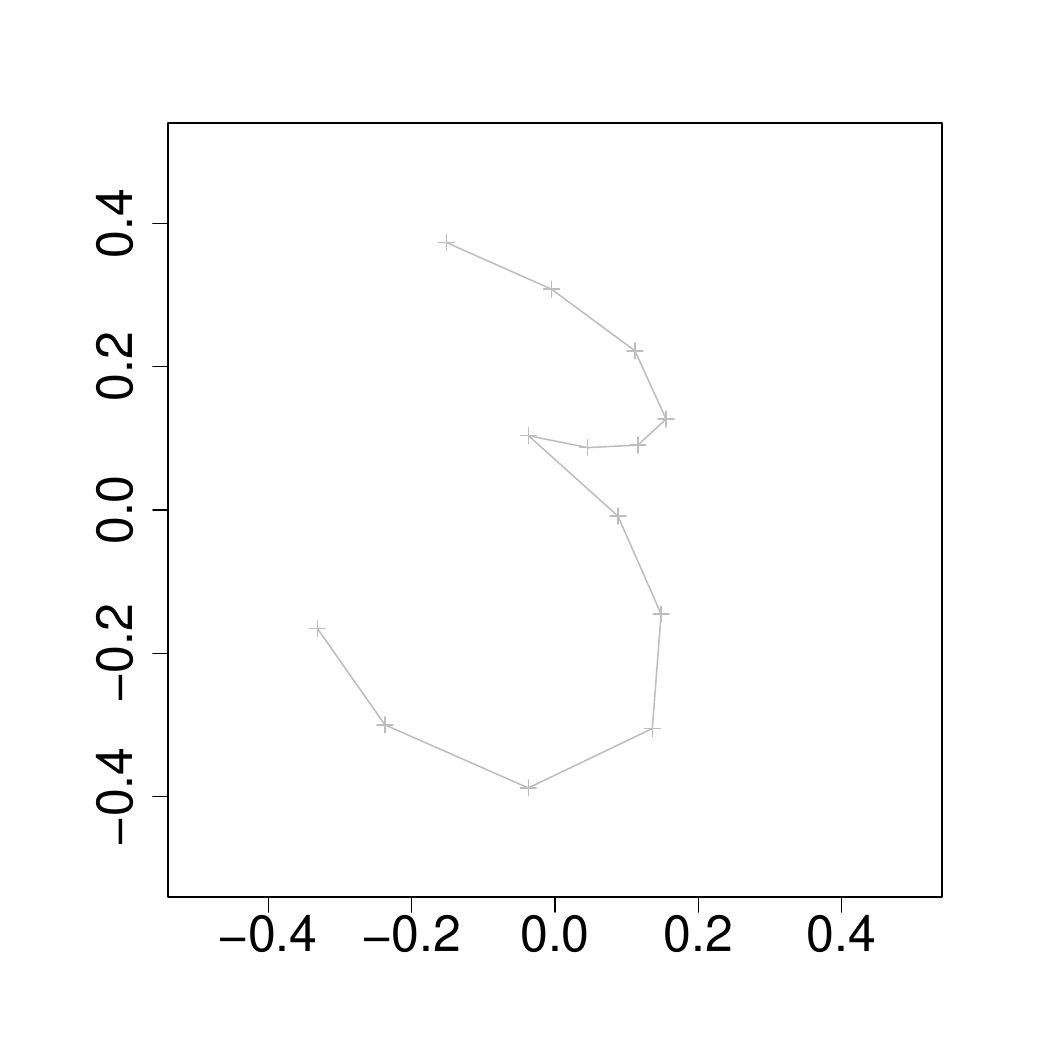}
  \includegraphics[width=0.5in]{PDF/grid-digit3-mean/empty}
  \includegraphics[width=0.5in]{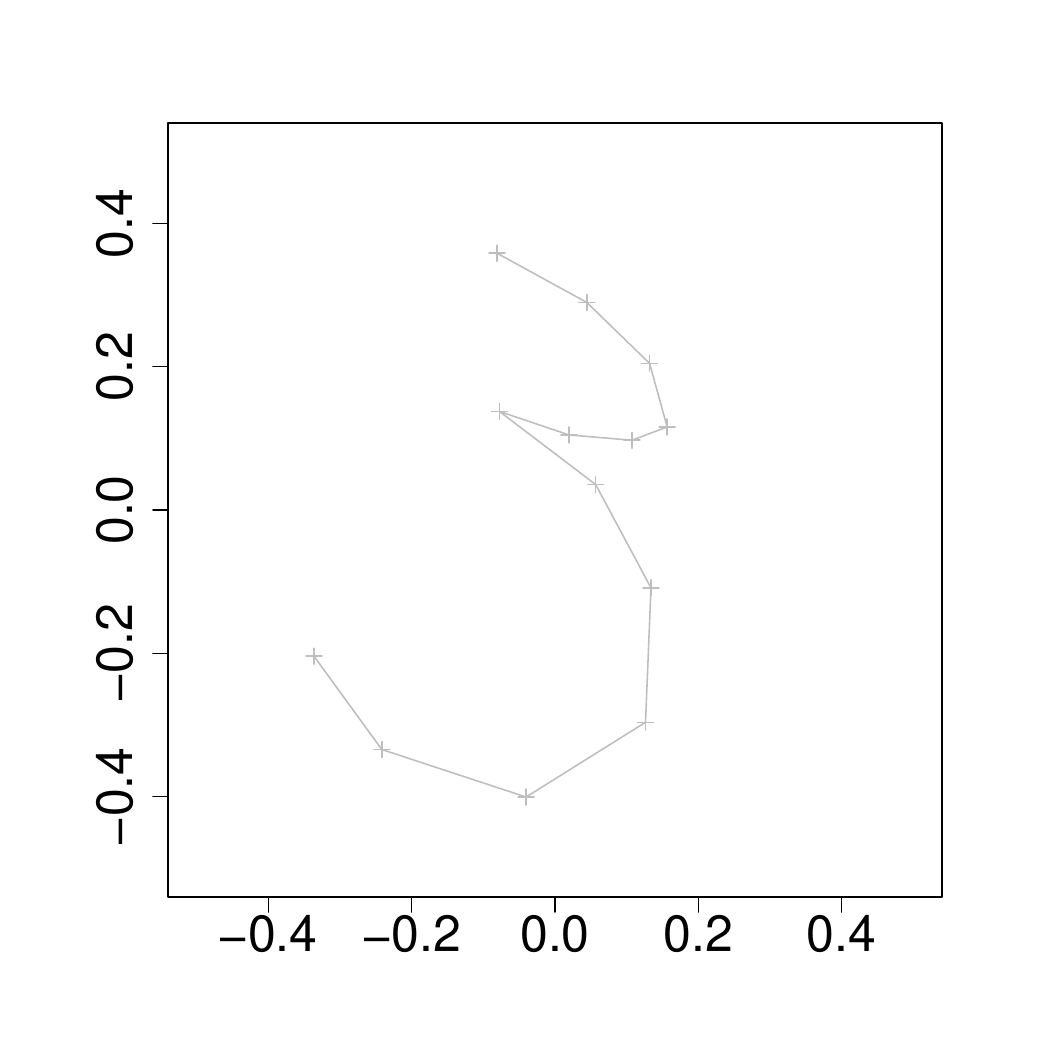}
  \includegraphics[width=0.5in]{PDF/grid-digit3-mean/empty}
  \includegraphics[width=0.5in]{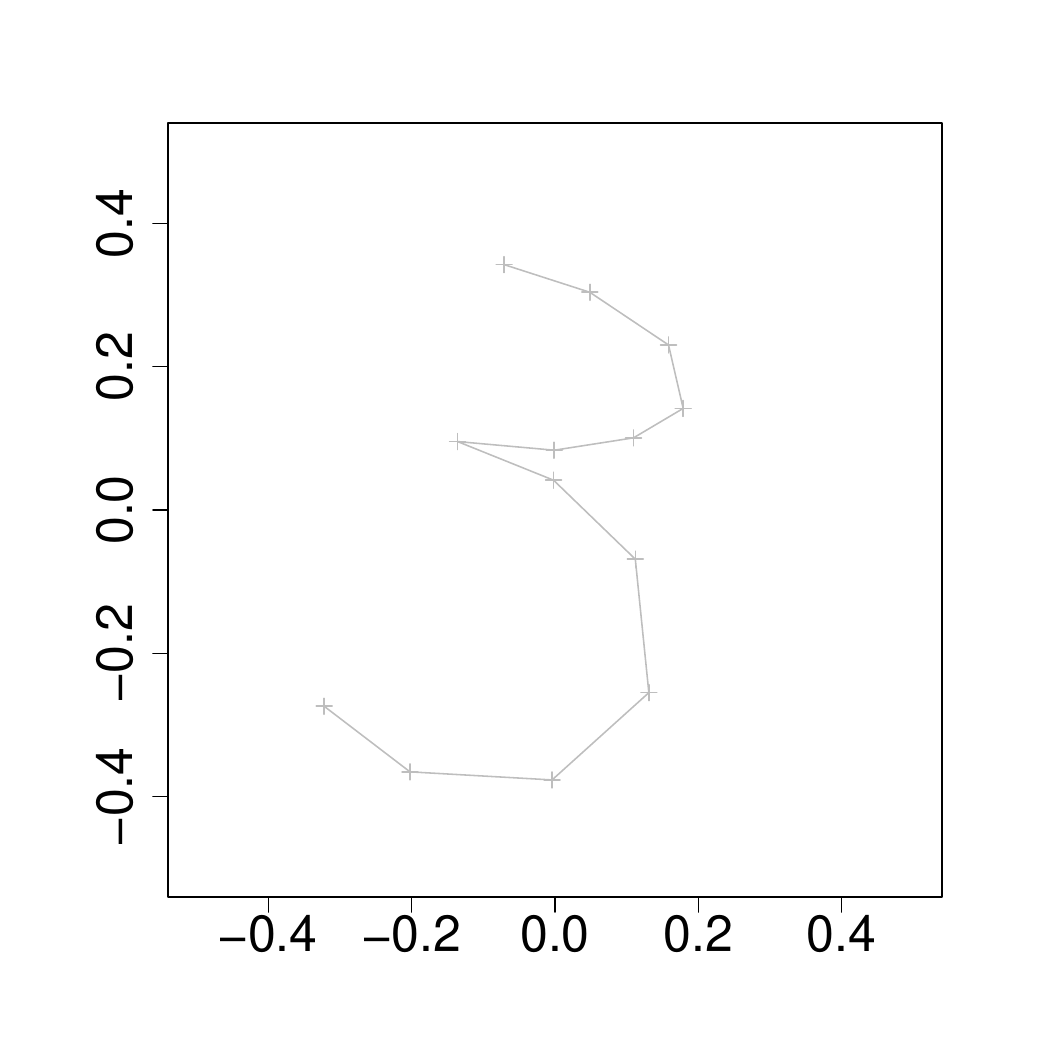}
  \includegraphics[width=0.5in]{PDF/grid-digit3-mean/empty}
  \includegraphics[width=0.5in]{PDF/grid-digit3-mean/empty}\\
  \includegraphics[width=0.5in]{PDF/grid-digit3-mean/empty}
  \includegraphics[width=0.5in]{PDF/grid-digit3-mean/empty}
  \includegraphics[width=0.5in]{PDF/grid-digit3-mean/empty}
  \includegraphics[width=0.5in]{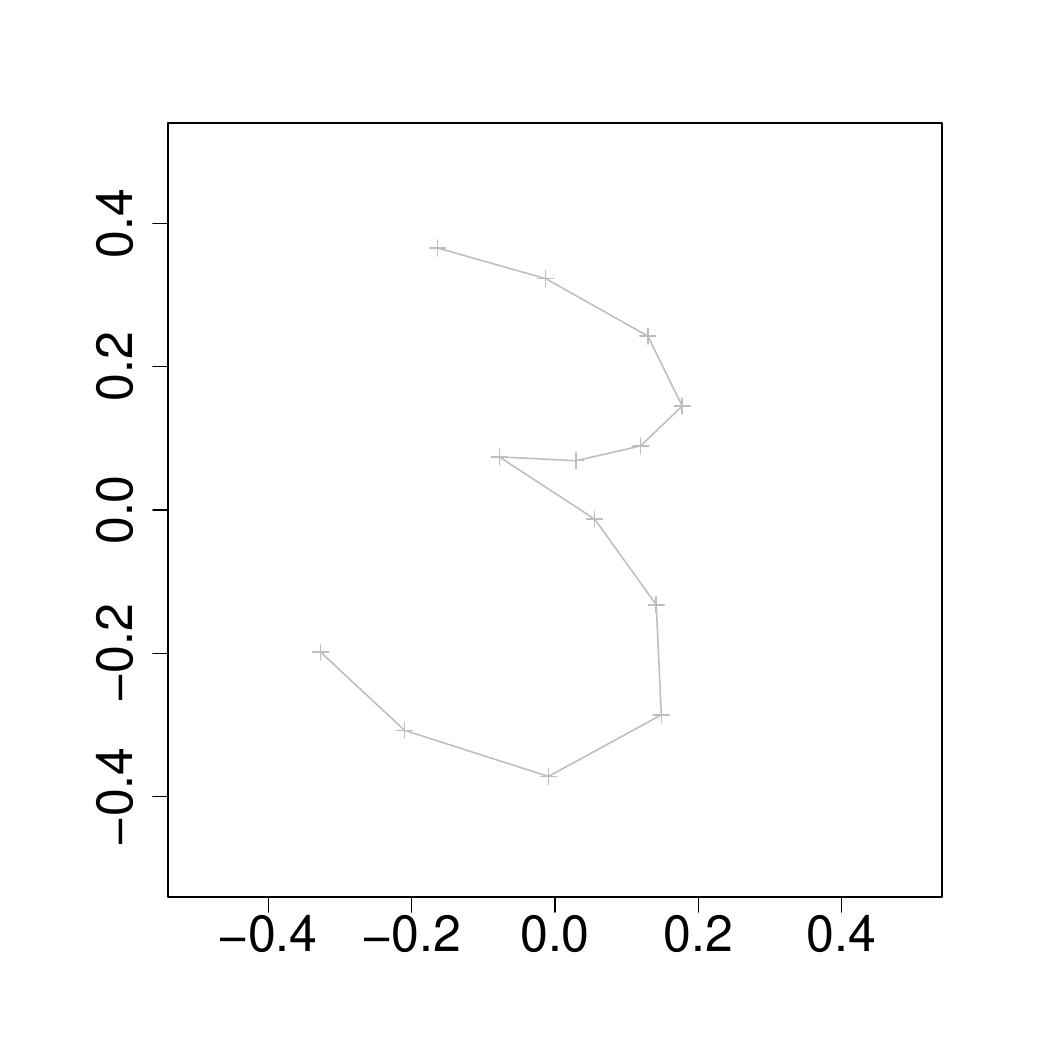}
  \includegraphics[width=0.5in]{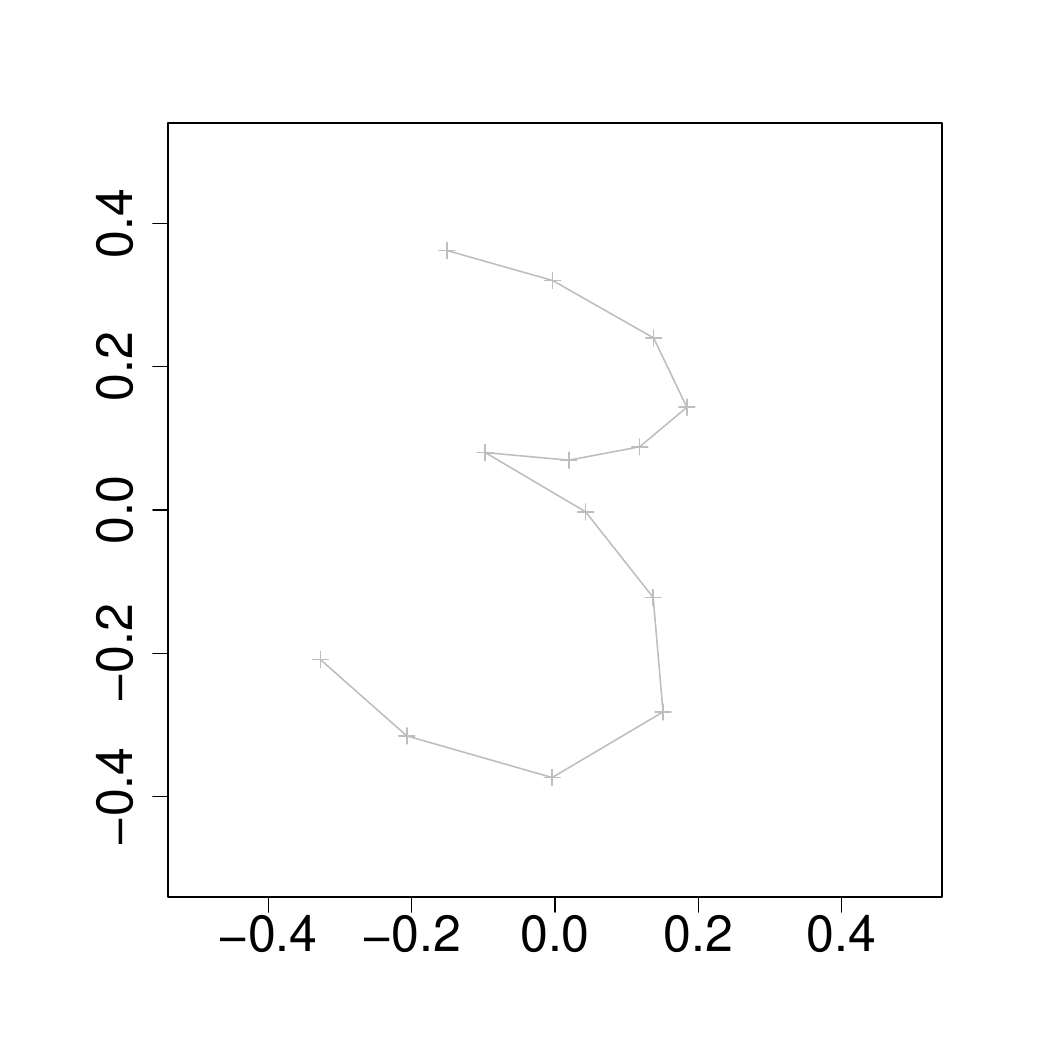}
  \includegraphics[width=0.5in]{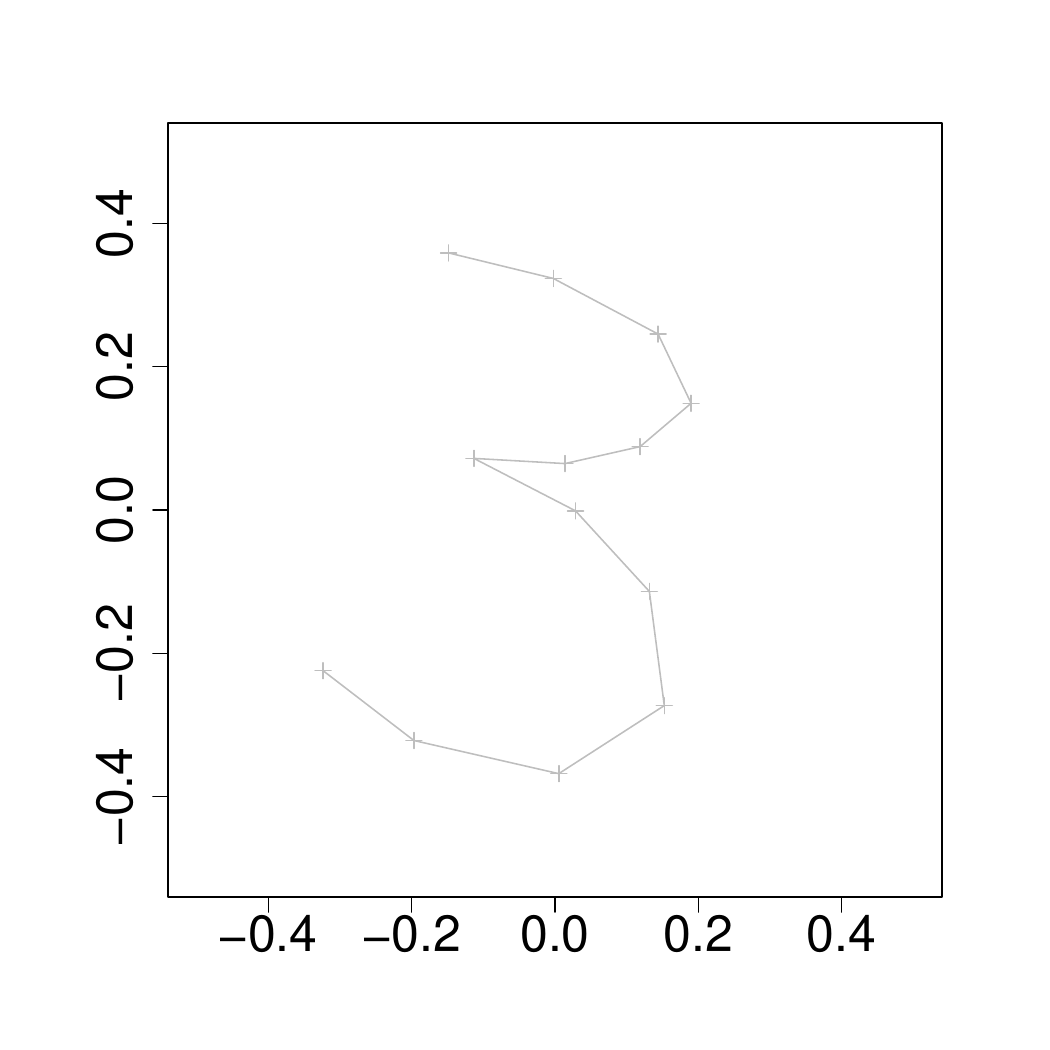}
  \includegraphics[width=0.5in]{PDF/grid-digit3-mean/empty}
  \includegraphics[width=0.5in]{PDF/grid-digit3-mean/empty}
  \includegraphics[width=0.5in]{PDF/grid-digit3-mean/empty}\\
  \includegraphics[width=0.5in]{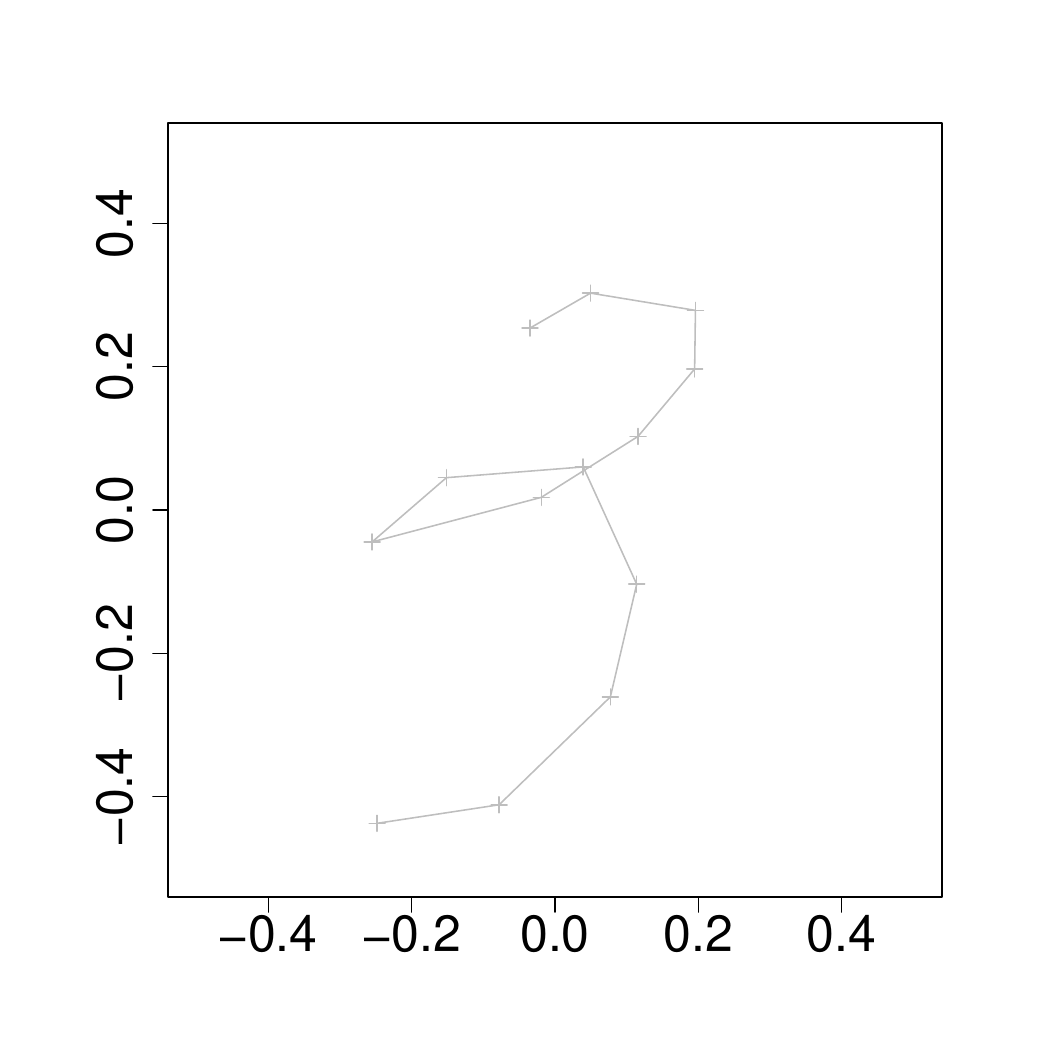}
  \includegraphics[width=0.5in]{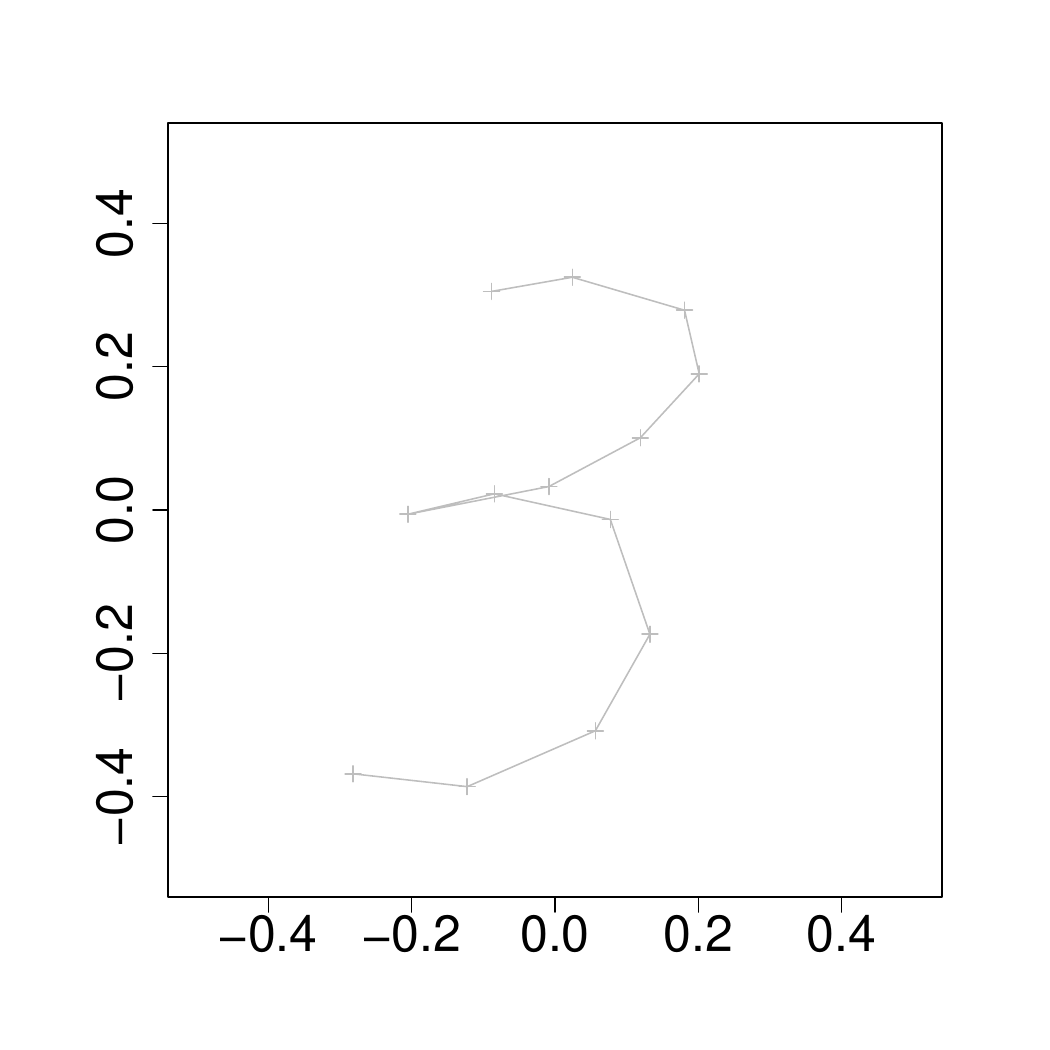}
  \includegraphics[width=0.5in]{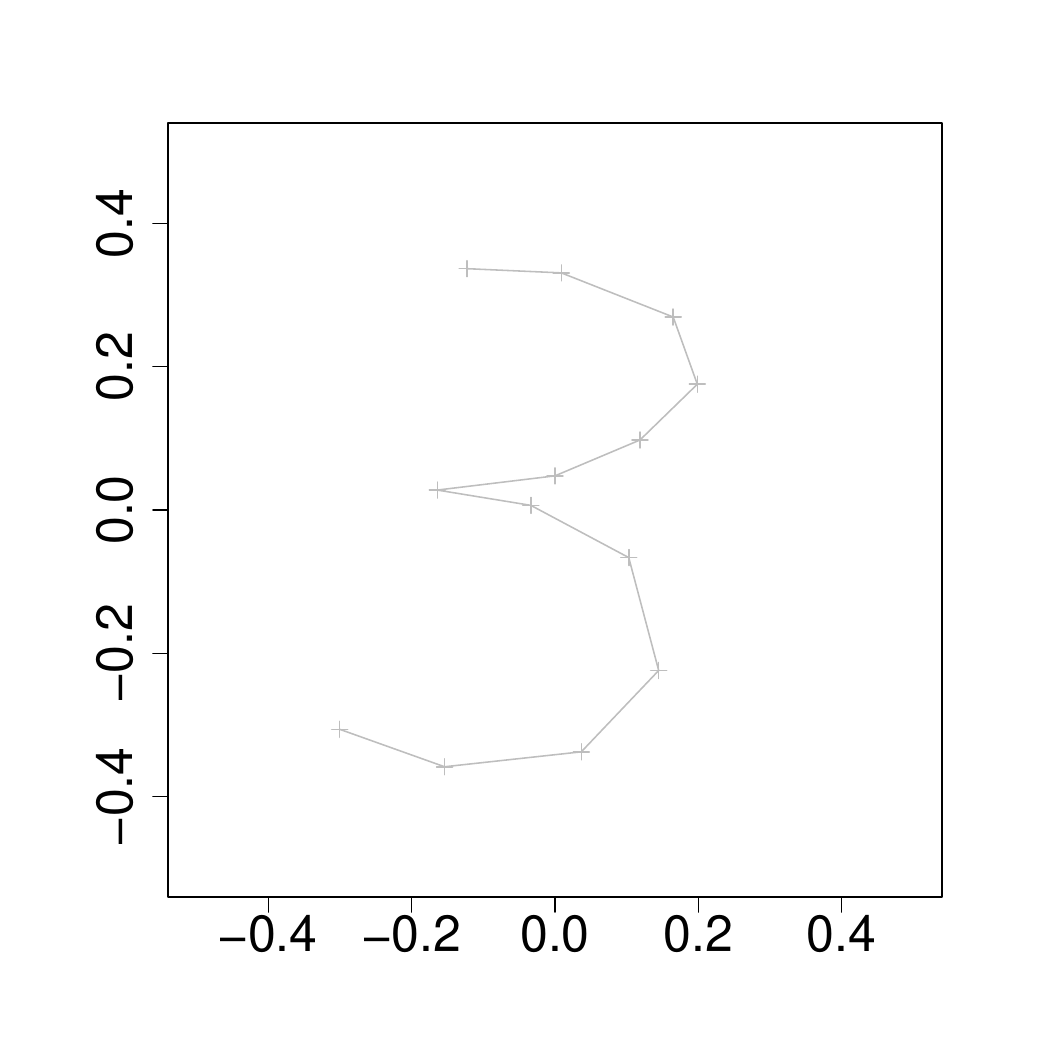}
  \includegraphics[width=0.5in]{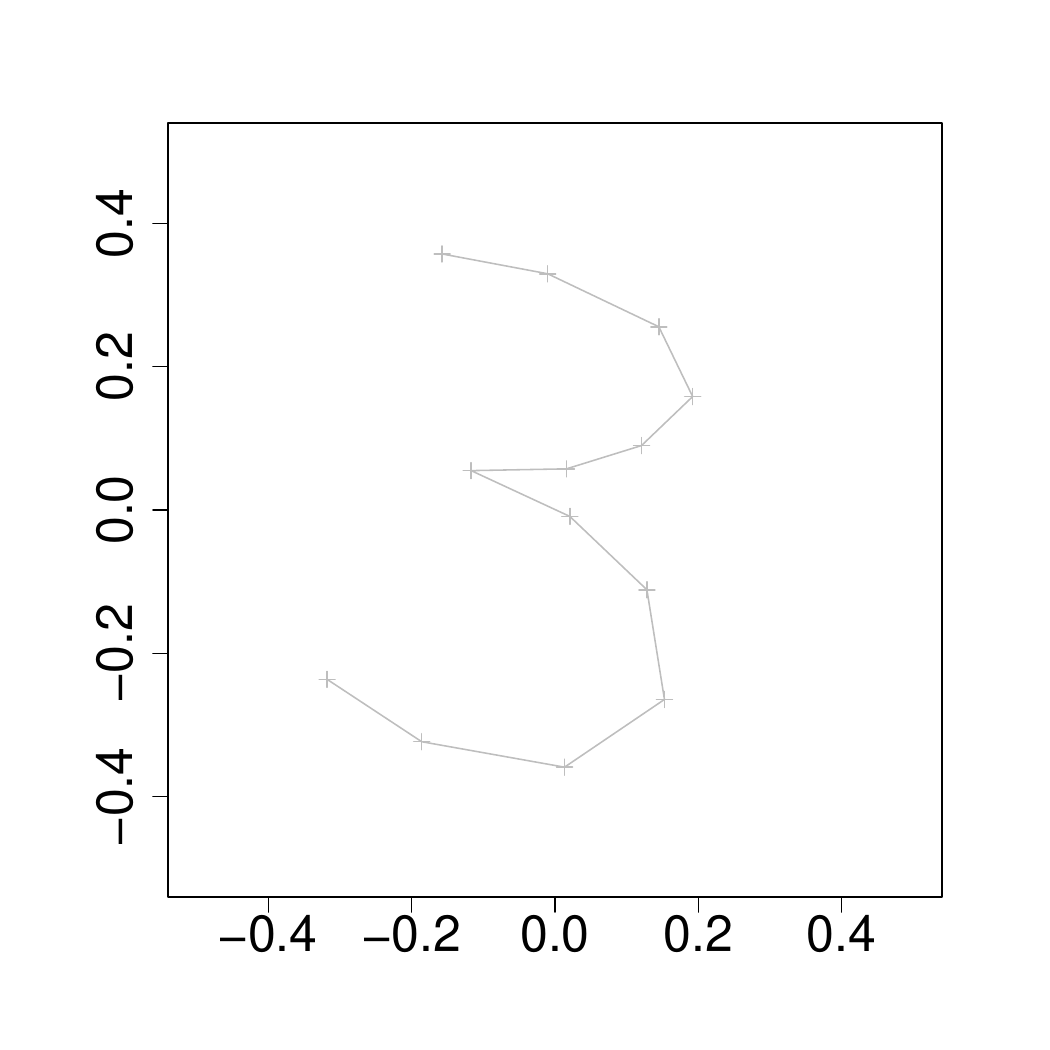}
  \includegraphics[width=0.5in]{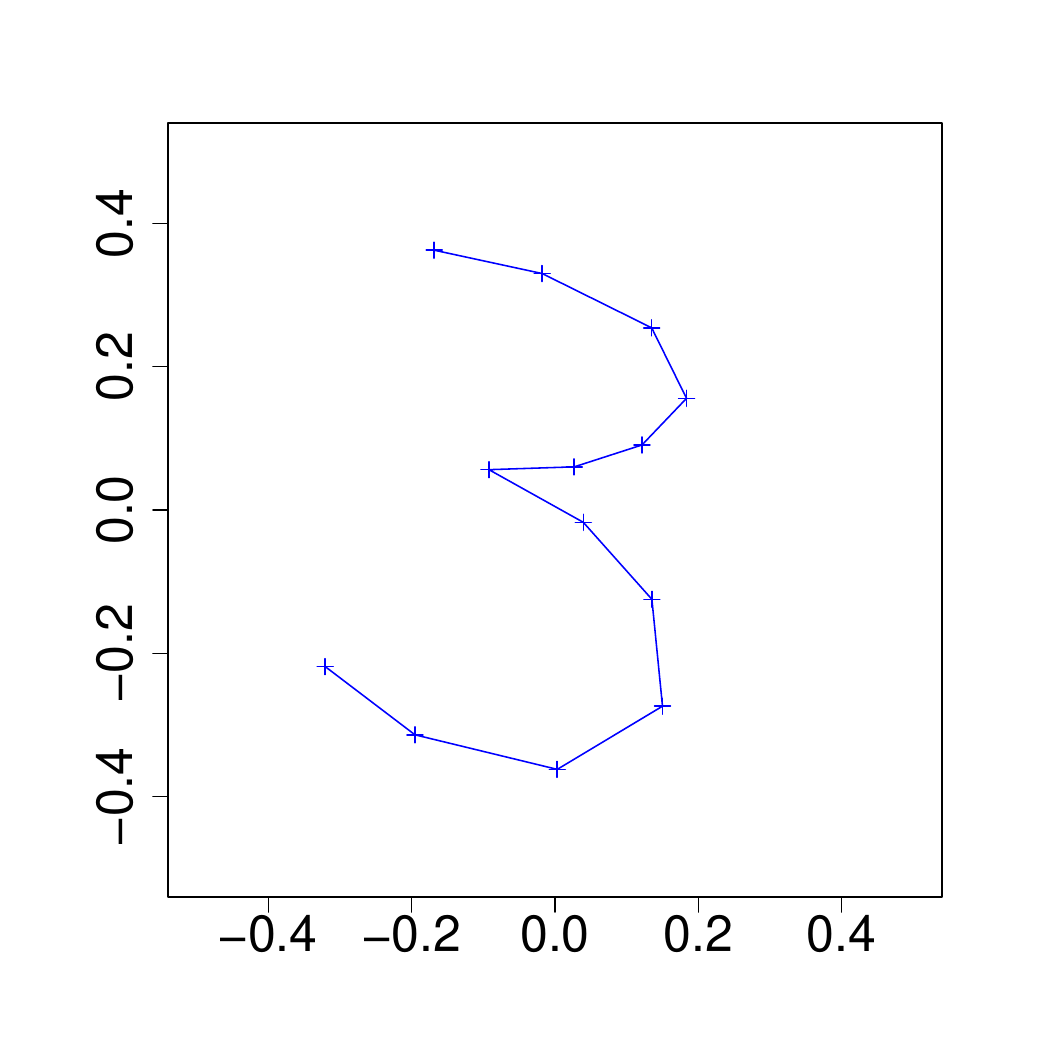}
  \includegraphics[width=0.5in]{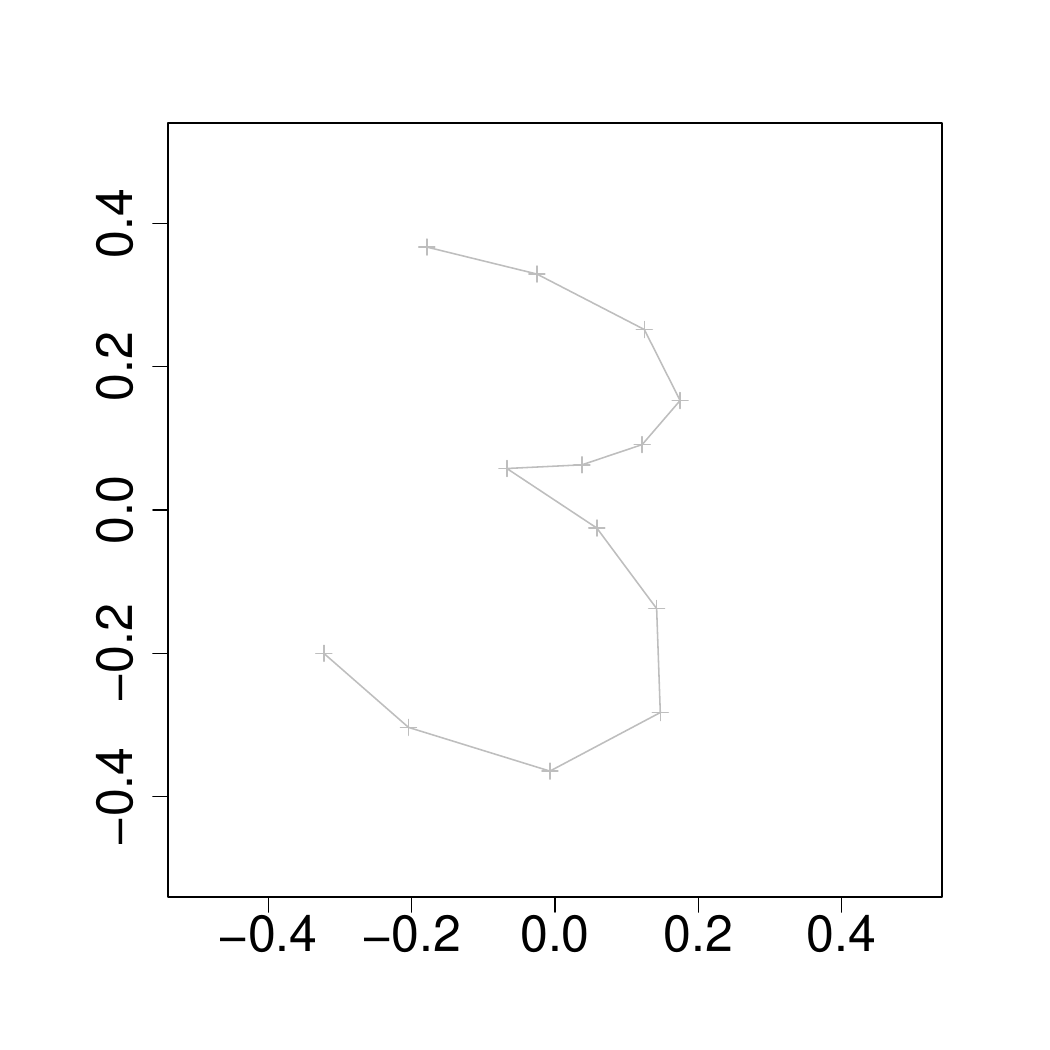}
  \includegraphics[width=0.5in]{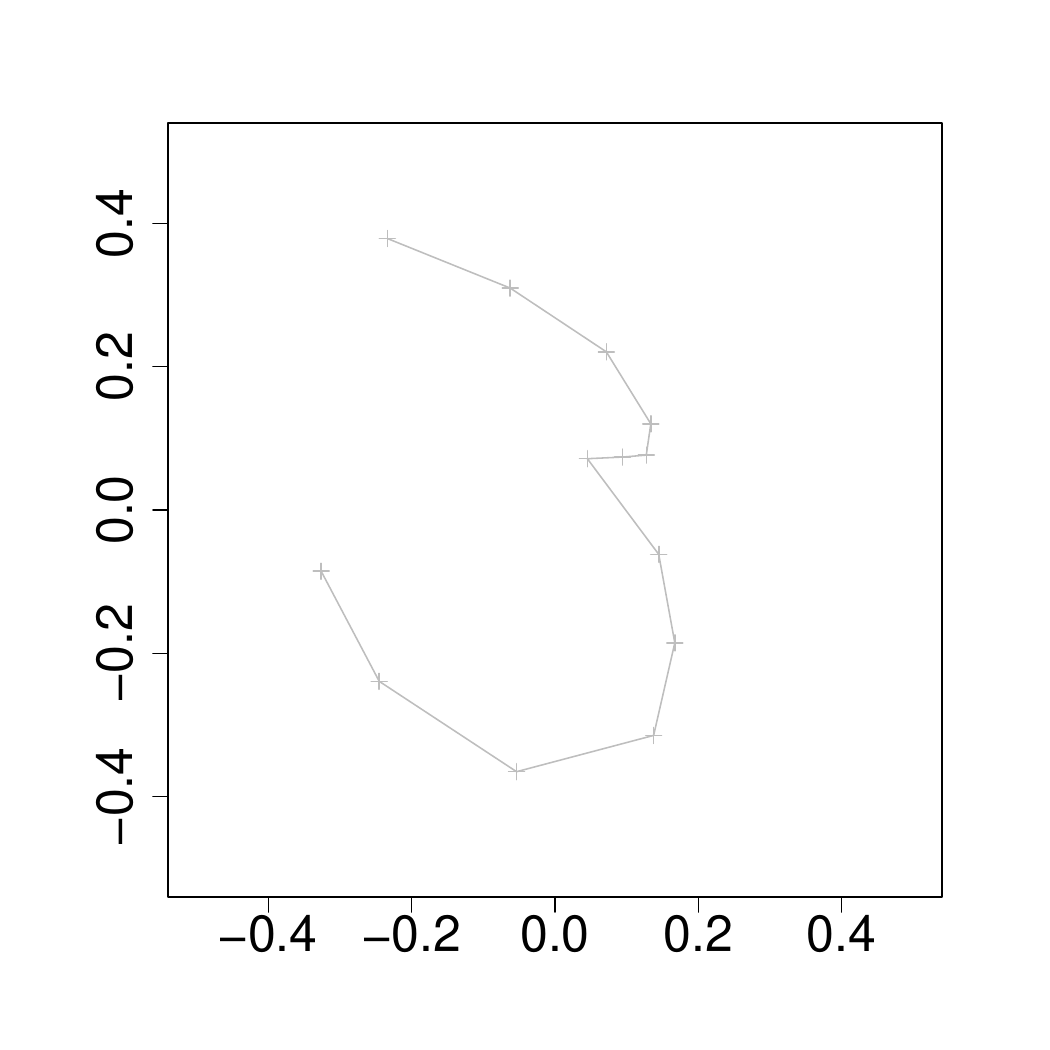}
  \includegraphics[width=0.5in]{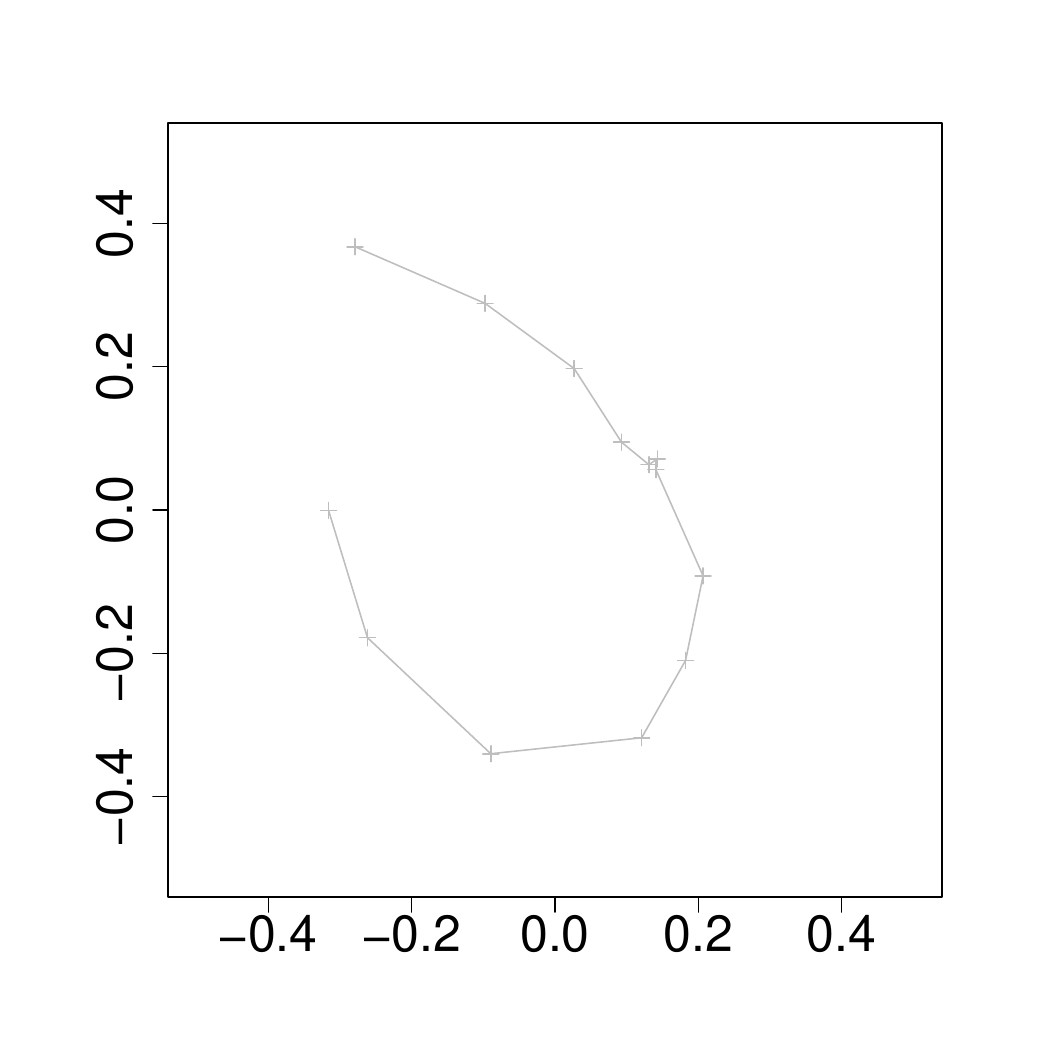}
  \includegraphics[width=0.5in]{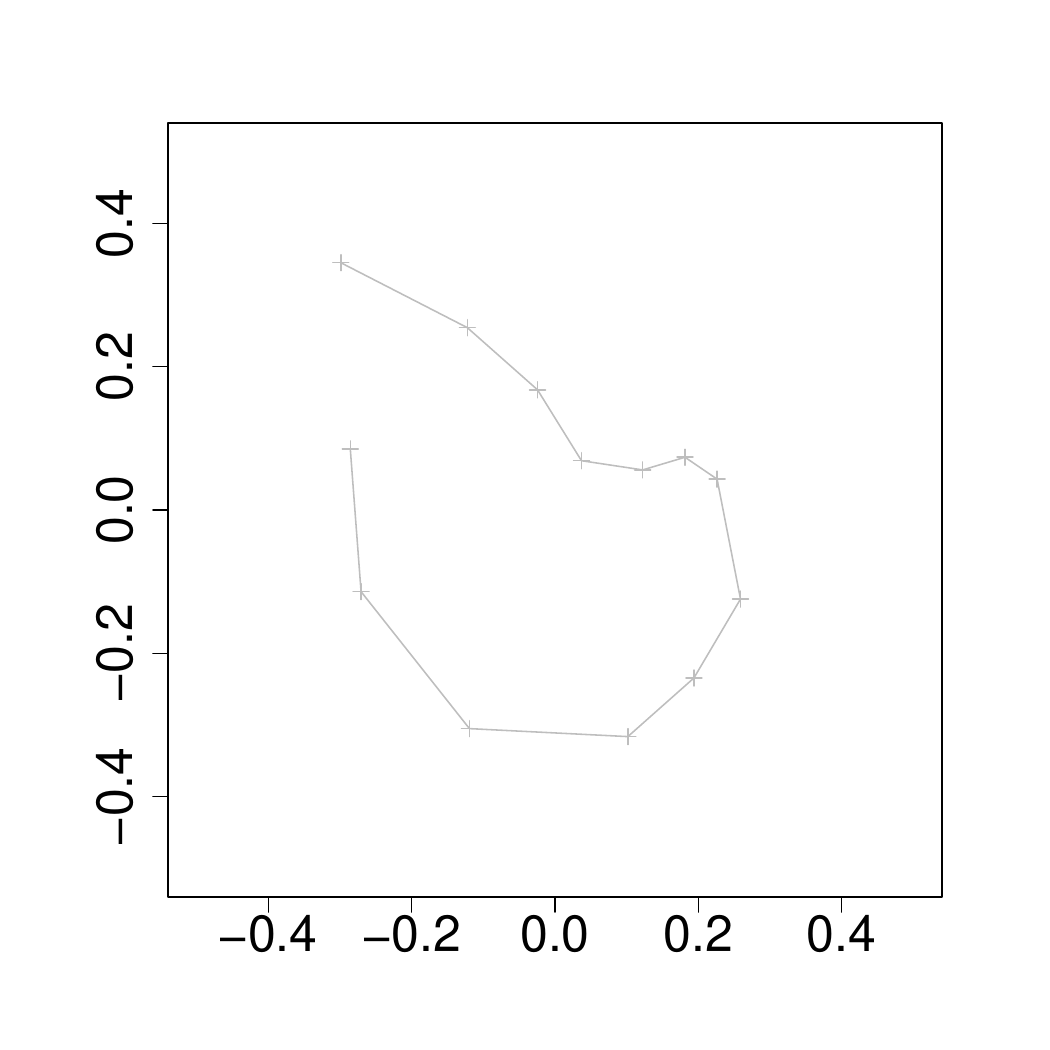}\\
  \includegraphics[width=0.5in]{PDF/grid-digit3-mean/empty}
  \includegraphics[width=0.5in]{PDF/grid-digit3-mean/empty}
  \includegraphics[width=0.5in]{PDF/grid-digit3-mean/empty}
  \includegraphics[width=0.5in]{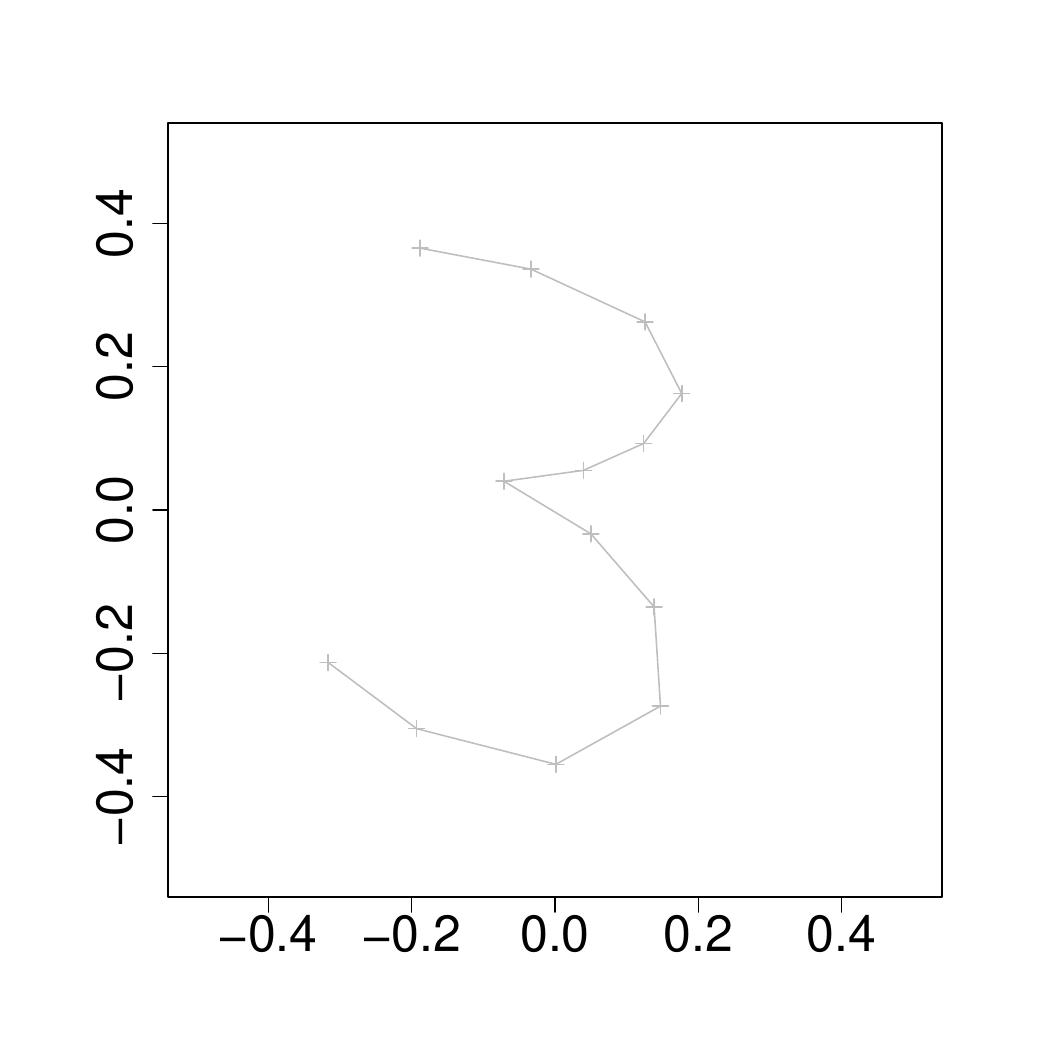}
  \includegraphics[width=0.5in]{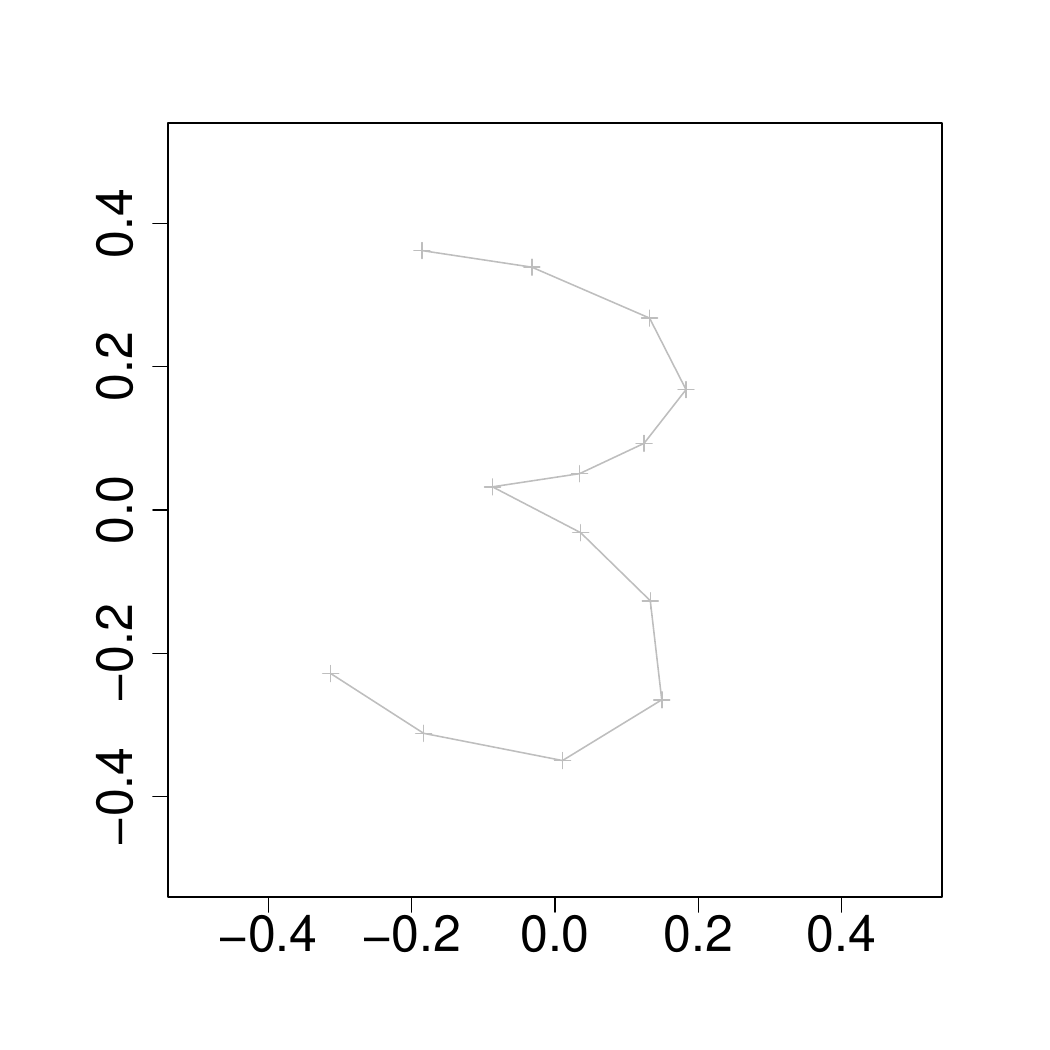}
  \includegraphics[width=0.5in]{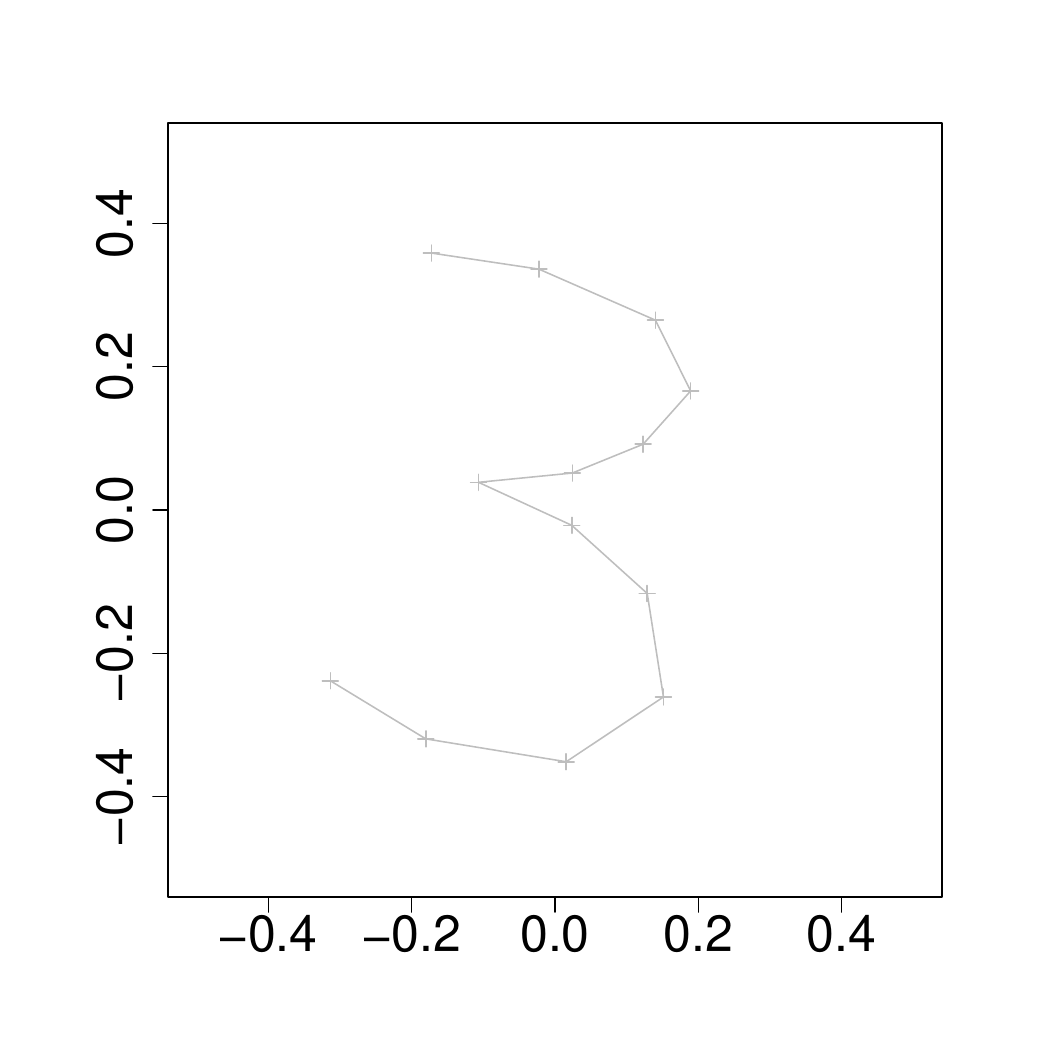}
  \includegraphics[width=0.5in]{PDF/grid-digit3-mean/empty}
  \includegraphics[width=0.5in]{PDF/grid-digit3-mean/empty}
  \includegraphics[width=0.5in]{PDF/grid-digit3-mean/empty}\\
  \includegraphics[width=0.5in]{PDF/grid-digit3-mean/empty}
  \includegraphics[width=0.5in]{PDF/grid-digit3-mean/empty}
  \includegraphics[width=0.5in]{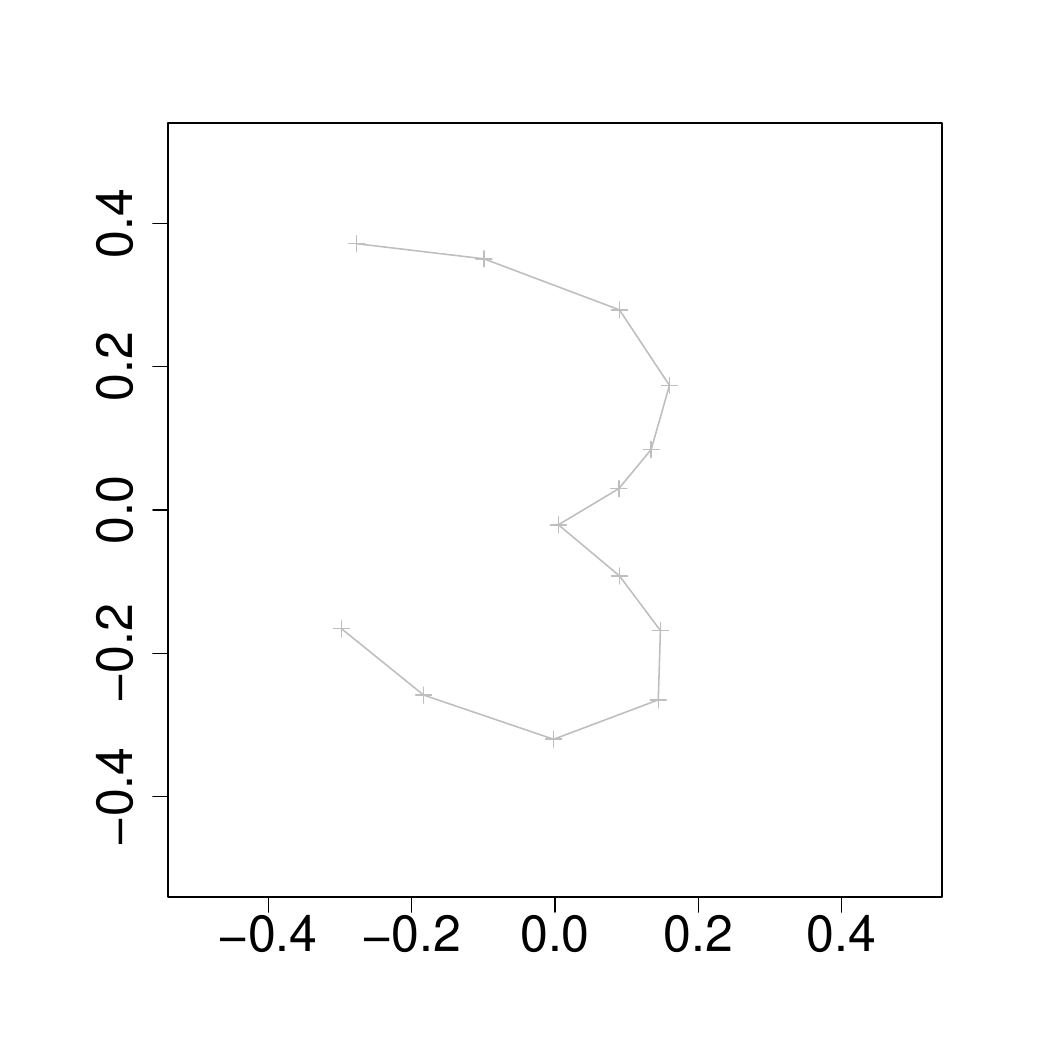}
  \includegraphics[width=0.5in]{PDF/grid-digit3-mean/empty}
  \includegraphics[width=0.5in]{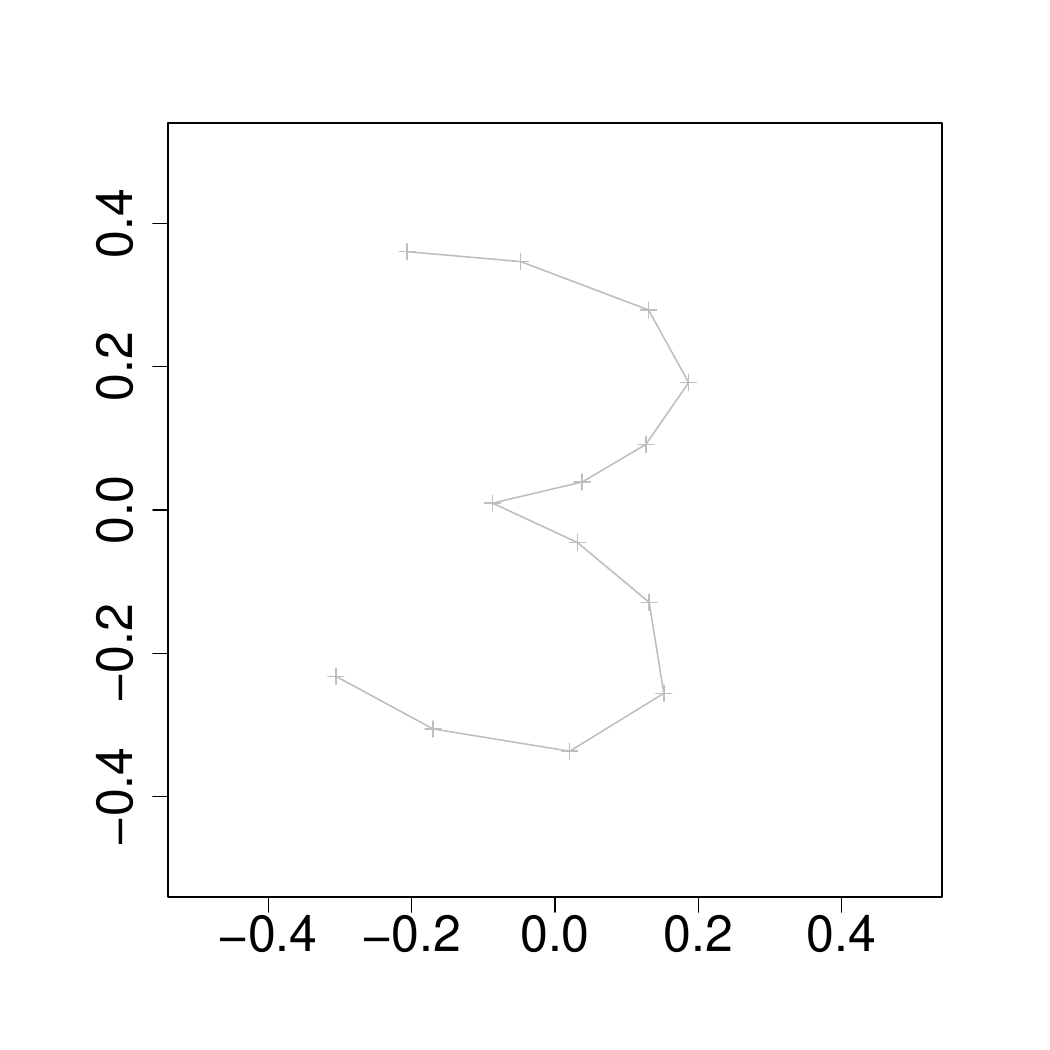}
  \includegraphics[width=0.5in]{PDF/grid-digit3-mean/empty}
  \includegraphics[width=0.5in]{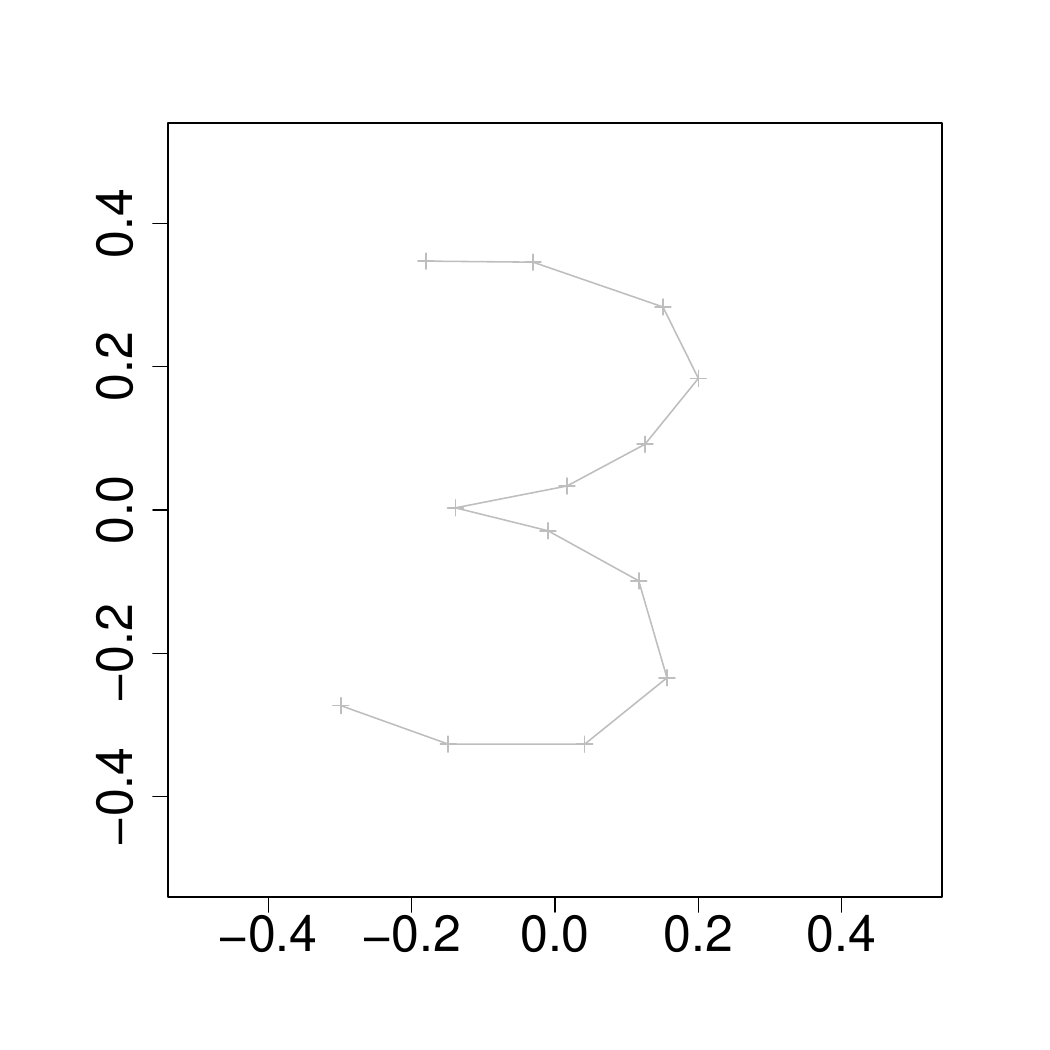}
  \includegraphics[width=0.5in]{PDF/grid-digit3-mean/empty}
  \includegraphics[width=0.5in]{PDF/grid-digit3-mean/empty}\\
  \includegraphics[width=0.5in]{PDF/grid-digit3-mean/empty}
  \includegraphics[width=0.5in]{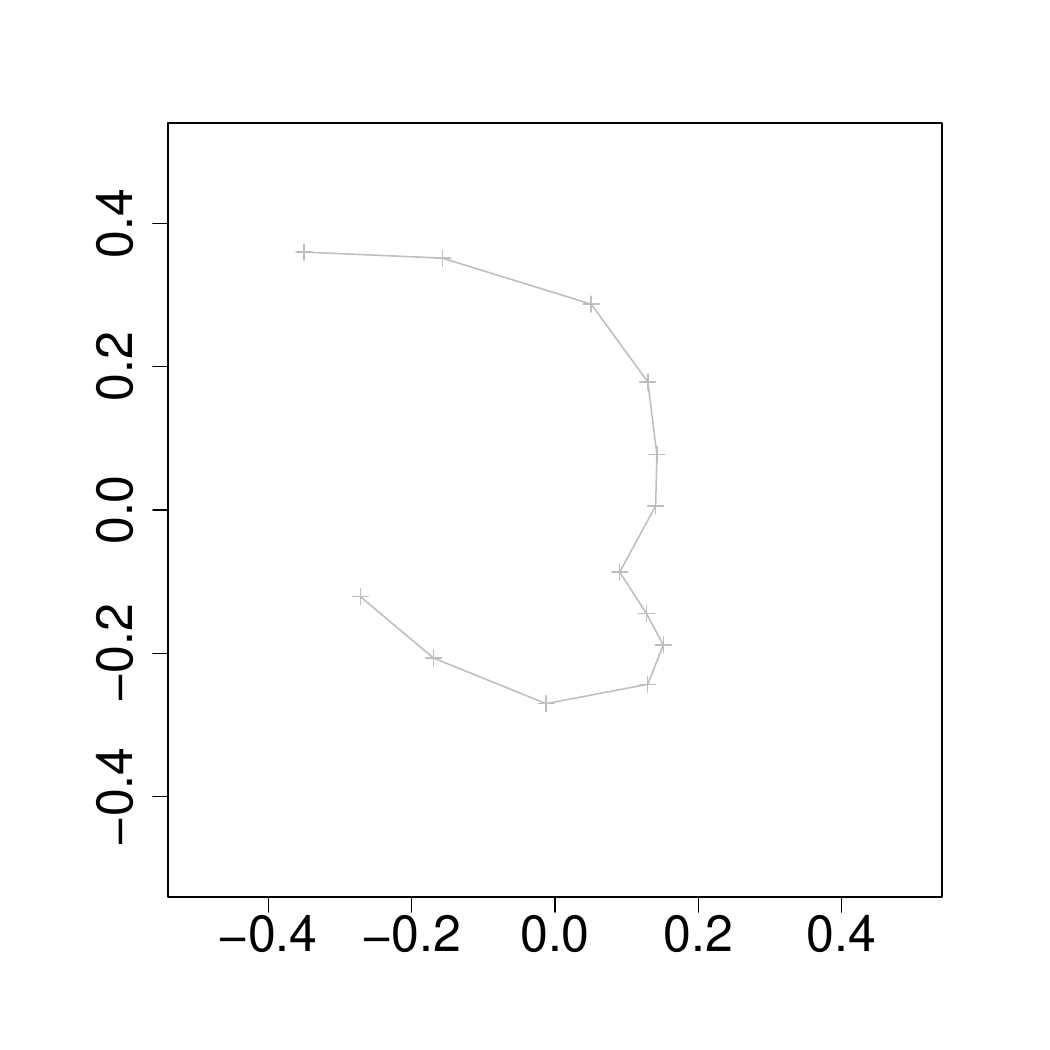}
  \includegraphics[width=0.5in]{PDF/grid-digit3-mean/empty}
  \includegraphics[width=0.5in]{PDF/grid-digit3-mean/empty}
  \includegraphics[width=0.5in]{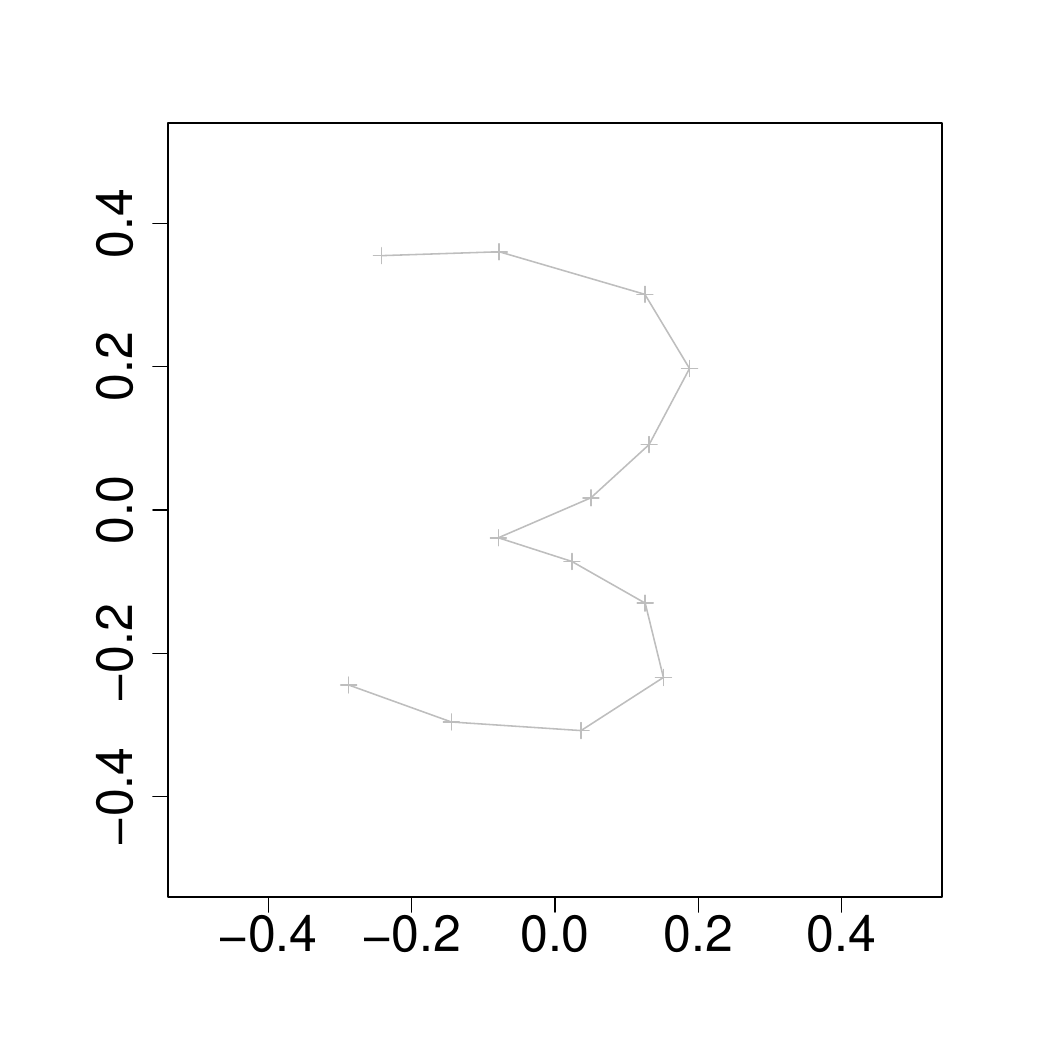}
  \includegraphics[width=0.5in]{PDF/grid-digit3-mean/empty}
  \includegraphics[width=0.5in]{PDF/grid-digit3-mean/empty}
  \includegraphics[width=0.5in]{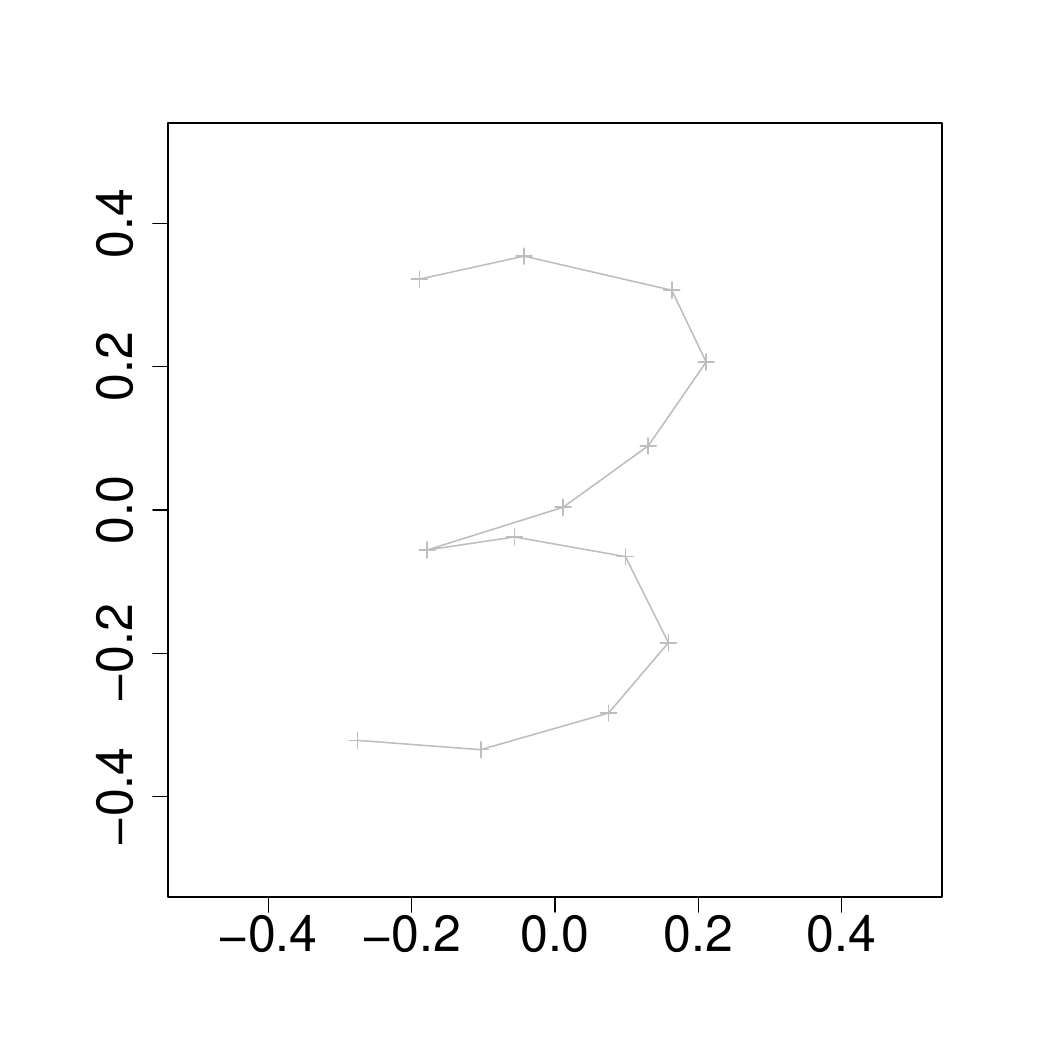}
  \includegraphics[width=0.5in]{PDF/grid-digit3-mean/empty}\\
  \includegraphics[width=0.5in]{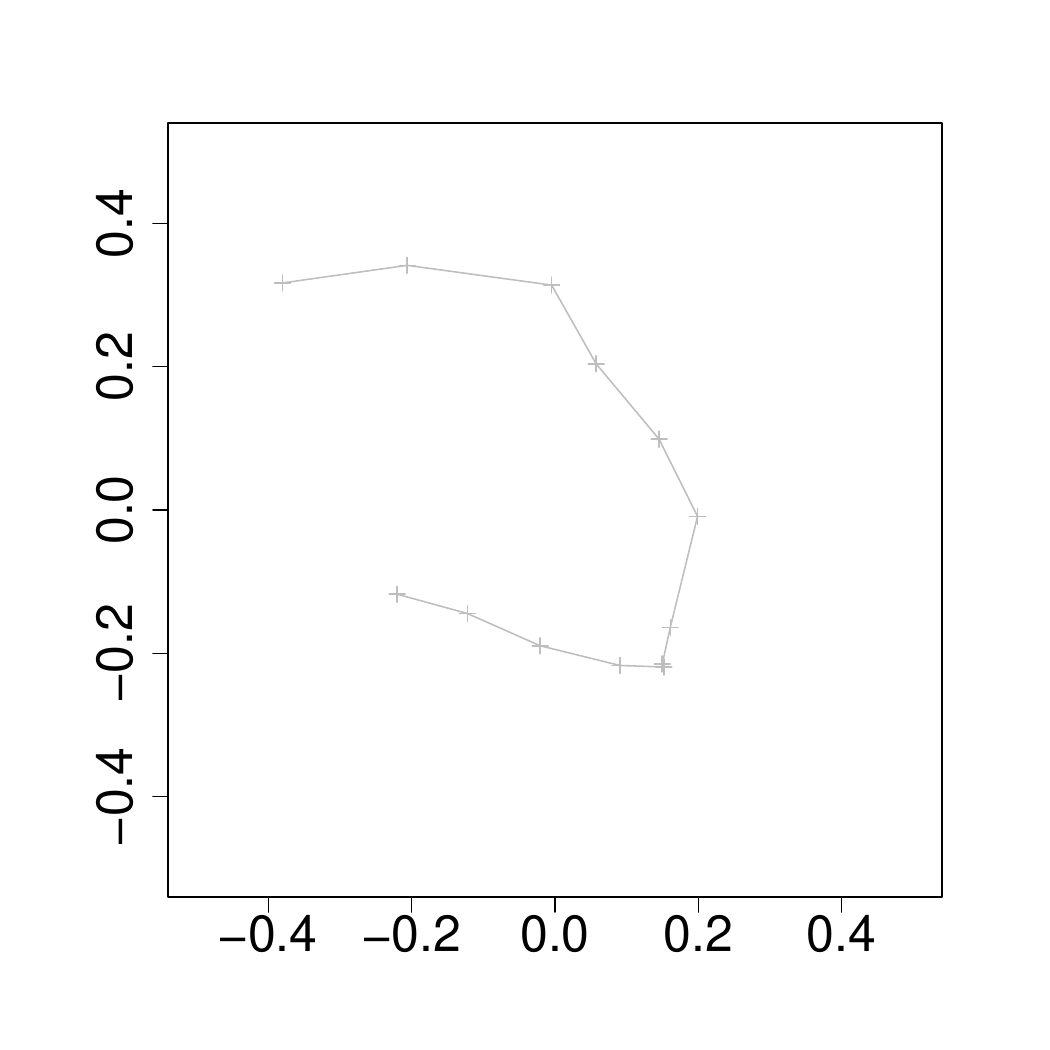}
  \includegraphics[width=0.5in]{PDF/grid-digit3-mean/empty}
  \includegraphics[width=0.5in]{PDF/grid-digit3-mean/empty}
  \includegraphics[width=0.5in]{PDF/grid-digit3-mean/empty}
  \includegraphics[width=0.5in]{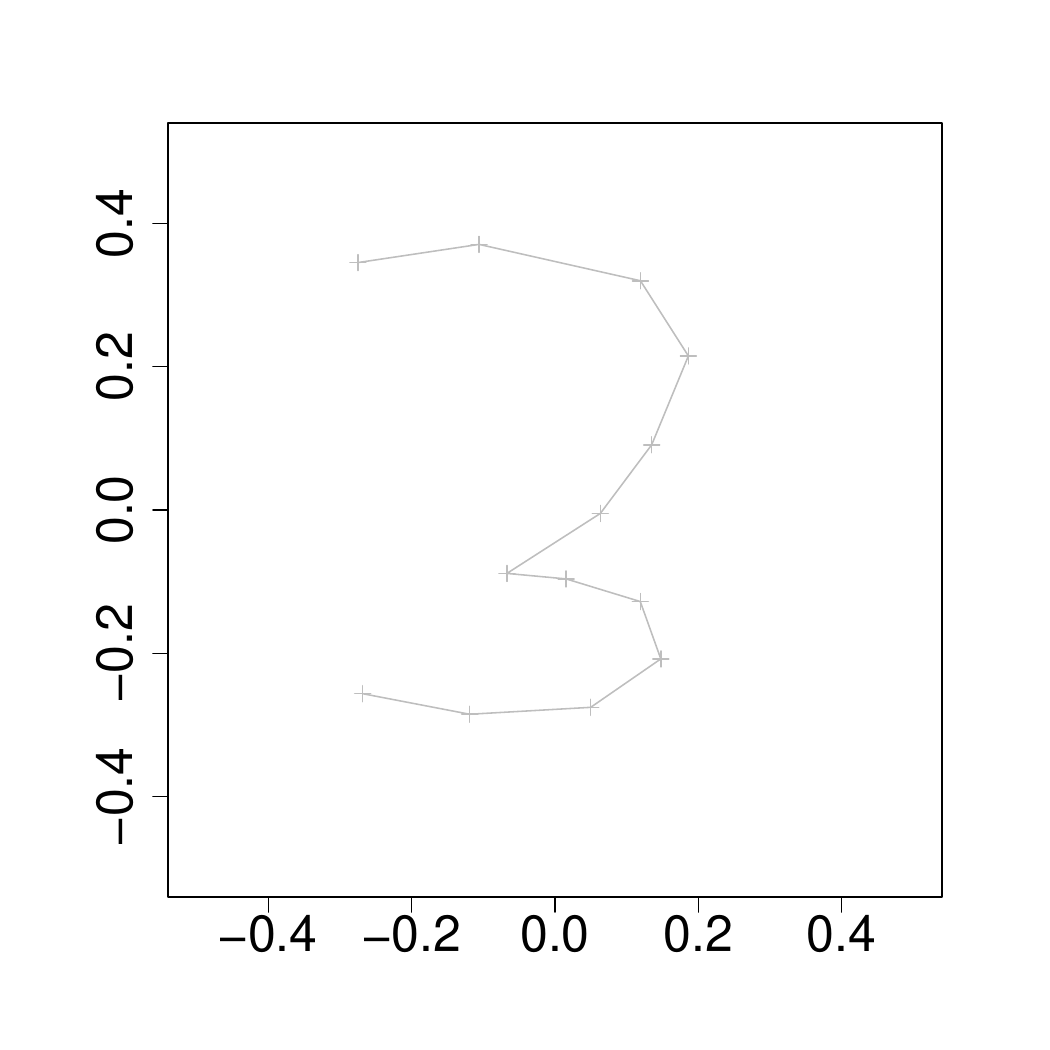}
  \includegraphics[width=0.5in]{PDF/grid-digit3-mean/empty}
  \includegraphics[width=0.5in]{PDF/grid-digit3-mean/empty}
  \includegraphics[width=0.5in]{PDF/grid-digit3-mean/empty}
  \includegraphics[width=0.5in]{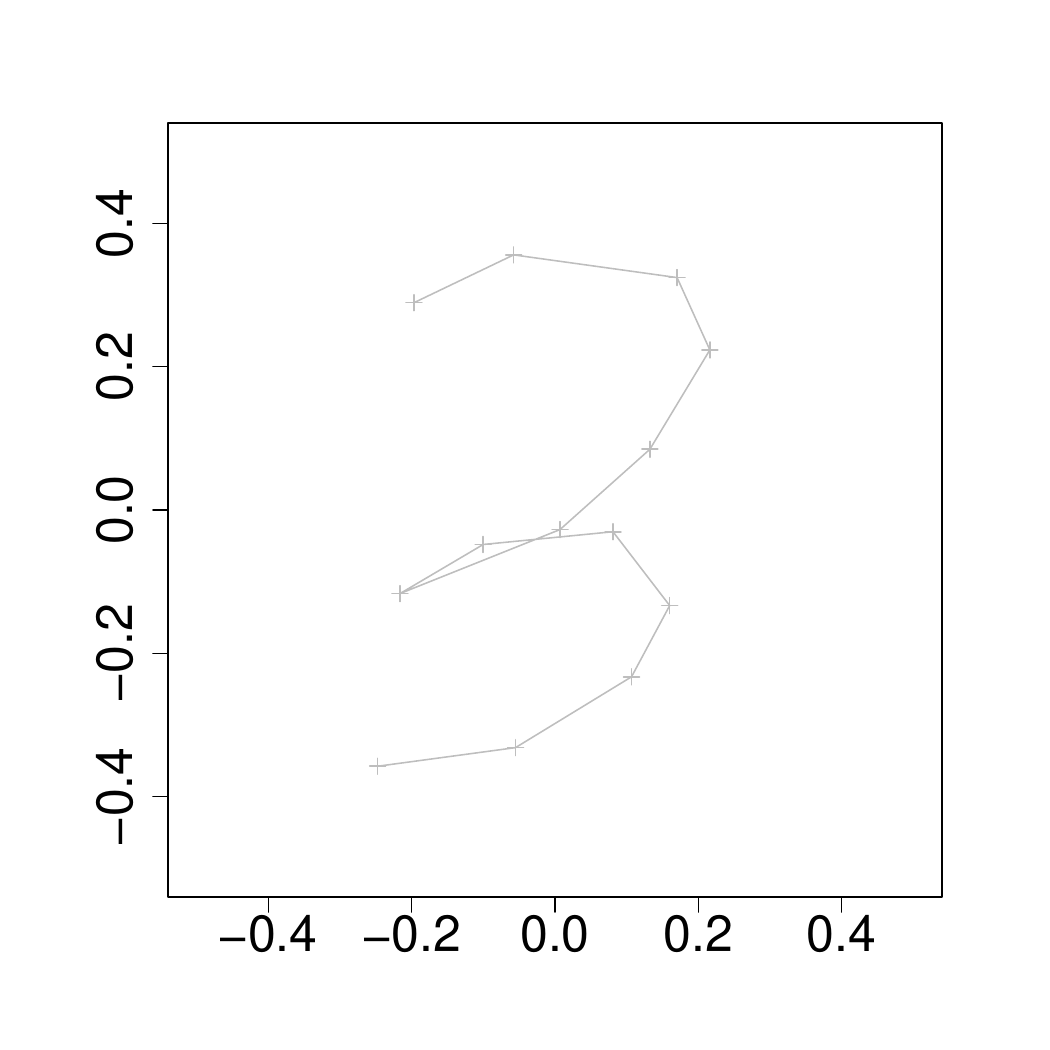}\\
  \caption{Principal sub-manifolds of the handwritten digits data, started from the mean. Among all the figures: the central figure (in blue) is the Fr\'{e}chet mean; the horizontal row contains images recovered from the first principal direction of the sub-manifold; the vertical column is the second principal direction; the main diagonal is the third principal direction; the other diagonal is the fourth principal direction.}
  \label{subman-digit3-mean}
\end{figure}

To illustrate the use of the principal sub-manifold in a concrete example, we consider a handwritten digit ``3'' data. The data, included in the GNU~R package \textit{shapes} \footnote{see \url{https://cran.r-project.org/web/packages/shapes/index.html}}, consists of 13 landmarks of a ``3'' in two dimensions, collected from 30 individuals. For visualization, we find a principal sub-manifold for the data and recover the shape variation of the ``3'' in four principal directions, started at two different shapes of the ``3''. In the first case (see Figure \ref{subman-digit3-mean}), the sub-manifold starts from the Fr\'{e}chet mean of the data. In each principal direction, the flow of images describes the shapes  of the ``3'' moving from one extreme to the other extreme. The horizontal set of the images represents the various shapes of ``3'' recovered from the first principal direction. From there, we can see that the most varying part is the middle part of the ``3''. The parts varying in the second principal direction are mainly the upper and lower parts of the ``3''. Those parts of the ``3'' have exhibited a significant shape change along the two principal directions. Both the main diagonal and the other diagonal show certain degrees of the shape change mostly in the middle part of the ``3'' but in an opposite direction. By observing the fact that there are two seemingly outlying individuals of ``3''s deviating from the rest in the data---the midpoint of the 3 having moved away from the center of the figure---a more sensible center of symmetry should be also considered. As in the second case (Appendix \ref{app:4-digits1}) serves to illustrate the slight effect of having a different choice (center of symmetry) of the starting point on the sub-manifold. However, no significant change in the representation of the sub-manifold is found.

\begin{figure}[ht]
  \centering
  \begin{subfigure}[b]{0.2\textwidth}
    \includegraphics[width=\textwidth]{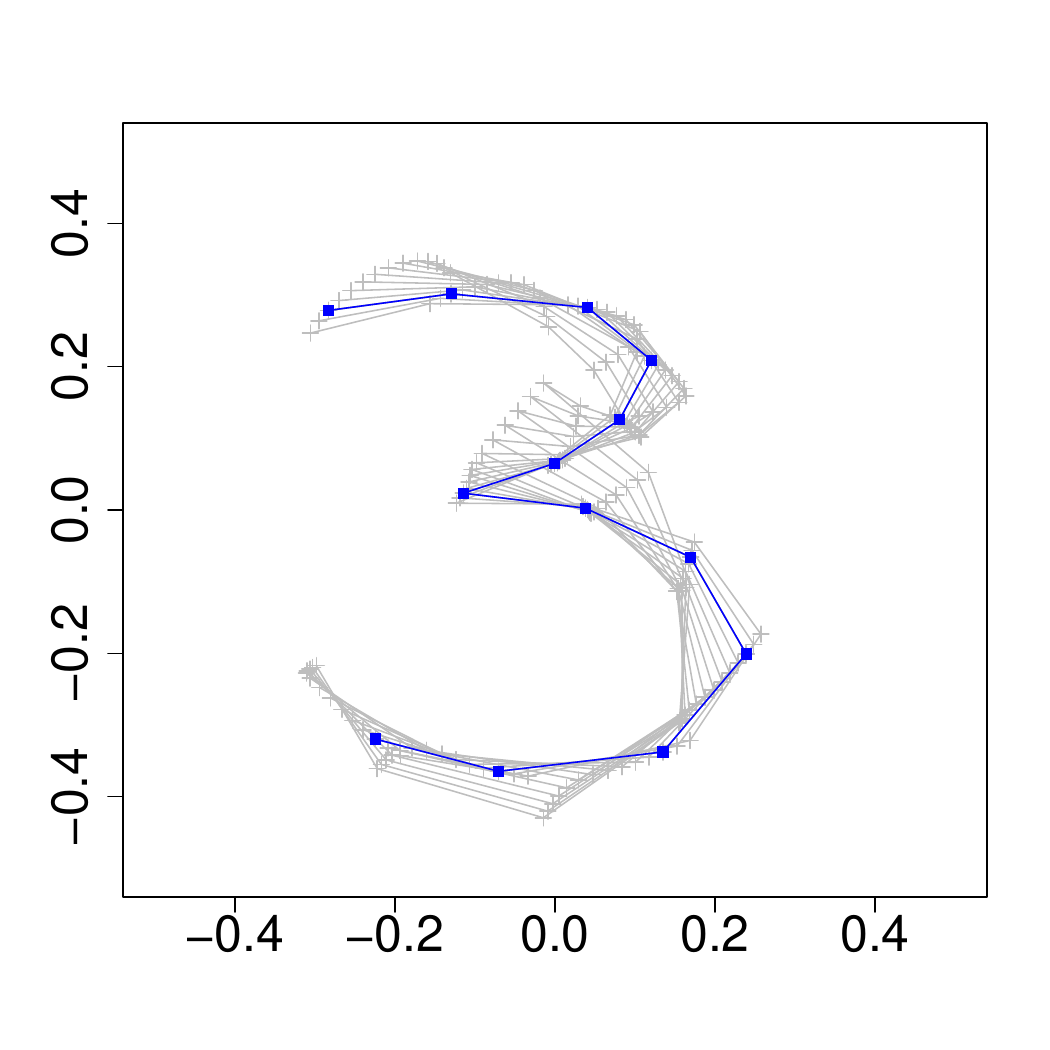}
    \caption{}
  \end{subfigure}
  \begin{subfigure}[b]{0.2\textwidth}
    \includegraphics[width=\textwidth]{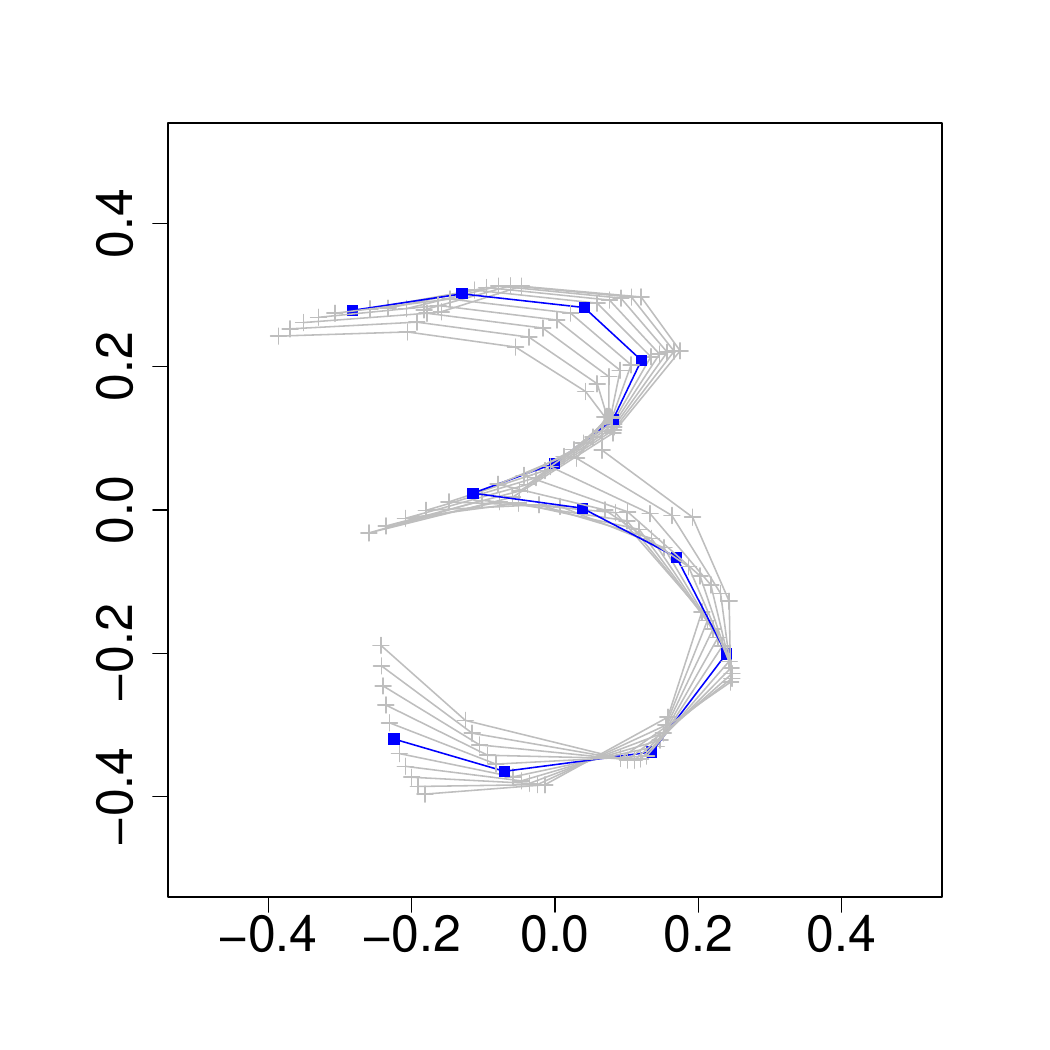}
    \caption{}
  \end{subfigure}
  \begin{subfigure}[b]{0.2\textwidth}
    \includegraphics[width=\textwidth]{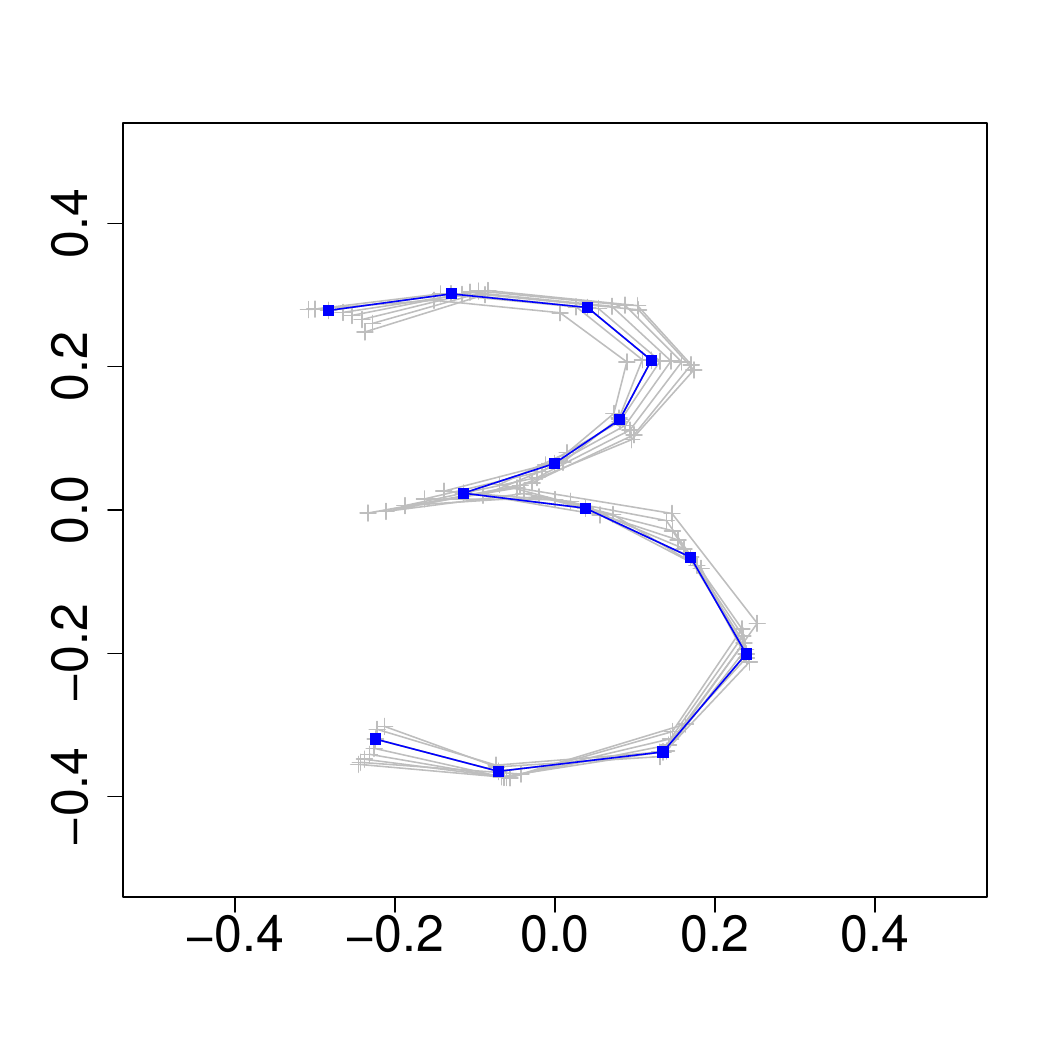}
    \caption{}
  \end{subfigure}
  \begin{subfigure}[b]{0.2\textwidth}
    \includegraphics[width=\textwidth]{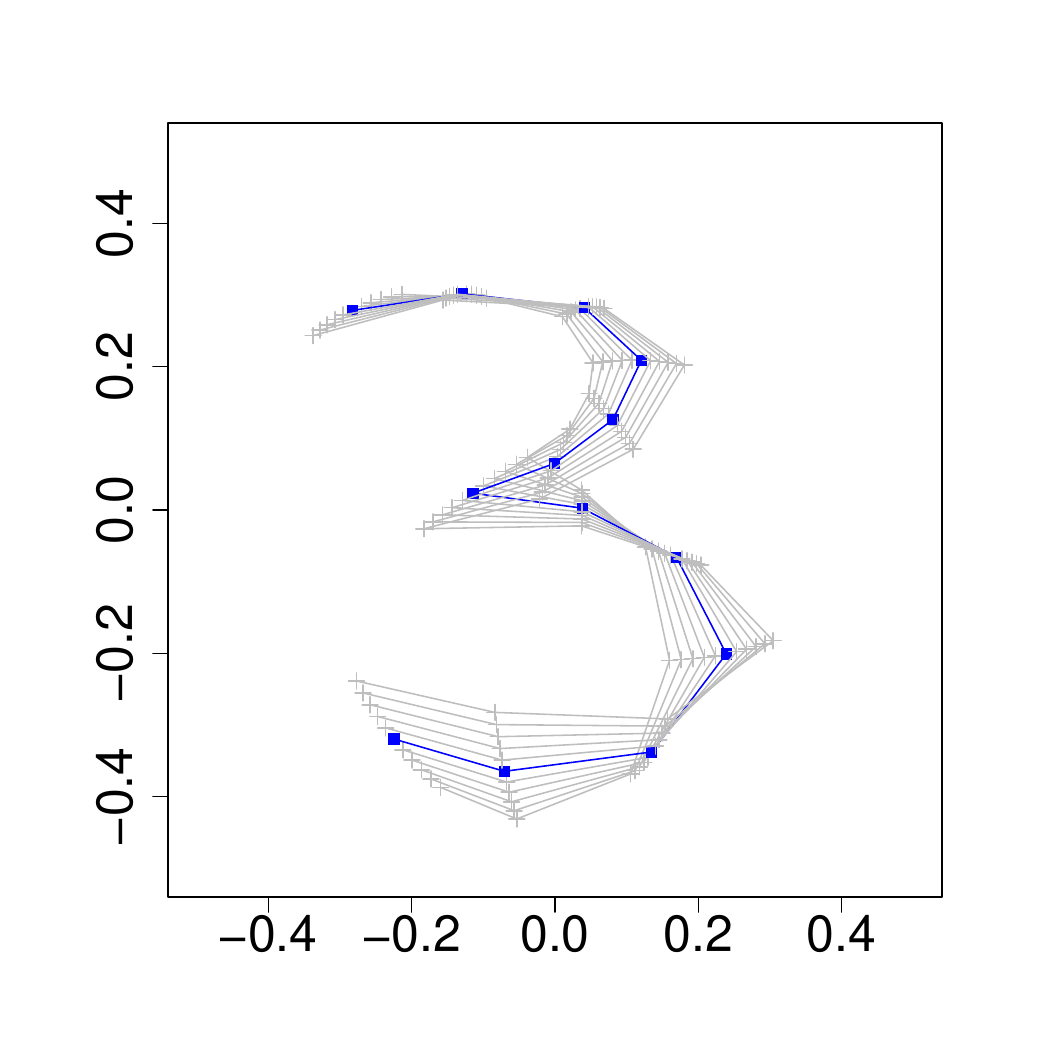}
    \caption{}
  \end{subfigure}
  \caption{Principal sub-manifolds and generalized Procrustes analysis on the handwritten digits data. Figures (a) and (b): the central figure (in blue) is the Procrustes mean; (a) contains images recovered from the first principal direction of the principal sub-manifold; (b) contains images recovered from the first principal component of generalized Procrustes analysis. Figures (c) and (d) give the same information for second principal direction (or principal component) of the principal sub-manifold (or generalized Procrustes analysis).}
  \label{GPA}
\end{figure}

To further understand the shape variation in configuration space, we contrast the results with that from the standard generalized Procrustes analysis (GPA). The profiling of shapes obtained from both methods along different principal directions (or principal components) in Figure \ref{GPA} has suggested quite different patterns. Not only does the variation differ at various parts of the ``3'', but the images of shapes recovered from the principal directions of the sub-manifold reveals an phenomenon of asymmetrical variation around the Procrustes mean, compared to the GPA: the principal sub-manifold tries to explain most of the variation by its first principal direction, while the GPA explains the variation almost equally along its first and second principal components. This is interesting to us, as this information is not available from standard procedures that profiles the images in the configuration space where they obey a standard PCA manner.

\section{Discussion}

The statistical analysis of  data on Riemannian manifold is a very challenging topic and it plays an increasingly important role in real-world problems. Conventional approaches, such as PCA in Euclidean space, are essentially helpful in neither learning the shape of the underlying manifold nor deciding its dimensionality. The main reason for this lies in the fact that those approaches simply do not use the intrinsic Riemannian manifold structures. 

With the aim of proposing a method that allows for finding a nonlinear manifold from the data, we introduced the notation of principal sub-manifold. We showed the importance of estimating a multi-dimensional sub-manifold, and its difference from finding only a one-dimensional curve. The principal sub-manifold was seen to be interpretable as a measure of non-geodesic variation of the data. Based on  a polar coordinate representation, the principal sub-manifold was constructed so that it coordinated with the local data variation. We illustrated that the principal sub-manifold is an extension of the principal flow, in the sense that it depicts a multi-dimensional manifold. When the manifold is linear, the $k$-dimensional principal sub-manifold reduces to the ball spanned by the principal component vectors corresponding to the $k$ largest eigenvalues.

We claim here that by definition, the implemented principal directions might or might not coincide with the principal flows that are defined in \citet{Panaretos2014}, although in practice, they appear to be close to or the same as the principal flows. Under the polar coordinate representation, we observe that the principal directions (these are plotted in green in Figure \ref{illustrate-flow-sub-manifold}) on the principal sub-manifold have presented the main modes of variation.

Regarding the issue of choosing the locality parameter $h$, or equivalently, which scale of the local covariance one should consider, we note that different sub-manifolds in this article have been fitted by choosing different parameters. Still, we refrain from making a strict statement on optimizing the $h$; rather, one should overview a sequence of $h$. Possible routes to approach this question are suggested by the criterion in \citet{Panaretos2014}, which could be adapted, and the scale space perspective \cite{Chaudhuri2000}. Simultaneously, we were able to define the principal sub-manifold to any dimension $k \leq d$, and this may also be seen as the development of a heuristic understanding of a backward stepwise principle of PCA on manifolds: in backward PCA, the best approximating affine subspaces are constructed from the highest dimension to the lowest one, see \citet{backward2010} for the case of spherical subspaces, while in the case of principal sub-manifolds, each ray of the principal sub-manifolds (i.e., the principal directions) corresponds to lower dimension sub-manifolds, compared to the entire sub-manifold.

Last but not least, the formulation of the principal sub-manifold opens the way to the generalization of many other statistical procedures. From the variance reduction perspective, one may categorize our proposed method as one of those competing methods that extend PCA on manifolds but not limited to only using lines or curves. This, potentially, can help us understand the data variation better and improve accuracy. From the classification point of view, this new method has been seen to be a useful tool to study shape changes. In the leaf growth example (Appendix \ref{app:12-digits-princ-var}), we studied the only two main modes of shape variation. This implies that one can extend a classification framework to manifolds. By projecting the new data points to any principal direction of the sub-manifold, one can calculate the distance and extend a classification rule based on all the distances. Surely, a successful classification also depends on 1) the data configuration; 2) how to define the local covariance matrix. If the data on the manifold is not too dense, one might consider using a kernel density estimation. The label information also needs to be considered in the local covariance matrix, in which one would account for both of the between class and within class effects. As this is one of our on-going works, we will investigate it in the future.

\section*{Acknowledgments}
Z. Yao has been supported by Singapore Ministry of Education Tier 2 grant (A-0008520-00-00, A-8001562-00-00) and Tier 1 grant (A-0004809-00-00, A8000987-00-00) at the National University of Singapore. B. Eltzner gratefully acknowledges funding by the DFG~CRC~803 project~Z02 and DFG~CRC~1456 project~B02. B. Eltzner is very grateful to the National University of Singapore for funding a two-month long term visit in February and March 2018 which enabled collaboration on this project. We thank S.~F.~Huckemann and J.~S.~Marron for helpful discussions.

\appendix
\section{Illustration referenced in the Introduction}\label{app:1-illustration}
%\label{Supplementary1}

The geometry plays an essential role for the shape of such a sub-manifold. To illustrate the concept of a principal sub-manifold, consider a set of data points $(x_{i,1}, x_{i,2}, x_{i,3}, x_{i,4})$ on $S^3 \subset \mathbb{R}^4$ , where the coordinates $(x_{i,1}, x_{i,2}, x_{i,3}) \subset \mathbb{R}^3$ of each data point form a rough sinusoid and the fourth coordinate $x_{i,4}$ is added so that every point lies on a sphere. The data is originally constructed by sampling the triplets $(x_{i,1}, x_{i,2}, x_{i,3}) \subset \mathbb{R}^3$ from the distribution% three coordinates ``lifted'' $S$-shape
\begin{equation*}
  \left(
  \begin{matrix}
    x_{i,1}  \\
    x_{i,2} \\
    x_{i,3} \\
  \end{matrix}
  \right)=\left(
  \begin{matrix}
    (i-n/2)/n  \\
    \sin (2 x_{i,1})/6+ 32U \\
    1+ 1/100 V \\
  \end{matrix}
  \right), \quad 1 \leq i \leq n
\end{equation*}
where $U$ and $V$ are independent normal variable $N(0, 1/10)$ and $N(0, 1/100)$.  To lift each point to $\mathbb{R}^4$, we add the fourth coordinate so that every point lies on a sphere satisfying
%such that the data points arrange themselves in the way as a part along a rough sinusoid in $\mathbb{R}^3$. 
\begin{equation*} 
  x_{i,1}^2+x_{i,2}^2+x_{i,3}^2+x_{i,4}^2=C,
\end{equation*}
where a shifting parameter $C$ (e.g., 0.45) is chosen such that 
\begin{equation*}
  \sqrt{C-x_{i,1}^2-x_{i,2}^2-x_{i,3}^2} \geq 0.
\end{equation*}
We still need to normalize the data to guarantee that the data are in $S^3$.

\begin{figure}[ht!] 
  \centering
  \begin{subfigure}[b]{0.35\textwidth}
    \centering
    \includegraphics[width=1\linewidth]{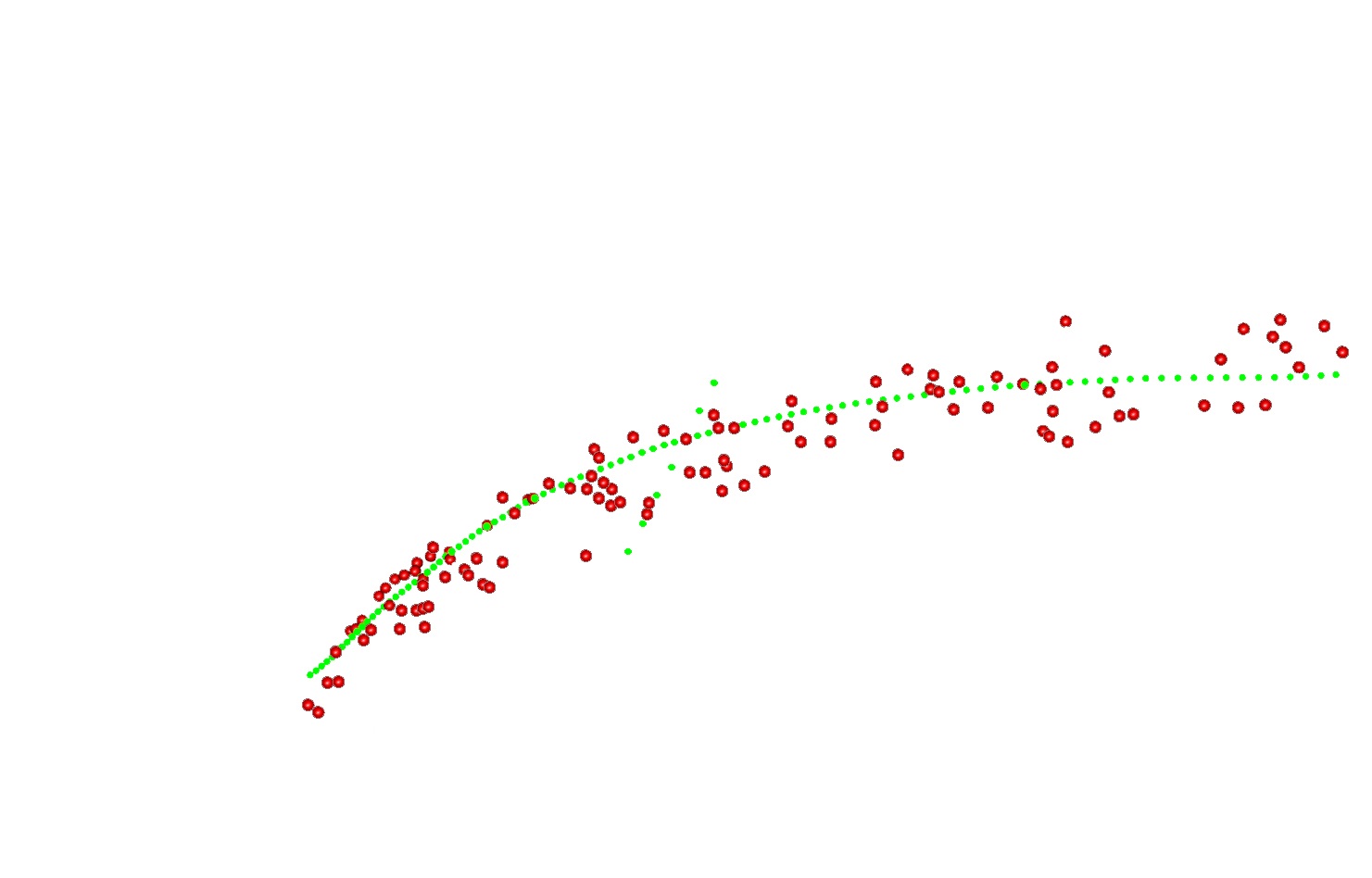}
    \caption{}
  \end{subfigure}%
  \hspace{0.15 in}
  \begin{subfigure}[b]{0.35\textwidth}
    \centering
    \includegraphics[width=1\linewidth]{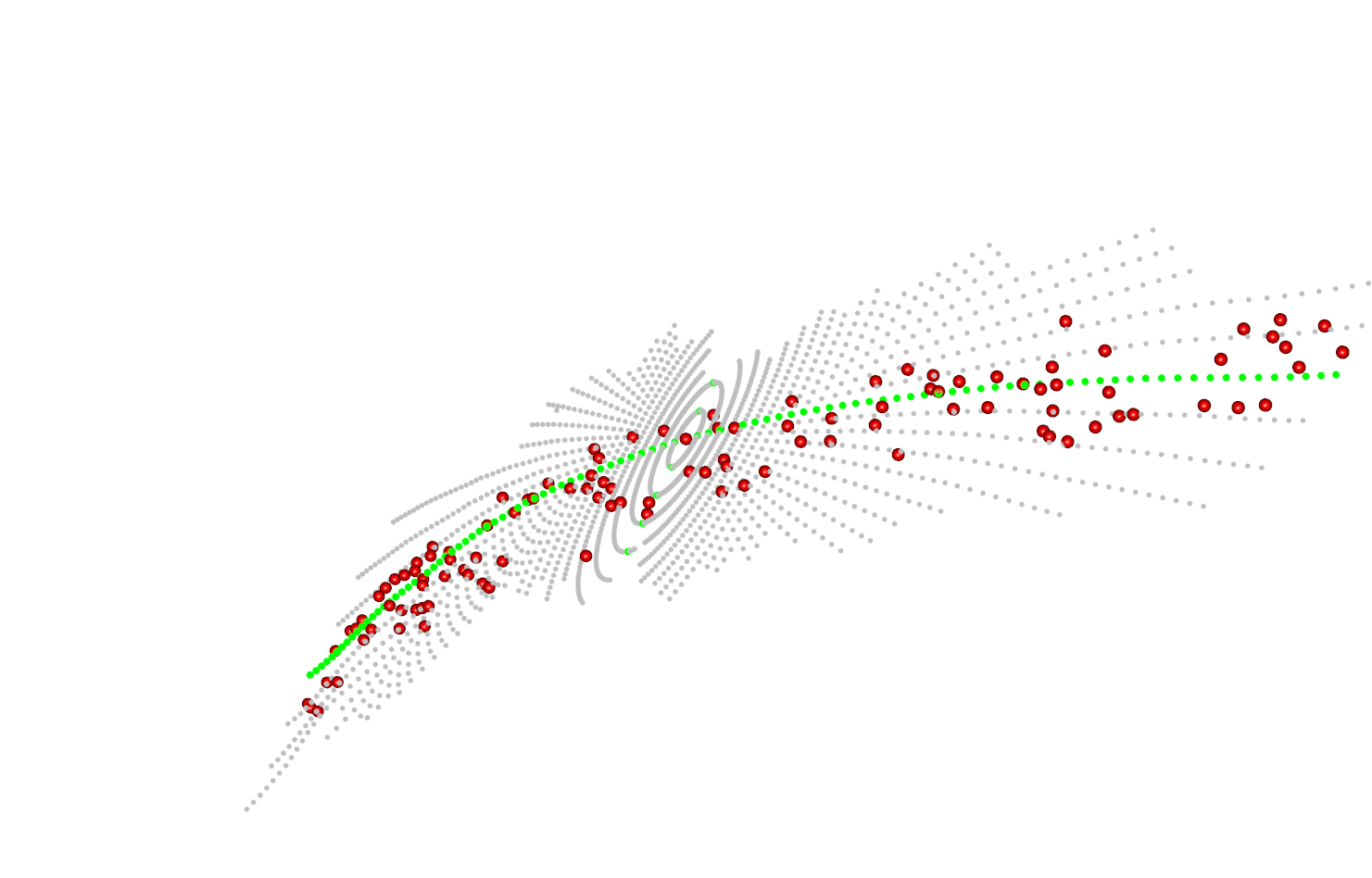}
    \caption{}
  \end{subfigure}% 
  \caption{Visualization of the projected two-dimensional sub-manifold for data on $S^3$. (a) Principal flow; (b) Principal sub-manifold. The data points are labeled in red, with the first and second principal flows (in green) going through the starting point. The sub-manifold (in gray) are the estimated principal sub-manifold.  For visualization purpose, the sub-manifold, the first and second principal direction and the data points have been projected to the first three eigenvectors of the covariance matrix at the starting point.}
  \label{illustrate-flow-submanifold}
\end{figure}

The ambient manifold determines some of the inherited geometry of the projected sub-manifold. In this case, the sub-manifold carries an $S$-pattern that mainly stems from the rough sinusoid of the first three coordinates. Figure \ref{illustrate-flow-submanifold} shows the data, the superimposed principal flow and the estimated principal sub-manifold. The starting point (in black) is the center of the data points (in red). The two green curves (Figure \ref{illustrate-flow-submanifold}(a)) are the first and second principal flows, while the estimated principal sub-manifold (Figure \ref{illustrate-flow-submanifold}(b)) is highlighted in gray, contrasting the two green principal flows lying on the surface of the estimated principal sub-manifold. All of them are projected to the first three eigenvectors of the covariance matrix at the starting point. The surface is able to bend wherever the curvature of the manifold changes rapidly. The first principal direction is towards the direction of maximum variance of the data points, while the surface extends in more directions that automatically account for more variance of the data points.

\section{Principal flows}\label{app:2-princ-flows}

Since the concept of a principal sub-manifold is strongly inspired by the principal flow, see \citet{Panaretos2014}, we will review this concept here. The principal flow yields a one-dimensional, not necessarily geodesic, approximation of a data set $\left\{x_1, \cdots, x_n\right\} \subset \mathcal{M}$. We parameterize the one dimensional sub-manifolds as a set of unit speed curves
\begin{multline} \label{candidate}
  \textnormal{SubM}(A, 1, v, \mathcal{M})= \Big\{\gamma:[0,r] \rightarrow \mathcal{M}, \gamma \in C^2(\mathcal{M}), \gamma(s) \neq \gamma(s') \mbox{ for } s \neq s',\\
  \gamma(0)=A, \dot{\gamma}(0)=v, \ell(\gamma[0,t])=t \mbox{  for all  } 0 \leq t \leq r \leq 1 \Big\},
\end{multline}
where $\gamma(0)=A$ and $\dot{\gamma}(0)=v$ are initial conditions for $\gamma$ and $\ell(\gamma)$ is the length of $\gamma$. The starting point $A$ can be chosen as the Fr\'{e}chet sample mean $\bar{x}$ or any other point of interest. Then $\textnormal{SubM}(A, 1, v, \mathcal{M})$ contains all smooth curves of length less than 1 with given initial speed and starting point.

The principal flow is defined by two curves
\begin{align}
  \gamma^+ &= \arg \sup_{\gamma \in \textnormal{SubM} \left(A, 1, v_1, \mathcal{M}\right)} \int_{0}^{\ell(\gamma)} \left\langle \dot{\gamma}(t), e_1(\gamma(t))\right\rangle dt \label{flow+}\\
  \gamma^- &= \arg \inf_{\gamma \in \textnormal{SubM} \left(A, 1, v_2, \mathcal{M}\right)} \int_{0}^{\ell(\gamma)} \left\langle \dot{\gamma}(t), e_1(\gamma(t))\right\rangle dt \label{flow-}
\end{align}
where $v_1=e_1(\gamma(t))$ , $v_2=-v_1$, $e_1(\gamma(t))$ is the first eigenvector of the covariance matrix $\Sigma_{\gamma(t)}$ at $\gamma(t)$. The integral for $\gamma^-$ is negative, therefore the infimum appears in its definition. At each point of $\gamma$, $\dot{\gamma}(t)$ is maximally compatible to the eigenvector to the largest eigenvalue of the local covariance matrix at scale $h$
\begin{align}\label{covar}
  \Sigma_{h,\gamma(t)}=\frac{1}{\sum_{i}\kappa_h(x_i,\gamma(t))} \sum_{i=1}^{n} \mbox{{\bf log}}_{\gamma(t)}(x_i) \otimes  \mbox{{\bf log}}_{\gamma(t)}(x_i) \kappa_h(x_i,\gamma(t)) \, ,
\end{align}
where $y \otimes y:=y y^{\tiny{\mbox{T}}}$ and $\kappa_h(x,\gamma(t))=K(h^{-1} d_\mathcal{M}(x,\gamma(t))$ with a smooth non-increasing kernel $K$ on $[0, \infty]$. All the above definitions are under the assumption that the first and second eigenvalues of $\Sigma_{h,\gamma(t)}$ are distinct.

Principal flows achieve higher data fidelity than geodesics and are more flexible than other curve-fitting approaches in trading off between data fidelity and avoiding too high curvature of the curve. The question whether non-linear variation can be captured in higher dimension has been discussed (for other assumptions on the embedding data space) under the names of \emph{principal curves} or \emph{principal surfaces} by \cite{hastie1989}. Note that principal surfaces are the extension of principal curves to higher dimensions in Euclidean space, restricted to a two-dimensional scenario. The present work is connected to both principal flow and principal surfaces, using a more general setting than \cite{hastie1989}.

\section{Proofs of Theorems 3 and 4}\label{app:3-proofs-thms-3+4}
%\label{Supplementary-asymptotics}

\begin{proof}[Proof of Theorem 3]
  First, since $\widehat{\mathcal{N}}_n$ is $C^2$, we can use a Taylor expansion around $A_n$ to see that for any $x \in \widehat{\mathcal{N}}_n$ we have some $w_x \in \widehat{W}_n(A_n)$ with $|w_x|=1$ such that
  \begin{align*}
    x = A_n + |A_n - x| w_x + \mathcal{O}(L_n^2) \, .
  \end{align*}
  Next, we note that since $W$ is $C^2$, we have a constant $C > 0$, such that
  \begin{align*}
    \angle \left( \widehat{W}_n(A_n), W(A)\right) \le&~ \angle \left( \widehat{W}_n(A_n), W(A_n)\right) + C d_{\mathcal{M}} (A_n, A)\\
    &+ \mathcal{O}(d_{\mathcal{M}} (A_n, A)^2) \, .
  \end{align*}
  Now, define the vector $w_y \in W(A)$ by
  \begin{align*}
    w_y := \mathop{\textnormal{argmin}}_{w_y \in W(A), |w_y|=1} \angle \left( w_x, w_y\right)
  \end{align*}
  and let $\gamma_{A,w_y}$ be the integral curve of $W$ starting at $A$ with tangent vector $w_y$ which stays closest to the straight line $c(t) := A + w_y t$. Then define $y := \gamma_{A,w_y}(|A_n - x|) \in \mathcal{N}$. By construction, we now have
  \begin{align*}
    y = A + |A_n - x| w_y + \mathcal{O}(L_n^2) \, .
  \end{align*}
  Thus we get the following bound
  \begin{align*}
    d_{\mathcal{M}} \left( x, y\right) \le&~ d_{\mathcal{M}} (A_n, A) + |A_n - x| \angle \left( w_x, w_y\right) + \mathcal{O}(L_n^2)\\
    \le&~ d_{\mathcal{M}} (A_n, A) + |A_n - x| \angle \left( \widehat{W}_n(A_n), W(A)\right) + \mathcal{O}(L_n^2)\\
    \le&~ d_{\mathcal{M}} (A_n, A) + |A_n - x| \angle \left( \widehat{W}_n(A_n), W(A_n)\right)\\
    &+ C |A_n - x| d_{\mathcal{M}} (A_n, A) + |A_n - x| \mathcal{O}(d_{\mathcal{M}} (A_n, A)^2) + \mathcal{O}(L_n^2)\\
    \le&~ (1 + C L_n) d_{\mathcal{M}} (A_n, A) + L_n \angle \left( \widehat{W}_n(A_n), W(A_n)\right)\\
    &+ \mathcal{O}(L_n d_{\mathcal{M}} (A_n, A)^2) + \mathcal{O}(L_n^2)\, .
  \end{align*}
  Using this bound, we see that
  \begin{align*}
    n^{1/2} d_{\mathcal{M}} \left( x, y\right) \le&~ (1 + C L_n) n^{1/2} d_{\mathcal{M}} (A_n, A) + L_n n^{1/2} \angle \left( \widehat{W}_n(A_n), W(A_n)\right)\\
    &+ \mathcal{O}(n^{1/2} L_n d_{\mathcal{M}} (A_n, A)^2) + \mathcal{O}(n^{1/2} L_n^2)\, .
  \end{align*}
  Now, using the assumptions $n^{1/2} d_{\mathcal{M}} (A_n, A) \to 0$ and $n^{1/4} L_n \to 0$, we get
  \begin{align*}
    n^{1/2} d_{\mathcal{M}} \left( x, y\right) \to&~  L_n n^{1/2} \angle \left( \widehat{W}_n(A_n), W(A_n)\right) \, ,
  \end{align*}
  which goes to $0$ in probability due to Theorem 2 as desired.
\end{proof}

\begin{proof}[Proof of Theorem 4]
  There are vectors $w_x \in W_n(A)$ with $|w_x|=1$ and $w_y \in W(A)$ with $|w_y|=1$ such that
  \begin{align*}
    x =&~ A + |A - x| w_x + \mathcal{O}(L_n^2) & y =&~ A + |A - y| w_y + \mathcal{O}(L_n^2) \, .
  \end{align*}
  In consequence,
  \begin{align*}
    d_{\mathcal{M}} \left( x, y\right) \le&~ \max(|A-x|, |A-y|) \angle \left( w_x, w_y\right) + \mathcal{O}(L_n^2)\\
    \le&~ L_n \angle \left( W_n(A), W(A)\right) + \mathcal{O}(L_n^2) \, ,
  \end{align*}
  so it remains to show that
  \begin{align*}
    n^{1/4} \angle \left( W_n(A), W(A)\right) \stackrel{\mathbb{P}}{\to} 0 \, .
  \end{align*}
  Because of $\sigma_n / h_n \to 0$ and $h_n \to 0$, there is a constant $C_1$ such that for $v \in T_A\mathcal{N}_0$ with $|v|=1$ one gets $v^T \Sigma_{n,A} v \to C_1 h_n^2$ and a constant $C_2$ such that for $v \in N_A\mathcal{N}_0$ normal to $\mathcal{N}_0$ with $|v|=1$
  Because of $\sigma_n / h_n \to 0$ and $h_n \to 0$, one gets $v^T \Sigma_{n,A} v \to h_n$ for $v \in T_x\mathcal{N}_0$ and $|v|=1$ while $v^T \Sigma_{n,A} v \to \sigma_n$ for $v \in N_x\mathcal{N}_0$ normal to $\mathcal{N}_0$ and $|v|=1$ one gets $v^T \Sigma_{n,A} v \to C_2 \sigma_n^2$.
  
  As a result, the eigenvectors of $\Sigma_{n,A}$ corresponding to the $k$ largest eigenvalues approach the tangent space $T_x\mathcal{N}_0$ with the maximal angle converging in probability
  \begin{align*}
    \angle \left( W_n(A), W(A)\right) =&~ \arctan \left( \frac{C_2 \sigma_n}{C_1 h_n} \right) \mathcal{O}_p (1)\\
    \Rightarrow \angle \left( W_n(A), W(A)\right) =&~ \left( \frac{C_2 \sigma_n}{C_1 h_n} + \mathcal{O} \left( \frac{\sigma_n^2}{h_n^2} \right) \right) \mathcal{O}_p (1) \, .
  \end{align*}
  The claim then follows from $n^{1/4} \sigma_n / h_n \to 0$.
\end{proof}

\newpage

\section{Principal sub-manifolds of the handwritten digits data, started from the center of symmetry}\label{app:4-digits1}
%\label{Supplementary2}

\begin{figure}[ht!]
  \centering
  \includegraphics[width=0.5in]{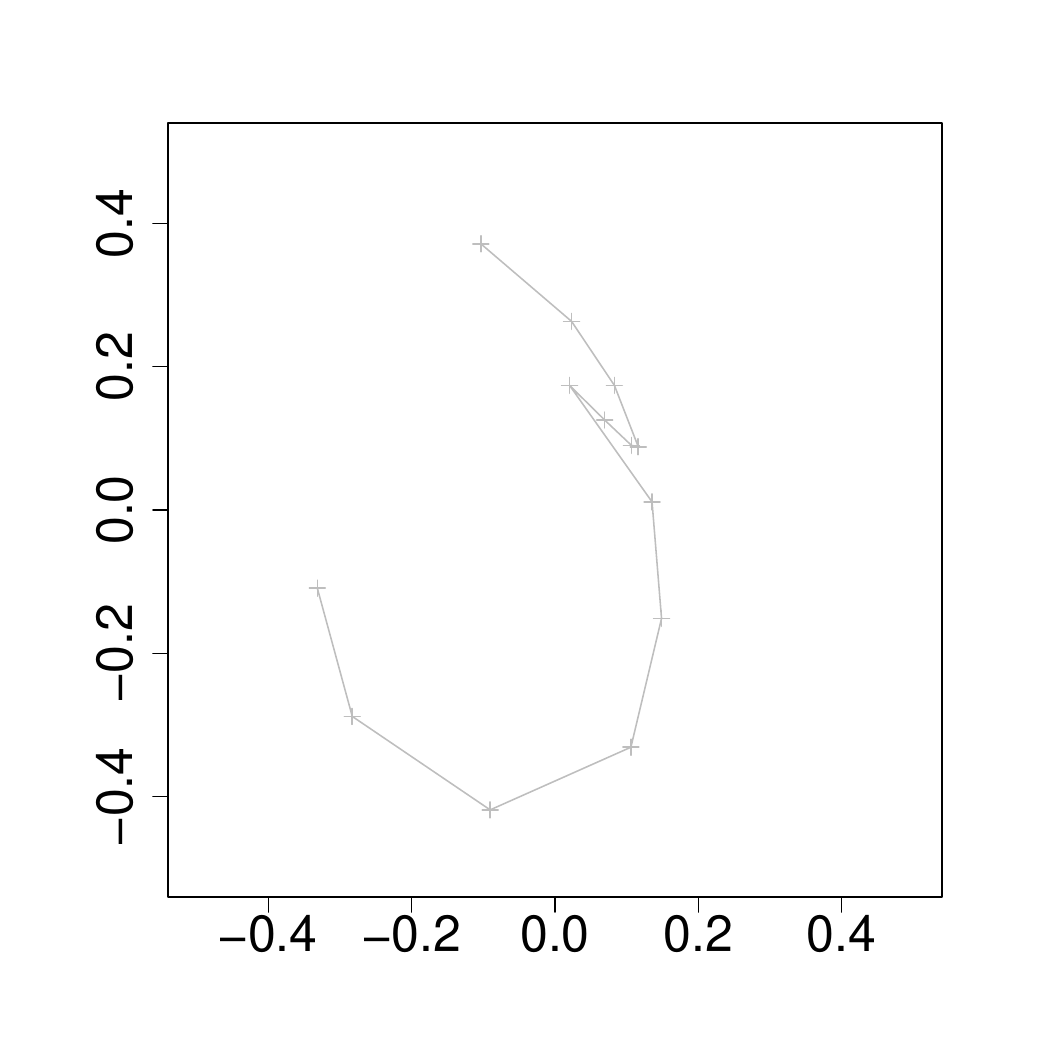}
  \includegraphics[width=0.5in]{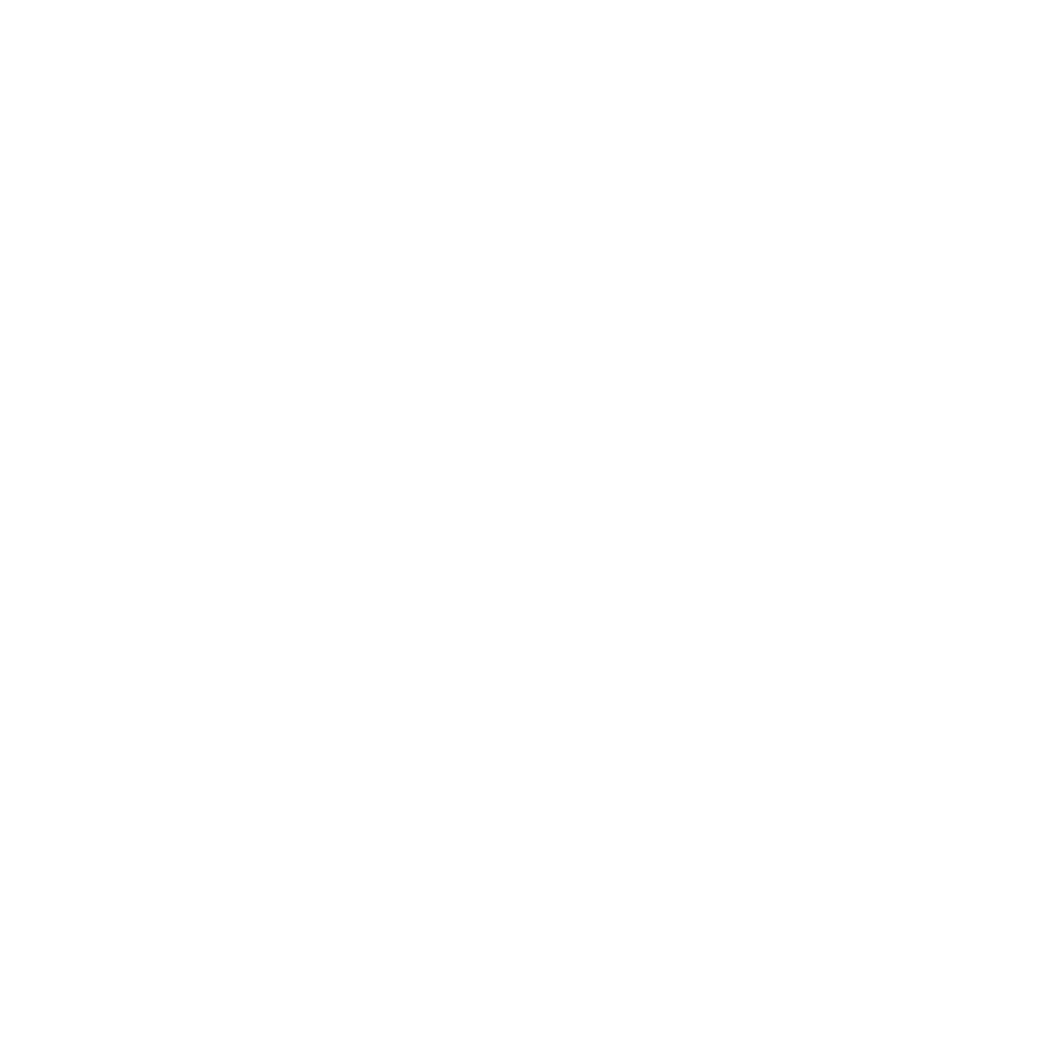}
  \includegraphics[width=0.5in]{PDF/grid-digit3/empty}
  \includegraphics[width=0.5in]{PDF/grid-digit3/empty}
  \includegraphics[width=0.5in]{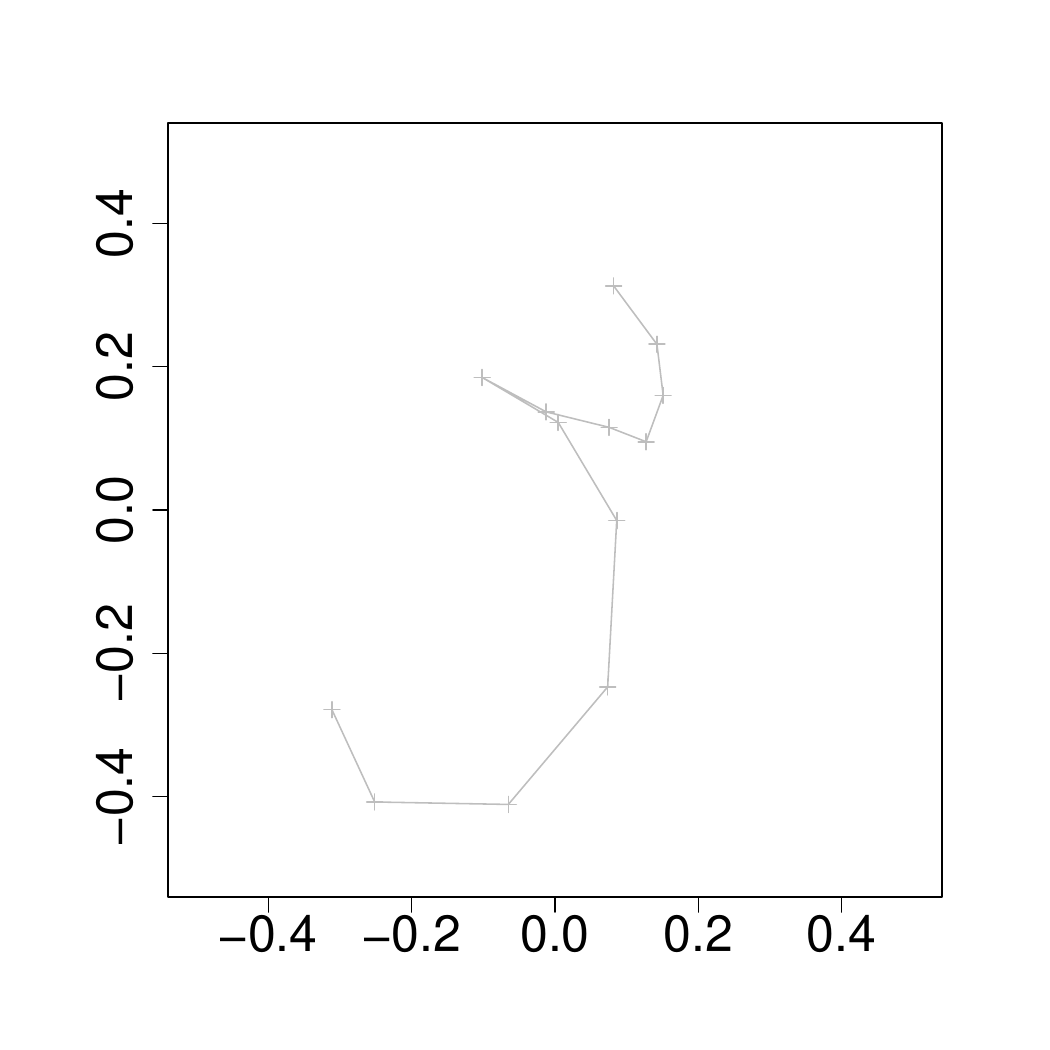}
  \includegraphics[width=0.5in]{PDF/grid-digit3/empty}
  \includegraphics[width=0.5in]{PDF/grid-digit3/empty}
  \includegraphics[width=0.5in]{PDF/grid-digit3/empty}
  \includegraphics[width=0.5in]{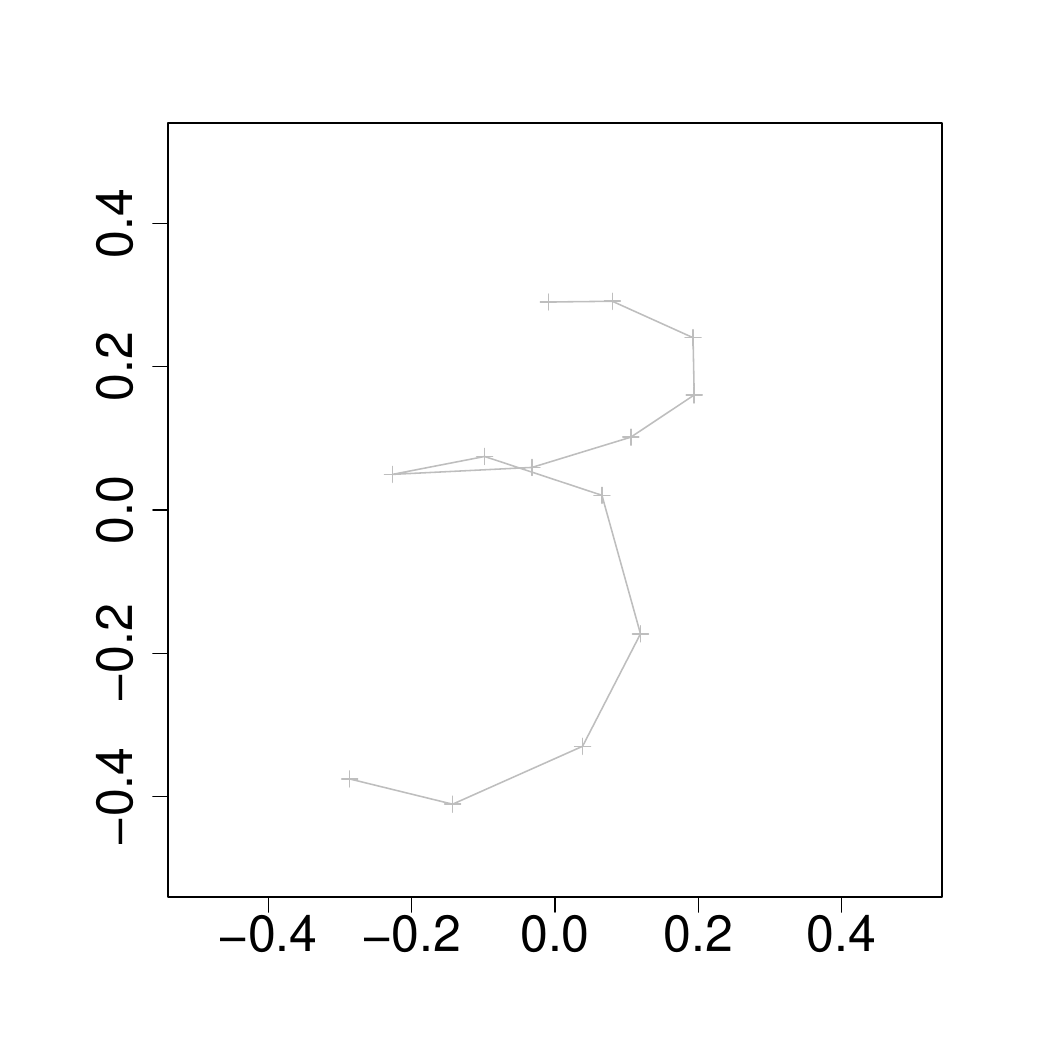}\\
  \includegraphics[width=0.5in]{PDF/grid-digit3/empty}
  \includegraphics[width=0.5in]{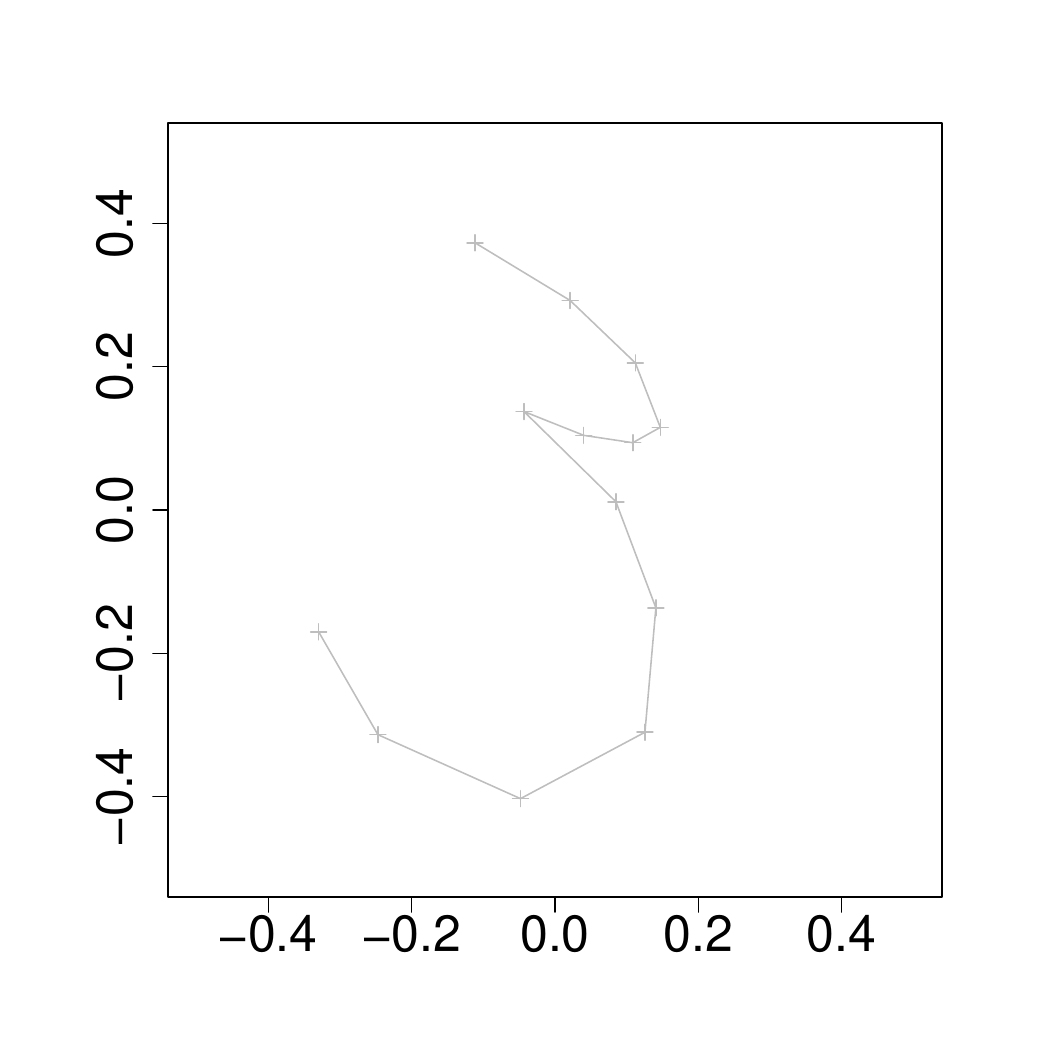}
  \includegraphics[width=0.5in]{PDF/grid-digit3/empty}
  \includegraphics[width=0.5in]{PDF/grid-digit3/empty}
  \includegraphics[width=0.5in]{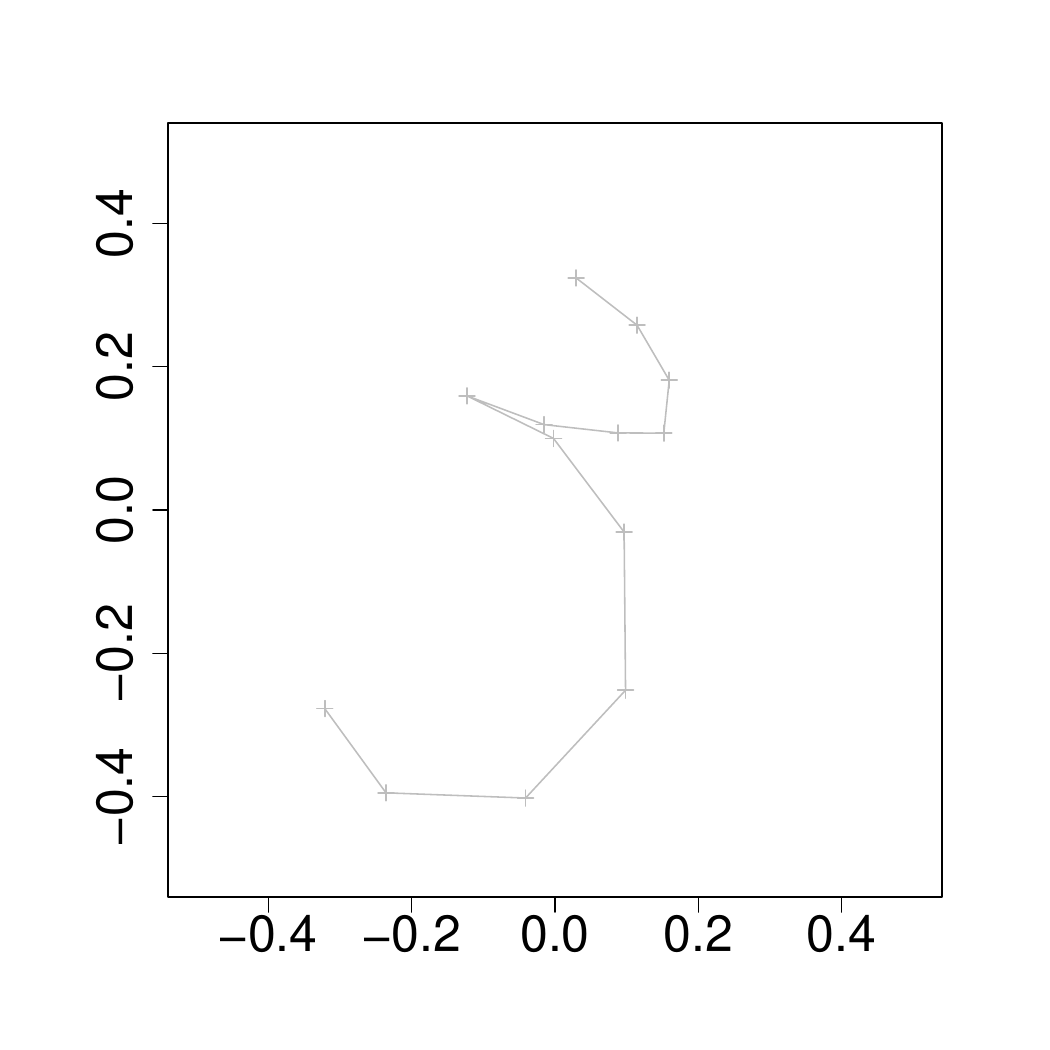}
  \includegraphics[width=0.5in]{PDF/grid-digit3/empty}
  \includegraphics[width=0.5in]{PDF/grid-digit3/empty}
  \includegraphics[width=0.5in]{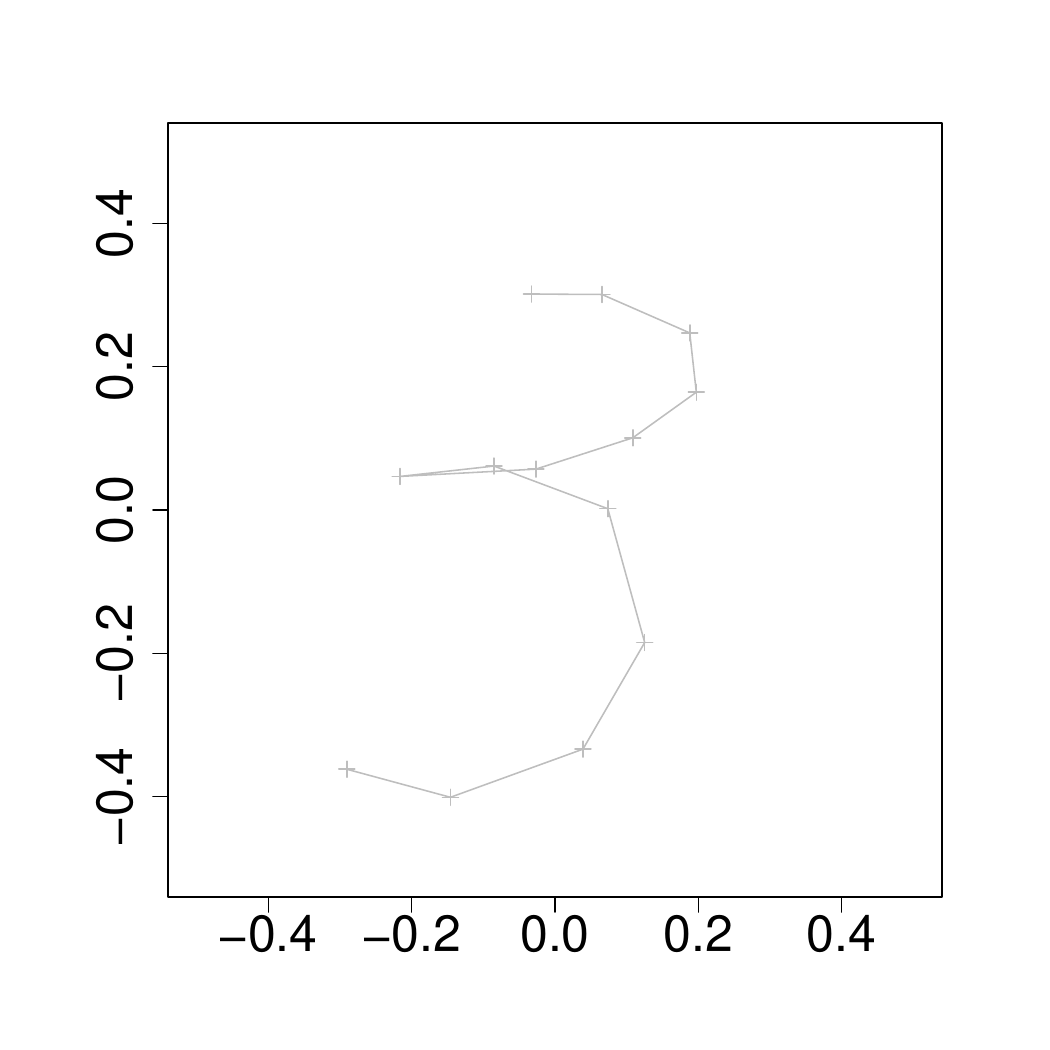}
  \includegraphics[width=0.5in]{PDF/grid-digit3/empty}\\
  \includegraphics[width=0.5in]{PDF/grid-digit3/empty}
  \includegraphics[width=0.5in]{PDF/grid-digit3/empty}
  \includegraphics[width=0.5in]{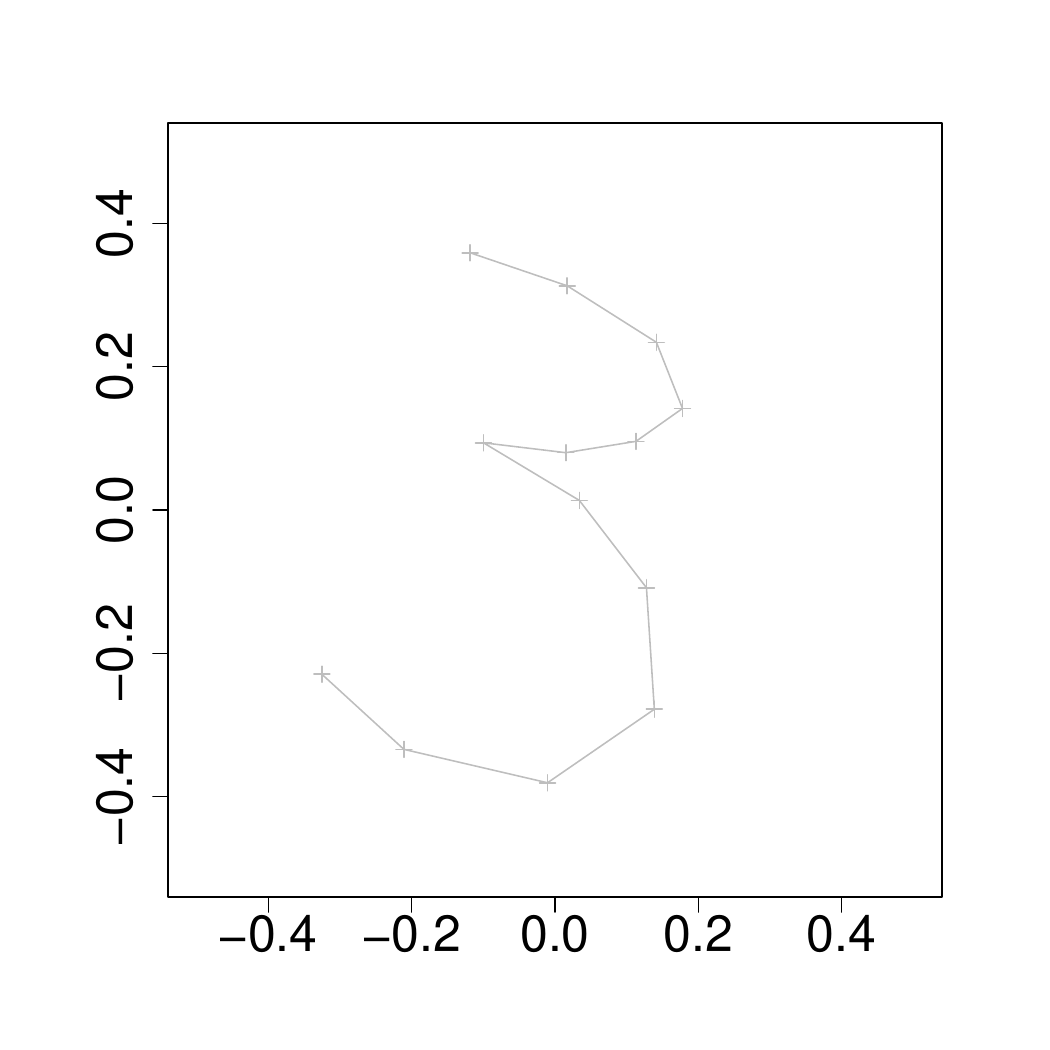}
  \includegraphics[width=0.5in]{PDF/grid-digit3/empty}
  \includegraphics[width=0.5in]{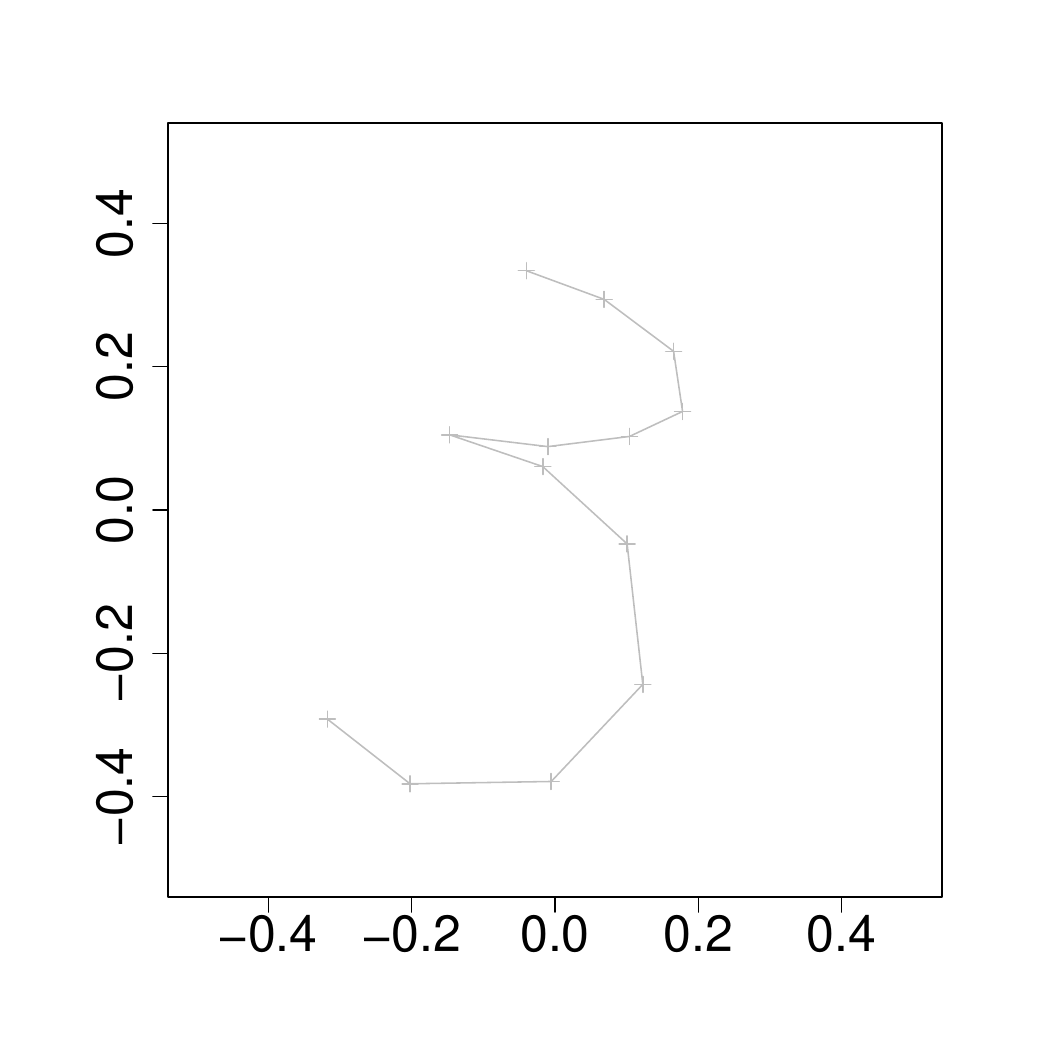}
  \includegraphics[width=0.5in]{PDF/grid-digit3/empty}
  \includegraphics[width=0.5in]{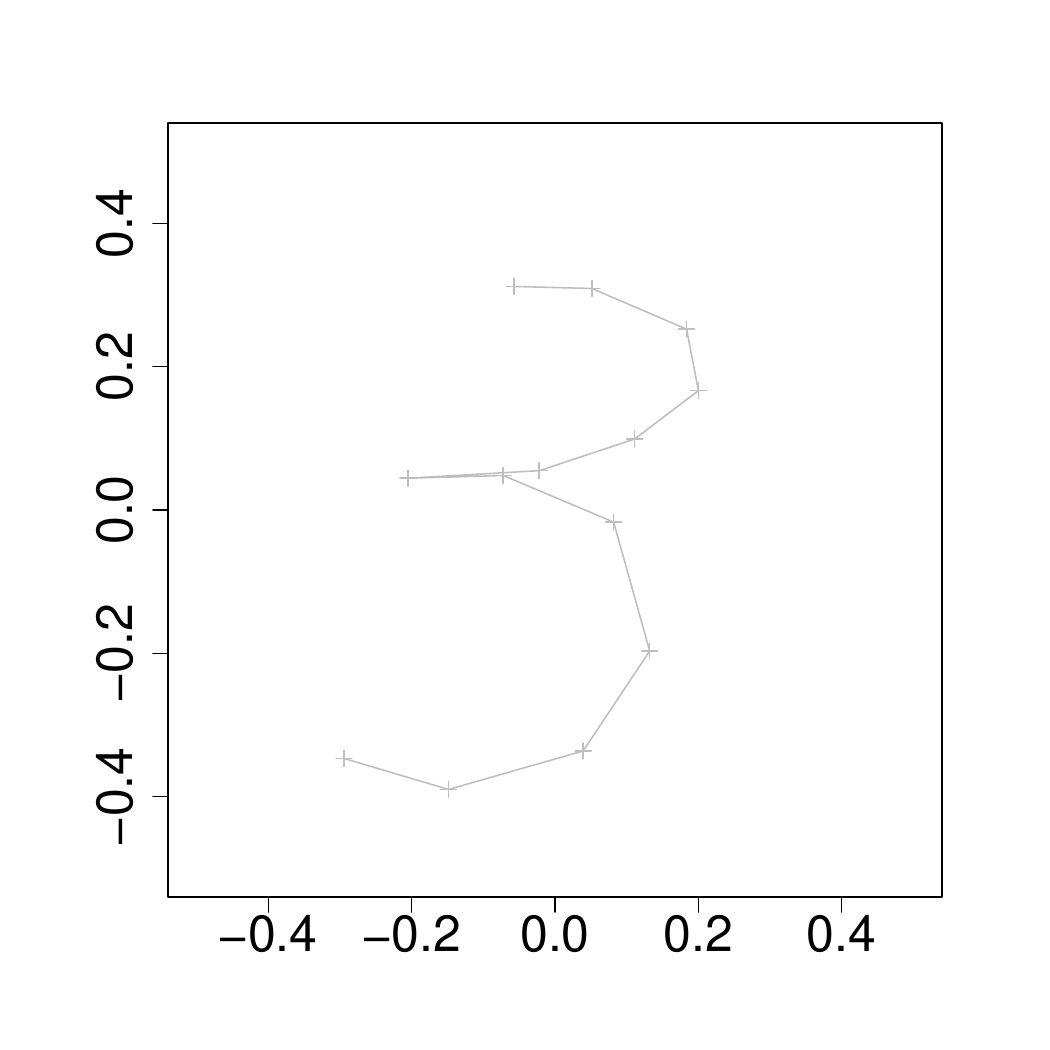}
  \includegraphics[width=0.5in]{PDF/grid-digit3/empty}
  \includegraphics[width=0.5in]{PDF/grid-digit3/empty}\\
  \includegraphics[width=0.5in]{PDF/grid-digit3/empty}
  \includegraphics[width=0.5in]{PDF/grid-digit3/empty}
  \includegraphics[width=0.5in]{PDF/grid-digit3/empty}
  \includegraphics[width=0.5in]{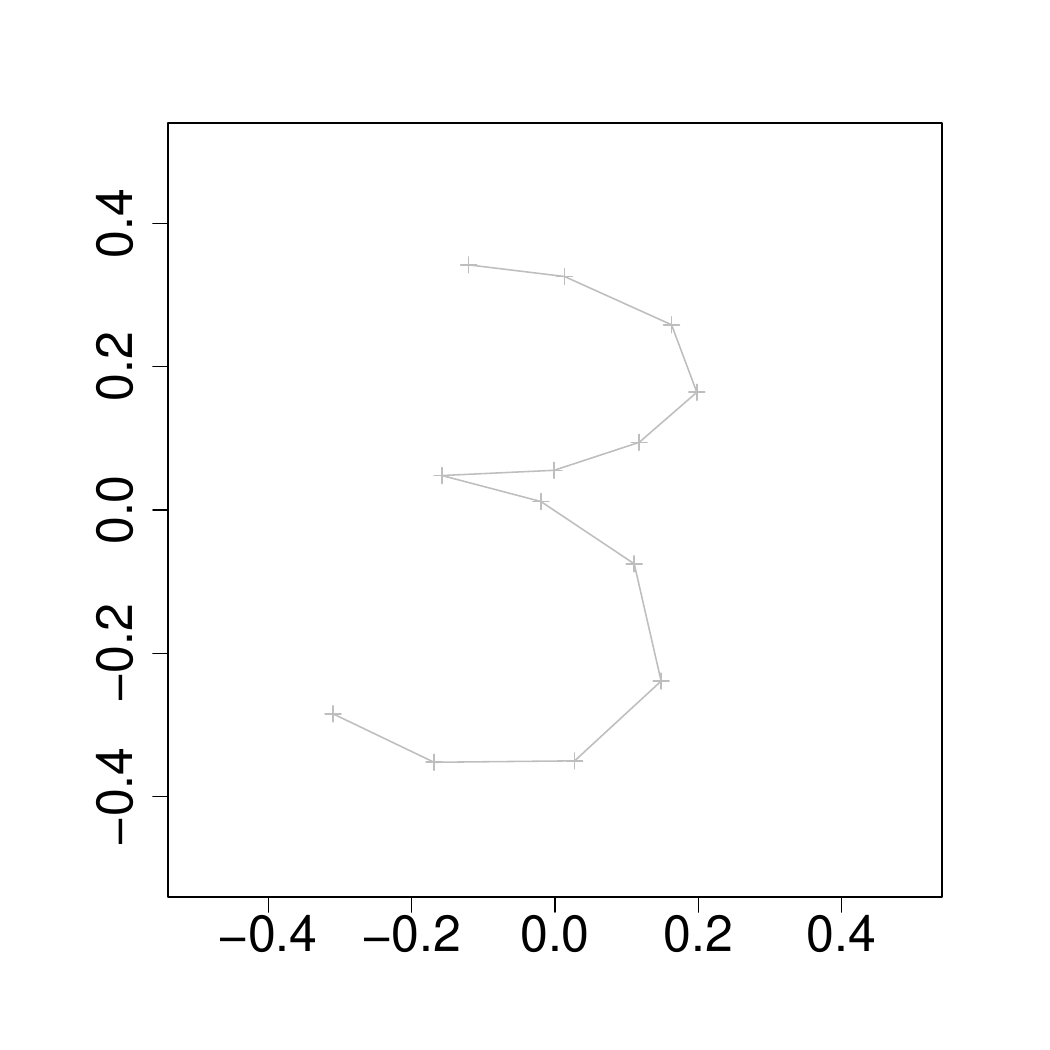}
  \includegraphics[width=0.5in]{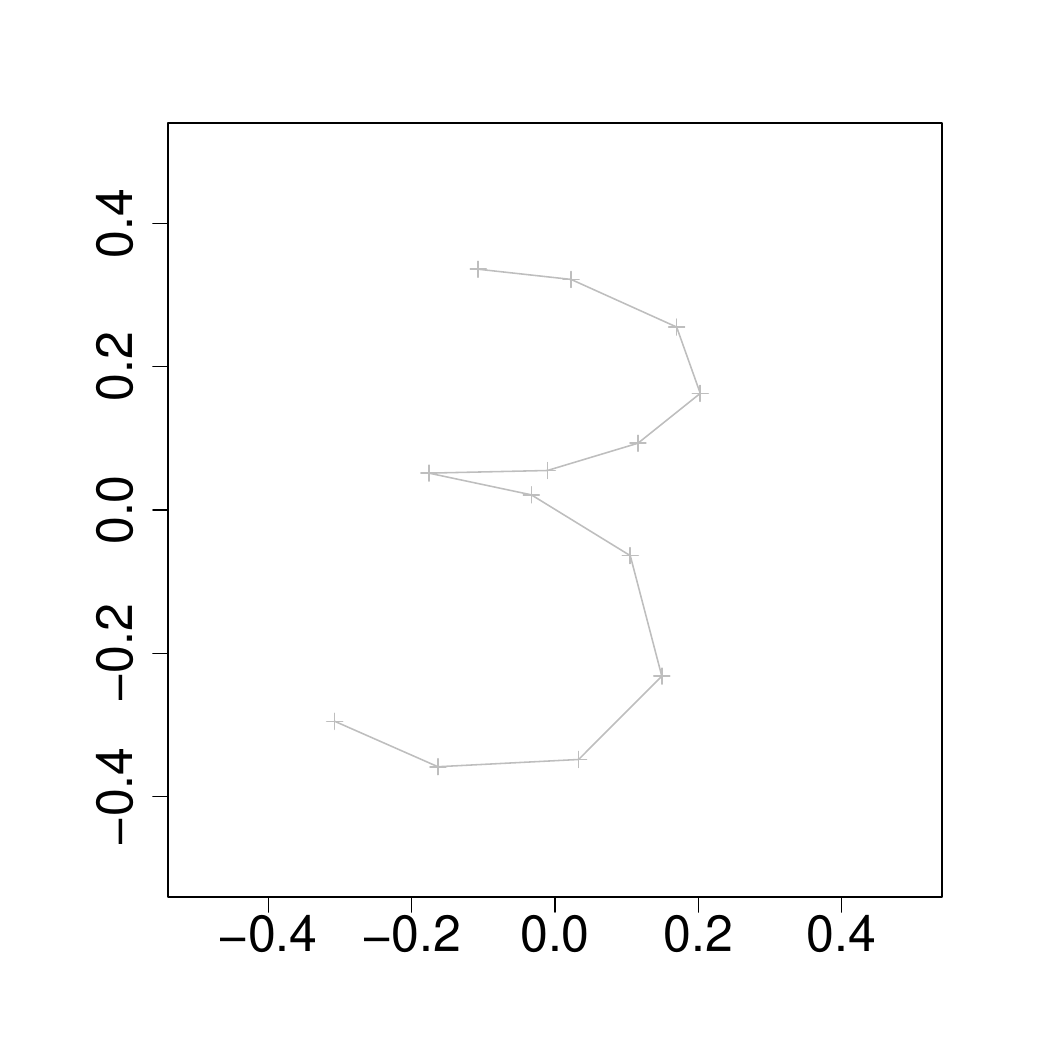}
  \includegraphics[width=0.5in]{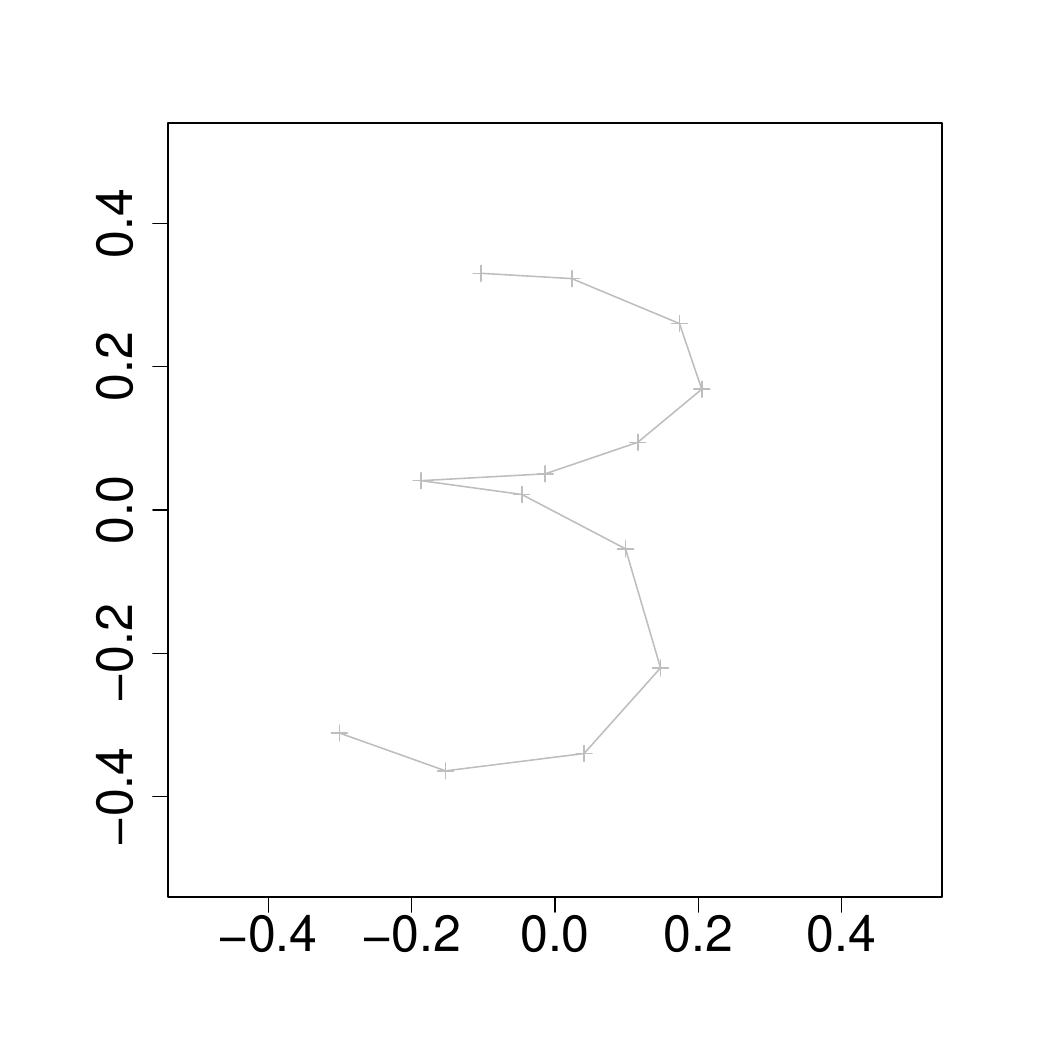}
  \includegraphics[width=0.5in]{PDF/grid-digit3/empty}
  \includegraphics[width=0.5in]{PDF/grid-digit3/empty}
  \includegraphics[width=0.5in]{PDF/grid-digit3/empty}\\
  \includegraphics[width=0.5in]{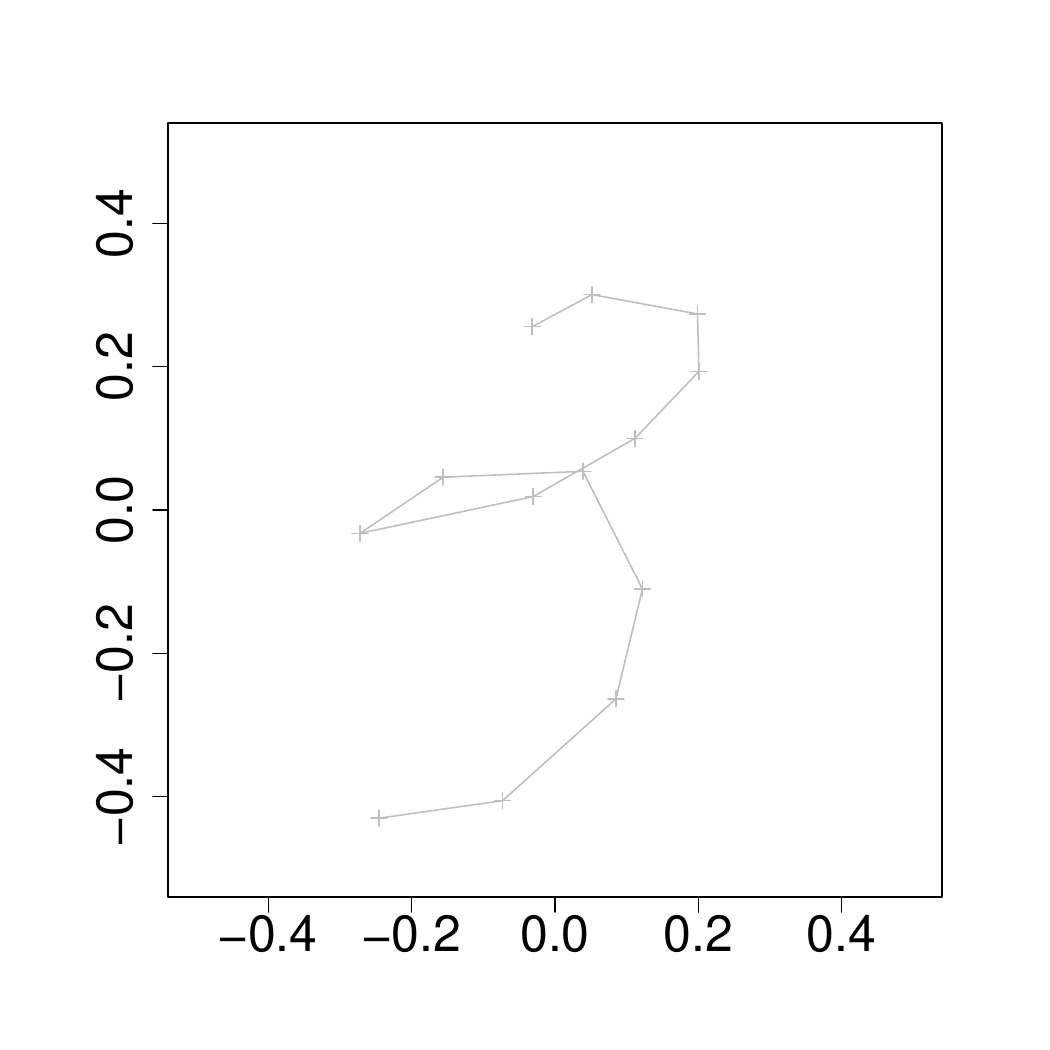}
  \includegraphics[width=0.5in]{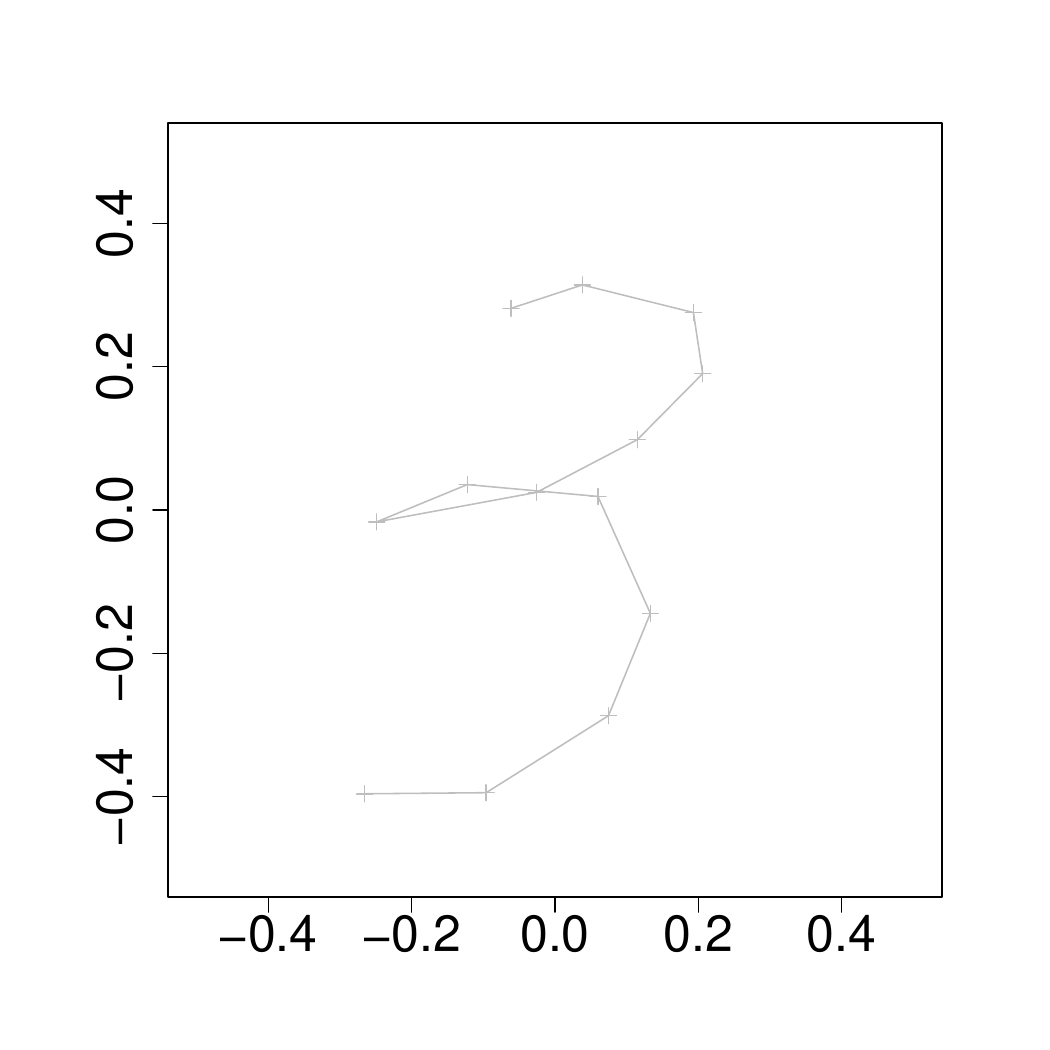}
  \includegraphics[width=0.5in]{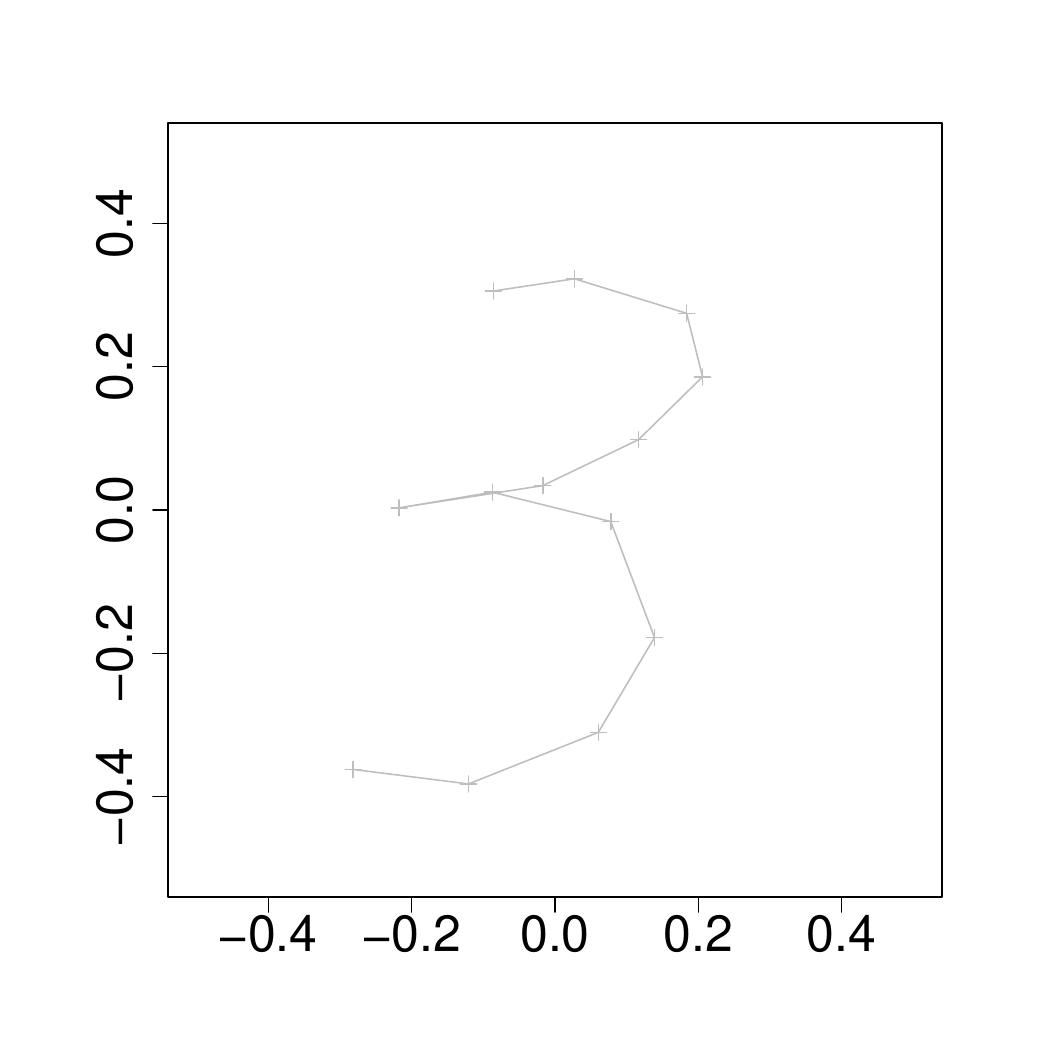}
  \includegraphics[width=0.5in]{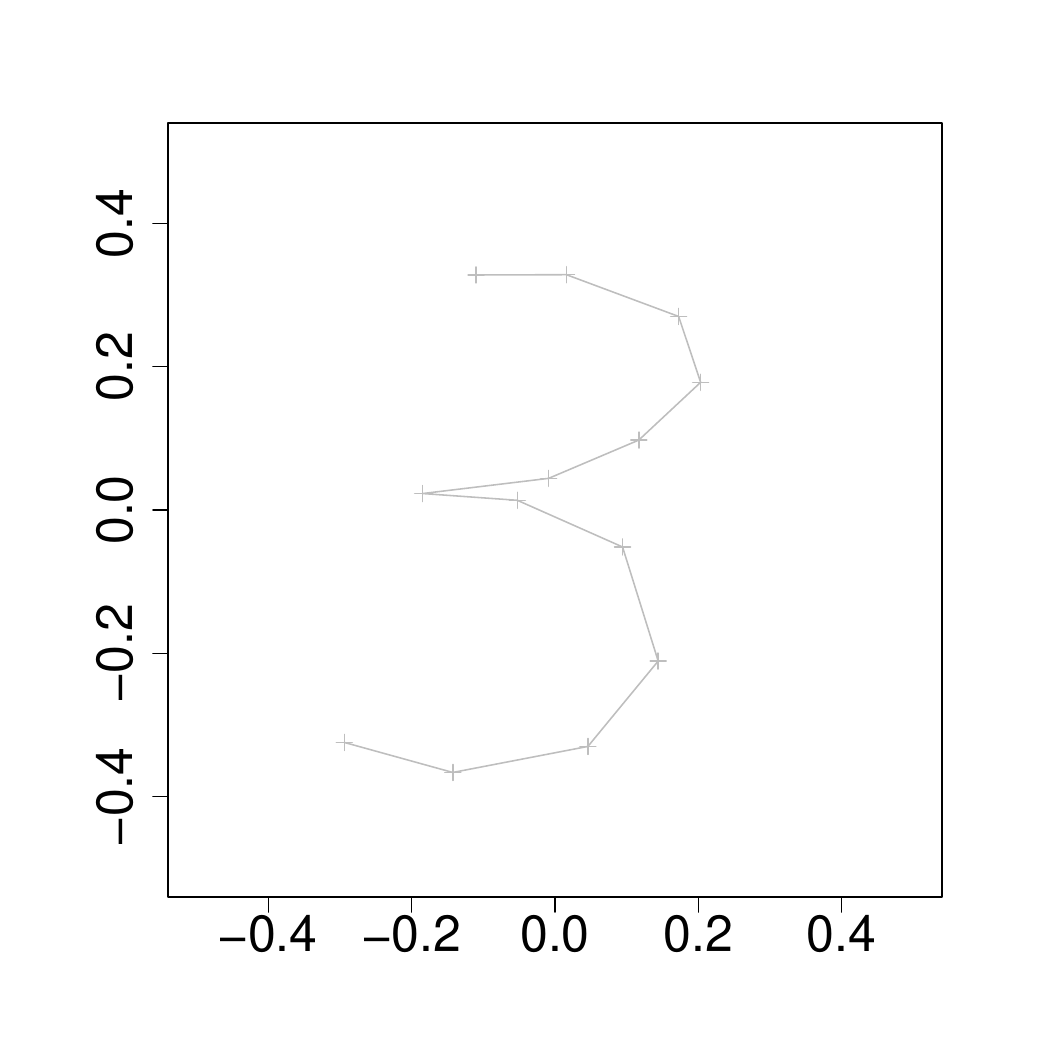}
  \includegraphics[width=0.5in]{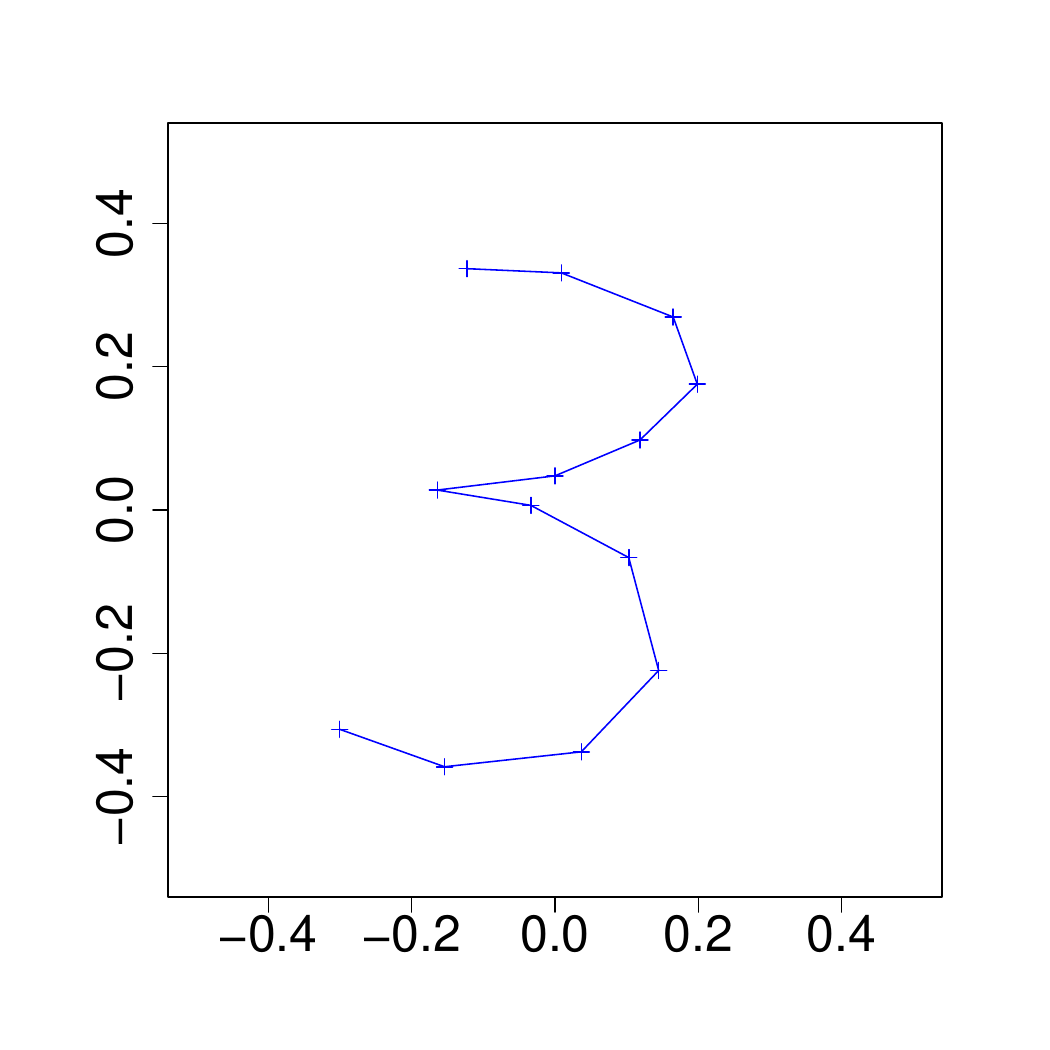}
  \includegraphics[width=0.5in]{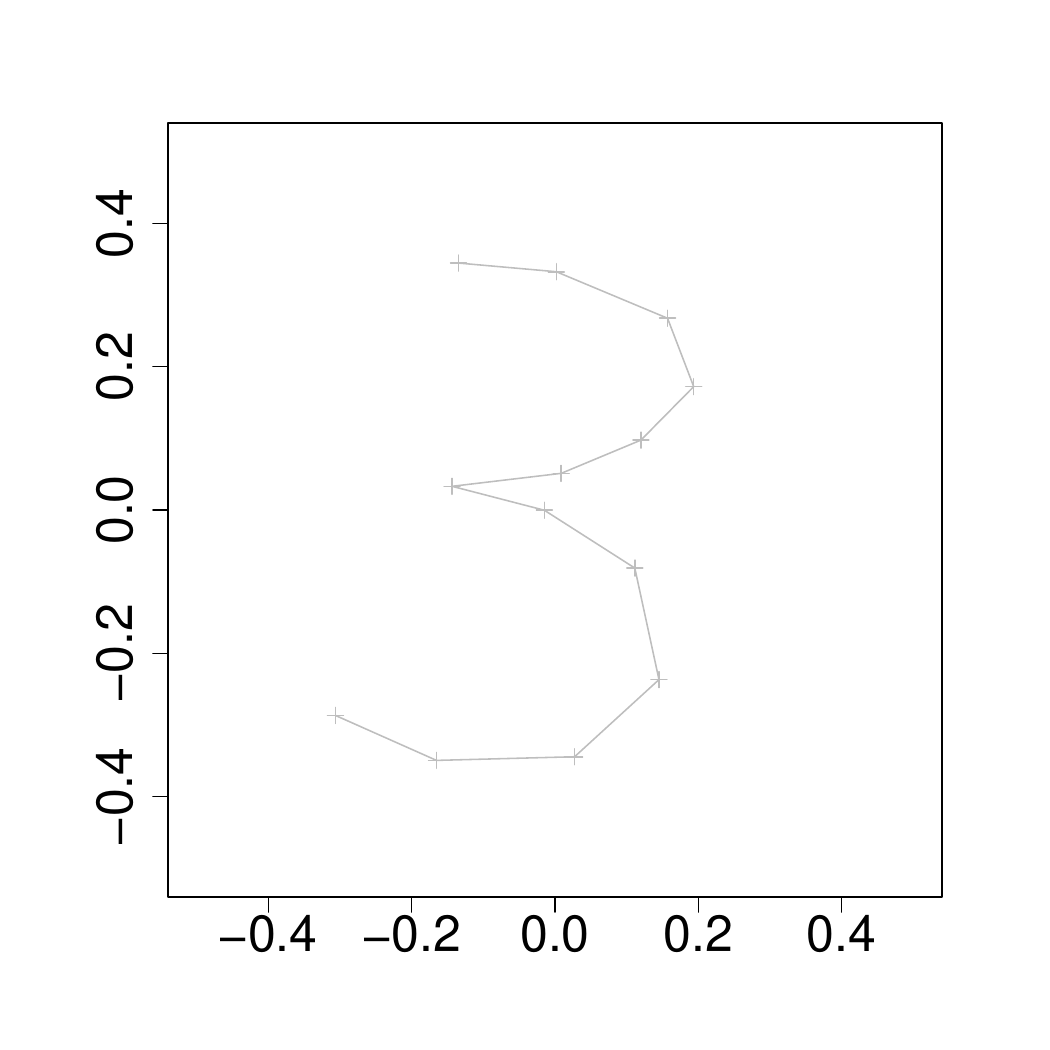}
  \includegraphics[width=0.5in]{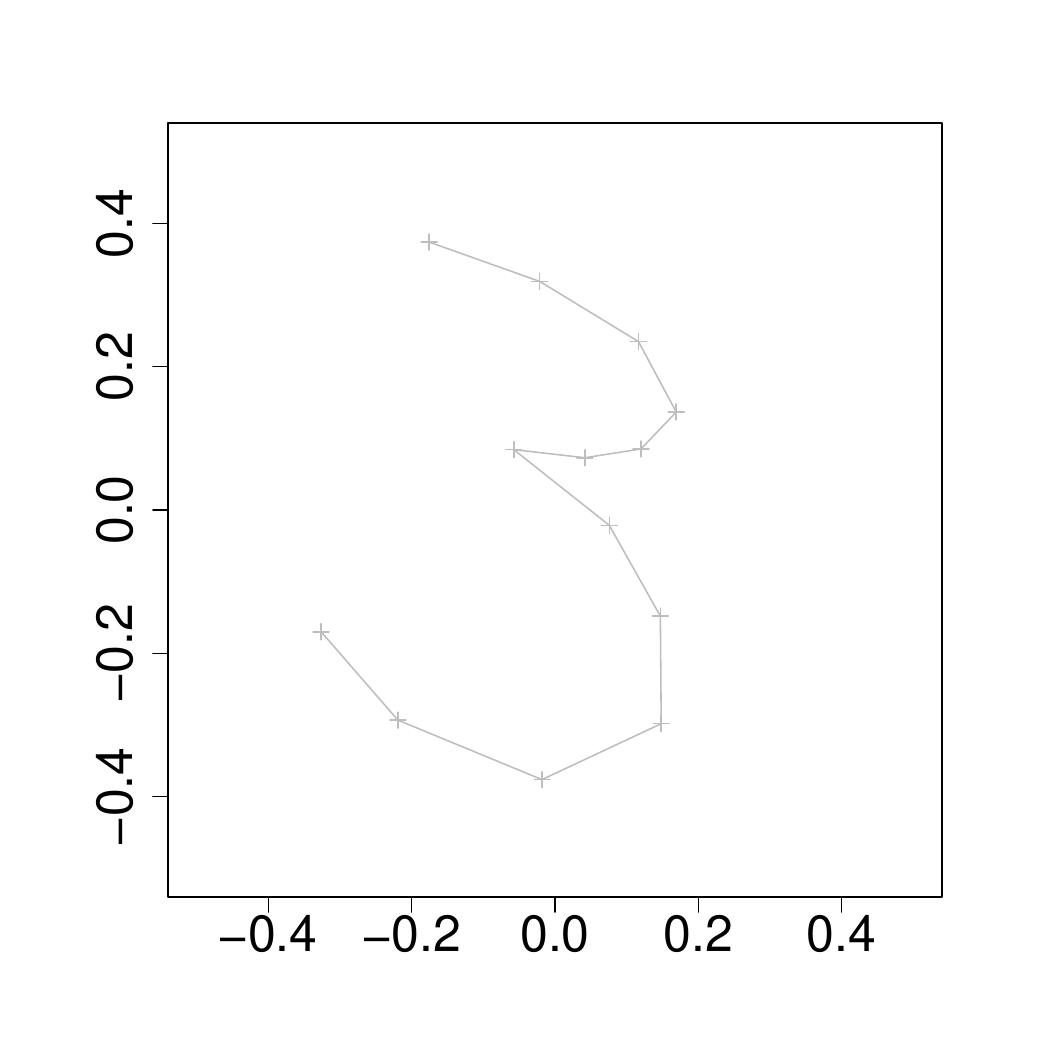}
  \includegraphics[width=0.5in]{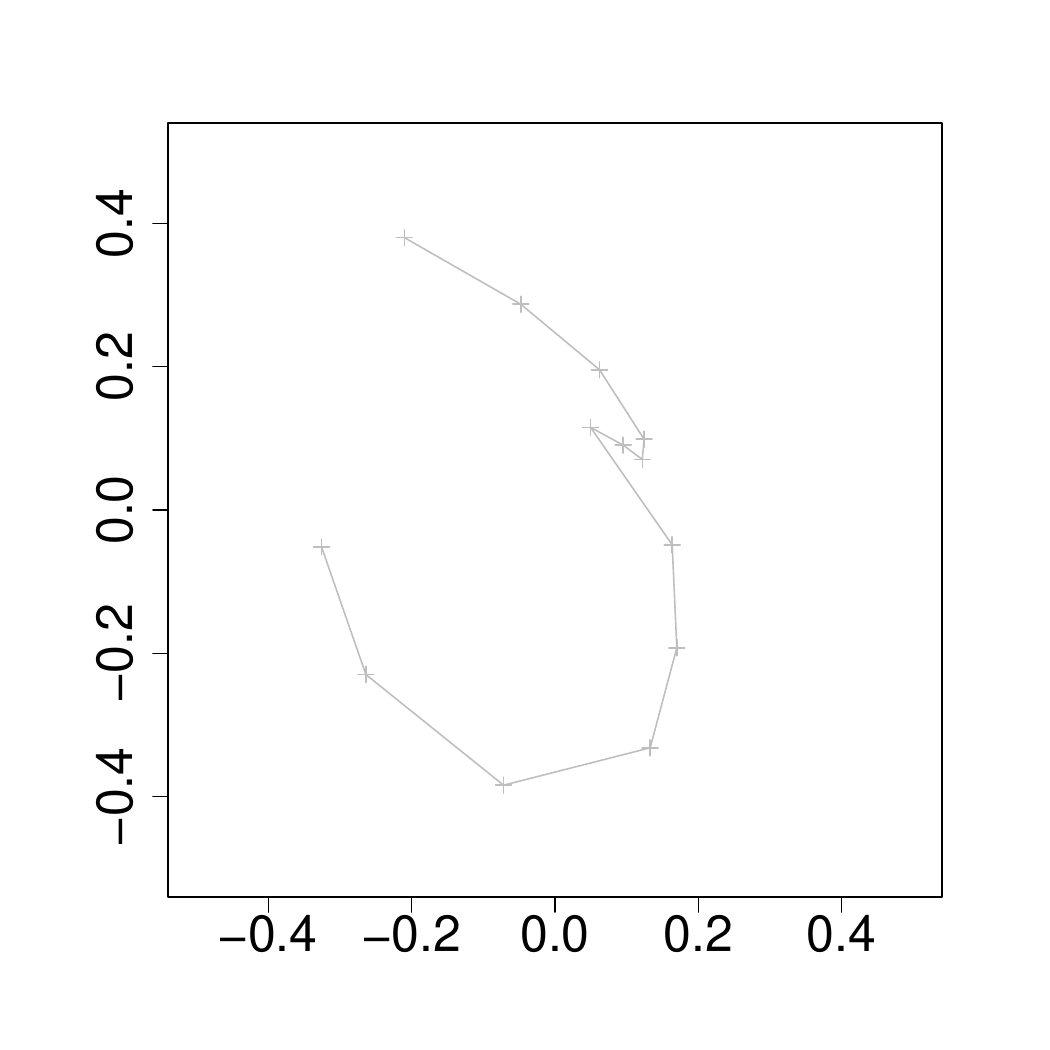}
  \includegraphics[width=0.5in]{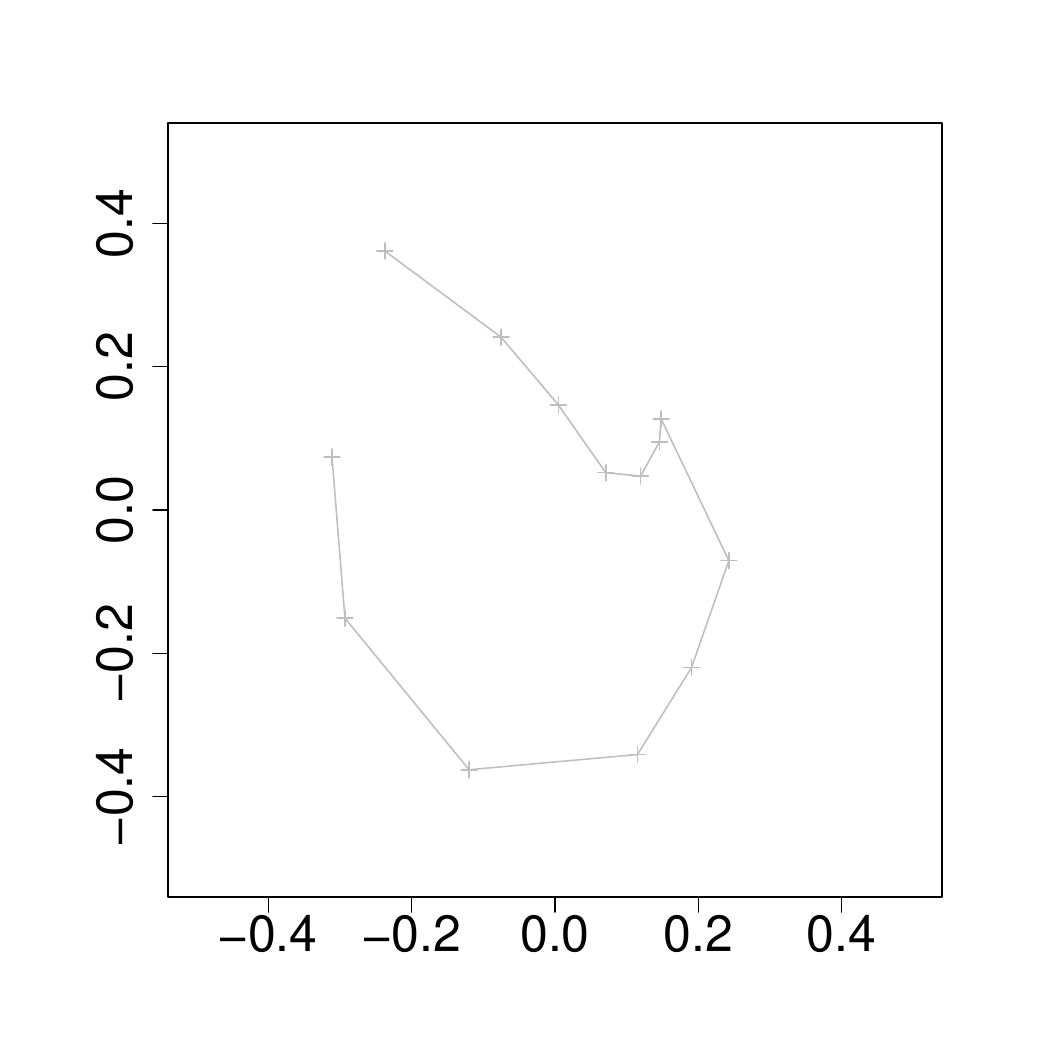}\\
  \includegraphics[width=0.5in]{PDF/grid-digit3/empty}
  \includegraphics[width=0.5in]{PDF/grid-digit3/empty}
  \includegraphics[width=0.5in]{PDF/grid-digit3/empty}
  \includegraphics[width=0.5in]{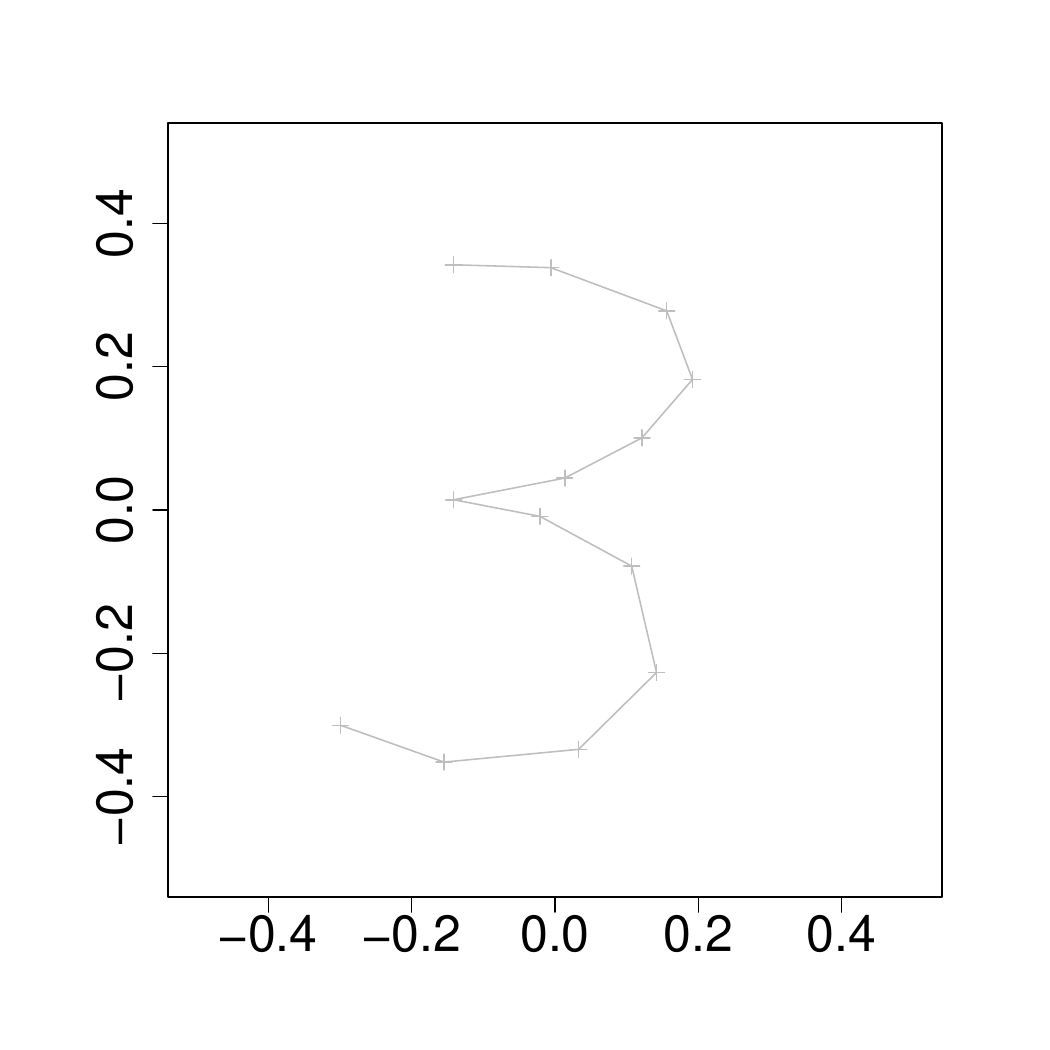}
  \includegraphics[width=0.5in]{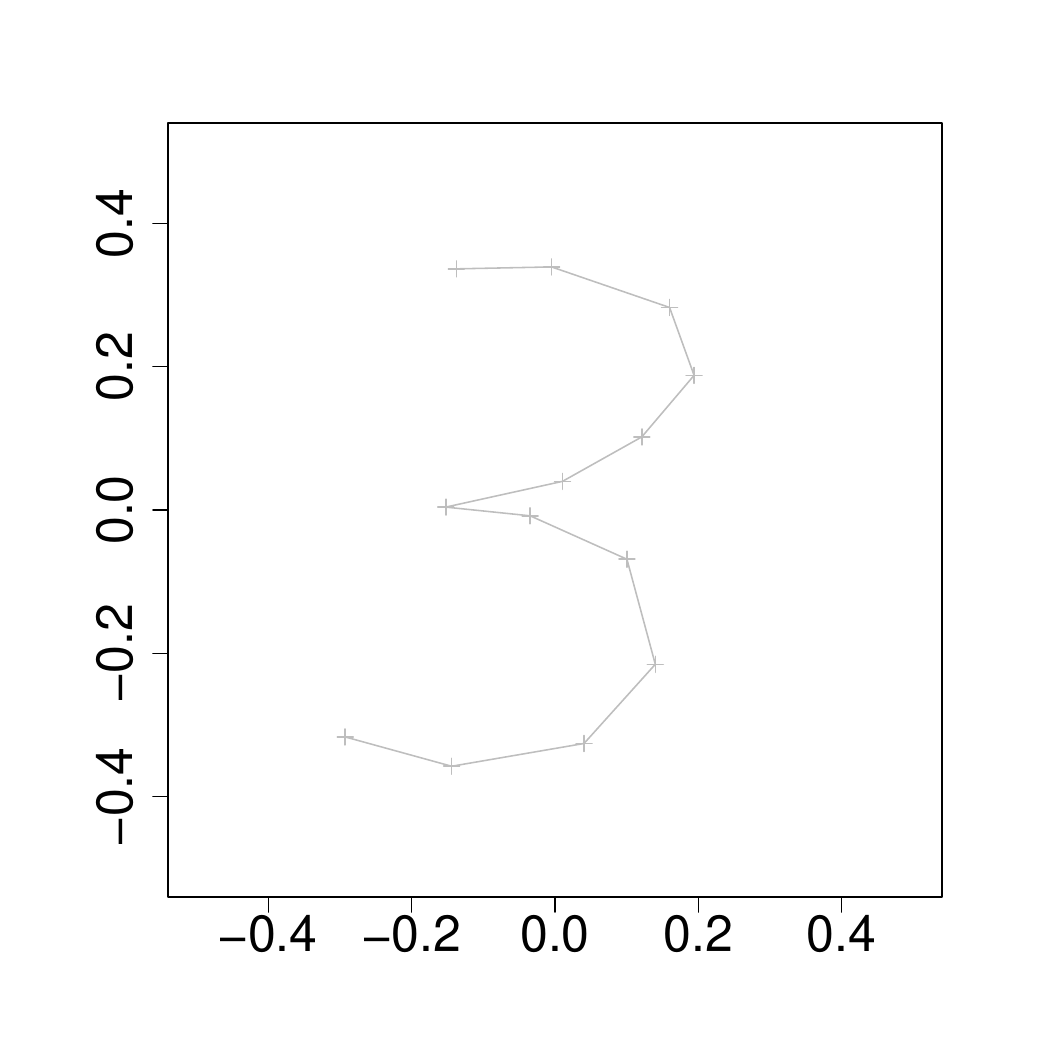}
  \includegraphics[width=0.5in]{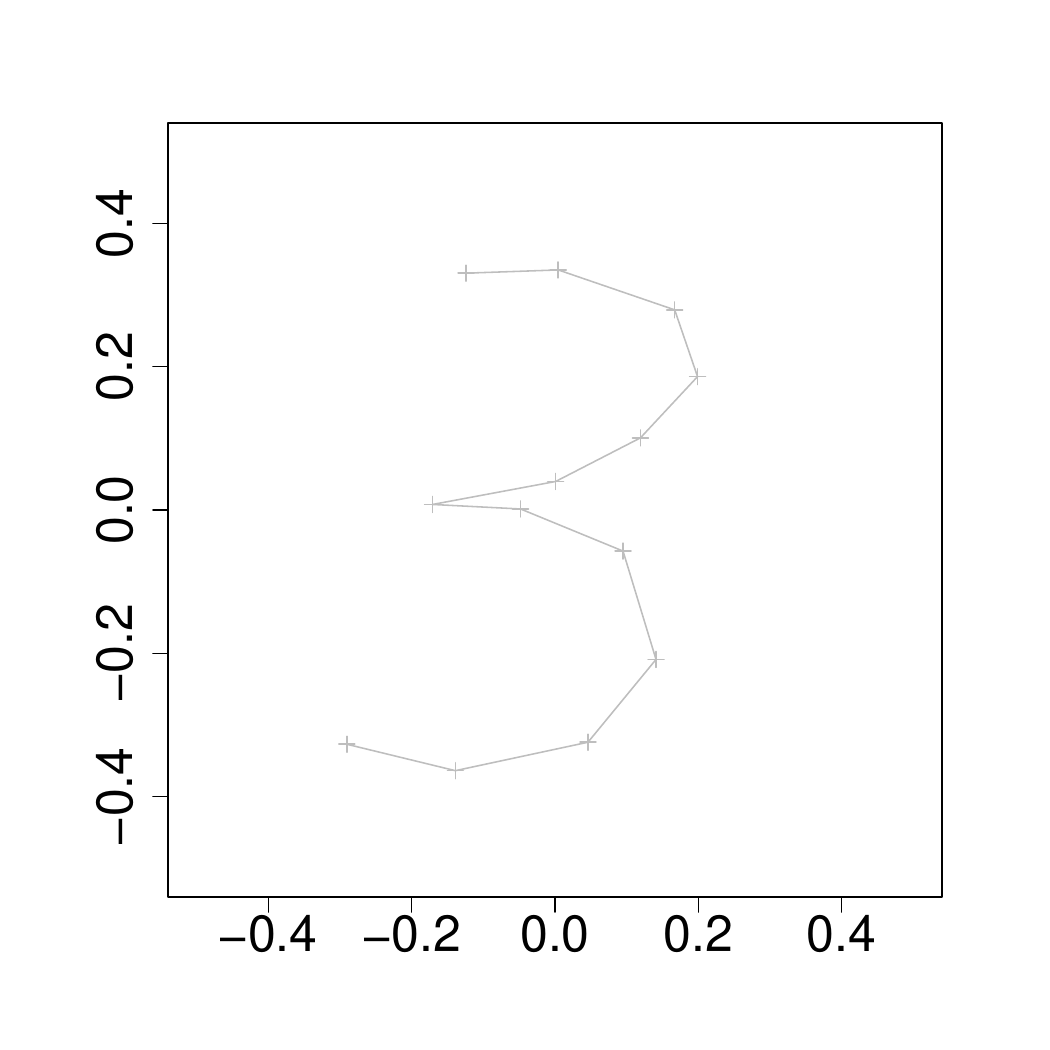}
  \includegraphics[width=0.5in]{PDF/grid-digit3/empty}
  \includegraphics[width=0.5in]{PDF/grid-digit3/empty}
  \includegraphics[width=0.5in]{PDF/grid-digit3/empty}\\
  \includegraphics[width=0.5in]{PDF/grid-digit3/empty}
  \includegraphics[width=0.5in]{PDF/grid-digit3/empty}
  \includegraphics[width=0.5in]{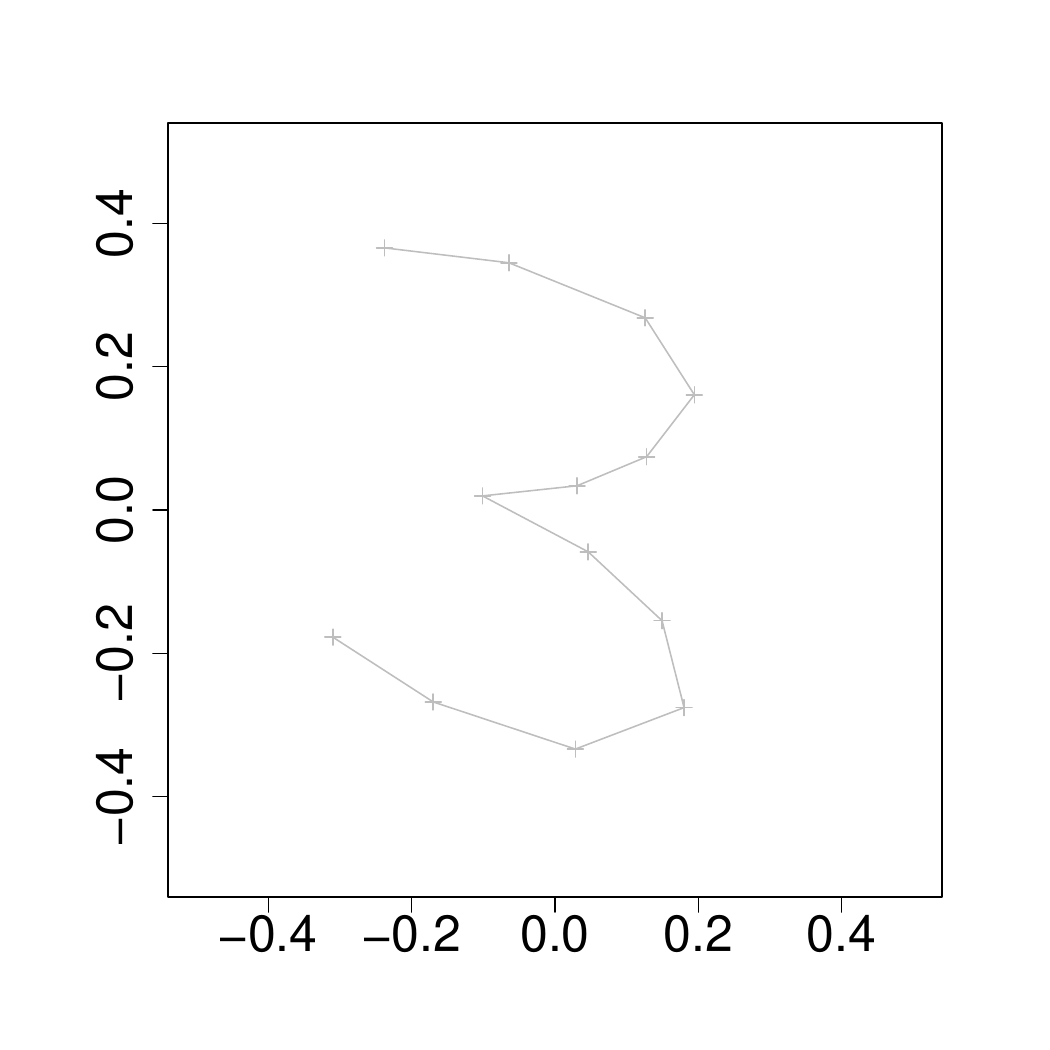}
  \includegraphics[width=0.5in]{PDF/grid-digit3/empty}
  \includegraphics[width=0.5in]{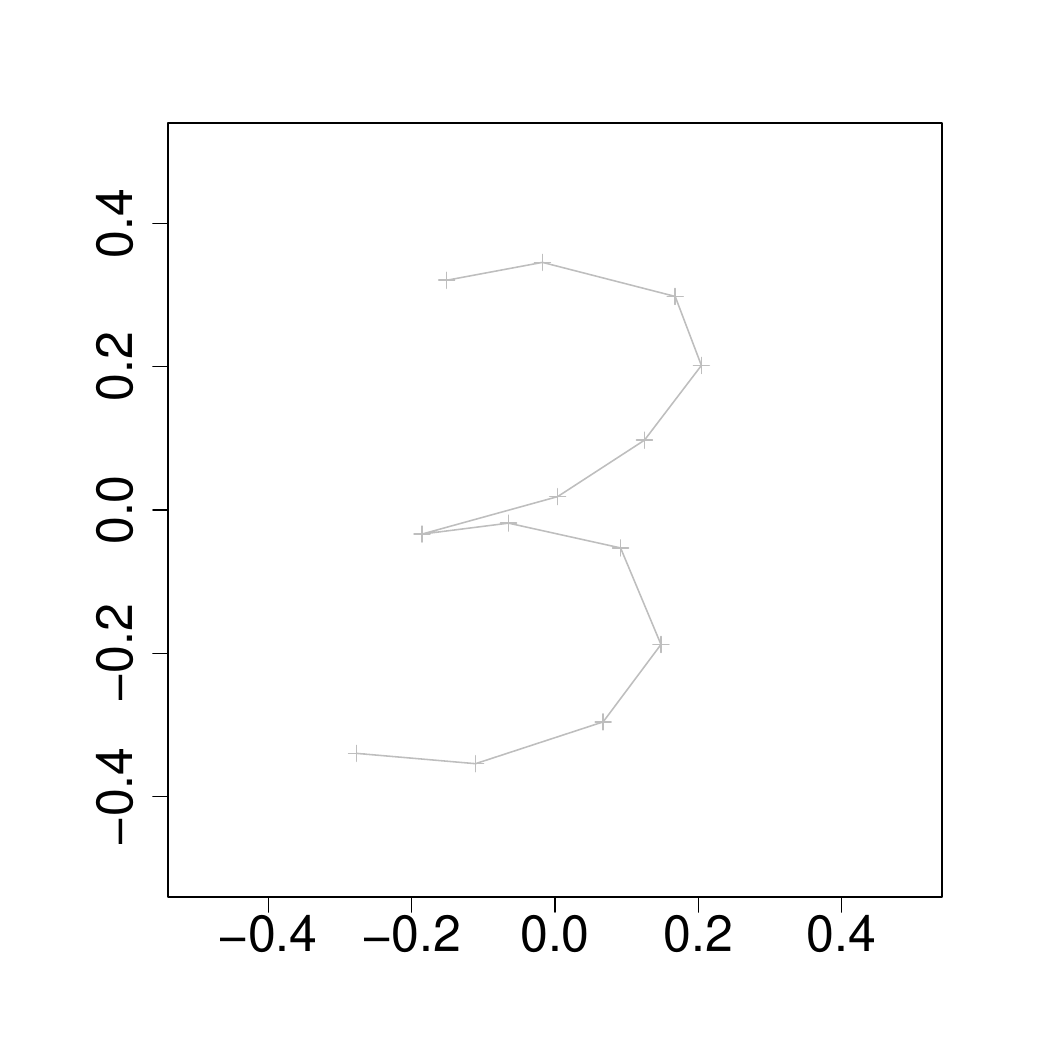}
  \includegraphics[width=0.5in]{PDF/grid-digit3/empty}
  \includegraphics[width=0.5in]{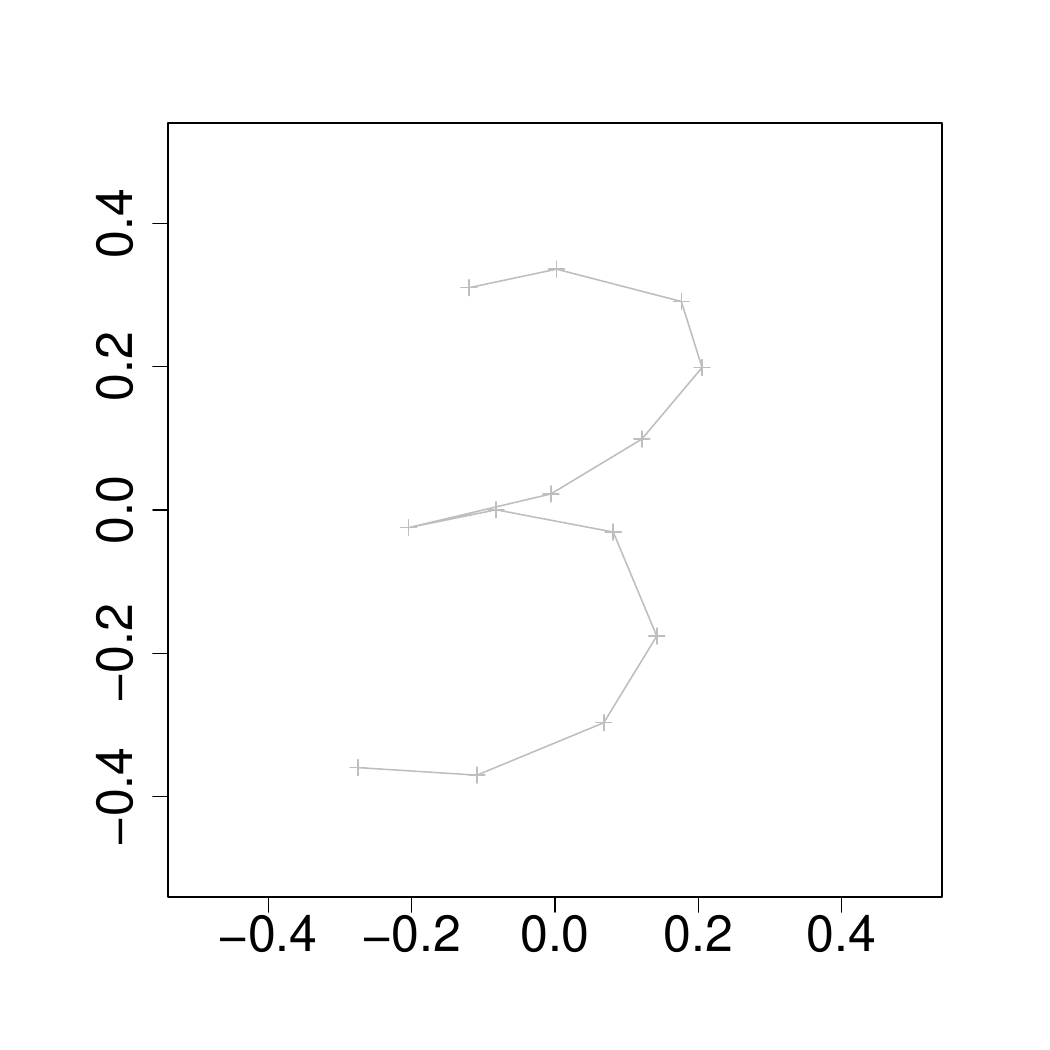}
  \includegraphics[width=0.5in]{PDF/grid-digit3/empty}
  \includegraphics[width=0.5in]{PDF/grid-digit3/empty}\\
  \includegraphics[width=0.5in]{PDF/grid-digit3/empty}
  \includegraphics[width=0.5in]{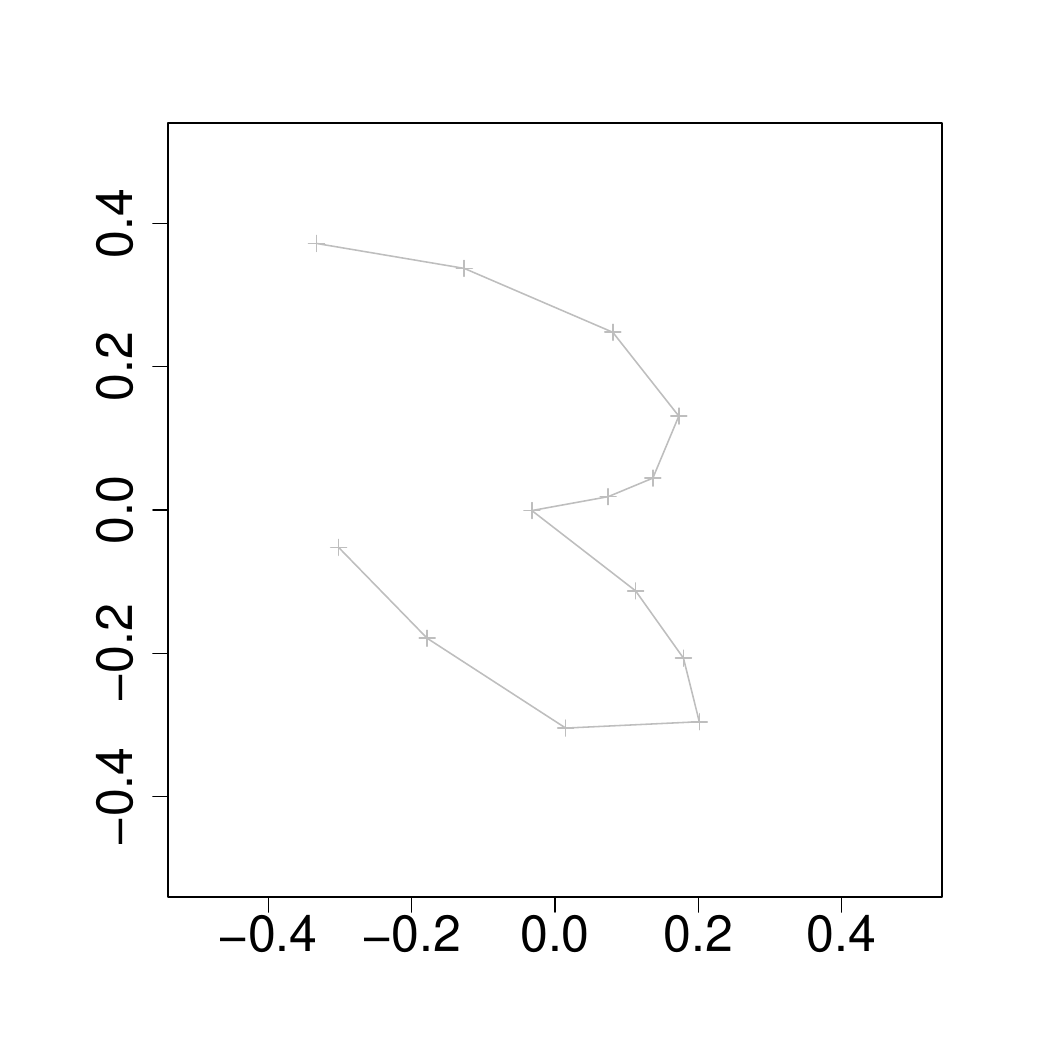}
  \includegraphics[width=0.5in]{PDF/grid-digit3/empty}
  \includegraphics[width=0.5in]{PDF/grid-digit3/empty}
  \includegraphics[width=0.5in]{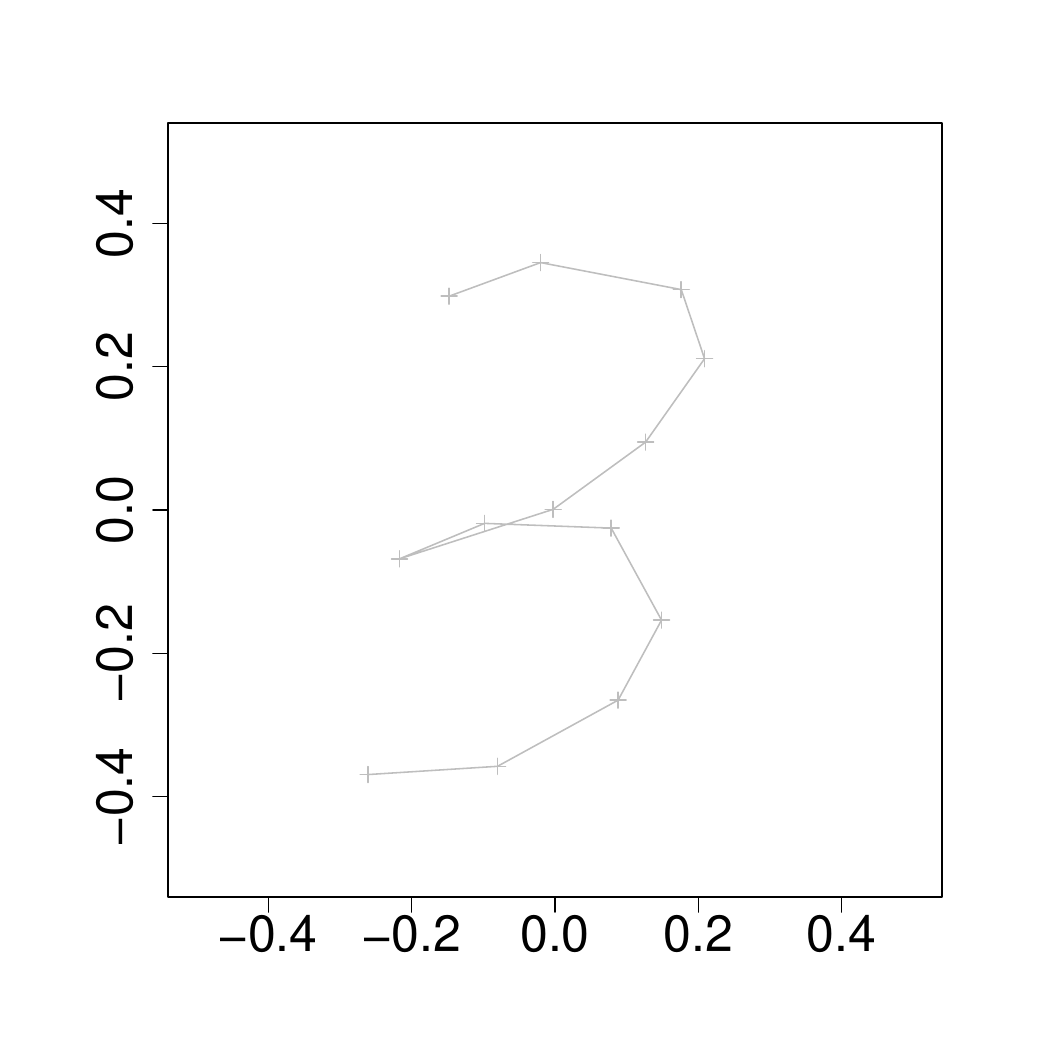}
  \includegraphics[width=0.5in]{PDF/grid-digit3/empty}
  \includegraphics[width=0.5in]{PDF/grid-digit3/empty}
  \includegraphics[width=0.5in]{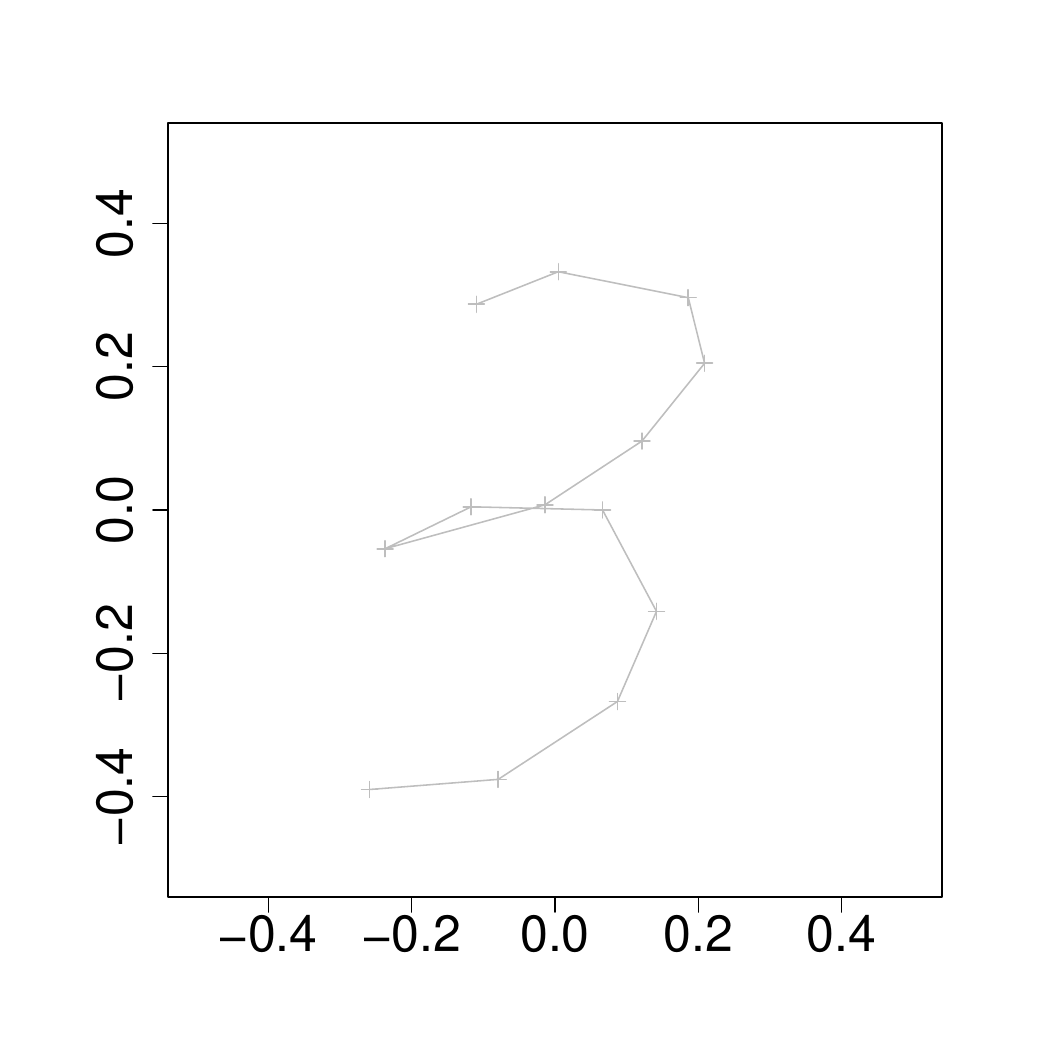}
  \includegraphics[width=0.5in]{PDF/grid-digit3/empty}\\
  \includegraphics[width=0.5in]{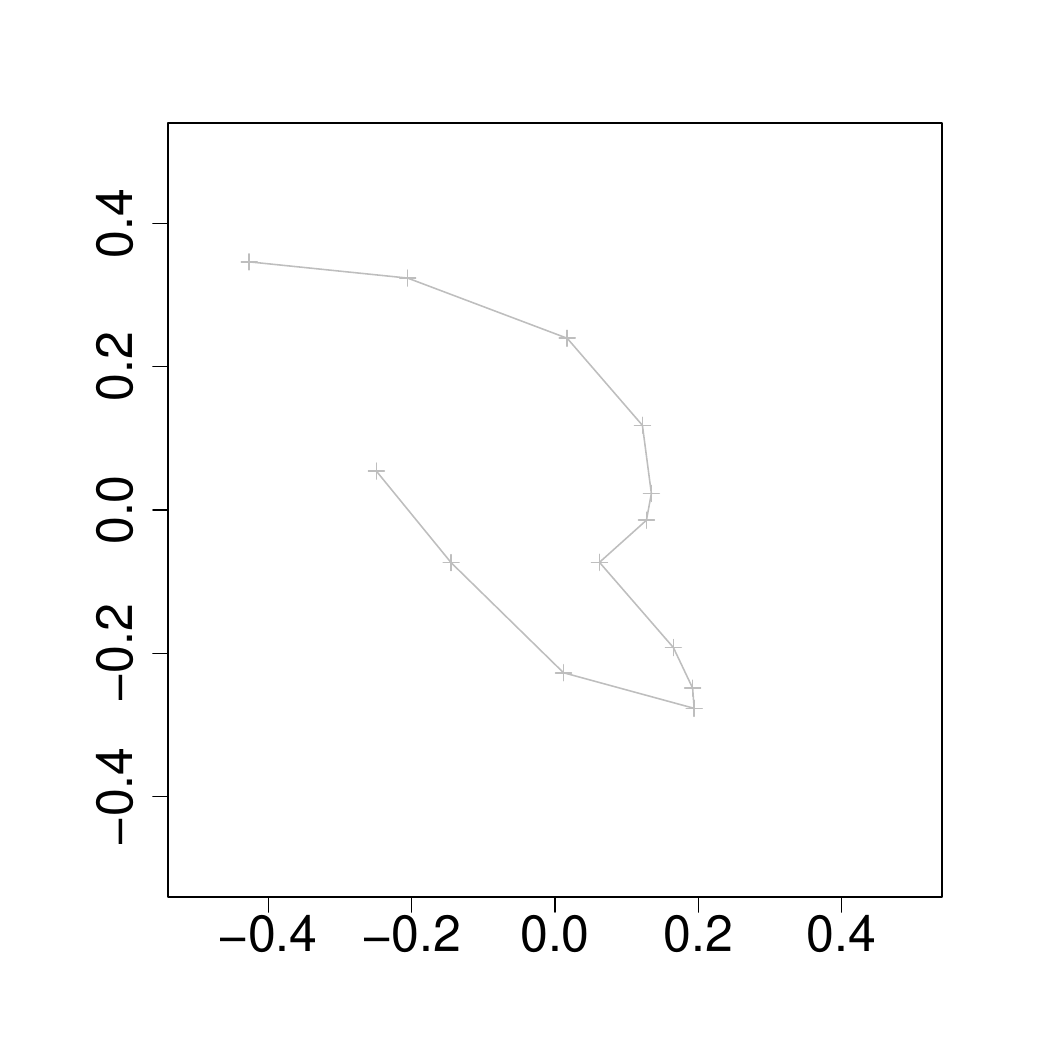}
  \includegraphics[width=0.5in]{PDF/grid-digit3/empty}
  \includegraphics[width=0.5in]{PDF/grid-digit3/empty}
  \includegraphics[width=0.5in]{PDF/grid-digit3/empty}
  \includegraphics[width=0.5in]{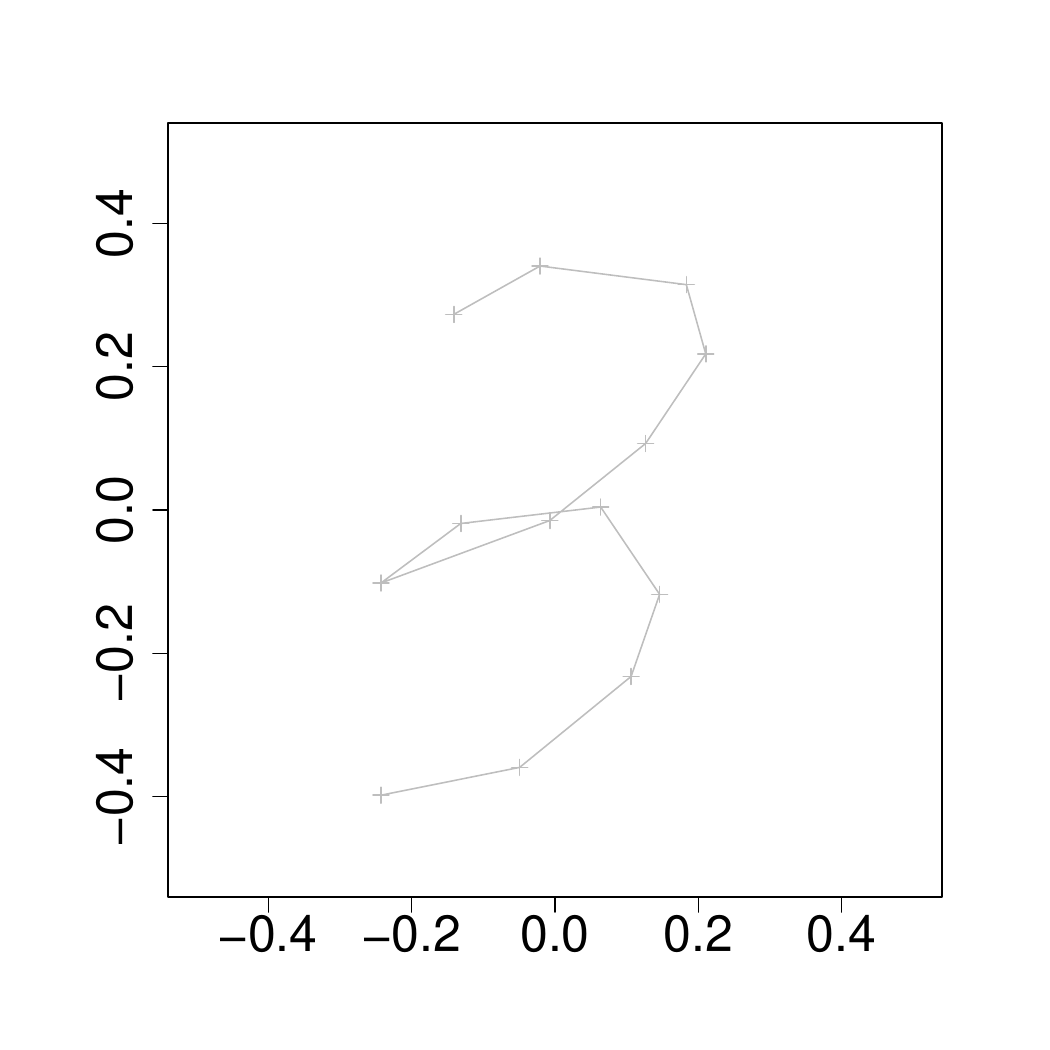}
  \includegraphics[width=0.5in]{PDF/grid-digit3/empty}
  \includegraphics[width=0.5in]{PDF/grid-digit3/empty}
  \includegraphics[width=0.5in]{PDF/grid-digit3/empty}
  \includegraphics[width=0.5in]{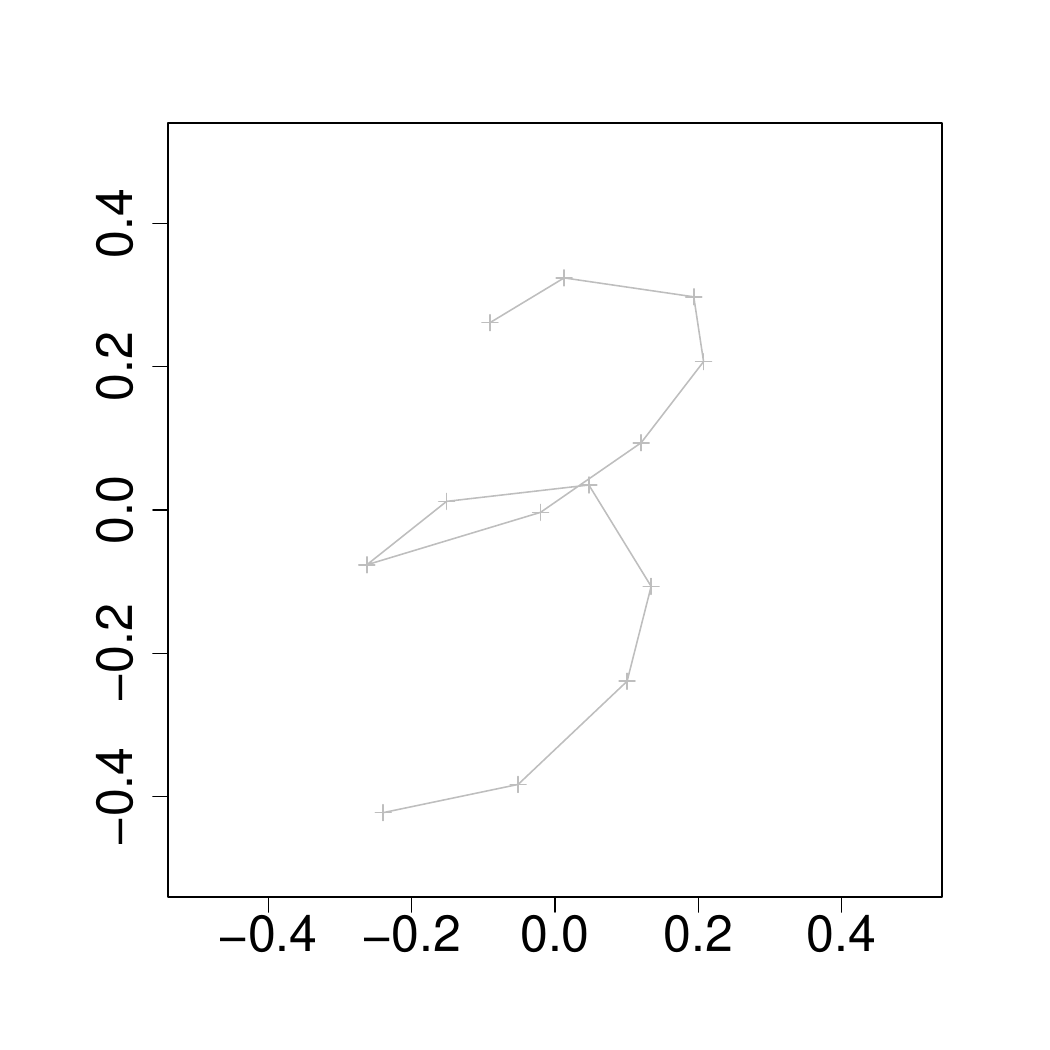}\\
  \caption{Principal sub-manifolds of the handwritten digits data, started from the center of symmetry. Among all the figures: the central figure (in blue) is the center of symmetry; the horizontal row contains images recovered from the first principal direction of the sub-manifold; the vertical column is the second principal direction; the main diagonal is the third principal direction; the other diagonal is the fourth principal direction.}
  \label{subman-digit3-nonmean}
\end{figure}

\section{Angle Between Linear Subspaces}\label{app:5-subspace-angles}%\label{sec:subspace-angle}

Assume two $k$-dimensional linear subspaces $A,B \subset \mathbb{R}^d$. Let $v_j := \sum_\alpha \limits A_{j\alpha} w_\alpha$ any normal vector in $A$, which means that $\|w\| = 1$. Then, the projection of this vector to $B$ is given by $\tilde{v}_j = \sum_{\alpha,\beta,l} \limits B_{j\beta} B_{l\beta} A_{l\alpha} w_\alpha$. Let $\phi(v,\tilde{v})$ denote the angle between the two vectors, then $\|\tilde{v}\| = \cos \phi(v,\tilde{v})$ and therefore
\begin{align*}
  \cos^2 \phi(A,B) = \min_v \cos^2 \phi(v,\tilde{v}) = \min_v \sum_j \limits v_j \cdot \tilde{v}_j = \min_w \sum_{\alpha,\beta,\gamma,j,l} \limits w_\gamma A_{j\gamma} B_{j\beta} B_{l\beta} A_{l\alpha} w_\alpha \, ,
\end{align*}
where minimization runs over unit vectors. Since the matrix $\left( \sum_{\beta,j,l} \limits A_{j\gamma} B_{j\beta} B_{l\beta} A_{l\alpha}\right)_{\gamma\alpha}$ is symmetric and positive semidefinite, the expression is minimized for $w$ being the eigenvector to the minimal eigenvalue $\lambda_\text{min}$ of this matrix. We therefore note
\begin{align*}
  \cos^2 \phi(A,B) = \lambda_\text{min} \left( \sum_{\beta,j,l} \limits A_{j\gamma} B_{j\beta} B_{l\beta} A_{l\alpha}\right)_{\gamma\alpha} \quad \Rightarrow \quad \cos \phi(A,B) = \lambda_\text{min} \left( \sum_l \limits B_{l\beta} A_{l\alpha} \right)_{\beta\alpha} \, .
\end{align*}
In the latter expression, we assume all eigenvalues of $\left( \sum_l \limits B_{l\beta} A_{l\alpha} \right)_{\beta\alpha}$ to be positive, which can be achieved by suitably choosing the signs of spanning vectors of $A$ and $B$.

Analogously, note that the angle between a $1$-dimensional linear subspace $A$ spanned by the unit vector $v$ and a $k$-dimensional linear subspace $B$ is given by
\begin{align*}
  \cos^2 \phi(A,B) = \cos^2 \phi(v,\tilde{v}) = \sum_j \limits v_j \cdot \tilde{v}_j = \sum_{\alpha,j,l} \limits v_j B_{j\alpha} B_{l\alpha} v_l \, .
\end{align*}

\section{A Lagrangian Formulation}\label{app:6-lagrangian}

In the following, we will denote components in $\mathbb{R}^d$ by latin letters from the middle of the alphabet and components in $\mathcal{N}$ and $W$ ranging from $1$ to $k$ by greek letter from the beginning of the alphabet.

We are aiming to describe a principal sub-manifold in terms of a Lagrangian problem, analogous to \citet{Panaretos2014}. To make the idea precise, we propose the following definition.

\begin{definition}\label{def:princ-submf}
  Assume a sub-manifold, described by the image of an injective smooth function
  \begin{align} \label{eq:princ-submf-map}
    N: \mathbb{R}^k \supset U \to \mathcal{N} \subset \mathcal{M} \subset \mathbb{R}^d\, .
  \end{align}
  In this expression, the image $\mathcal{N} := \{N(t)\}$ is the principal sub-manifold. We denote the space of local $k$-dimensional sub-manifolds containing some point $A \in \mathcal{M}$ and satisfying $\forall N \in \mathcal{N} \, : \, d_{\mathcal{N}}(N,A) < \epsilon$ by $\mbox{SubM}(A, \epsilon, k, \mathcal{M})$. Here $d_{\mathcal{N}}$ is the metric on $\mathcal{N}$ induced by the metric on $\mathcal{M}$.
\end{definition}

We denote a generic point in the principal sub-manifold by $N$, suppressing the argument $t$, and write $\dot{N}_{j\alpha} := \nabla_\alpha N_j$ in a slight abuse of notation.

The main term of the Lagrangian characterizes the compatibility of the principal sub-manifold with the tensor field $W$. A simple measure for goodness of fit at some point $N \in \mathcal{N}$ is given by the angle between $T_N\mathcal{N}$ and $W(N)$, which we will denote by $\phi(T_N\mathcal{N}, W(N))$. Furthermore, the Lagrangian will contain two sets of constraints. One set will enforce that the tangent vectors of $T\mathcal{N}$ will always be orthonormal and the second set consists of the algebraic equations $F (N) = 0$, restricting to $\mathcal{M}$.

Using the result of Appendix \ref{app:3-proofs-thms-3+4}, we can formulate the Lagrangian for the principal sub-manifold as
\begin{align}\label{eq:lagrange1}
  \mathscr{L}_1 (N, \dot{N}) := \lambda_\text{min} \left( \sum_l \limits W_{l\beta} (N) \dot{N}_{l\alpha} \right)_{\beta\alpha} + \sum_{\alpha\beta} \kappa_{\alpha\beta} \left( \sum_l \dot{N}_{l\alpha} \dot{N}_{l\beta} - \delta_{\alpha\beta} \right) + \sum_\nu \limits z_\nu F_\nu(N) \, .
\end{align}

For the solution of the dynamic system resulting from this Lagrangian we have the following result.
\begin{theorem} \label{thm:linearspace1}
  Assume that $\mathcal{M}=\mathbb{R}^d$, which means that $F \equiv 0$, and let $h=\infty$. Assume that the first $k+1$ eigenvalues of $\Sigma_{x, \mathcal{M}}$ are distinct for any point $x \in \mathcal{M}$. Then the solution of the dynamic system corresponding to $\mathscr{L}_1$ with initial conditions $N(0) = A$ and $\forall \alpha \, : \, \dot{N}_{\alpha}(0) = e_\alpha(A, \mathbb{R}^d)$ is the affine space containing $A$ and spanned by $\big\{ e_1(A,\mathbb{R}^d), e_2(A,\mathbb{R}^d), \ldots,  e_k(A,\mathbb{R}^d) \big\}$, the first $k$ eigenvectors of $\Sigma_{A, \mathcal{M}}$.
\end{theorem}

\noindent\textit{Proof.}
%\begin{proof}
Since $h = \infty$, we have for any point $x \in \mathbb{R}^d$ that $W_{j\beta} (x) = \Big( e_\beta(A, \mathbb{R}^d) \Big)_j$. Assuming the constraint $\forall \alpha, \beta \, : \, \sum_l \dot{N}_{l\alpha} \dot{N}_{l\beta} = \delta_{\alpha\beta}$, it is clear that
\begin{align*}
  \lambda_\text{min} \left( \sum_l \limits W_{l\beta} (N) \dot{N}_{l\alpha} \right)_{\beta\alpha} \le 1
\end{align*}
where equality holds if and only if $\forall \alpha, j \, : \, W_{j\alpha} (N) = \dot{N}_{j\alpha}$. Integrating the equation $\dot{N}_{\alpha} = e_\alpha(A, \mathbb{R}^d)$ with the initial conditions yields the desired affine subspace. \QED
%\end{proof}

However, many complications arise when trying to solve the dynamic system resulting from this Lagrangian. In particular, the solution strategy outlined in \citet{Panaretos2014} cannot be applied in this generalized setting. We will therefore turn to a simpler approach, which aims at constructing a principal sub-manifold in terms of a spherical mesh, centered at the reference point $A$. This means, that we are looking for curves $\gamma$ starting at $A$, whose tangent vectors $\dot{\gamma}$ stay as close as possible to $W(\gamma)$.

Consider the following set of curves, which encodes the constraints
\begin{align}
  \Gamma_{1, \mathcal{M}} := \left\{ \gamma : [a,b] \to \mathbb{R}^d \,\Bigg|\, \sum_j \dot{\gamma}_j \dot{\gamma}_j = 1 \mbox{ and } F(\gamma) = 0 \right\} \, .
\end{align}
We then maximize
\begin{align} \label{eq:minimize2}
  \mathop{\rm argmax}_{\gamma \in \Gamma_{1, \mathcal{M}}} \limits \sum_{\alpha,j,l} \limits \dot{\gamma}_j W_{j\alpha} (\gamma) W_{l\alpha} (\gamma) \dot{\gamma}_l \, .
\end{align}

We can write this in terms of a Lagrangian, where we square the main term of the Lagrangian to achieve a simpler formulation
\begin{align}\label{eq:lagrange2}
  \mathscr{L}_2 (\gamma, \dot{\gamma}) := \sum_{\alpha,j,l} \limits \dot{\gamma}_j W_{j\alpha} (\gamma) W_{l\alpha} (\gamma) \dot{\gamma}_l + \kappa \left( \sum_j \dot{\gamma}_j \dot{\gamma}_j - 1 \right) + \sum_\nu \limits z_\nu F_\nu(\gamma) \, .
\end{align}
We get the following Theorem for this Lagrangian
\begin{theorem} \label{thm:linearspace2}
  Assume that $\mathcal{M}=\mathbb{R}^d$, which means that $F \equiv 0$, and let $h=\infty$. Assume that the first $k+1$ eigenvalues of $\Sigma_{x, \mathcal{M}}$ are distinct for any point $x \in \mathcal{M}$. Then the solutions $\gamma_\alpha$ of the dynamic system corresponding to $\mathscr{L}_2$ with initial conditions $\gamma (0) = A$ and $\dot{\gamma}(0) = e_\alpha(A, \mathbb{R}^d)$ span the affine space containing $A$ and spanned by $\big\{ e_1(A,\mathbb{R}^d), e_2(A,\mathbb{R}^d), \ldots,  e_k(A,\mathbb{R}^d) \big\}$, the first $k$ eigenvectors of $\Sigma_{A, \mathcal{M}}$.
\end{theorem}

The Lagrangian $\mathscr{L}_2$ is much simpler than $\mathscr{L}_1$ and is in fact very similar to the principal flow Lagrangian. %Standard solution techniques as presented in \cite{Panaretos2014} could be applied to solve the dynamic systems. However, since a large number of curves with different initial conditions have to be determined, this approach would be very slow.
However, the solution technique used in \citet{Panaretos2014} is not suitable here, therefore we provide a simpler ``greedy'' algorithm to approximate principal submanifolds in Section 3.1 of the article.

\begin{figure}[ht!]
  \centering
  \begin{subfigure}[b]{0.33\textwidth}
    \centering
    \includegraphics[width=.85\linewidth]{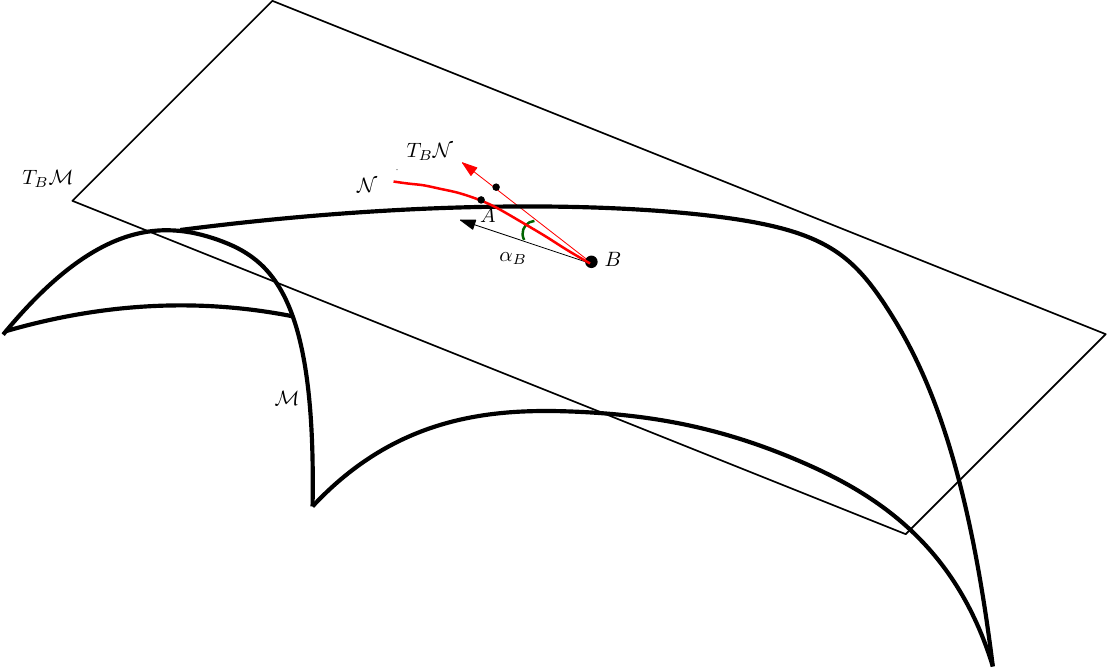}
    \caption{}
  \end{subfigure}%
  %\hspace{0.15 in}
  \begin{subfigure}[b]{0.33\textwidth}
    \centering
    \includegraphics[width=.85\linewidth]{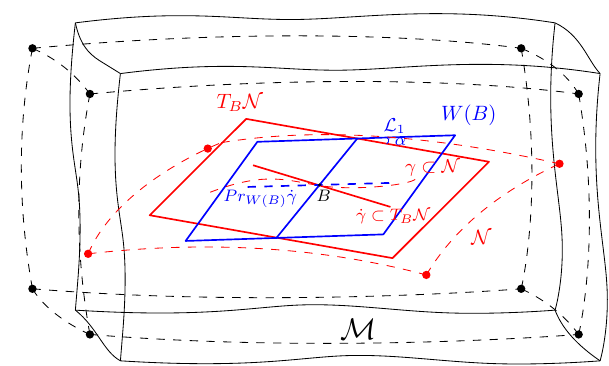}
    \caption{}
  \end{subfigure}%
  %\hspace{0.15 in}
  \begin{subfigure}[b]{0.33\textwidth}
    \centering
    \includegraphics[width=.85\linewidth]{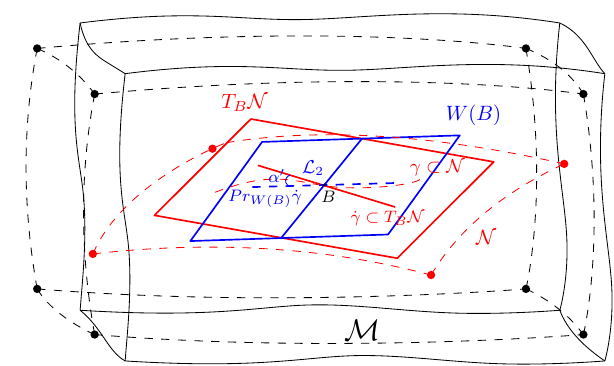}
    \caption{}
  \end{subfigure}% 
  \caption{Principal sub-manifolds. (a) Principal flow, $\alpha_B$ is the angle between $T_B \mathcal{N}$ ($k=1$) and $W(B)$ ($k=1$); (b) Principal sub-manifold (according to $\mathscr{L}_1$), $\alpha_B$ is the angle between $T_B \mathcal{N}$ ($k \geq 2$) and $W(B)$ ($k \geq 2$); (c) Principal sub-manifold (according to $\mathscr{L}_2$), $\alpha'_B$ is the angle between $\dot{\gamma}|_B \in T_B \mathcal{N}$ ($k=1$) and $W(B)$ ($k \geq 2$).}
  \label{fig:manifolds}
\end{figure}

For geometric intuition on the Lagrangians, denote the angle between the $k$-dimensional tangent space $\dot{N}$ from $\mathscr{L}_1$ and the tensor field $W(N)$ at point $B$ as $\alpha_B$ and denote the angle between the $1$-dimensional tangent vector $\dot{\gamma}$ and $W(\gamma)$ at point $B$ as $\alpha'_B$ and note that $\alpha'_B \le \alpha_B$. Figure \ref{fig:manifolds} illustrates the two angles and compares them to the situation of the principal flow.

We conclude this session with a remark on the interpretation of Theorem \ref{thm:linearspace1}. Theorem \ref{thm:linearspace1} shows that, in a flat space, the principal sub-manifold reduces to the $k$-dimensional space spanned by the $k$ eigenvectors of $\Sigma_{A, \mathcal{M}}$ when $h=\infty$. In this sense, the approach reduces to linear PCA. In connection with the principal flow, recall the similar result (Proposition 5.1, \citet{Panaretos2014}) where the first order of the principal flow on a flat space has been shown to coincide with the first principal direction if the locality parameter $h$ of the tangent covariance matrix is chosen to be infinity.

\section{A Greedy Algorithm}\label{app:7-greedy-algorithm}%\label{sec:algorithm}

%\scalebox{0.85}{
  %[ht!]
  %\caption
  \begin{algorithm}{two-dimensional principal sub-manifold}
    \label{algo1}
    \begin{enumerate}
      \item At a point $A$ (mean or other point), use the logarithm  map: ${\bf log}_A(x_i) = y_i$.
      \item Find the covariance matrix $\Sigma_{A,  \mathcal{M}}$ from $y_1,\ldots,y_n$ by (2.9).
      %using equivalently the covariance in \eqref{covar} with kernel $\quad K(y)=\mathbf{1}(y\leq 1)$
      %$$\Sigma_{A}= \frac{1}{n}\sum_{i=1}^{n}(y_i -A)^{T} (y_i-A).$$
      %$$\Sigma_{A, \mathcal{M}}= \frac{1}{\sum_{i} (\left\|y_i -A \right\| \leq h)}(y_i -A)^{T} (y_i-A).$$
      \item Let $e_1(A)$ and $e_2(A)$ be the first and second eigenvector of $\Sigma_{A, \mathcal{M}}$. Define
      $$Z_l = \epsilon' \times \Big[\cos \big(2l\pi/L\big) e_1(A) + \sin\big(2l  \pi/L\big) e_2(A)\Big], $$
      with $l = 1,\ldots,L$.
      \item Use exponential map to map $Z_l$ onto $\mathcal{M}$ so we get a set of new points ${\bf exp}_{A}(Z_l) = A_l$.
      \item Assume that we stay at point $A_{l,i}$, we are going to find $A_{l,i+1}$ ($A_{l,0} = A$ and $A_{l,1} = A_l$) via steps (a)-(g)
      \begin{enumerate}
        \item find $\Sigma_{A_{l,i}, \mathcal{M}}$.
        \item find $e_1\big(A_{l,i}\big)$ and $e_2\big(A_{l,i}\big)$.
        \item find $v_{l,i}={\bf log}_{A_{l,i}}\big(A_{l,i-1}\big)$. 
        \item find 
        $$  \tilde{v}_{l,i} =  \Big\langle v_{l,i}, e_1\big(A_{l,i}\big) \Big\rangle  e_1\big(A_{l,i}\big)  +\Big\langle v_{l,i}, e_2\big(A_{l,i}\big) \Big\rangle  e_2\big(A_{l,i}\big) $$
        where $\langle a,b\rangle  = \sum_{i=1}^n a_i b_i$ with $a = (a_1,\ldots,a_n)$ and $b = (b_1,\ldots,b_n)$.
        \item let $u_{l,i}=  \tilde{v}_{l,i}$ (or $u_{l,i}=2\tilde{v}_{l,i}-v_{l,i}$). 
        \item calculate 
        $$r_{l,i} = -\epsilon' \times  \frac{u_{l,i}}{\big\|u_{l,i} \big\|} .$$
        \item update 
        $$ A_{l,i+1} = {\bf exp}_{A_{l,i}}\big(r_{l,i} \big)  .$$
        \item stop at $A_{l,i+1}$ when 
        $$ \big\|{\bf log}_{A_{l,i+1}}(x_j) \big\|> \delta \mbox{ or }  \Big \langle {\bf log}_{A_{l,i+1}}(A_{l,i}) , {\bf log}_{A_{l,i+1}}(x_j) \Big \rangle \geq 0.$$
        for all $j = 1,\ldots,n$.
      \end{enumerate}
      \item For every $l = 1,\ldots, L$, connect $A_{l,i}$ with $A_{l,i+1}$ by $i$ we get $\mathcal{A}_l$, a ray of principal sub-manifold.
      \item {\bf Output}: all $\mathcal{A}_l$'s as in (3.15), where $1 \leq l \leq L$. 
    \end{enumerate}
  \end{algorithm}
  %}

\begin{remark}
  In Step 3, there is no difference in either forming a circle or an ellipse for small $\epsilon'$. In case of an ellipse, the axes of ellipse would be proportional to the first and second eigenvalue of $\Sigma_{A, \mathcal{M}}$. In case of a $k$-dimensional sub-manifold, Step 3 and Step 5(b), (d) will need to be updated with the first $k$ eigenvectors.
\end{remark}

\section{Convergence of the Greedy Algorithm}\label{app:8-greedy-convergence}% \label{sec:greedy-proofs}

First we simplify the algorithm from Appendix \ref{app:7-greedy-algorithm} to only represent abstract calculation steps. We call curves defined by this algorithm \textit{principal spokes}.

%[ht!]
%\caption
\begin{algorithm}
  \label{algo:greedy}{Algorithm for principal spokes of fixed length $L$.}
  \begin{enumerate}
    \item Start with a point $p^{(0)} \in \mathcal{M}$ and a tangent vector $v^{(0)} \in W (p^{(0)})$,
    \item for $i \ge 0$ and writing an orthonormal basis of $W (p^{(i+1)})$ by vectors $W_\alpha(p^{(i+1)})$ for $1 \le \alpha \le k$ let
    \begin{align*}
      p^{(i+1)} &= {\bf exp}_{p^{(i)}}\big(\epsilon' v^{(i)} \big)\\
      \widetilde{u}^{(i+1)} &= - {\bf log}_{p^{(i+1)}}(p^{(i)})\\
      u^{(i+1)} &= \sum_{\alpha =1}^{k} \sum_{j=1}^{m} \widetilde{u}^{(i+1)}_{j} W_{\alpha j}(p^{(i+1)}) W_\alpha(p^{(i+1)})\\
      %\widetilde{v}^{(i+1)} &= 2 u^{(i+1)} - \widetilde{u}^{(i+1)}\\
      v^{(i+1)} &= \frac{u^{(i+1)}}{\big\|u^{(i+1)} \big\|}
    \end{align*}
    \item stop when $(j+1) \epsilon' \ge L$.
  \end{enumerate}
\end{algorithm}

\noindent Assume a sequence of points $(p_j)_j \in \mathcal{M}$ which converges to a point $p \in \mathcal{M}$. Then smoothness of $\mathcal{M}$ yields that $d_g(p_j, p) \to \| p_j - p \|$, where $d_g$ is the geodesic distance. Thus proving any convergence statement in terms of euclidean distance immediately yields the same convergence statement in terns of geodesic distance in $\mathcal{M}$.

\begin{lemma} \label{lem:distribution-rotation}
  Assume $W$ to be $C^1$ and assume a sequence of points $(p_j)_j \in \mathcal{M}$ which converges to a point $p \in \mathcal{M}$. For any vector $v \in W(p)$ let $v_j \in W(p_j)$ define the sequence of its projections. Then the angle between $v_j$ and $v$ decreases as $\mathcal{O}(\|p-p_j\|)$.
\end{lemma}

\noindent\textit{Proof.}
%\begin{proof}
The claim follows immediately by applying the Taylor expansion in $\|p-p_j\|$ to the local spanning vector fields $X_j$ of $W$. The linear order dominates for $p_j \to p$. \QED
%\end{proof}

\begin{lemma} \label{lem:v-step}
  Assume $W$ to be $C^1$ and assume a sequence of points $p^{(j)}$ constructed by algorithm \ref{algo:greedy} with fixed $\epsilon'$. Then there is a constant $K_0$ such that
  \begin{align}
    \| v^{(j+1)} - v^{(j)} \| \le K_0 \epsilon' + o(\epsilon'). \label{eq:v-step-bound}
  \end{align}
\end{lemma}

\noindent\textit{Proof.}
%\begin{proof}
The points $p^{(j+1)}$ and $p^{(j)}$ are connected by an arc length parametrized geodesic $\gamma$, which is $C^2$, since $\mathcal{M}$ is $C^2$. Let $\gamma(0) = p^{(j)}$ and $\gamma(\epsilon') = p^{(j+1)}$ then $v^{(j)} = \dot\gamma(0)$ and $\widetilde{v}^{(j+1)} := \dot\gamma(\epsilon') = \frac{1}{\epsilon'} \widetilde{u}^{(j+1)}$ by construction. Since $\dot\gamma$ is $C^1$, we can use the Taylor expansion to note that there is a constant $A_1$ such that we have for the angle
\begin{align*}
  \angle \left( \widetilde{v}^{(j+1)}, v^{(j)} \right) &\le A_1 \epsilon' + o(\epsilon')\\
  \angle \left( v^{(j+1)}, P_{W(p^{(j+1)})} v^{(j)} \right) = \angle \left( P_{W(p^{(j+1)})} \widetilde{v}^{(j+1)}, P_{W(p^{(j+1)})} v^{(j)} \right) &\le A_1 \epsilon' + o(\epsilon') \, .
\end{align*}
From Lemma \ref{lem:distribution-rotation} we conclude that there is a constant $A_2$ such that
\begin{align*}
  \angle \left( P_{W(p^{(j+1)})} v^{(j)}, v^{(j)} \right) &\le A_2 \epsilon' + o(\epsilon')\\
  \angle \left( v^{(j+1)}, v^{(j)} \right) \le \left( v^{(j+1)}, P_{W(p^{(j+1)})} v^{(j)} \right) + \left( P_{W(p^{(j+1)})} v^{(j)}, v^{(j)} \right) &\le (A_1 + A_2) \epsilon' + o(\epsilon') \, .
\end{align*}
The claim follows immediately since $v^{(j)}$, $\widetilde{v}^{(j+1)}$ and $v^{(j+1)}$ are unit vectors. \QED

\begin{theorem}
  Assume $W$ to be $C^2$ and assume there is a $C^2$ integral submanifold $\mathcal{N}$ of $W$ through $p^{(0)}$. From the fact that $\mathcal{M}$ and $\mathcal{N}$ are $C^2$ and from Lemma \ref{lem:distribution-rotation} we can conclude the following uniform bounds in a ball of radius $L$ around $p^{(0)}$, where $P_X$ denotes orthogonal projection onto $X$ and $v$ is always normalized:
  \begin{align}
    \forall \, p \in \mathcal{M}, \, v \in T_p\mathcal{M} &: \, \| {\bf exp}_{p}\big(\epsilon' v \big) - p - \epsilon' v \| \le \frac{K_2}{2} (\epsilon')^2 + o((\epsilon')^2), \label{eq:bound1}\\
    \forall \, p \in \mathcal{N}, \, v \in T_p\mathcal{N} &: \, \| P_\mathcal{N} (p + \epsilon' v) - p - \epsilon' v \| \le \frac{K_2}{2} (\epsilon')^2 + o((\epsilon')^2), \label{eq:bound2}\\
    \forall \, p, q \in \mathcal{M}, \, v \in T_p\mathcal{N} &: \, \| P_{\mathcal{W}_k(q)} v - v \| \le K_1 \| p - q \| + o(\| p - q \|) . \label{eq:bound3}
  \end{align}
  Then the curves of length $L$ starting at $\gamma (0) = p^{(0)}$ constructed by algorithm \ref{algo:greedy} converge for step size $\epsilon' \to 0$ to curves that lie within the integral submanifold of $W$ through $p^{(0)}$.
\end{theorem}

\noindent\textit{Proof.}
%\begin{proof}
For every point $q \in \mathcal{N}$ we have $T_q \mathcal{N} = W(q)$. We show that the distance of $p^{(s)}$ from $\mathcal{N}$ is of order $\mathcal{O}\left((\epsilon')^2 \sum_{j=0}^{s-1} (1 + K_1 \epsilon')^{j}\right)$. The proof is done by induction. Note that we will simply bound the distance between the points $p^{(j)}$ and some corresponding points $q^{(j)} \in \mathcal{N}$. Since these $q^{(j)}$ need not be the closest points in $\mathcal{N}$ to the $p^{(j)}$, we derive, strictly speaking, upper bounds for the distances of the $p^{(s)}$ from $\mathcal{N}$.

As above let $v^{(0)} \in W (p^{(0)}) = T_{p^{(0)}} \mathcal{N}$ the initial direction of the curve. Then, using that $v^{(0)}$ is tangent to $\mathcal{N}$ and $p^{(1)} = {\bf exp}_{p^{(0)}}\big(\epsilon' v^{(0)} \big)$, we get from equation \eqref{eq:bound1}
\begin{align*}
  \| p^{(1)} - p^{(0)} - \epsilon' v^{(0)} \| = \frac{K_2}{2} (\epsilon')^2 + o((\epsilon')^2) .
\end{align*}
Let $q^{(1)}$ be the orthogonal projection of $p^{(0)} + \epsilon' v^{(0)}$ to $\mathcal{N}$, which is unique for small enough $\epsilon'$. Then by equation \eqref{eq:bound2} we get $\| q^{(1)} - p^{(0)} - \epsilon' v^{(0)} \| = \frac{K_2}{2} (\epsilon')^2 + o((\epsilon')^2)$ and thus $\|q^{(1)} - p^{(1)} \| \le K_2 (\epsilon')^2 + o((\epsilon')^2)$. This concludes the beginning of the induction.

Now assume $\|q^{(s)} - p^{(s)} \| \le K_2 (\epsilon')^2 \sum_{j=0}^{s-1} (1 + K_1 \epsilon')^{j} + o\left((\epsilon')^2 \sum_{j=0}^{s-1} (1 + K_1 \epsilon')^{j}\right)$ where $q^{(s)}$ is some point on $\mathcal{N}$. As before we have
\begin{align*}
  \| p^{(s+1)} - p^{(s)} - \epsilon' v^{(s)} \| = \frac{K_2}{2} (\epsilon')^2 + o((\epsilon')^2) .
\end{align*}
Let $w^{(s)}$ be the projection of $v^{(s)}$ to $T_{q^{(s)}} \mathcal{N}$, then equation \eqref{eq:bound3} yields
\begin{align*}
  \epsilon' \| w^{(s)} - v^{(s)} \| \le K_1 \epsilon' \| q^{(s)} - p^{(s)} \| + o(\epsilon' \| q^{(s)} - p^{(s)} \|) .
\end{align*}
Let $q^{(s+1)}$ be the orthogonal projection of $q^{(s)} + \epsilon' w^{(s)}$ to $\mathcal{N}$. As above
\begin{align*}
  \| q^{(s+1)} - q^{(s)} - \epsilon' w^{(s)} \| = \frac{K_2}{2} (\epsilon')^2 + o((\epsilon')^2).
\end{align*}
From the above considerations we can thus conclude
\begin{align*}
  \| p^{(s+1)} - q^{(s+1)} \| &\le \| p^{(s)} + \epsilon' v^{(s)} - q^{(s)} - \epsilon' w^{(s)}\| + K_2 (\epsilon')^2 + o((\epsilon')^2)  \\
  &\le \|q^{(s)} - p^{(s)} \| + K_1 \epsilon' \|q^{(s)} - p^{(s)} \| + K_2 (\epsilon')^2 + o(\epsilon' \| q^{(s)} - p^{(s)} \|) + o((\epsilon')^2)\\
  &\le K_2 (\epsilon')^2 \sum_{j=1}^{s} (1 + K_1 \epsilon')^{j} + K_2 (\epsilon')^2 + o\left((\epsilon')^2 \sum_{j=1}^{s} (1 + K_1 \epsilon')^{j}\right) + o((\epsilon')^2\\
  &= K_2 (\epsilon')^2 \sum_{j=0}^{s} (1 + K_1 \epsilon')^{j} + o\left((\epsilon')^2 \sum_{j=0}^{s} (1 + K_1 \epsilon')^{j}\right) .
\end{align*}
This concludes the induction step.

Now, note that
\begin{align*}
  \sum_{j=0}^{s-1} (1 + K_1 \epsilon')^{j} &= \sum_{j=0}^{s-1} \sum_{l=0}^{j} \binom{j}{l} (K_1 \epsilon')^l = \sum_{l=0}^{s-1} (K_1 \epsilon')^l \sum_{j=l}^{s-1} \binom{j}{l} = \sum_{l=0}^{s-1} \binom{s}{l+1} (K_1 \epsilon')^l\\
  &\le \sum_{l=0}^{s-1} \frac{s^{l+1}}{(l+1)!} (K_1 \epsilon')^l = \frac{1}{K_1 \epsilon'} \sum_{l=1}^{s} \frac{(K_1 s \epsilon')^l}{l!} \le \frac{\exp (K_1 s \epsilon') - 1}{K_1 \epsilon'},
\end{align*}
where we use the hockey-stick identity in the first line.

To achieve a curve of length at least $L$, we need $s = \ceil{\frac{L}{\epsilon'}}$ steps of length $\epsilon'$. The maximum distance of the curve from $\mathcal{N}$ is thus bounded by $d_{\text{max}} = \frac{K_2(\exp (K_1 L) - 1)}{K_1} \epsilon' + o(\epsilon')$ which goes to $0$ as $\epsilon'\to 0$. \QED
%\end{proof}

\begin{remark}
  Note that, by definition of an integral manifold, $\dot{N}_{j\alpha} = W_{j\alpha}$ at every point in $\mathcal{N}$ and therefore $\mathcal{N}$ can be understood as a principal submanifold in our setting.
\end{remark}

\begin{corollary}
  Assume $W$ to be $C^2$ and involutive, then there is a $C^2$ integral submanifold $\mathcal{N}$ of $W$ through $p^{(0)}$. Then the curves of length $L$ starting at $\gamma (0) = p^{(0)}$ constructed by algorithm \ref{algo:greedy} converge for step size $\epsilon' \to 0$ to curves that lie within the integral submanifold of $W$ through $p^{(0)}$.
\end{corollary}

\section{Reflection versus Projection}\label{app:9-reflection}%\label{app:reflection}

To illustrate the properties of the projection and reflection algorithms we consider the simpler case of the principal flow to make illustration easier. However, the qualitative results can be generalized to the principal sub-manifold setting. In general, the projection algorithm will incur an error proportional to the local curvature of true integral curves of the vector field $W$ at each step. If the vector field changes strongly, this leads to an increasingly bad fit. The reflection algorithm does not incur an error when following a field whose integral curves have constant curvature, which means they are circles. In the general case, it will incur an error at each step which is proportional to the change of curvature. In this sense, the error is of ``higher order'' and can be expected to be smaller in general.

\begin{figure}[ht!]
  \centering
  \begin{subfigure}[b]{0.333\textwidth}
    \centering
    \includegraphics[width=.85\linewidth]{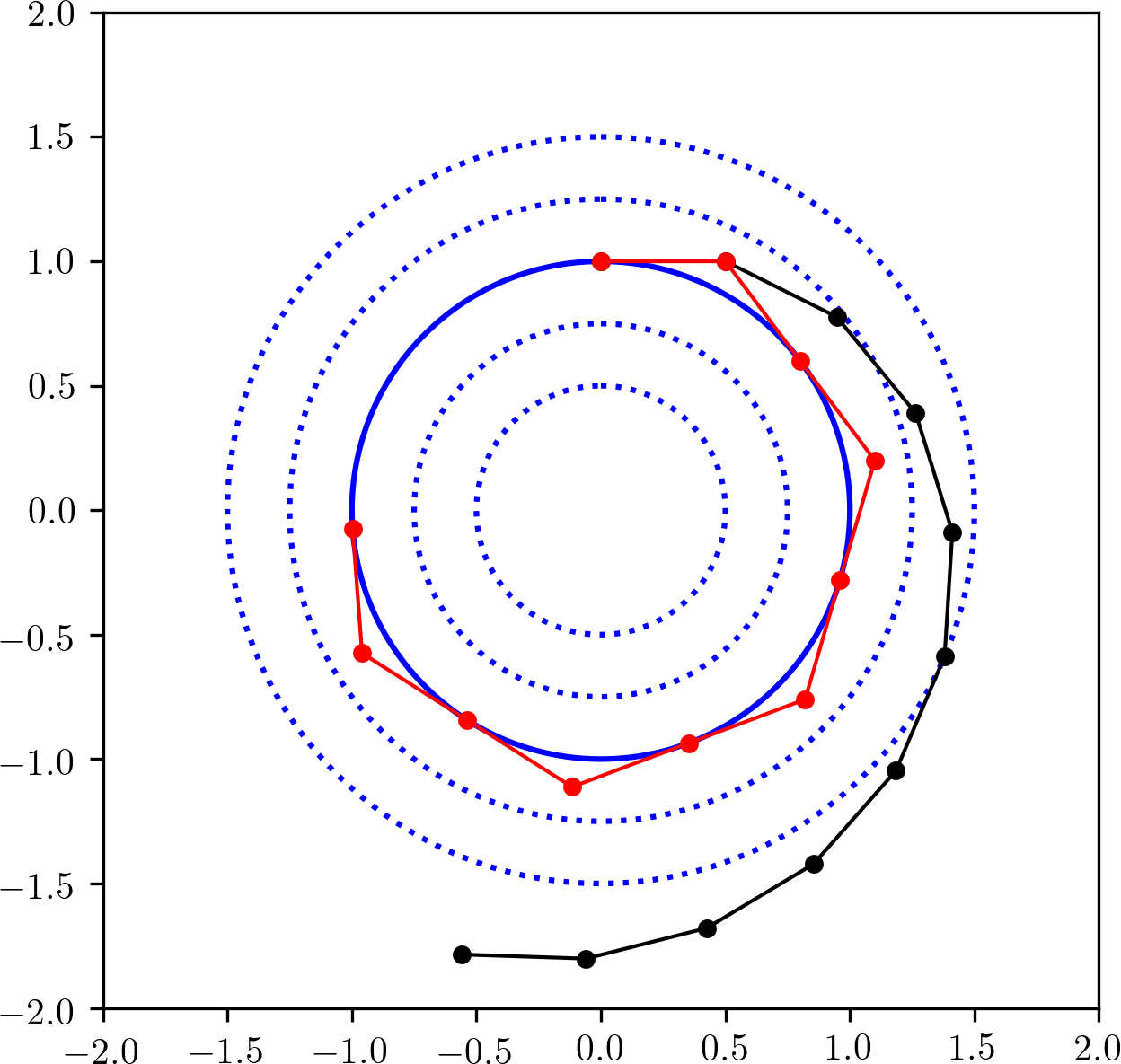}
    \caption{}
  \end{subfigure}%
  %\hspace{0.15 in}
  \begin{subfigure}[b]{0.32\textwidth}
    \centering
    \includegraphics[width=.85\linewidth]{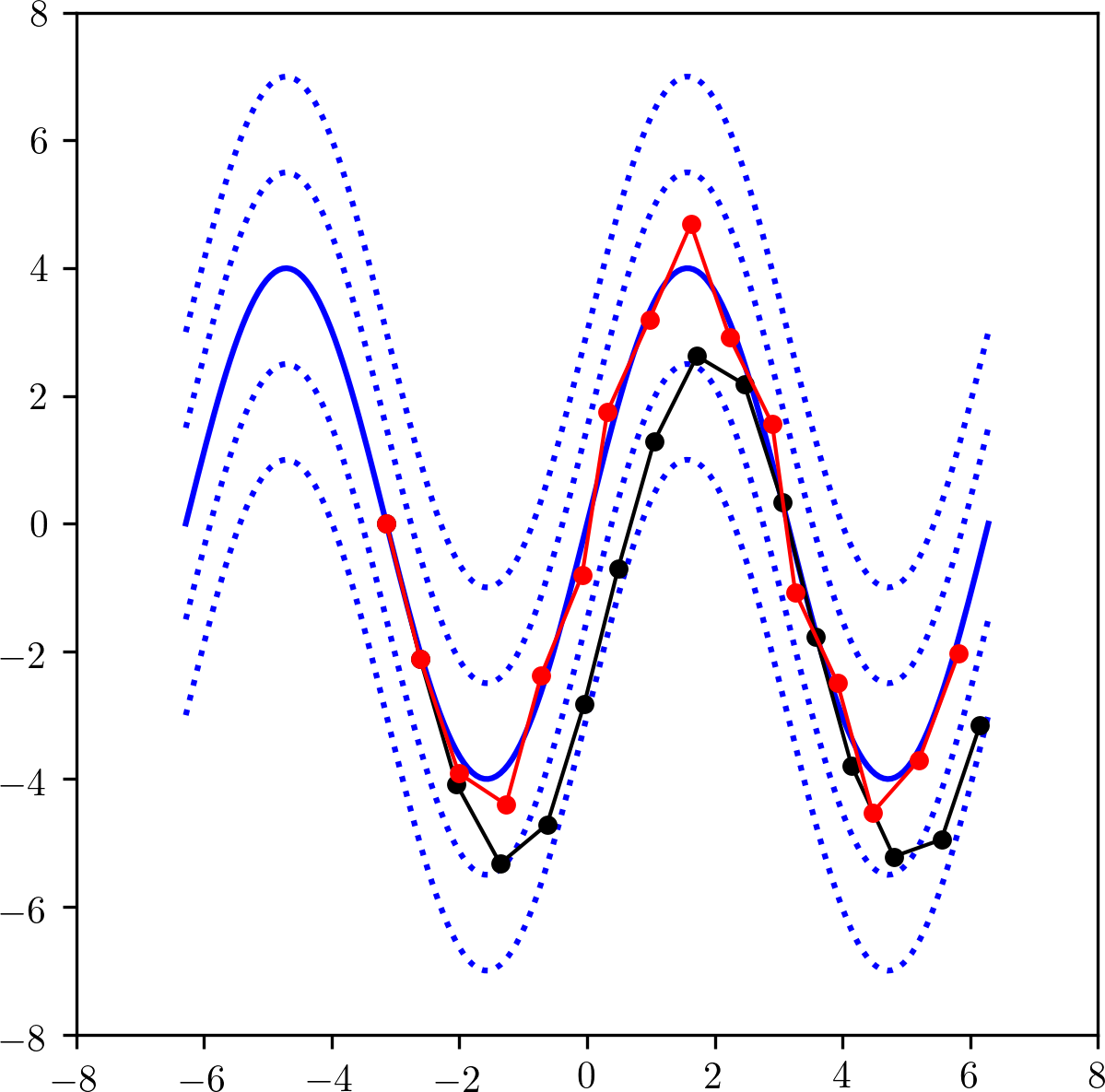}
    \caption{}
  \end{subfigure}%
  \begin{subfigure}[b]{0.32\textwidth}
    \centering
    \includegraphics[width=.85\linewidth]{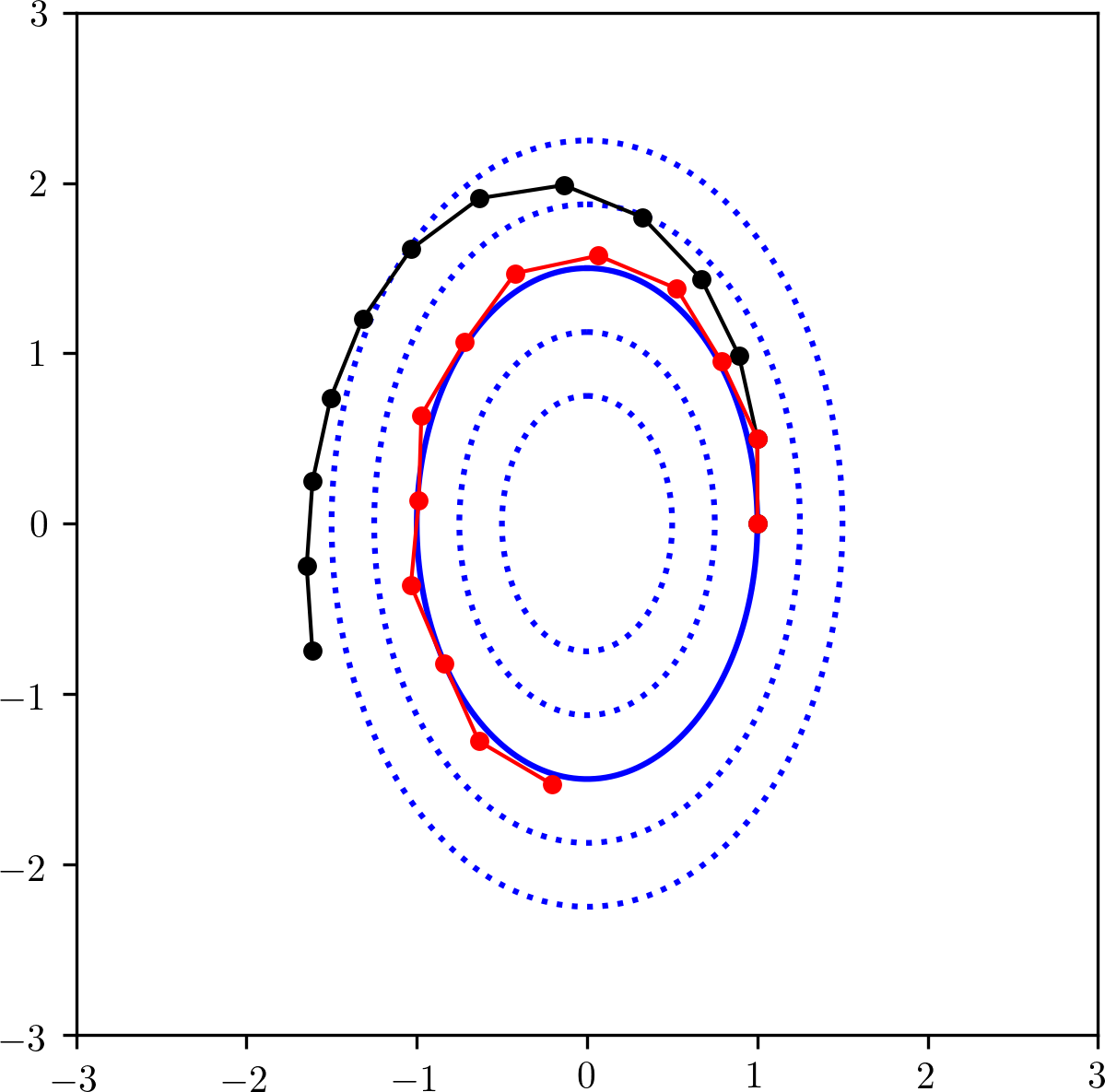}
    \caption{}
  \end{subfigure}\\
  \begin{subfigure}[b]{0.333\textwidth}
    \centering
    \includegraphics[width=.85\linewidth]{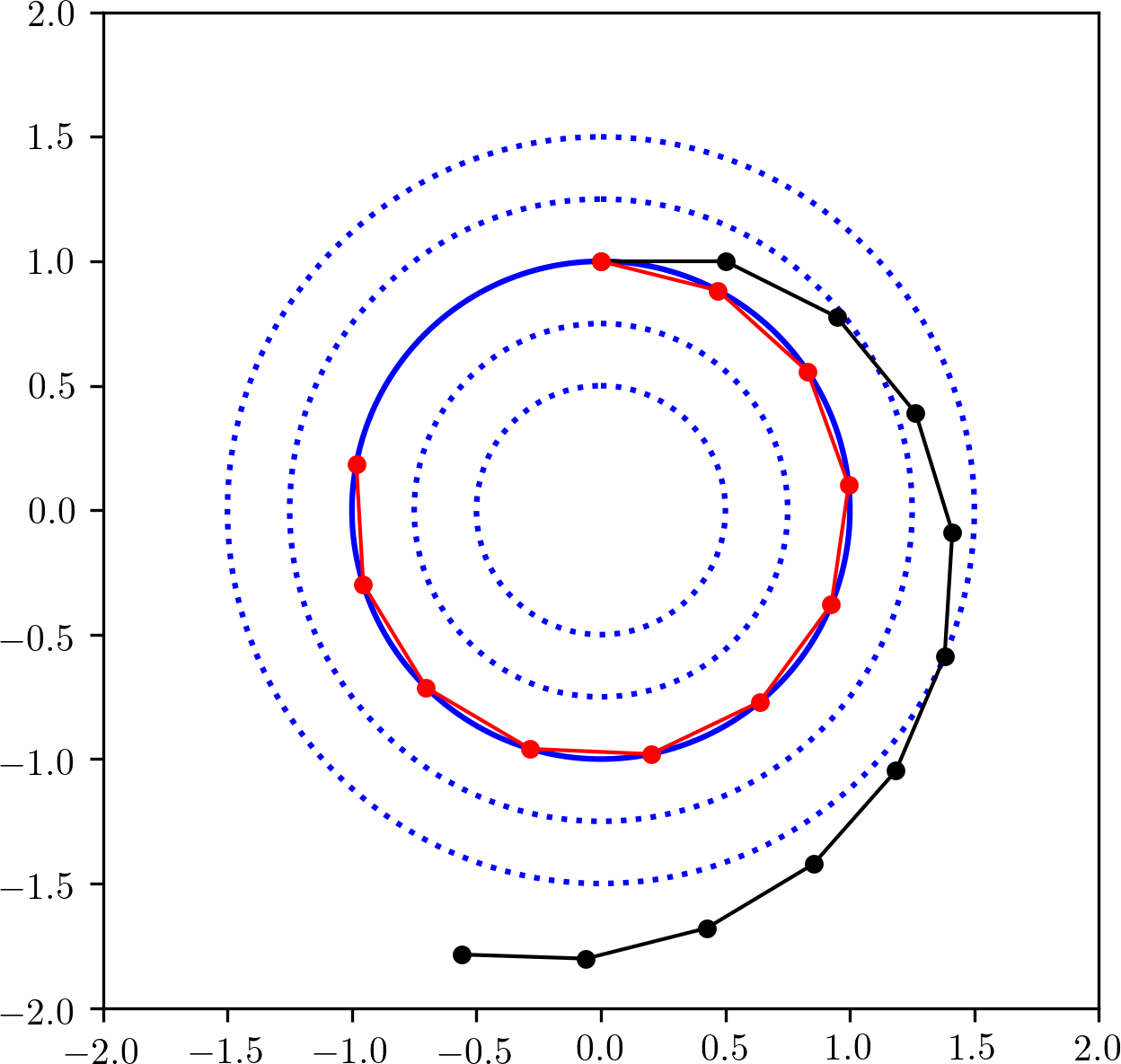}
    \caption{}
  \end{subfigure}%
  %\hspace{0.15 in}
  \begin{subfigure}[b]{0.32\textwidth}
    \centering
    \includegraphics[width=.85\linewidth]{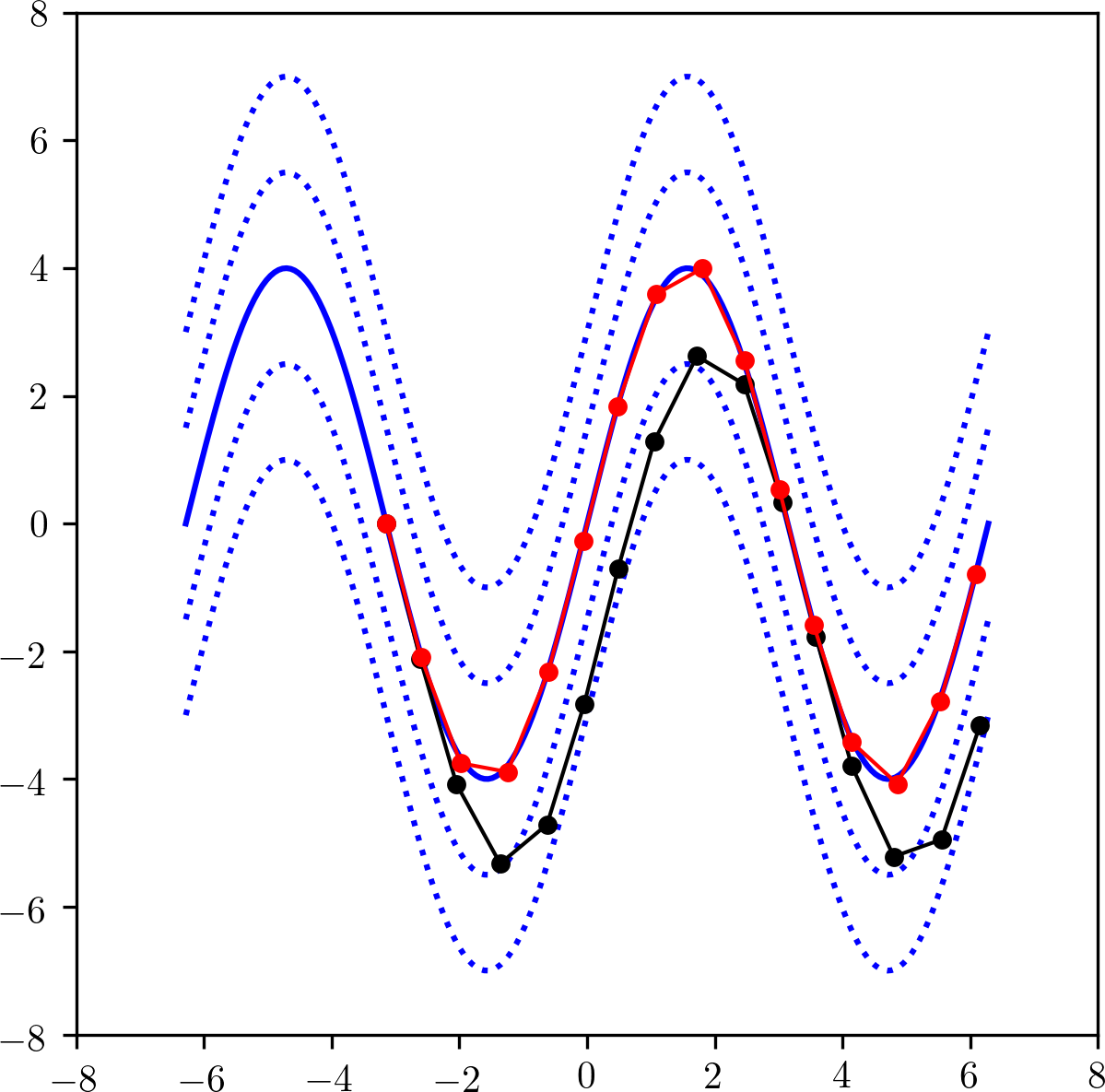}
    \caption{}
  \end{subfigure}%
  \begin{subfigure}[b]{0.32\textwidth}
    \centering
    \includegraphics[width=.85\linewidth]{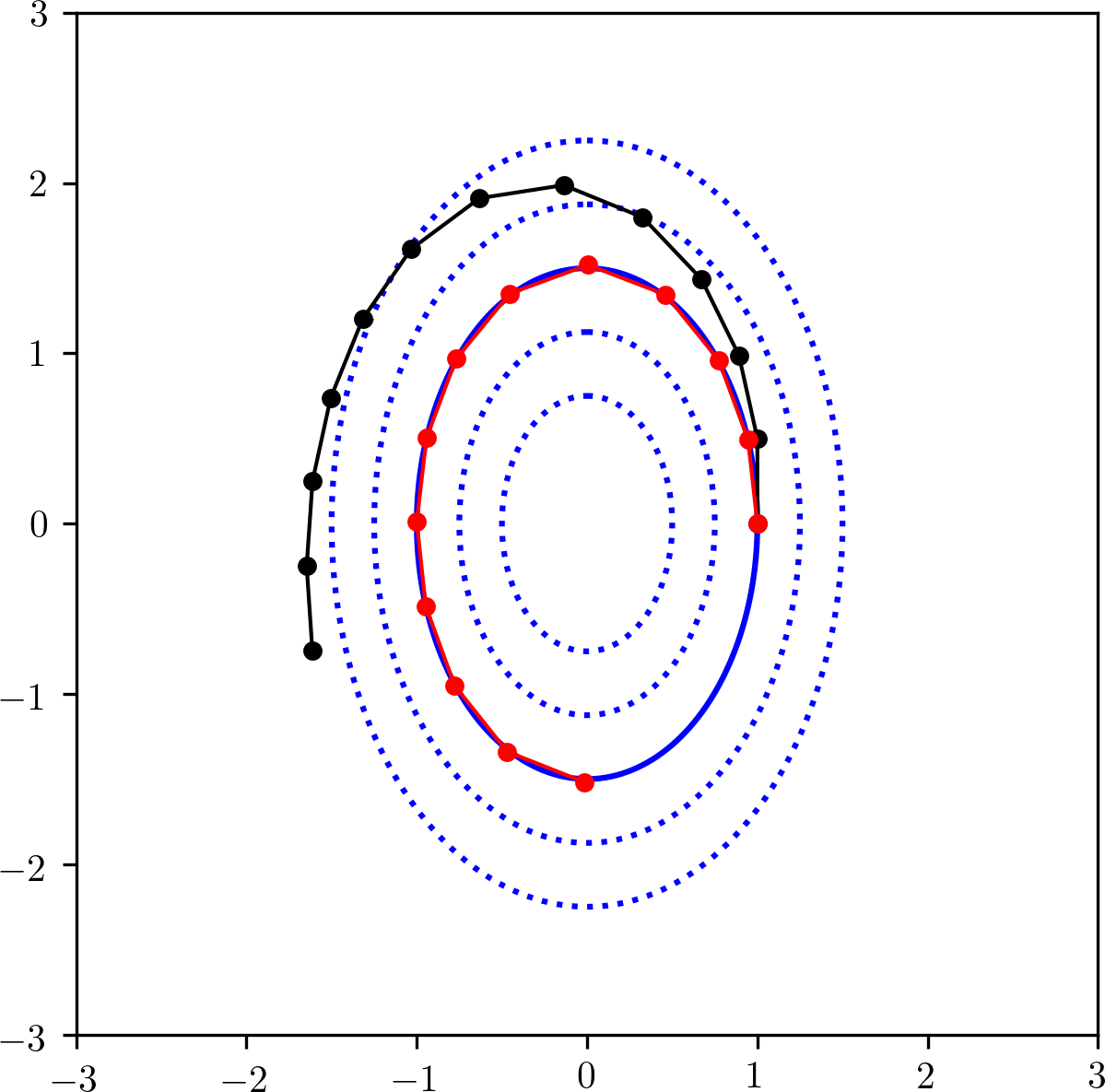}
    \caption{}
  \end{subfigure}% 
  \caption{Comparing the results of the projection and reflection algorithms for a simple principal flow. The blue lines indicate true integral curves of the principal flow, thus the local PCA vector field $W$ is always tangential to these lines. The solid blue line is the curve on which the algorithm starts and which it should ideally follow. The black points and lines represent steps of the algorithm with projection and the red points and lines represent steps of the algorithm with reflection. In Figures (a)-(c) all steps of the reflection algorithm are shown, whereas in Figures (d)-(f) the reflection algorithm uses halved step size and only every second point is used.}
  \label{fig:reflections}
\end{figure}

Figure \ref{fig:reflections} illustrates three example cases: spherical, elliptical and sinus integral curves. To highlight the qualitative behaviour, the step sizes of the algorithms are strongly exaggerated. One can clearly see that the projection algorithm determines a curve which successively departs from the true integral curve. The reflection algorithm, in contrast, stays close to the true curve. It is thus also much less susceptible to perturbations in the vector field $W$.

%\begin{figure}[ht!]
%  \centering
%  \begin{subfigure}[b]{0.5\textwidth}
  %    \centering
  %    \includegraphics[width=.85\linewidth]{PDF/ellipse1.png}
  %    \caption{}
  %  \end{subfigure}%
%  %\hspace{0.15 in}
%  \begin{subfigure}[b]{0.5\textwidth}
  %    \centering
  %    \includegraphics[width=.85\linewidth]{PDF/alt_ellipse1.png}
  %    \caption{}
  %  \end{subfigure}% 
%  \caption{Comparing the results of the projection and reflection algorithms for a simple principal flow. The blue lines indicate true integral curves of the principal flow, thus the local PCA vector field $W$ is always tangential to these lines. The solid blue line is the curve on which the algorithm starts and which it should ideally follow. The black points and lines represent steps of the algorithm with projection and the red points and lines represent steps of the algorithm with reflection. In Figure (a) all steps of the reflection algorithm are shown, whereas in Figure (b) the reflection algorithm uses halved step size and only every second point is used.}
%  \label{fig:reflections}
%\end{figure}

Figure \ref{fig:reflections} also shows that the integral curves for the reflection algorithm are much smoother, if only every second step is taken into account. This is due to the fact that the curve starts out tangential to the true integral curve, changes direction after the first step and is then again close to tangential after the second step, if curvature is only slowly varying. Therefore, the change of direction after the second step and indeed every even numbered step is much smaller than after an odd numbered step, if curvature changes slowly.

\section{Simulated data}\label{app:10-simulations-s3}
To further investigate the behavior of the principal sub-manifold as dependent on the configuration of the data points and the choice of scale parameter, we considered three sets of examples on $S^3$. We chose this manifold as a ``test manifold'' since it represents one of the most natural spaces from which the projected sub-manifold can be well understood, and since it provides a manifold for which we can compare the principal sub-manifolds with the principal geodesic. We observe here that the full manifold variation of the sub-manifold from the data can be very complicated; hence, we do not look at them on a quantitative basis, but rather investigate them qualitatively.

The first set of examples involves five data clouds in $S^3$, each presenting a different curvature. As the curvature is non-constant, the Fr\'{e}chet mean is no longer a good starting point for the principal sub-manifold. Instead, we choose the center of symmetry for each data set as a starting point. The first and second data cloud are constructed in a way that the first three coordinates of each point are concentrated around a one dimensional curve; the configuration of the third and fourth are such that the points are on a two-dimensional surface/plane; the fifth one is much more diffuse: the points lie on a sea-wave-like surface.      

For each one of them, a two-dimensional principal sub-manifold was fitted using three different bandwidths $h$ and the results are presented. The results indicate that the corresponding sub-manifolds perform well in capturing the local and global variation. We note that the sub-manifold fits well for data Cloud 1 no matter what scale of $h$ is used (Figure 6(a)-(c)); the sub-manifold seems to capture a finer structure with a reduced value for $h$ for data Cloud 2 (Figure 6(d)-(f)): this can be also seen as the first principal direction evolves with the scale of $h$. When the surface becomes two-dimensional for data Clouds 3 and 4, the principal sub-manifolds also excel: the fitted sub-manifold remains unchanged for different $h$ as the surface is flat (Figure 6(g)-(i)), while it picks up the appropriate structure with a reduced $h$ for the bent surface (Figure 6(j)-(l)); for data Cloud 5 (Figure 6(m)-(o)), it is more obvious that using a sub-manifold is more appealing than using only a curve or its equivalent; the sub-manifold fits the data points surprisingly well even with a surface of high curvature.

To probe how a sub-manifold performs with a noisy surface, we created four sets of data by blurring the sea-wave-like surface aforementioned with increasing levels of noise. Although the data reside in $S^3$, most of the variation originates around but not exactly on the surface. By knowing how the data points lie around the surface, we can get a sense of such variability. As we should no longer look at the local scale when the points tend to have large variability, we found a two-dimensional sub-manifold by choosing an appropriate scale parameter $h$, potentially a larger one, for each data set. In Figure \ref{example-subman2}(a), when there is no noise, it is expected that the sub-manifold would capture total variation of the data in the projected space. When the noise increases (Figure \ref{example-subman2}(b), (c), and (d)), where all points are more diffused away from the underlying projected surface, the fitted sub-manifold is, although not a perfect sub-manifold, still well explaining for total data variability.  

%Since the curvature varies, thus the variation is global. 
\begin{figure}
  \begin{center}
    \begin{tabular}{|c|c|c|c|c|} 
      \hline
      \addheight{\includegraphics[width=0.17\textwidth]{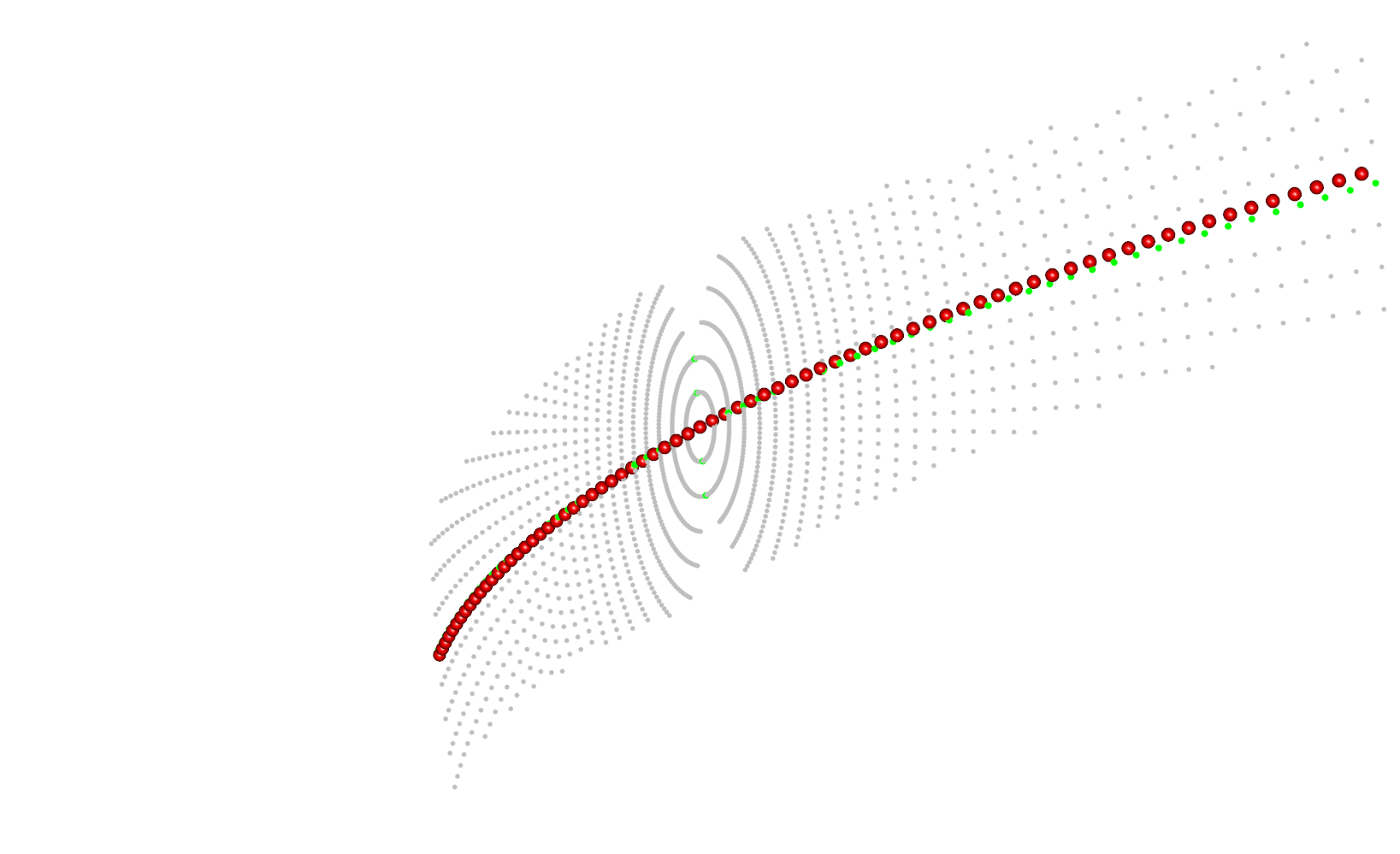}} &
      \addheight{\includegraphics[width=0.17\textwidth]{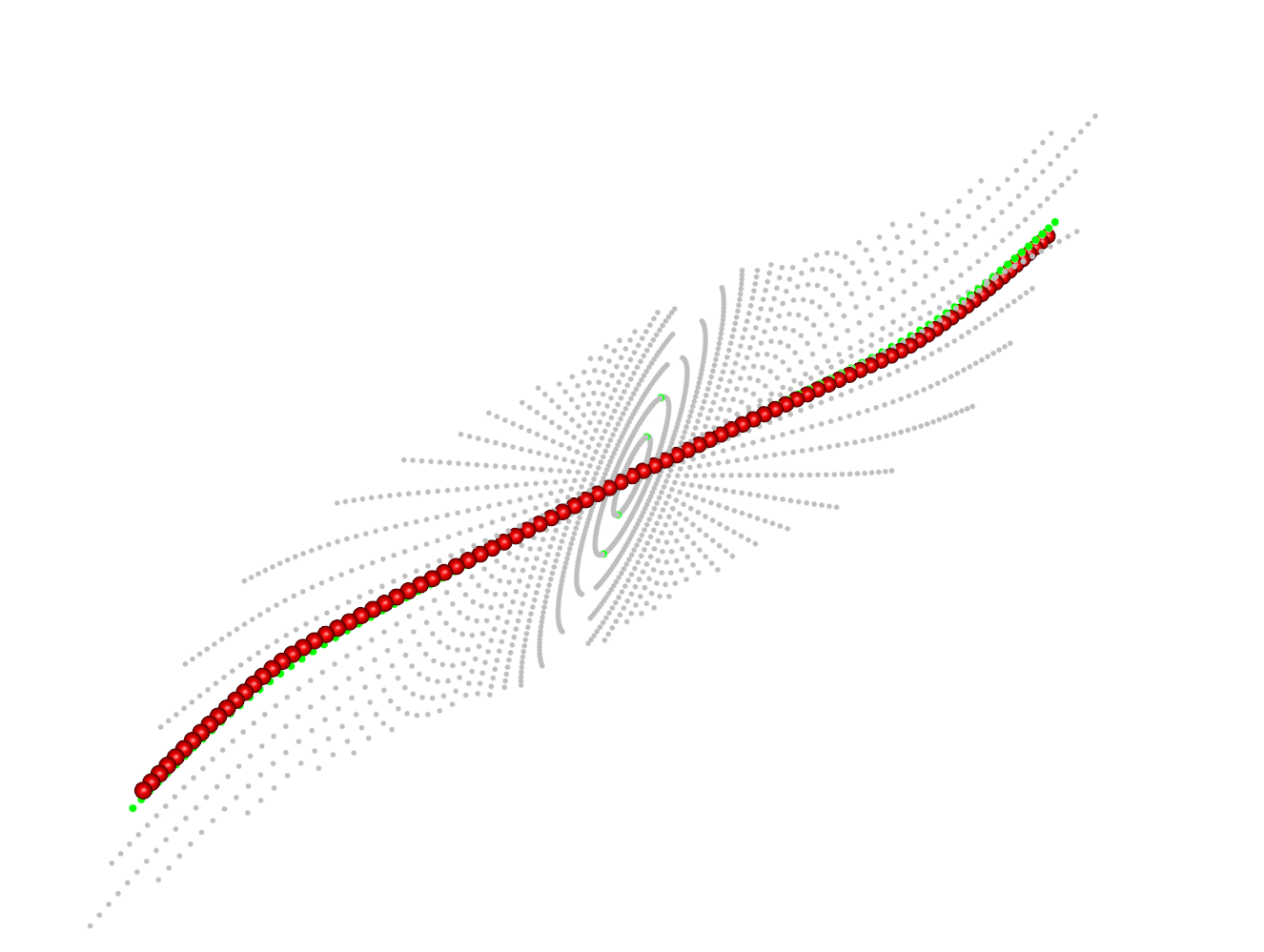}} & 
      \addheight{\includegraphics[width=0.17\textwidth]{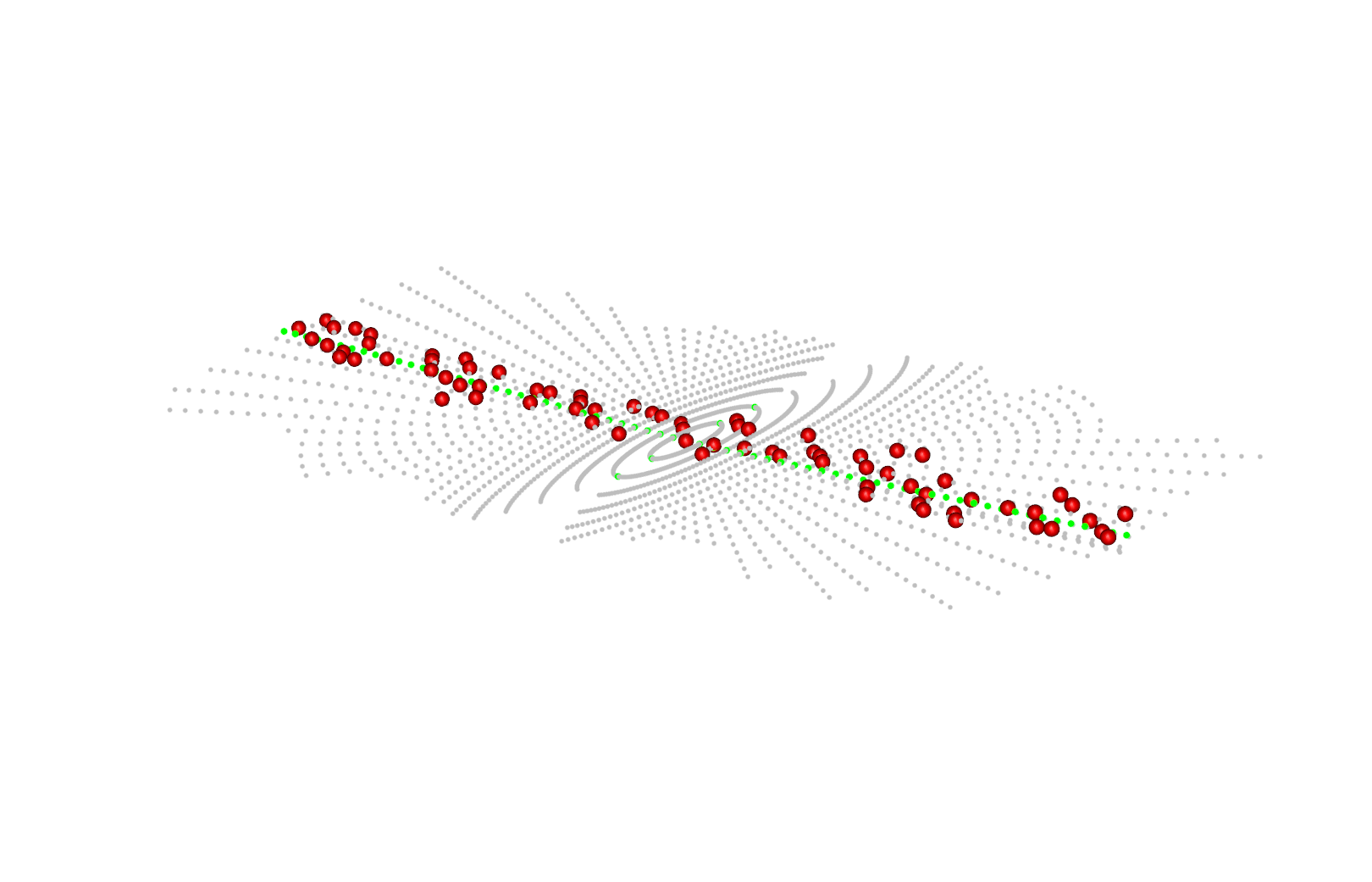}} &
      \addheight{\includegraphics[width=0.17\textwidth]{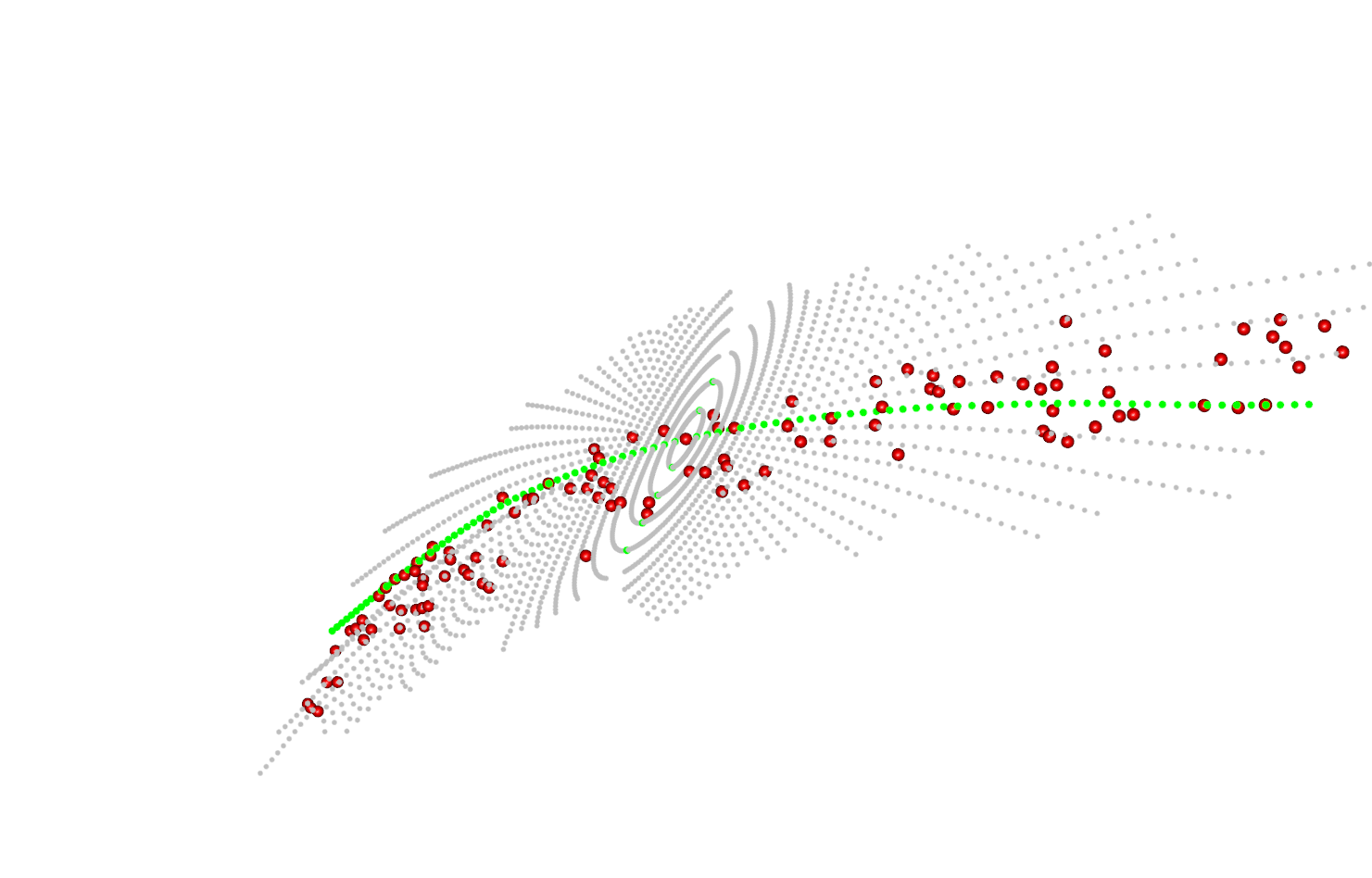}} & 
      \addheight{\includegraphics[width=0.17\textwidth]{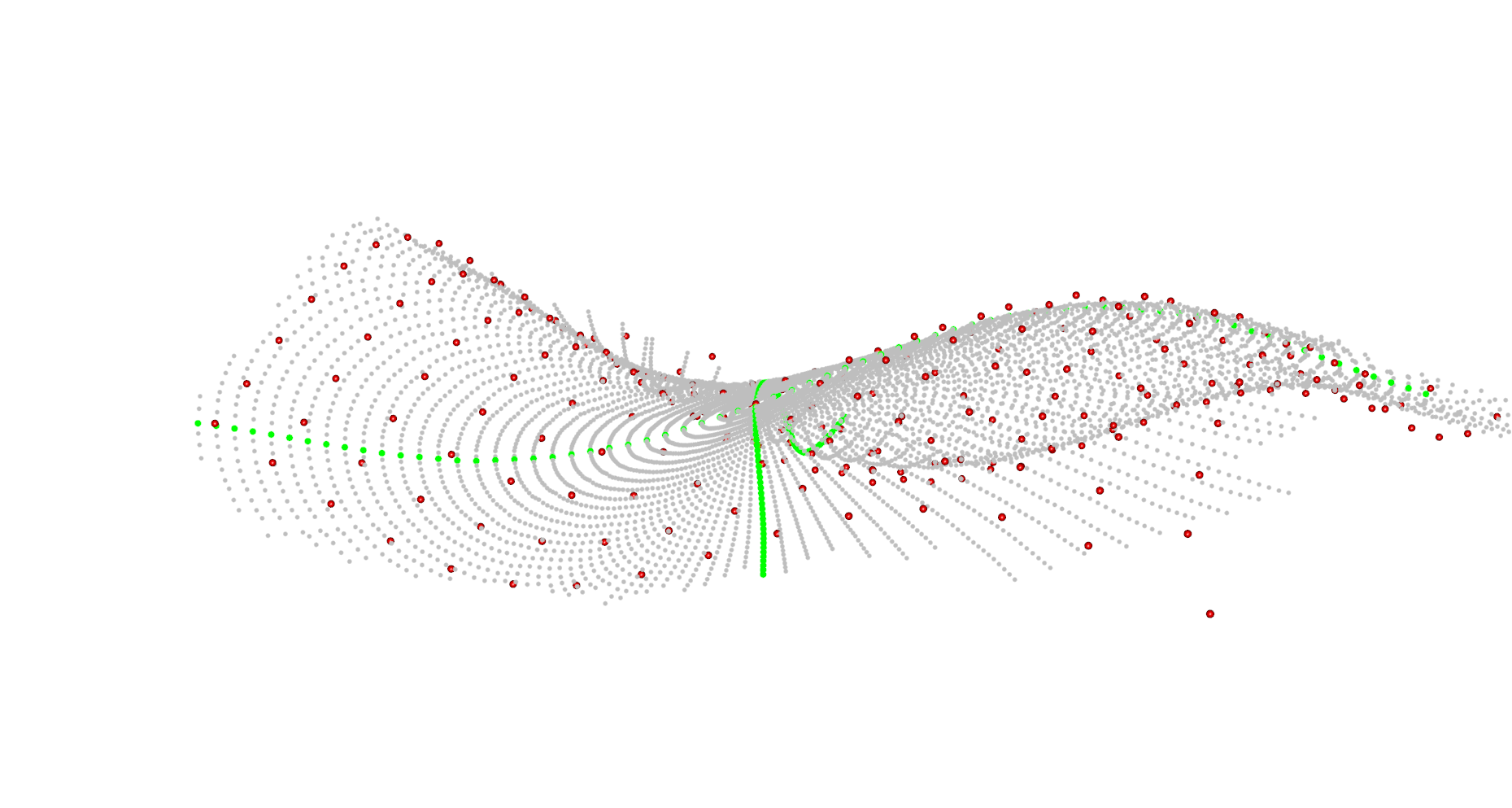}} \\
      \small (a) & (d) &  (g) & (j) & (m) \\
      \hline
      \addheight{\includegraphics[width=0.17\textwidth]{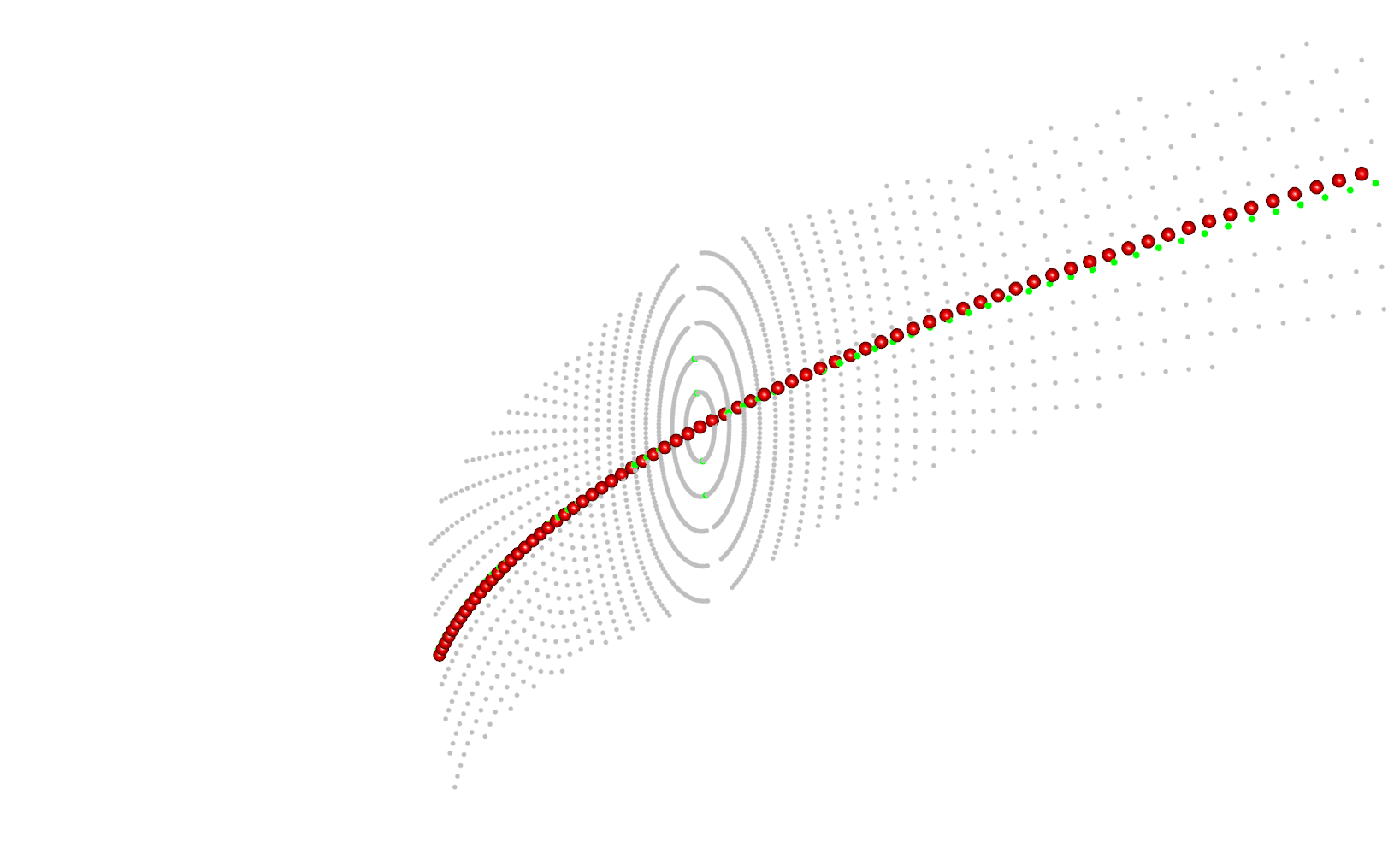}} &
      \addheight{\includegraphics[width=0.17\textwidth]{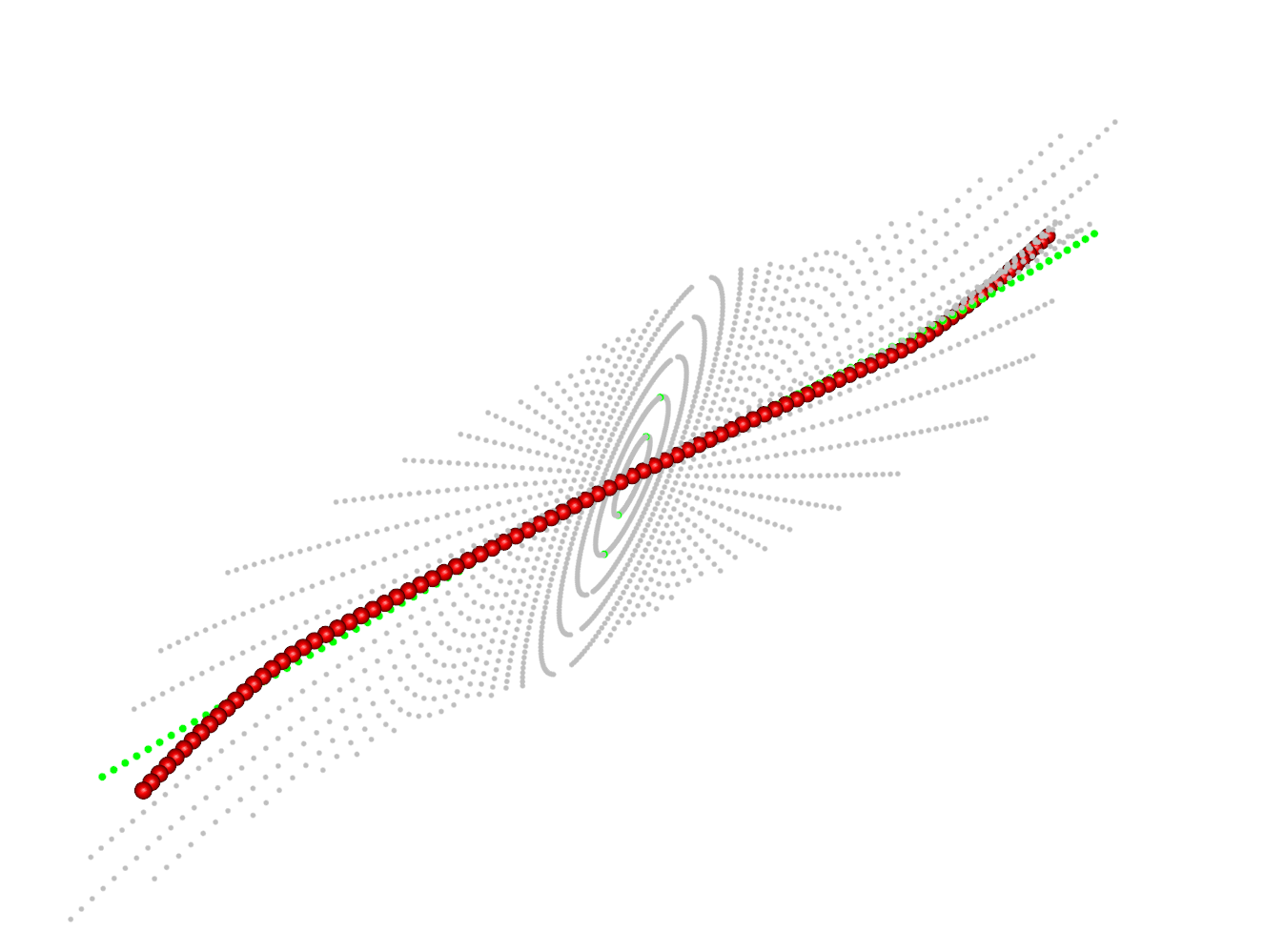}} & 
      \addheight{\includegraphics[width=0.17\textwidth]{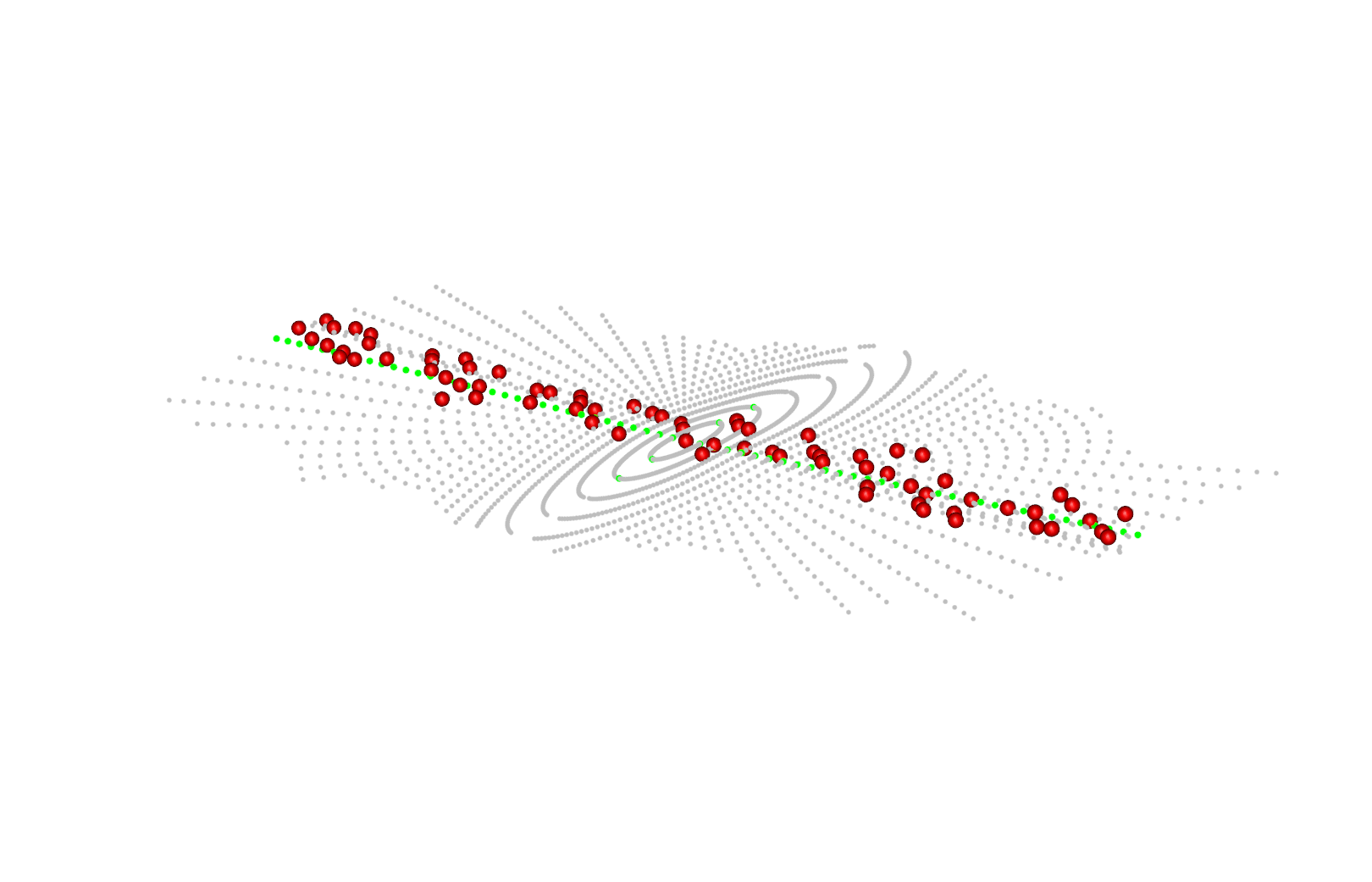}}&
      \addheight{\includegraphics[width=0.17\textwidth]{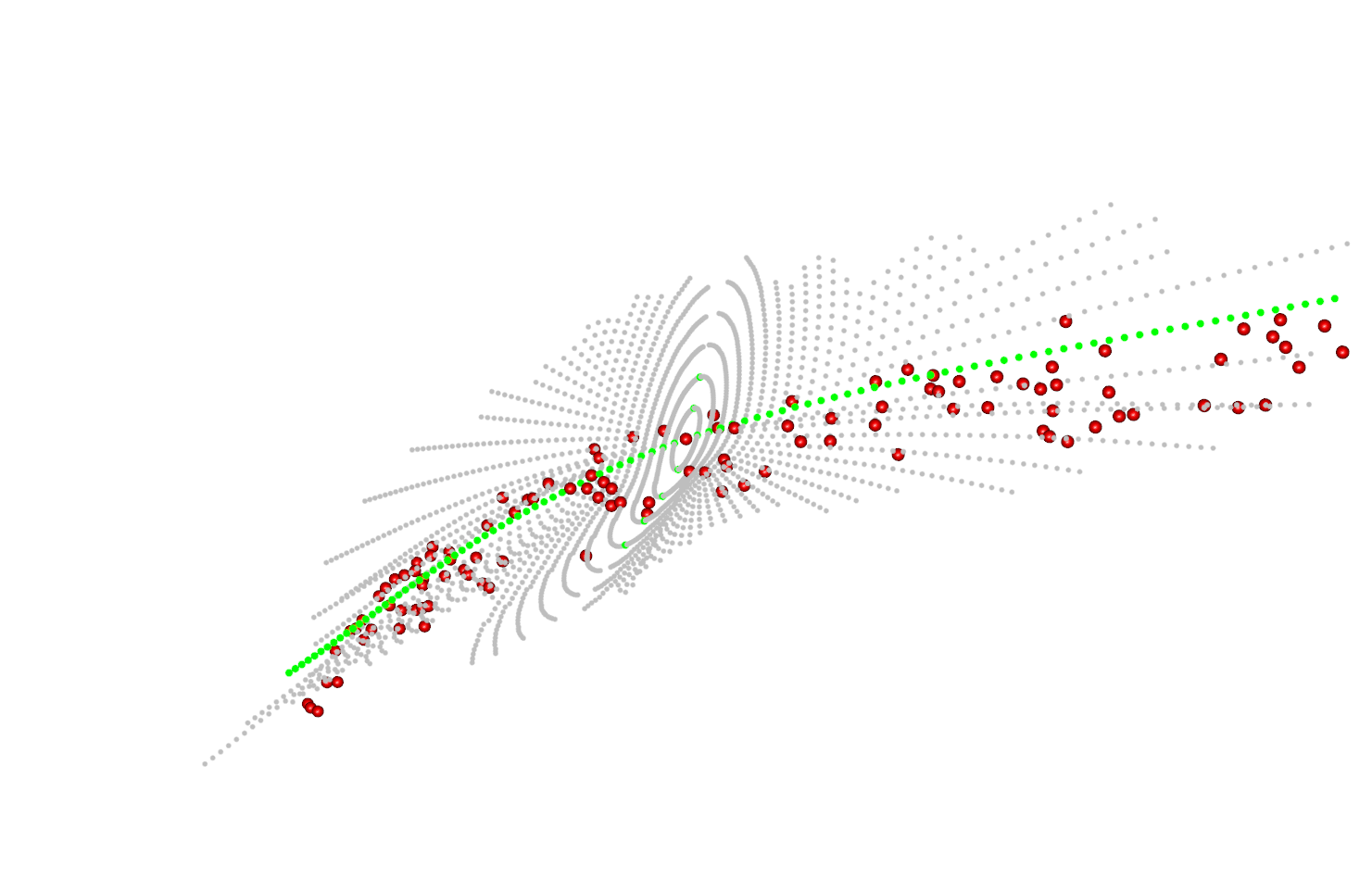}} & 
      \addheight{\includegraphics[width=0.17\textwidth]{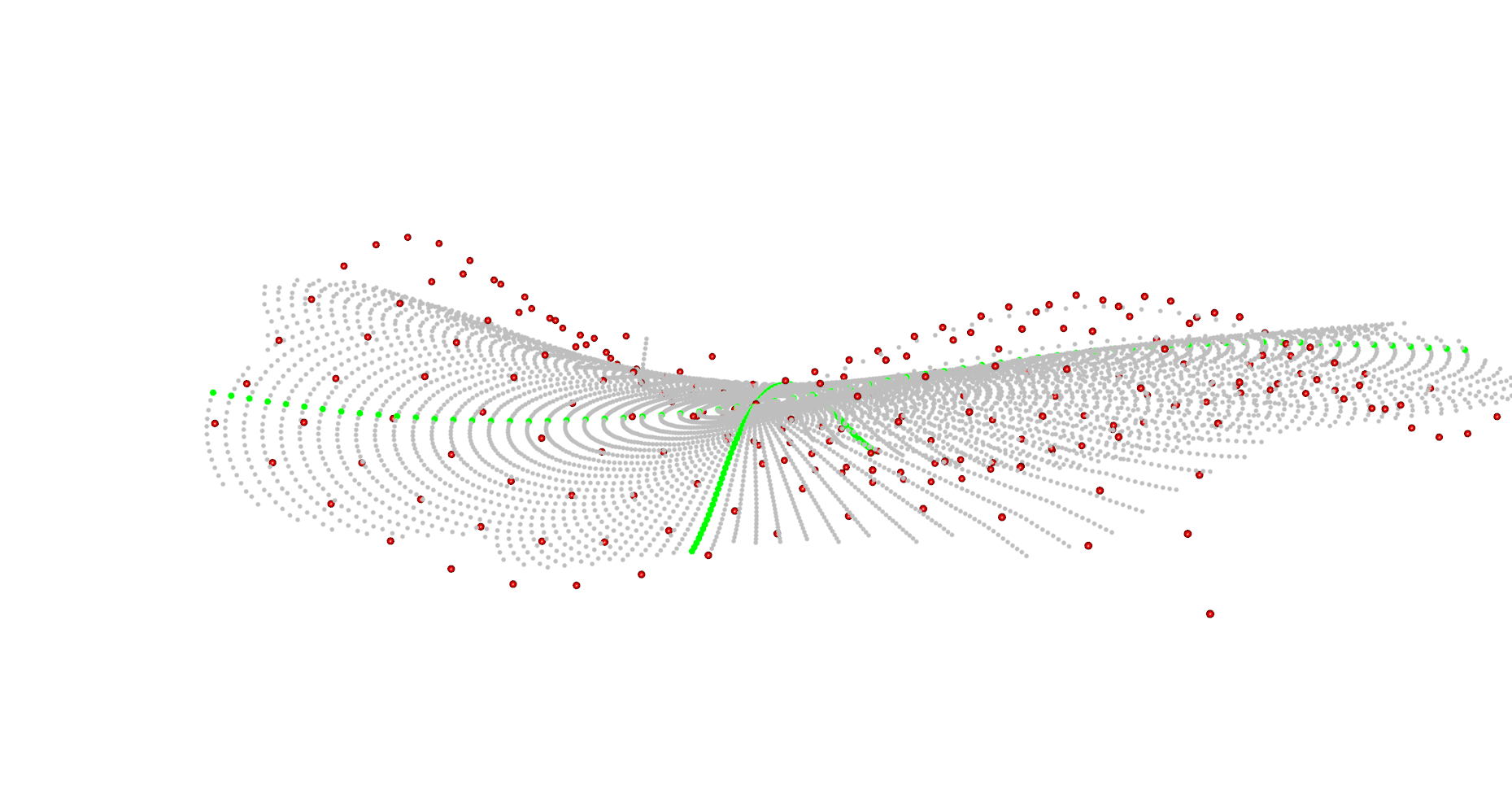}}\\
      \small (b) & (e) &  (h) &  (k) & (n) \\
      \hline
      \addheight{\includegraphics[width=0.17\textwidth]{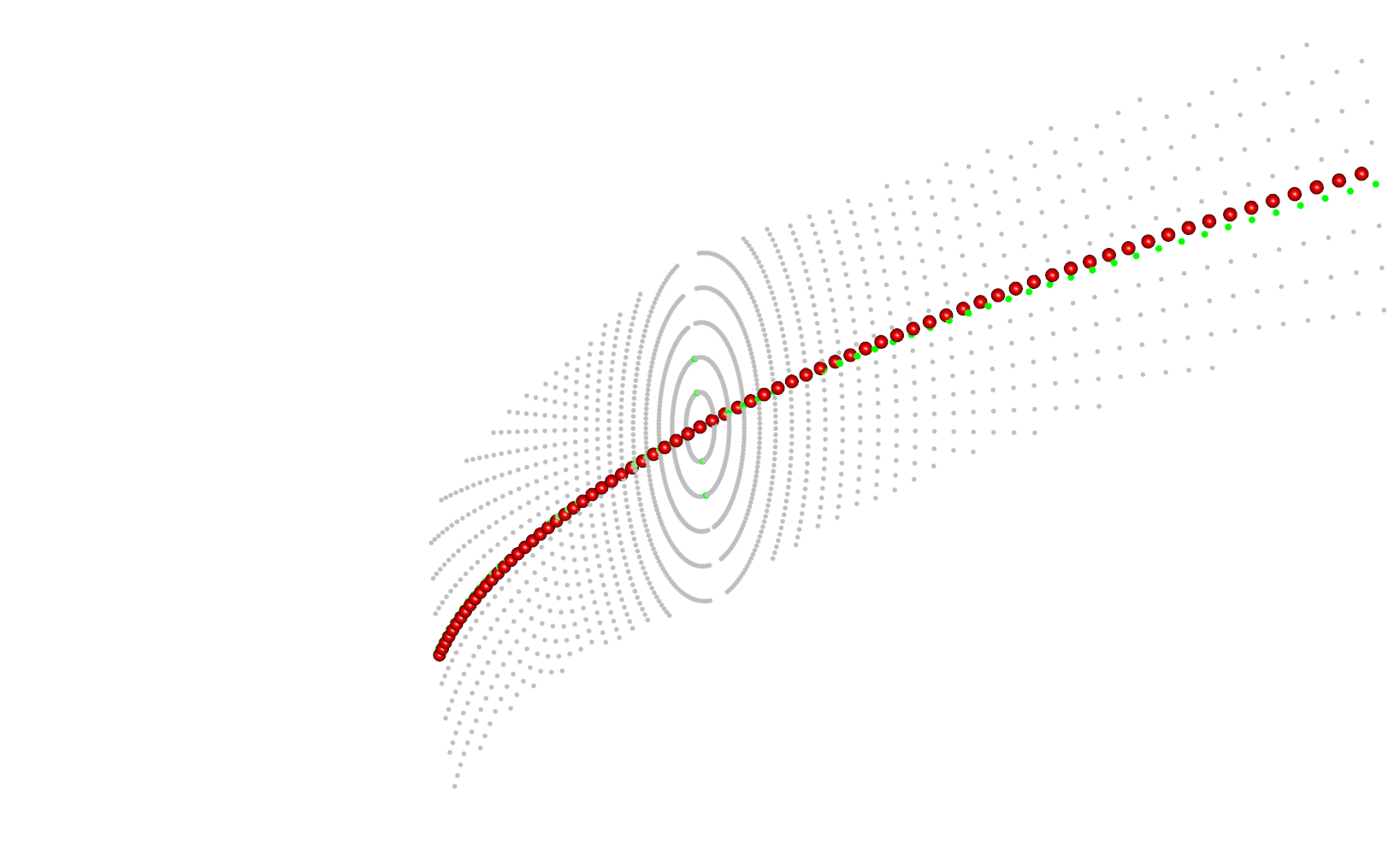}} &
      \addheight{\includegraphics[width=0.17\textwidth]{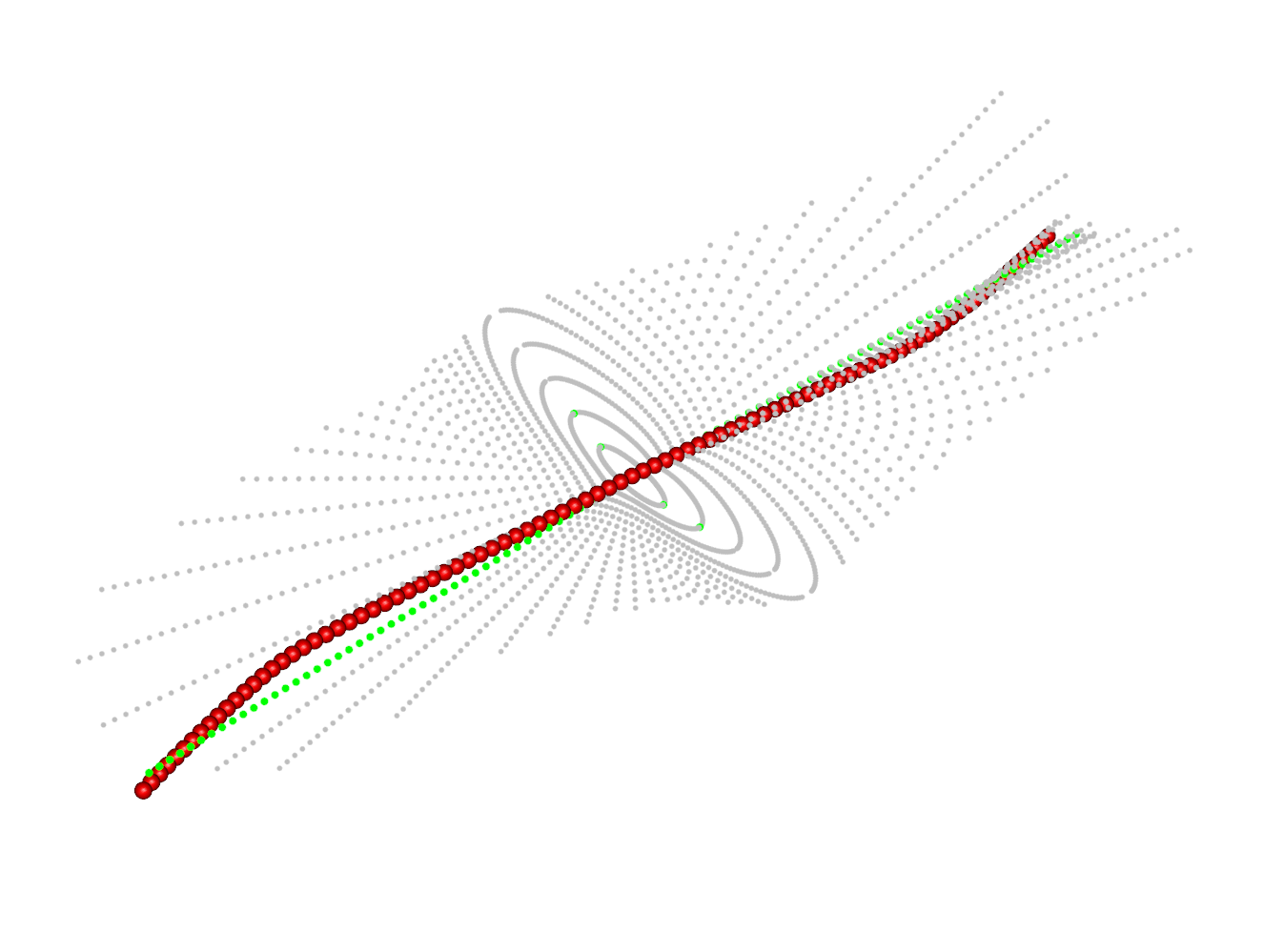}} & 
      \addheight{\includegraphics[width=0.17\textwidth]{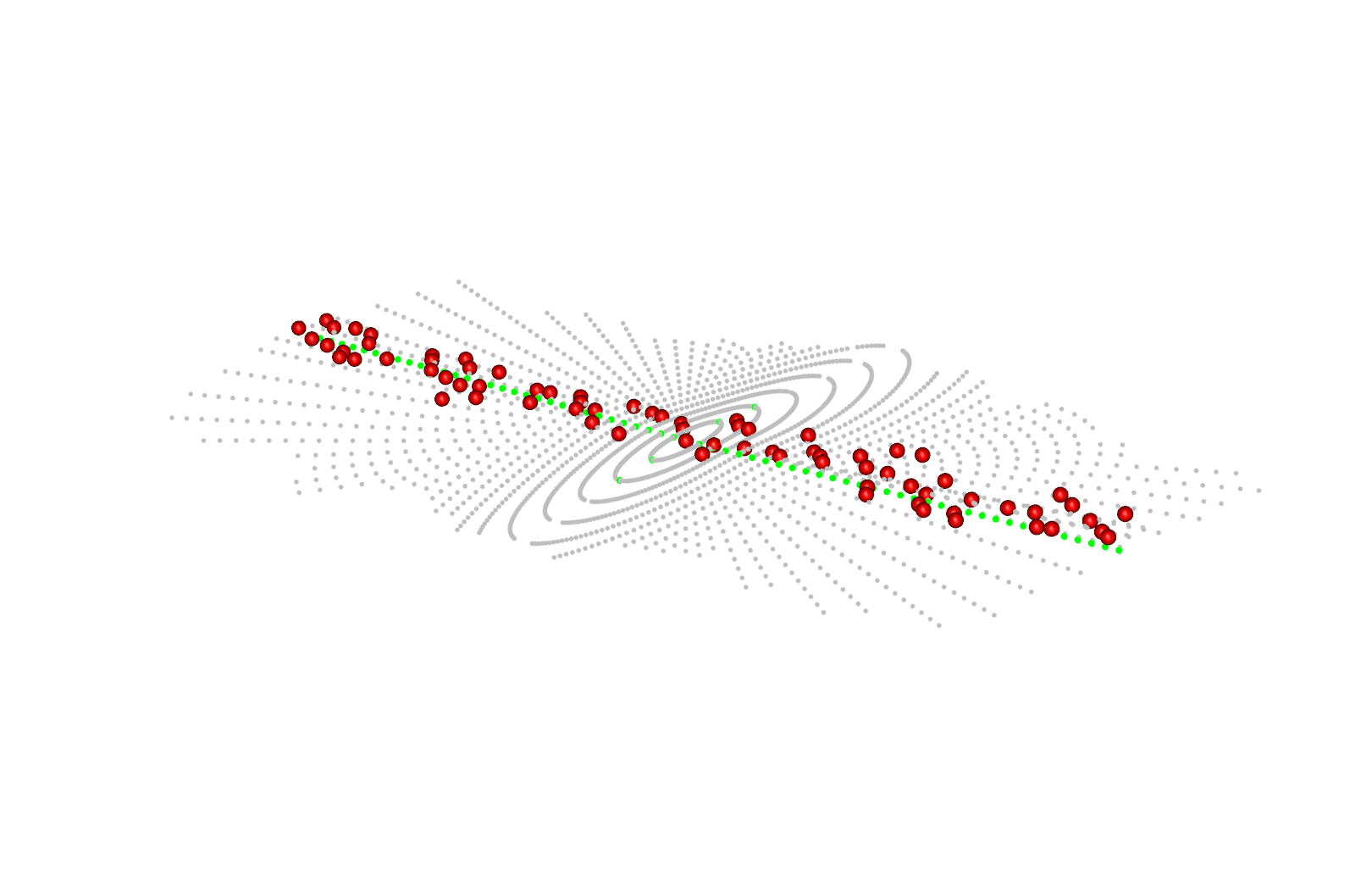}}&
      \addheight{\includegraphics[width=0.17\textwidth]{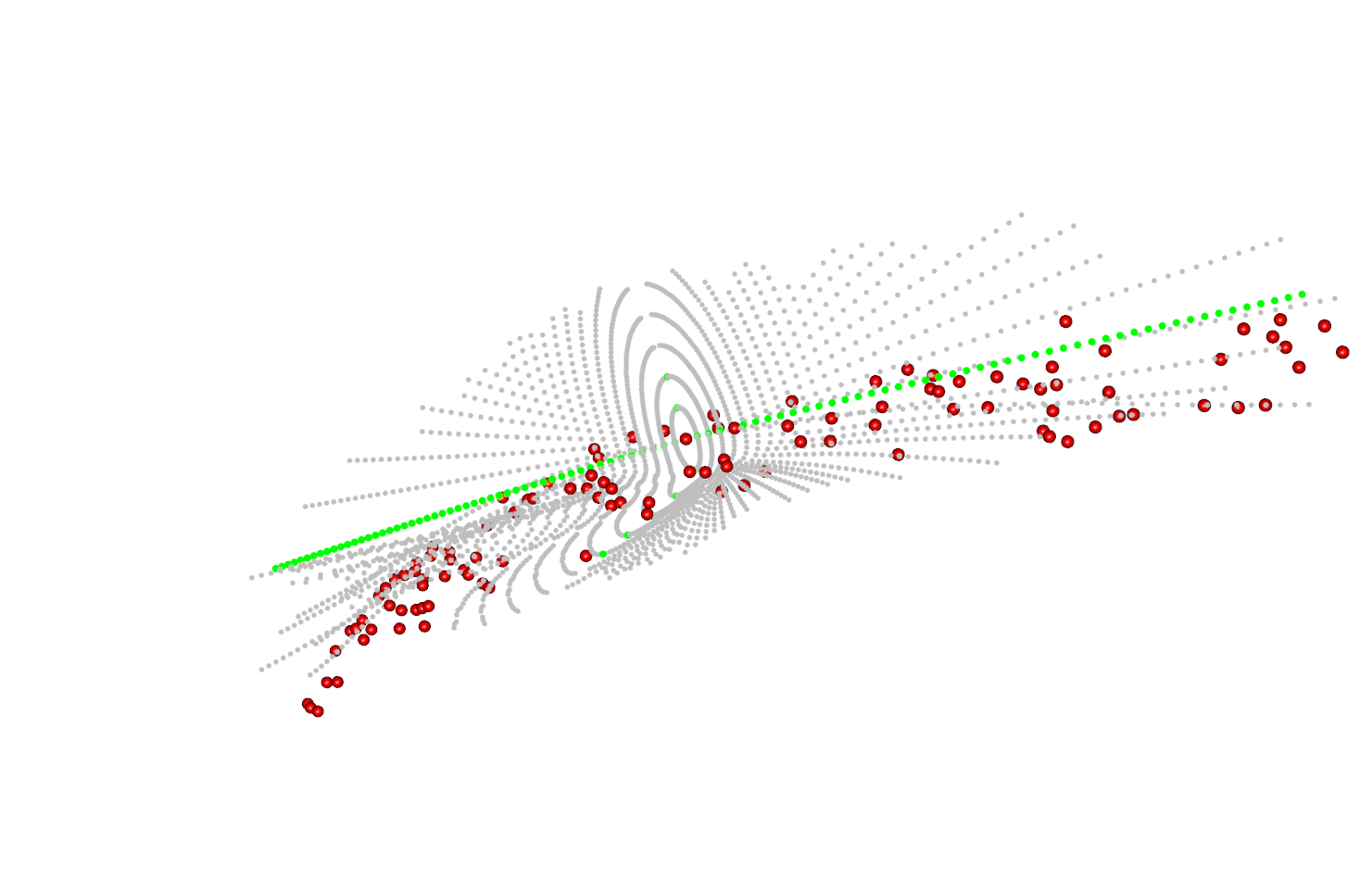}} & 
      \addheight{\includegraphics[width=0.17\textwidth]{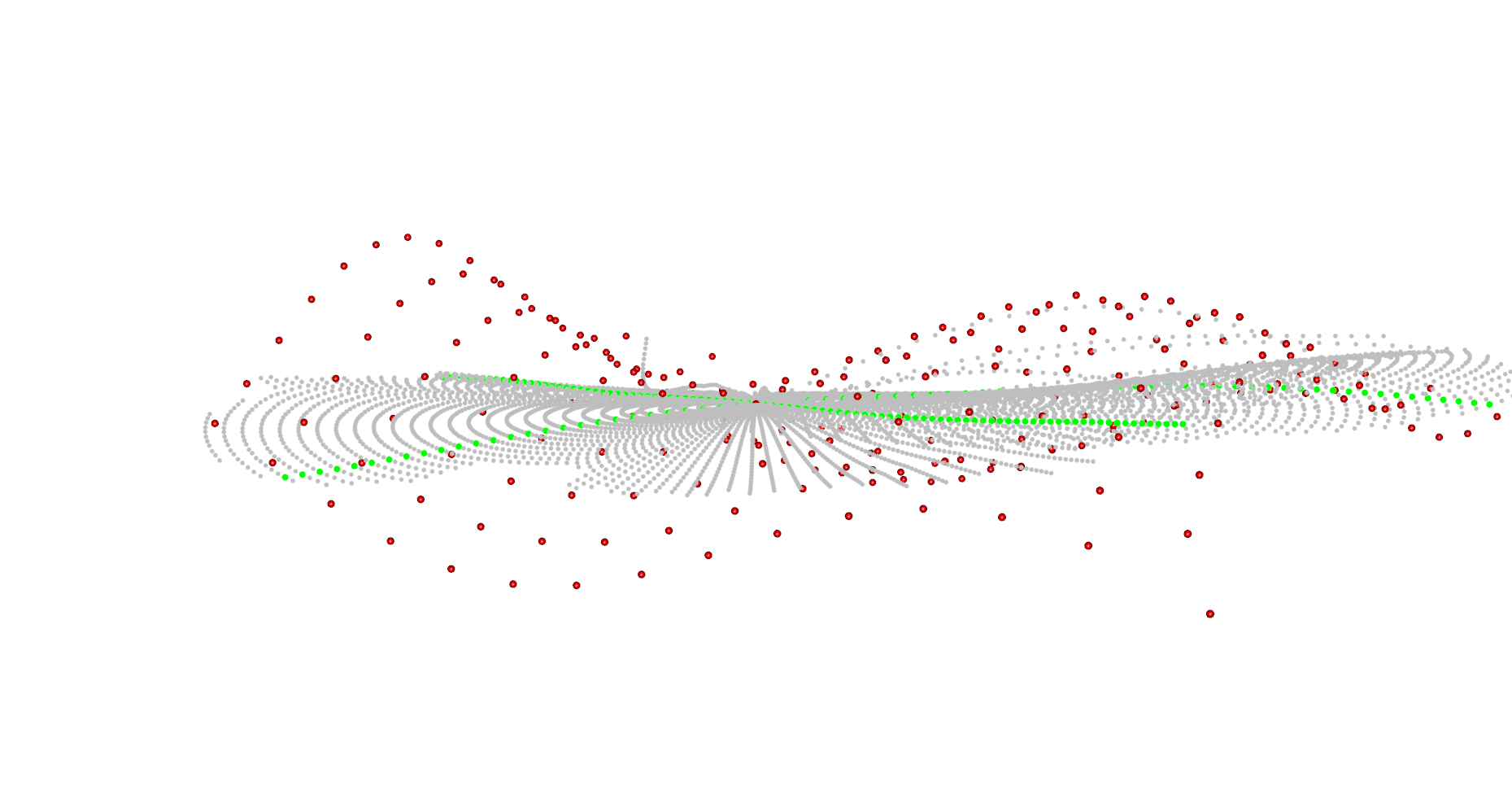}}\\
      \small (c) & (f) &  (i) &  (l) & (o) \\
      \hline
    \end{tabular}
    \captionof{figure}{Principal sub-manifolds (with superimposed principal directions) for five data clouds in $S^3$, with different scale parameters. (a)-(c) Principal sub-manifolds (in gray) and principal directions (in green) for data Cloud 1 (in red) for different values of $h$ (small, middle, large). (d)-(f), (g)-(i), (j)-(l) and (m)-(o) provide the same information for data Clouds 2,  3, 4 and 5.}
  \end{center}
\end{figure}

\begin{figure}[ht]
  \centering
  \begin{subfigure}[b]{0.4\textwidth}
    \includegraphics[width=1.8in]{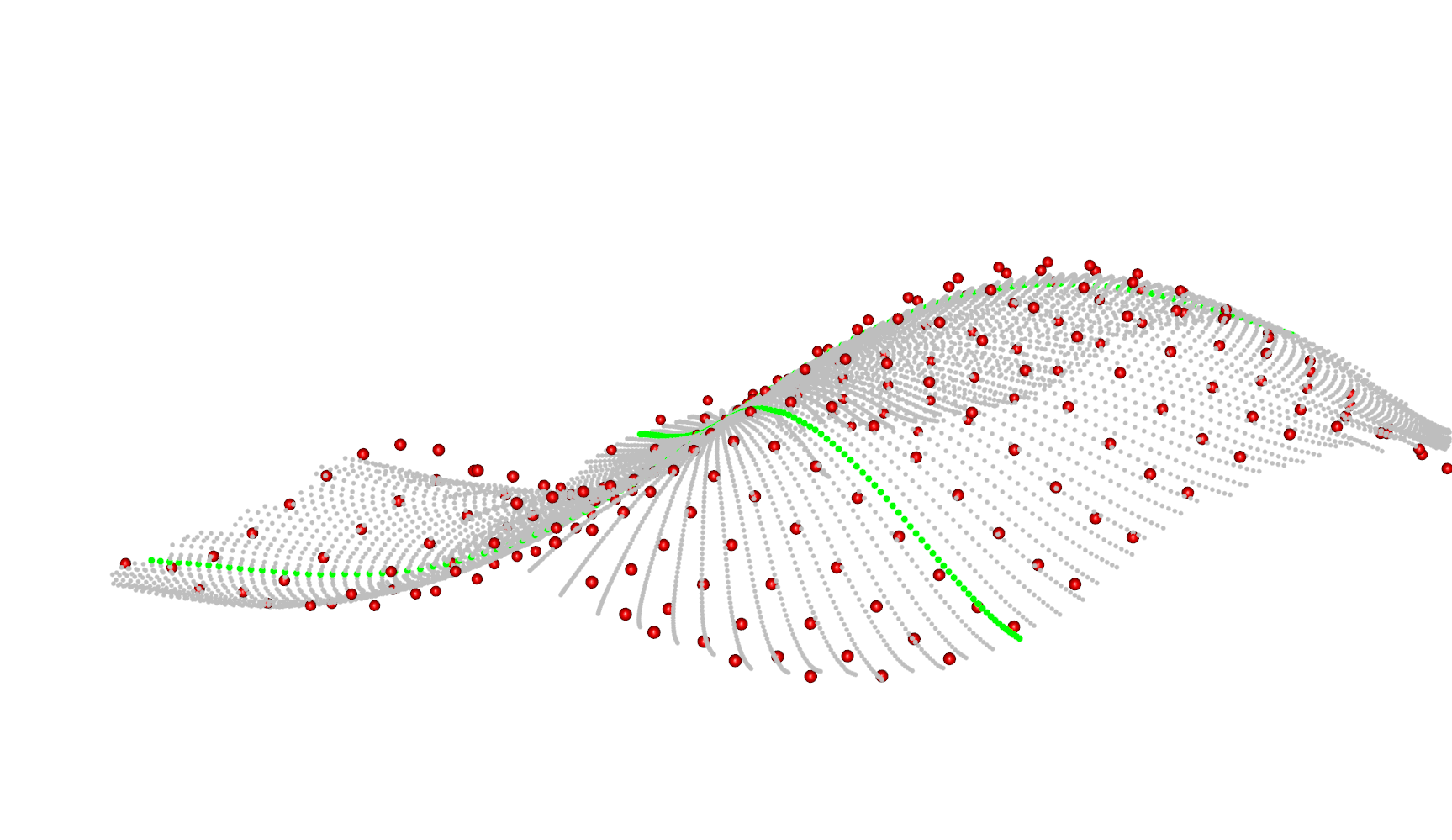} 
    \caption{no noise}
    \label{figure:noisye1}
  \end{subfigure}
  \begin{subfigure}[b]{0.4\textwidth}
    \includegraphics[width=1.8in]{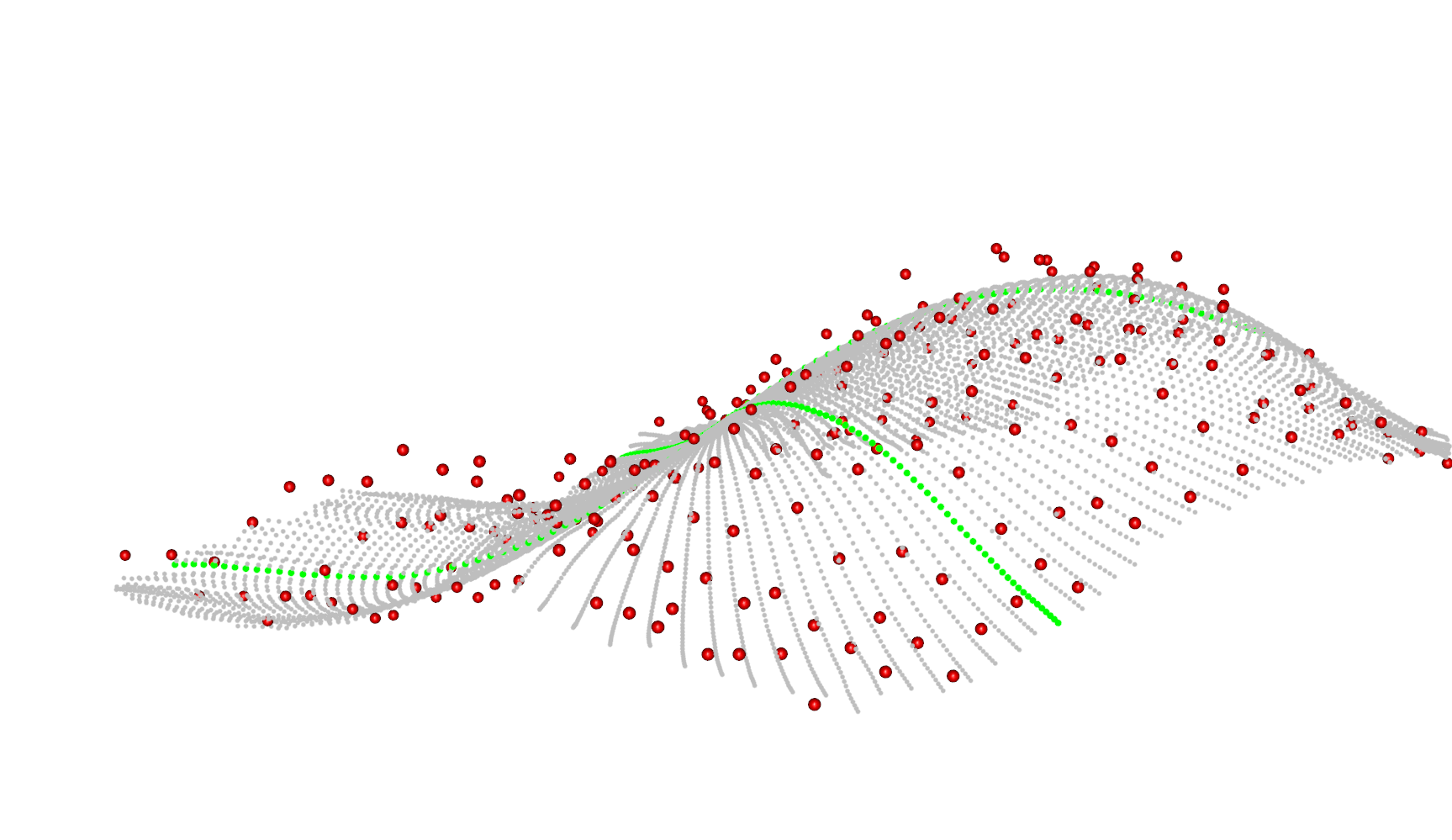}
    \caption{noise level 1}
    \label{figure:noisye2}
  \end{subfigure}    
  \begin{subfigure}[b]{0.4\textwidth}
    \includegraphics[width=1.8in]{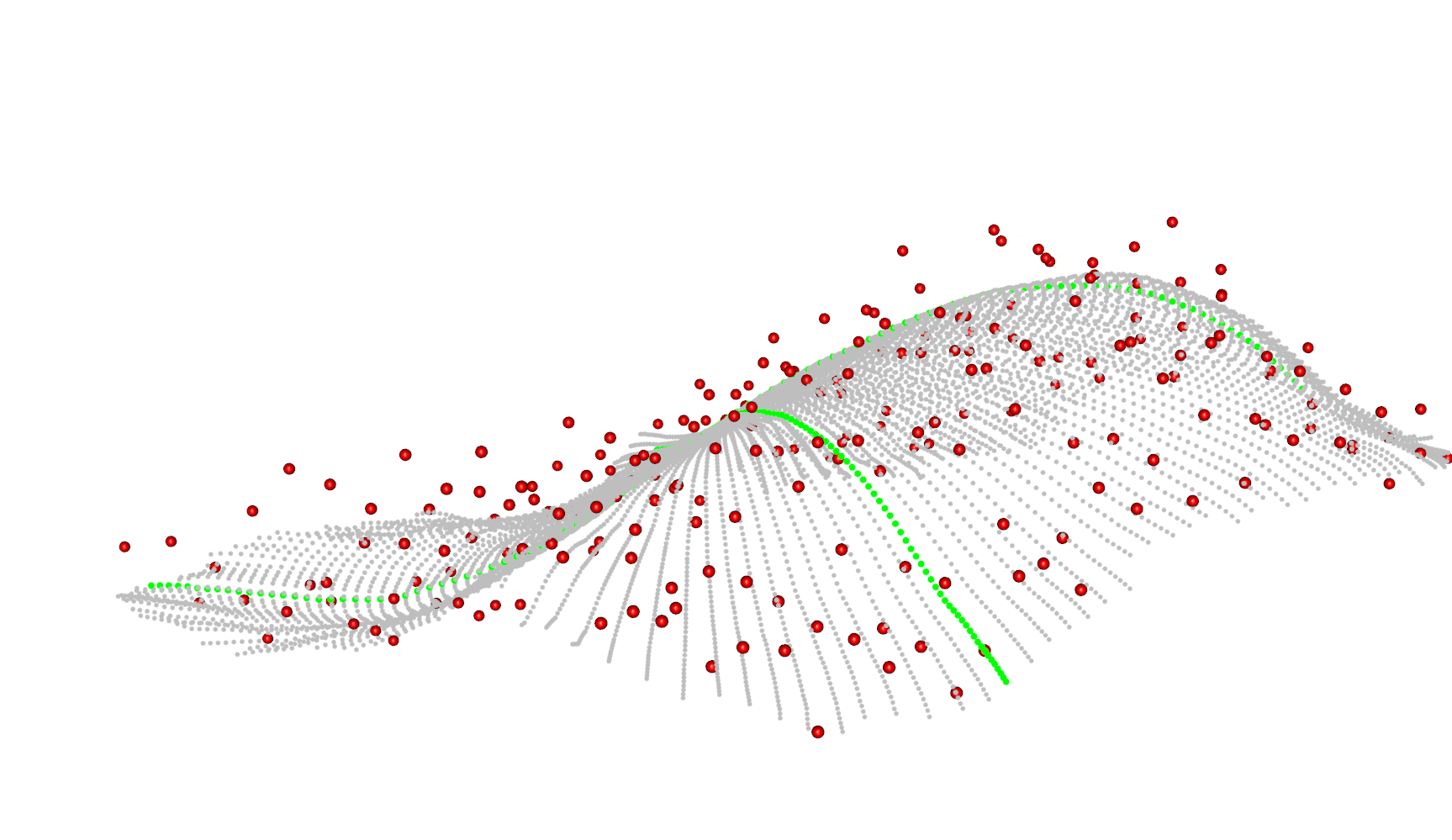}
    \caption{noise level 2}
    \label{figure:noisye3}
  \end{subfigure}
  \begin{subfigure}[b]{0.4\textwidth}
    \includegraphics[width=1.8in]{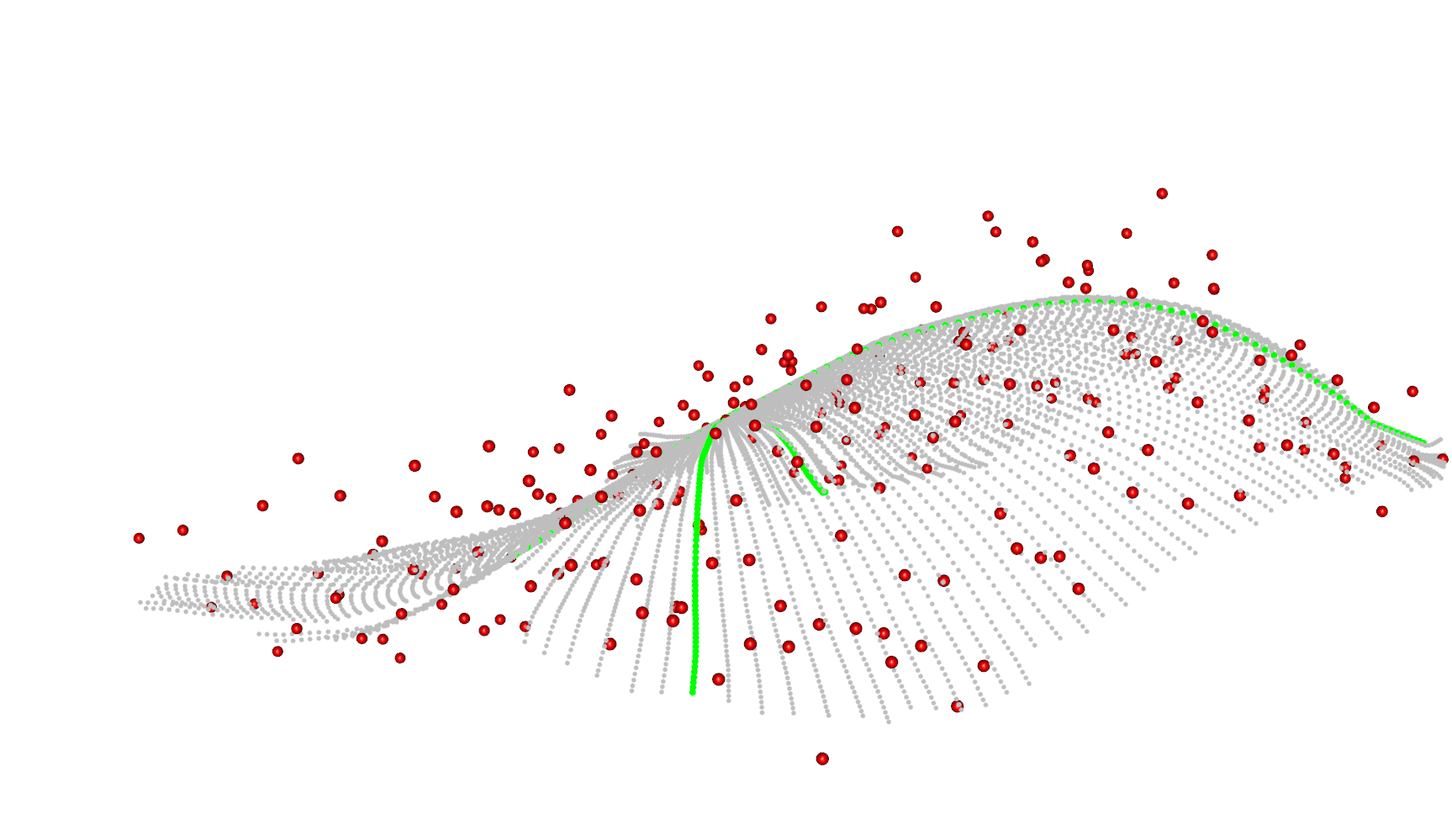}
    \caption{noise level 3}
    \label{figure:noisye4}
  \end{subfigure}
  \caption{Principal sub-manifolds (with superimposed principal directions) for four sea wave sets of data with noise on $S^3$. (a) Principal sub-manifolds with no noise added. (b), (c) and (d) provide the same information for three different levels of noise.}
  \label{example-subman2}
\end{figure}

\begin{figure}
  \centering
  \begin{subfigure}[b]{0.4\textwidth}
    \includegraphics[width=1.8in]{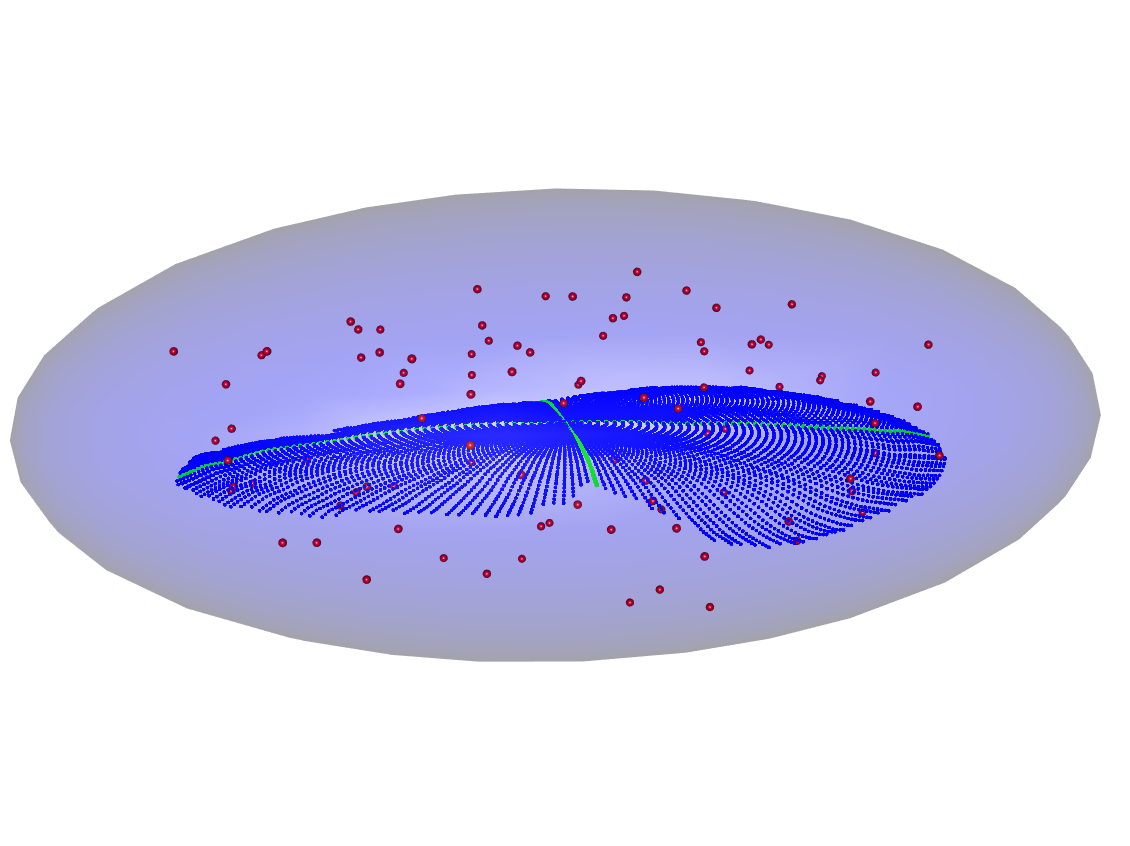} 
    \caption{$\frac{x_1^2}{2.5^2} + \frac{x_2^2}{2} + \frac{x_3^2}{1} \leq 1$}
    \label{fig:case1}
  \end{subfigure}
  \begin{subfigure}[b]{0.4\textwidth}
    \includegraphics[width=1.8in]{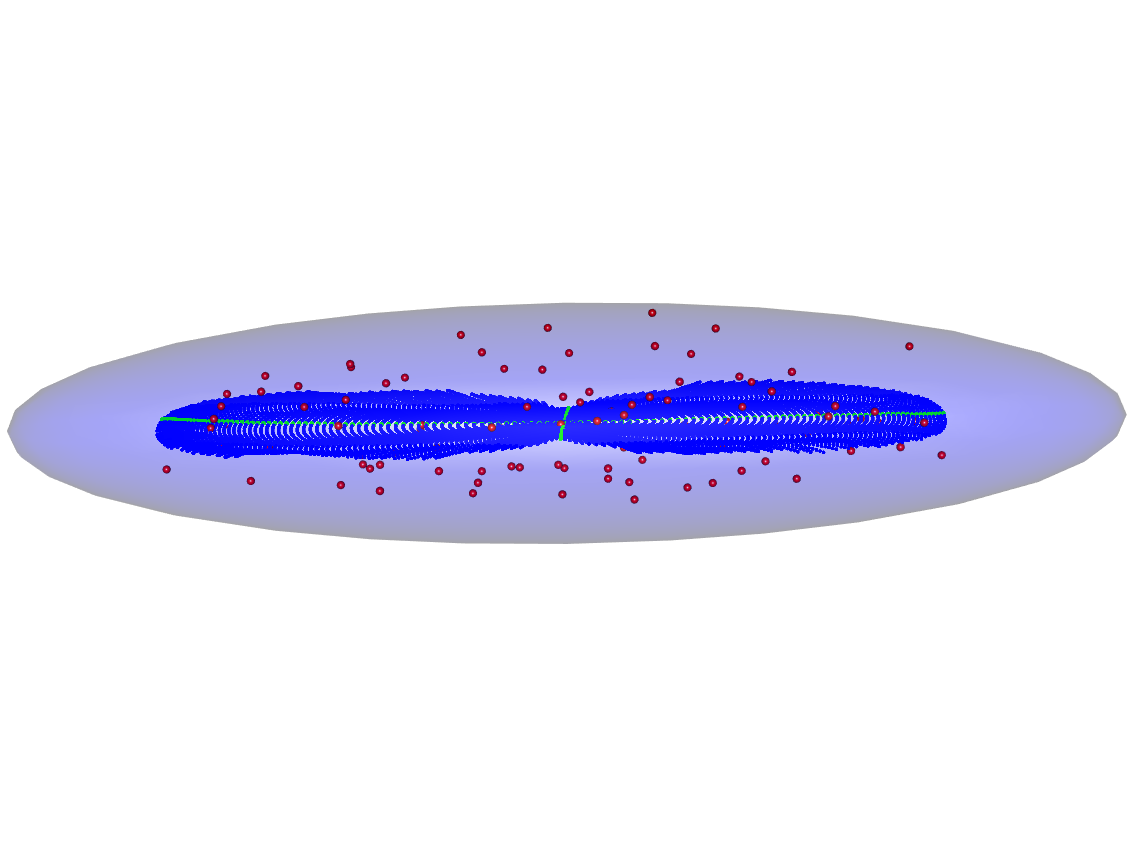}
    \caption{$\frac{x_1^2}{5^2} + \frac{x_2^2}{2} + \frac{x_3^2}{1} \leq 1$}
    \label{fig:case2}
  \end{subfigure}    
  \begin{subfigure}[b]{0.4\textwidth}
    \includegraphics[width=1.8in]{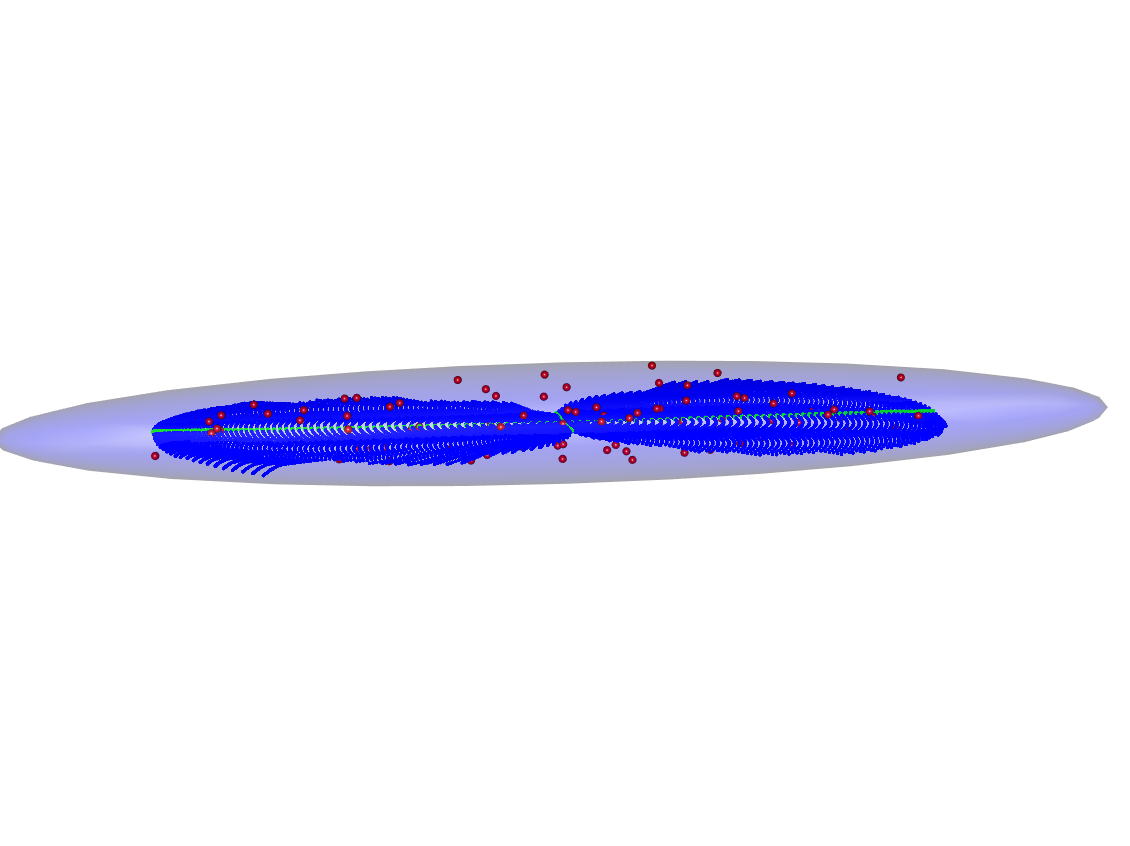}
    \caption{$\frac{x_1^2}{10^2} + \frac{x_2^2}{2} + \frac{x_3^2}{1} \leq 1$}
    \label{fig:case3}
  \end{subfigure}
  \begin{subfigure}[b]{0.4\textwidth}
    \includegraphics[width=1.8in]{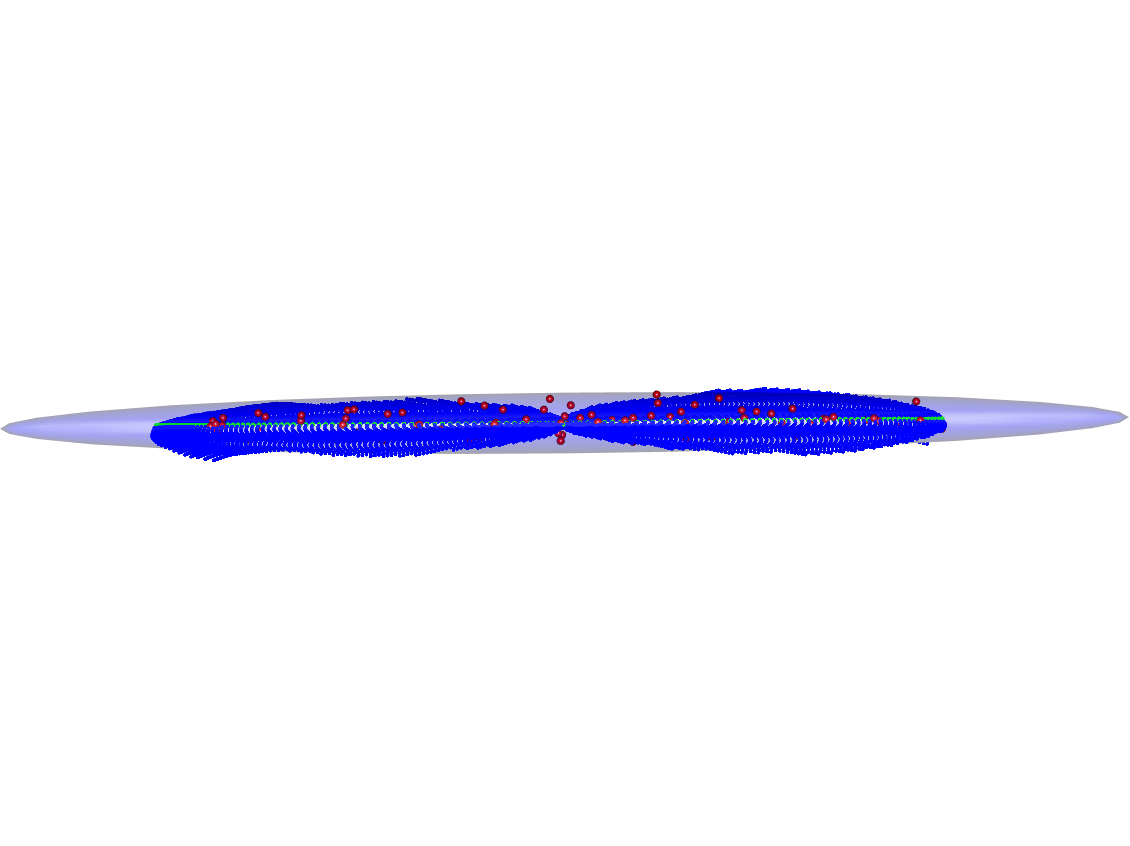}
    \caption{$\frac{x_1^2}{20^2} + \frac{x_2^2}{2} + \frac{x_3^2}{1} \leq 1$}
    \label{fig:case4}
  \end{subfigure}
  \caption{Principal sub-manifolds (with superimposed principal directions) for four ellipsoid sets of data on $S^3$. (a) Principal sub-manifolds (in blue) and principal directions (in green) for data set (in red) of case (1). %started at the Fr\'{e}chet mean. 
    (b), (c) and (d) provide the same information for case (2), (3) and (4).}
  \label{example-subman3}
\end{figure}

The last sets of examples are from a ``lifted'' ellipsoid in $S^3$. Intuitively, the four data sets we generated represent different but inter-connected types of situation: (1) the triplets are well spread out inside the ellipsoid; (2)-(3) the triplets are mostly being concentrated in the middle of a more flatter ellipsoid; (4) the triplets are chosen nearly on the diameter of the ellipsoid (potentially around an ellipse). %As we have purposely created the above data in the projected space with a lifting process followed, the four data sets are more or less symmetric and therefore the mean of the data provides a good starting point for the principal sub-manifold. 
For case (1) (Figure \ref{example-subman3}(a)), where most points are inside the ellipsoid, neither one-dimensional nor two-dimensional sub-manifold would be a perfect sub-manifold. As the diffusion decreases, such as in case (2) (Figure \ref{example-subman3}(b)) and (3) (Figure \ref{example-subman3}(c)), the sub-manifold of dimension two appears to be more and more appropriate. In case (4) (Figure \ref{example-subman3}(d)), the sub-manifold provides the best fit such that all the projected data points lie on the sub-manifold. As one has already observed, the benefit of using a two-dimensional sub-manifold in this example is only marginal.
%the two principal directions out of all the four cases provide only the marginal description of the data variation, provided that the total variations have rooted in more than one dimension. The sub-manifold can thus be interpreted as an extension of the one-dimensional representation of the variability. 
Arguably though, one can go further, for instance, having a higher dimensional sub-manifold in case (1) or case (2). Such an extension of the algorithm would be very natural, but the details of implementing the algorithm are quite subtle and we choose not to pursue this further.

\begin{figure}
  \centering
  \begin{subfigure}[b]{0.4\textwidth}
    \includegraphics[width=2.3in]{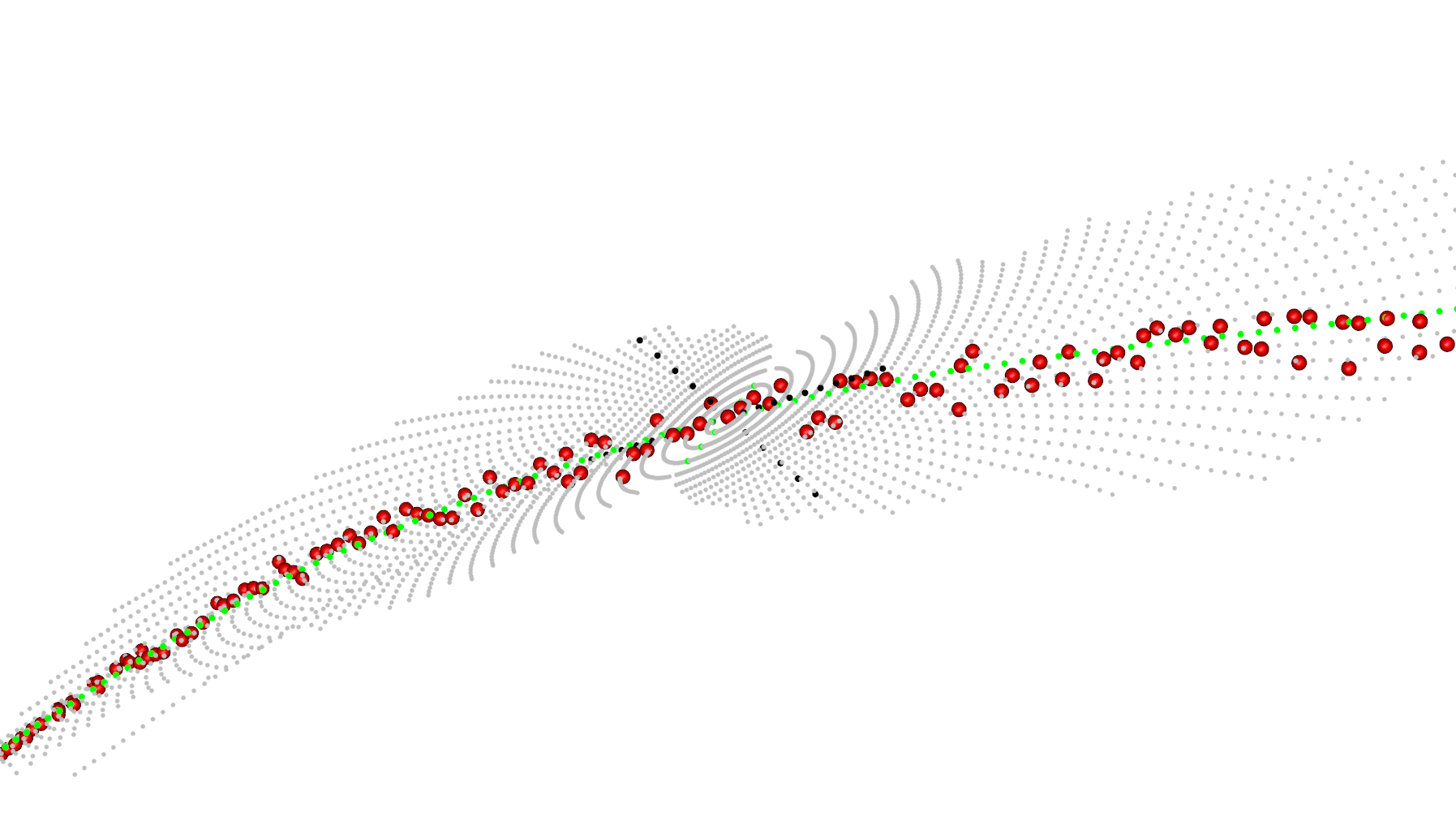} 
    \caption{}
    %\label{figure:redo2}
  \end{subfigure}
  \begin{subfigure}[b]{0.4\textwidth}
    \includegraphics[width=2.3in]{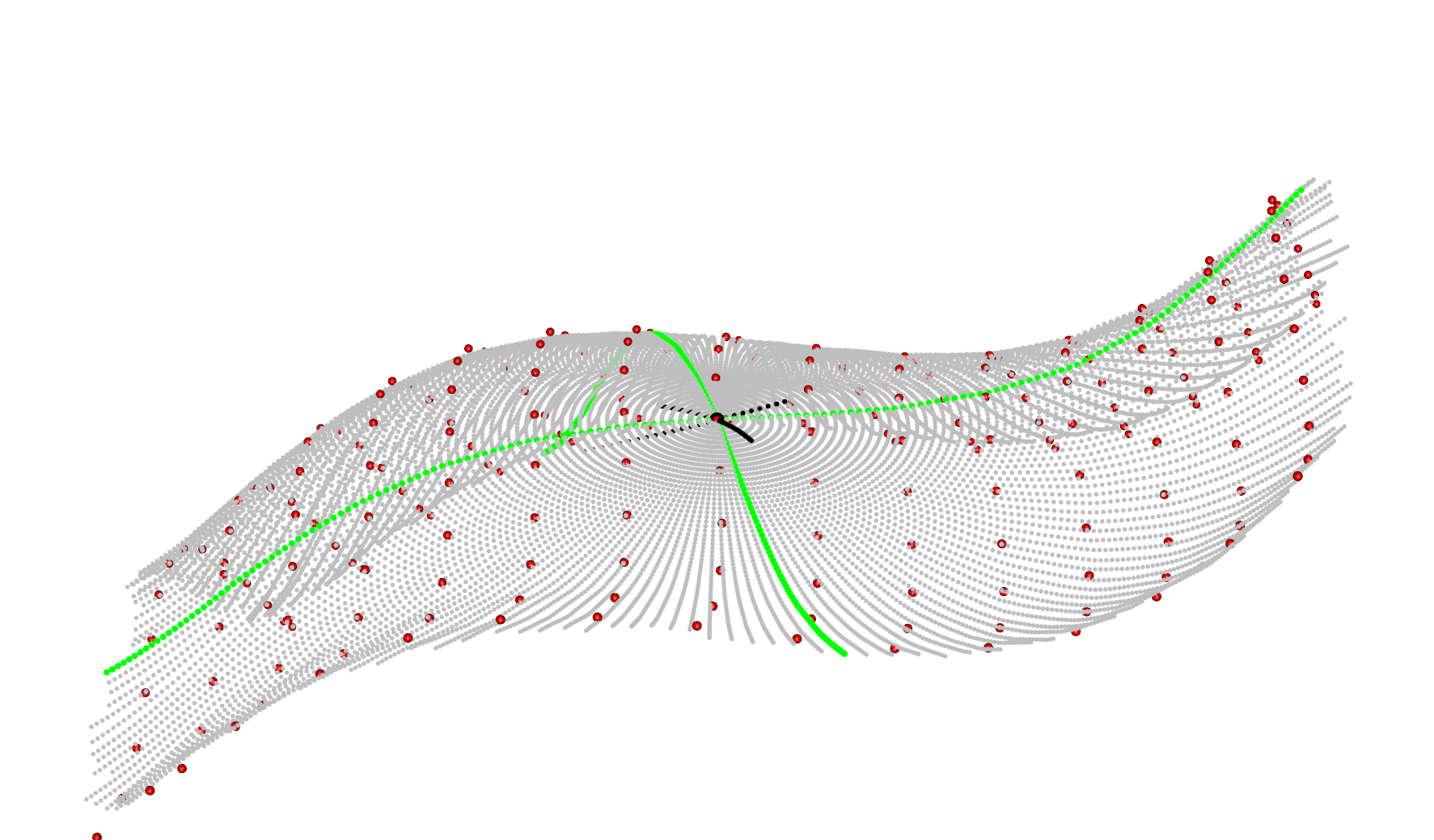}
    \caption{}
    %\label{figure:redo1}
  \end{subfigure}    
  \caption{Comparison of principal sub-manifolds and principal geodesics. The principal geodesics (in black) are superimposed to the principal directions (in green) in the projected space. Only segments of the principal geodesics are highlighted for visualization purpose.}
  \label{example-compare}
\end{figure}

%\subsection{Comparison of principal sub-manifolds and principal geodesic analysis}
To contrast the principal sub-manifold with the standard principal geodesic, we include the results of principal geodesics adjusted to its 2d version, for the case of Figure 6(j) and Figure 6(m). Specifically, the best $h$ has been chosen for either method to perform appropriately. It is expected that the principal geodesic, essentially a principal great circle along its first and second principal component, is not capable of capturing the curvature of the manifold; that is, the two principal geodesics (in black) for both cases (Figure \ref{example-compare}(a)) and (Figure \ref{example-compare}(b)) tend to deviate from the principal directions (in green) shortly after the starting point, thus not lying on the surface. In contrast, the principal sub-manifold handles the curvature well in both cases.

%\begin{figure}[ht]
%\centering
%\includegraphics[width=1.8in]{PDF/e1} 
%\includegraphics[width=1.8in]{PDF/e2} \\
%\includegraphics[width=1.8in]{PDF/e3} 
%\includegraphics[width=1.8in]{PDF/e4} 
%\includegraphics[width=2.4in]{PDF/submanifold2} 
%\caption{Principal sub-manifolds (with principal directions) for three different sets of data on $S^3$ embedded in $\mathbb{R}^4$. (a) Principal sub-manifolds (blue) and principal directions (green) for data set %(red) of case (1) started at the Fr\'{e}chet mean; (b) and (c) provide the same information for data set of case (2) and (3).}A
%\label{example-subman2}
%\end{figure}

\section{Simulations for Different Kernel Bandwidths}\label{app:11-simulations-s2}

In this subsection, we present some simulation results highlighting the influence of the kernel bandwidth on the results of the principal submanifold estimation. We sample $n$ points uniformly for the surface of a sphere of unit radius $S^2 \subset \mathbb{R}^3$ and add Gaussian white noise with a standard deviation of $\sigma$. Then we apply the principal submanifold algorithm with 16 initial directions with a Gaussian kernel of bandwidth $h$.

\begin{figure}[ht!]
  \centering
  \begin{subfigure}[b]{0.32\textwidth}
    \centering
    \includegraphics[width=.85\linewidth]{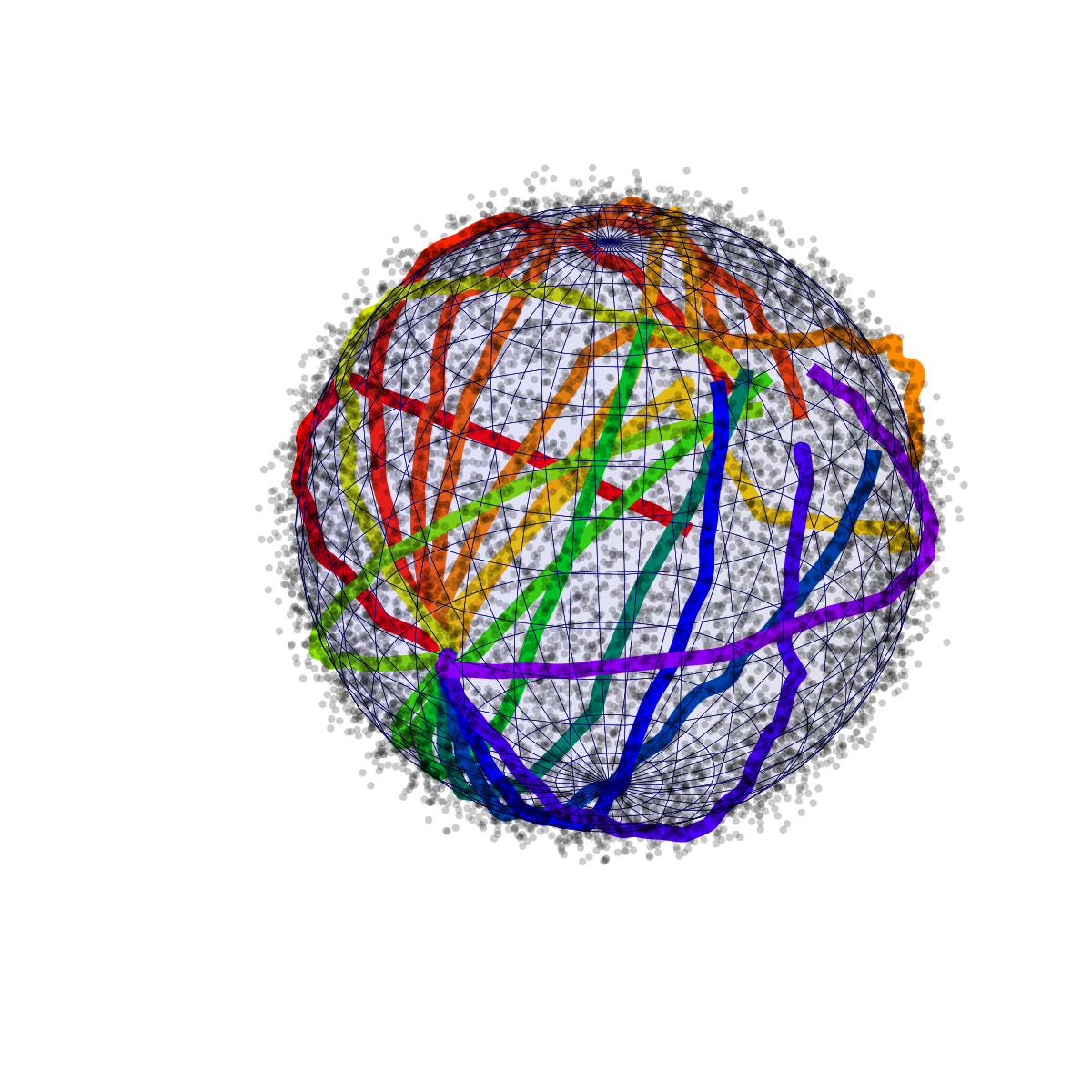}
    \caption{$h = 0.05$}
  \end{subfigure}%
  %\hspace{0.15 in}
  \begin{subfigure}[b]{0.32\textwidth}
    \centering
    \includegraphics[width=.85\linewidth]{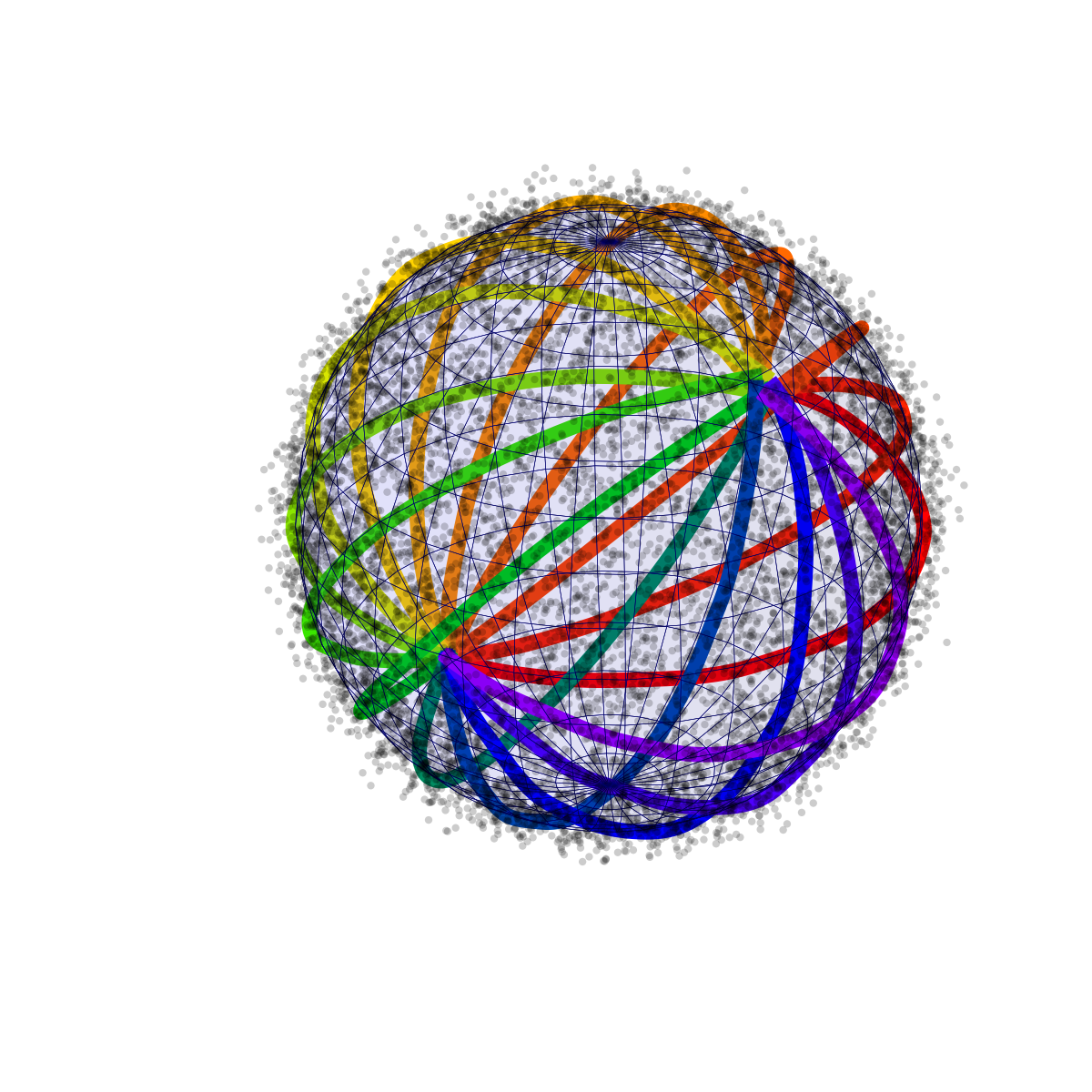}
    \caption{$h = 0.1$}
  \end{subfigure}%
  \begin{subfigure}[b]{0.32\textwidth}
    \centering
    \includegraphics[width=.85\linewidth]{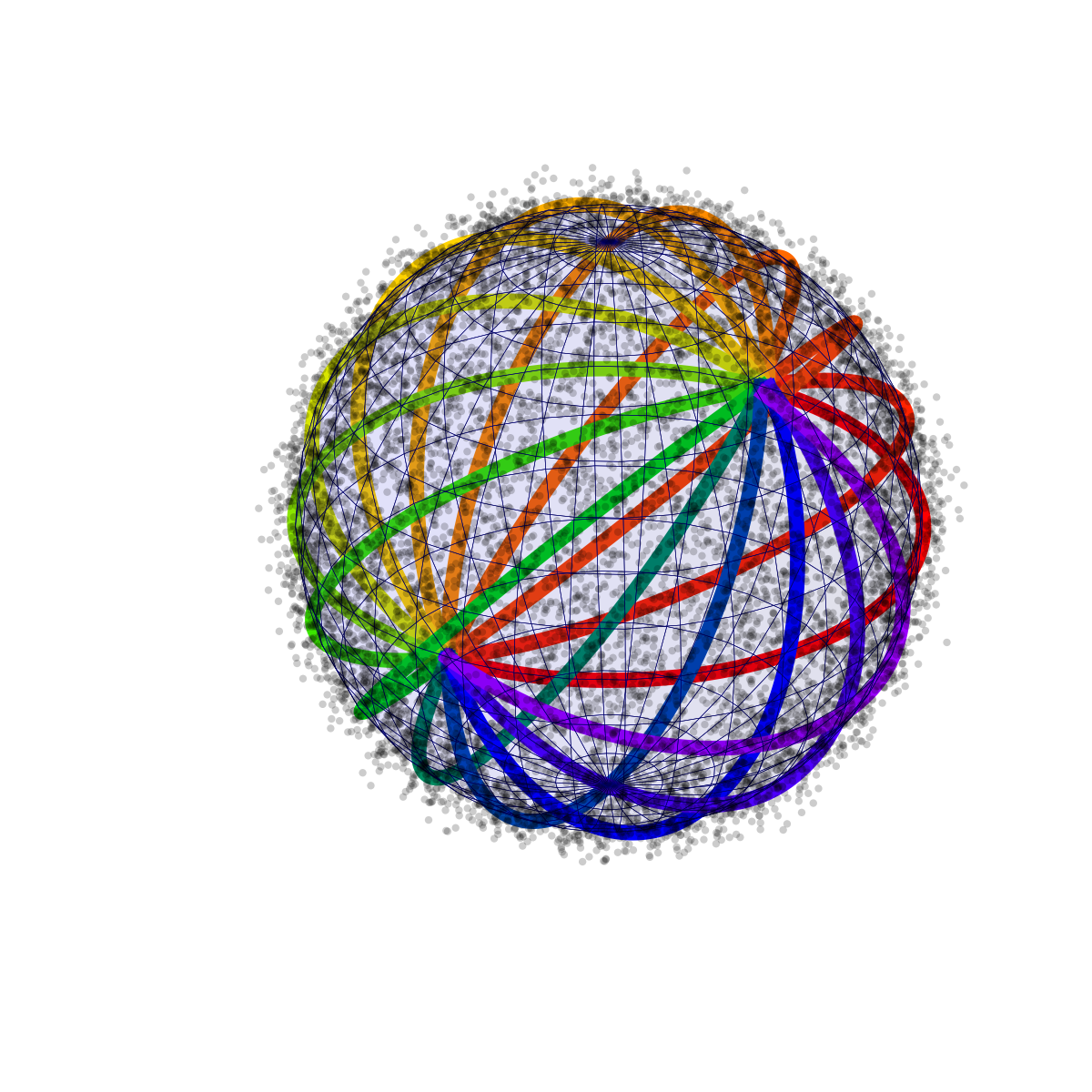}
    \caption{$h = 0.2$}
  \end{subfigure}\\
  \begin{subfigure}[b]{0.32\textwidth}
    \centering
    \includegraphics[width=.85\linewidth]{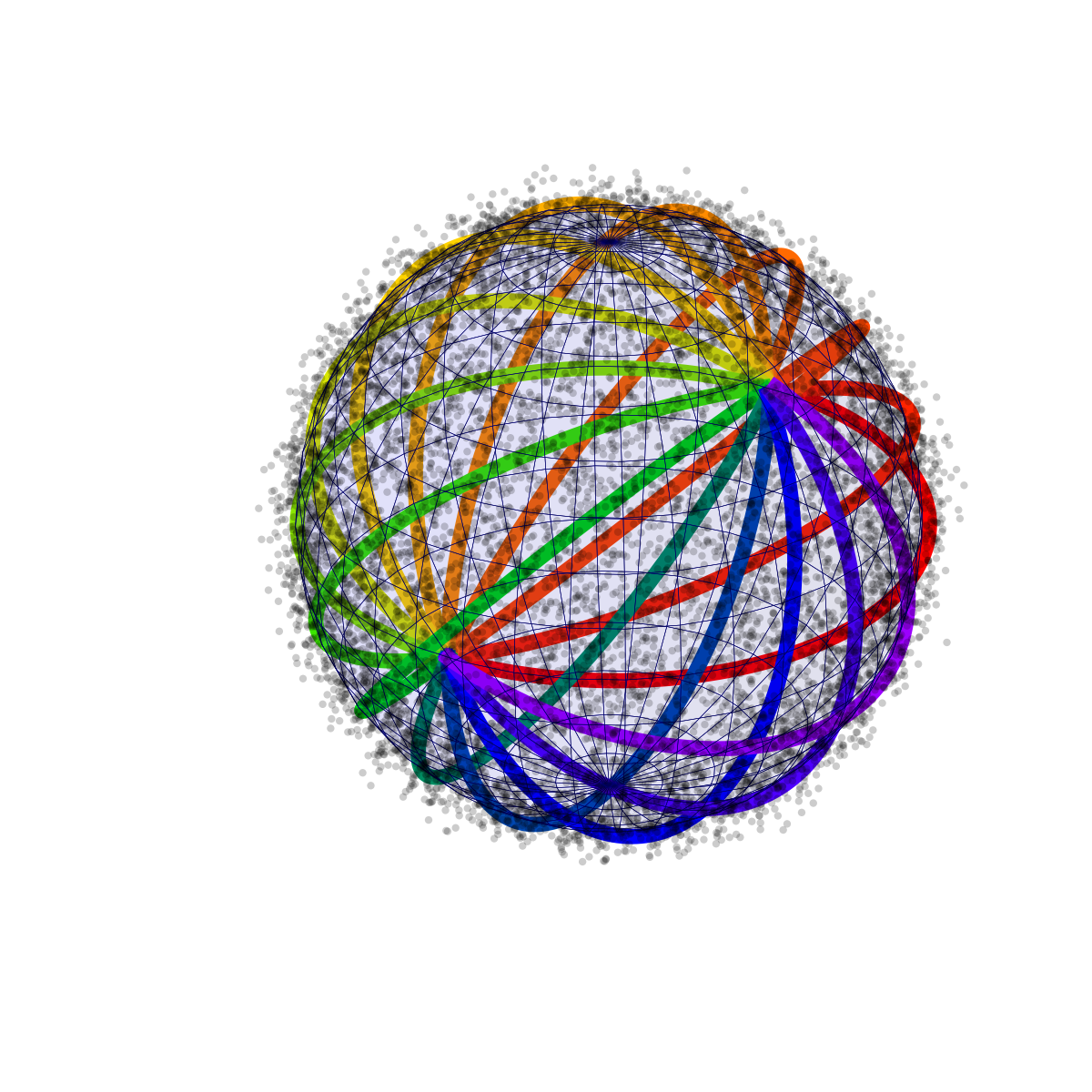}
    \caption{$h = 0.5$}
  \end{subfigure}%
  %\hspace{0.15 in}
  \begin{subfigure}[b]{0.32\textwidth}
    \centering
    \includegraphics[width=.85\linewidth]{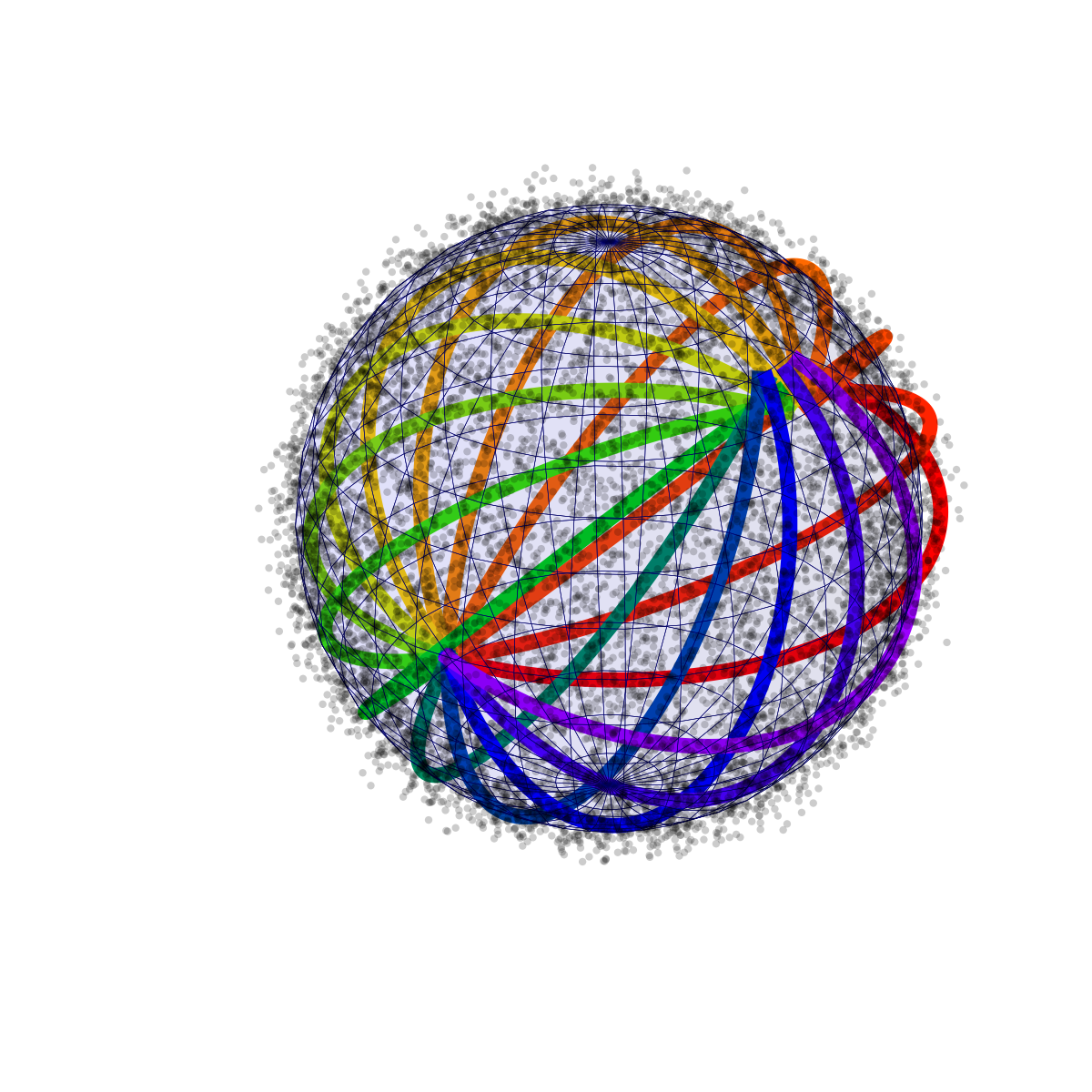}
    \caption{$h = 1$}
  \end{subfigure}%
  \begin{subfigure}[b]{0.32\textwidth}
    \centering
    \includegraphics[width=.85\linewidth]{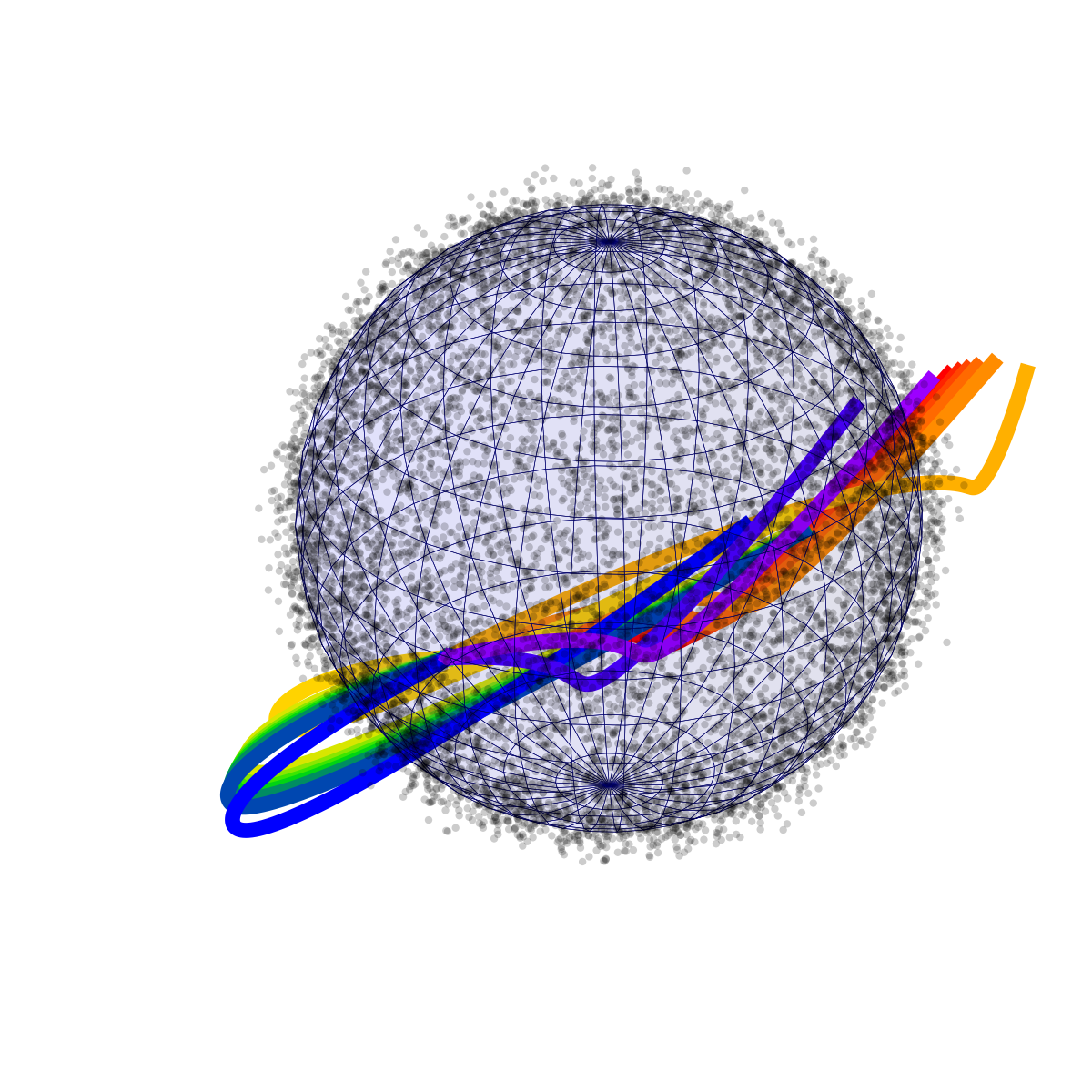}
    \caption{$h = 2$}
  \end{subfigure}% 
  \caption{Principal submanifold results for noise level $\sigma = 0.05$ and sample size $n = 10\,000$. One can clearly see that a bandwidth much larger than the noise level but smaller than the diameter of the sphere, i.e. $\sigma \ll h \ll 2$ one gets a good fit to the sphere. For small bandwidths the eigenvectors to the two largest eigenvalues are in some cases not tangential to the sphere, leading to erratic direction changes of the rays. For large bandwidths, the opposite side of the sphere starts to contribute substantially to the local PCA, leading to curves that do not follow the curvature of the sphere, but are instead too straight.}
  \label{fig:bandwidth_10000}
\end{figure}

\begin{figure}[ht!]
  \centering
  \begin{subfigure}[b]{0.32\textwidth}
    \centering
    \includegraphics[width=.85\linewidth]{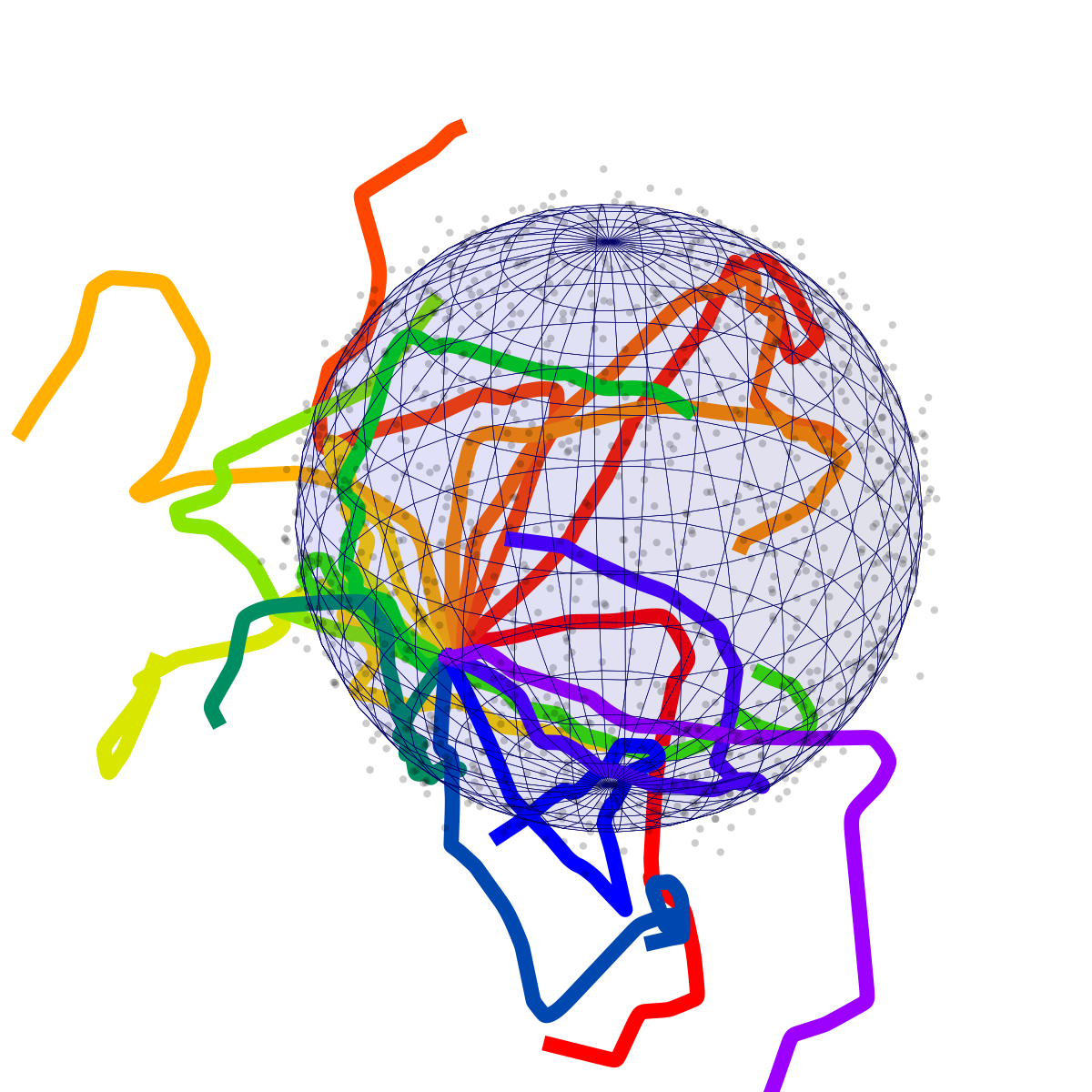}
    \caption{$h = 0.05$}
  \end{subfigure}%
  %\hspace{0.15 in}
  \begin{subfigure}[b]{0.32\textwidth}
    \centering
    \includegraphics[width=.85\linewidth]{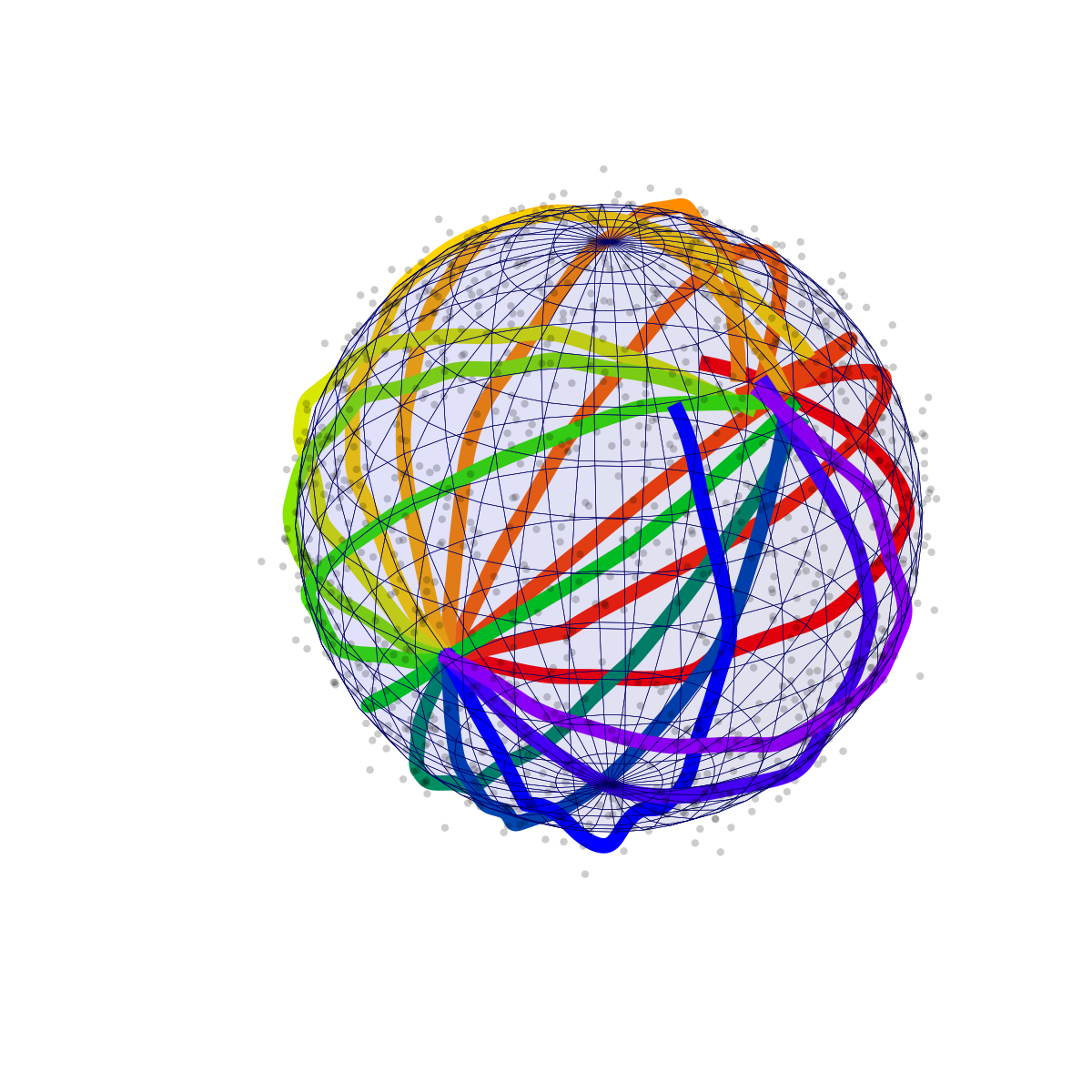}
    \caption{$h = 0.1$}
  \end{subfigure}%
  \begin{subfigure}[b]{0.32\textwidth}
    \centering
    \includegraphics[width=.85\linewidth]{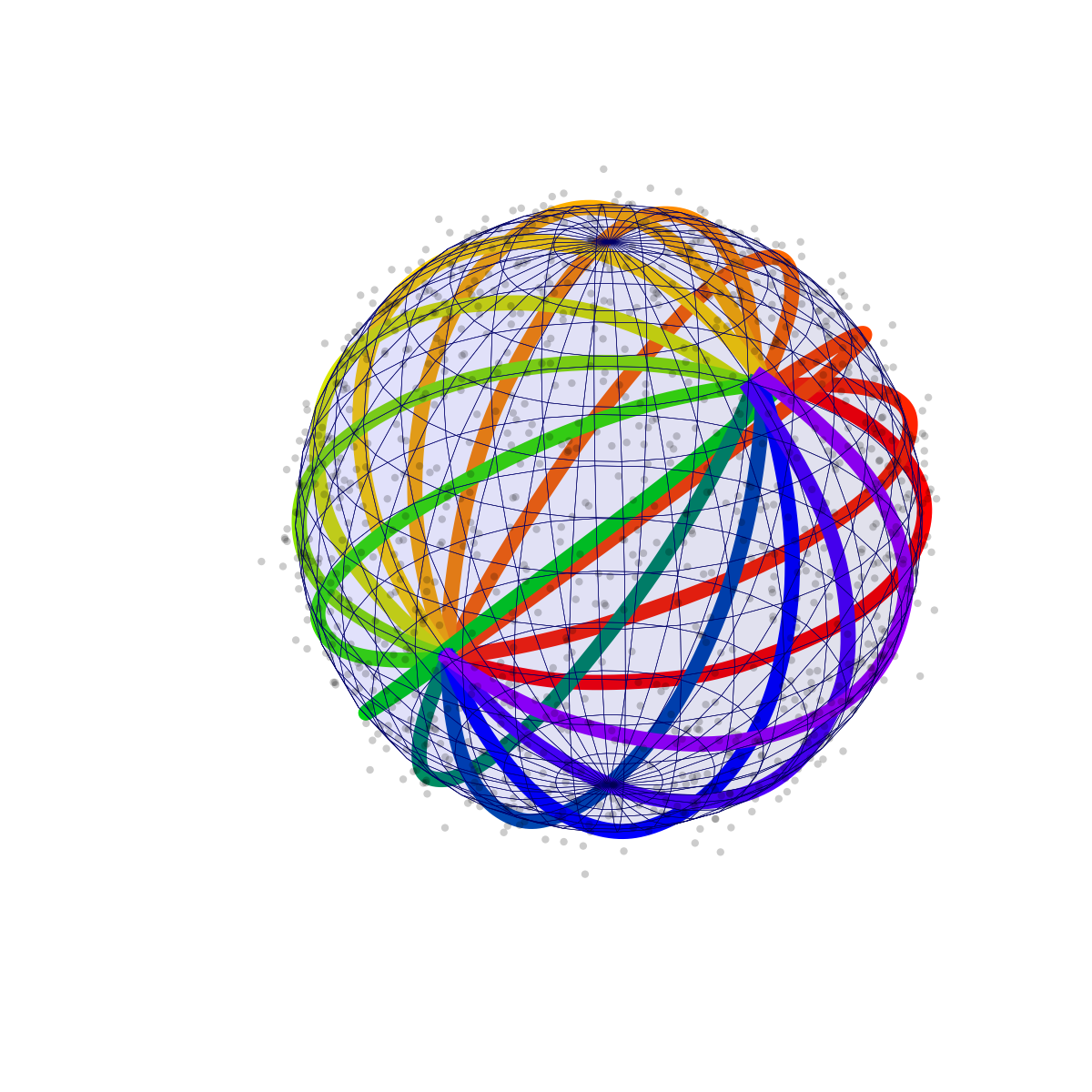}
    \caption{$h = 0.2$}
  \end{subfigure}\\
  \begin{subfigure}[b]{0.32\textwidth}
    \centering
    \includegraphics[width=.85\linewidth]{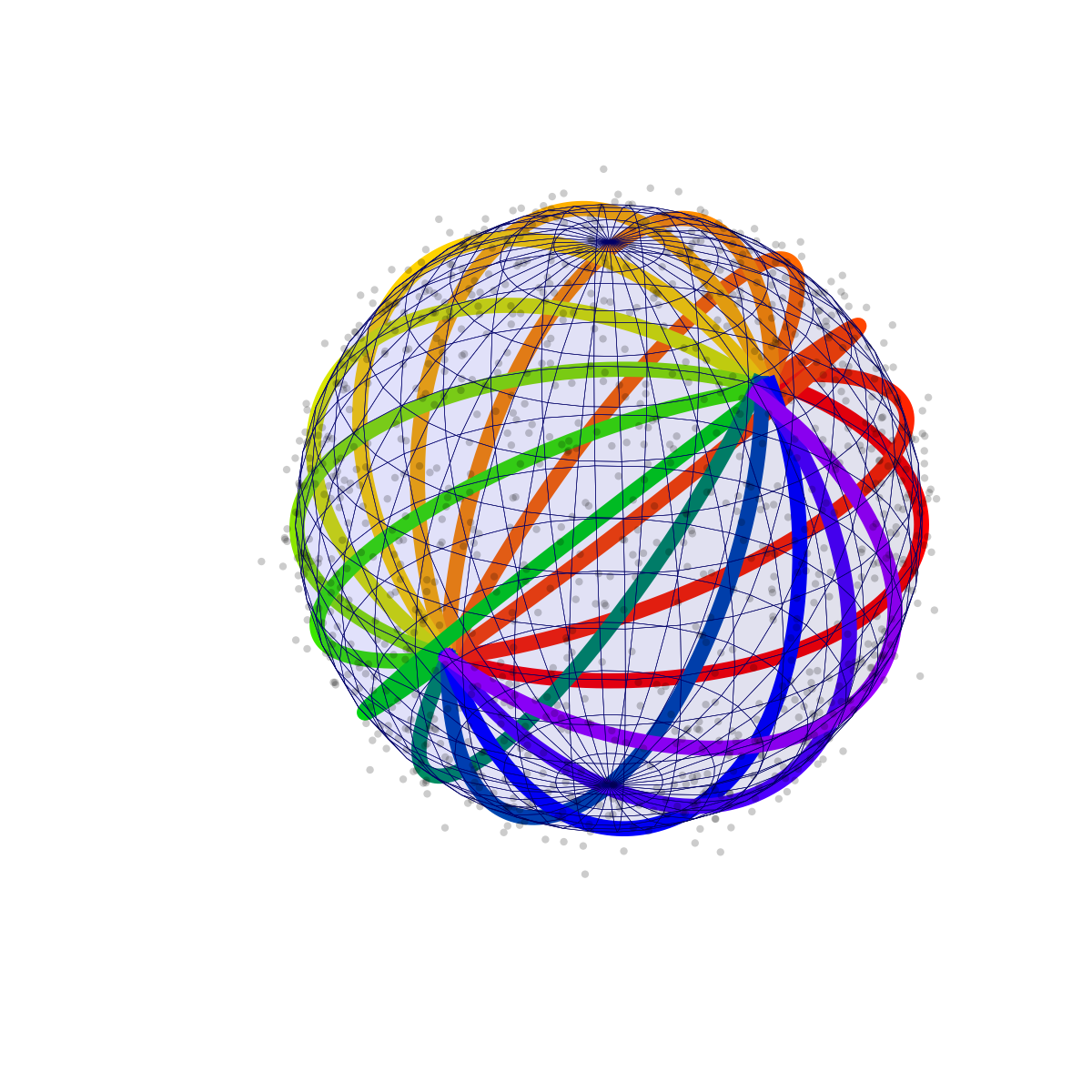}
    \caption{$h = 0.5$}
  \end{subfigure}%
  %\hspace{0.15 in}
  \begin{subfigure}[b]{0.32\textwidth}
    \centering
    \includegraphics[width=.85\linewidth]{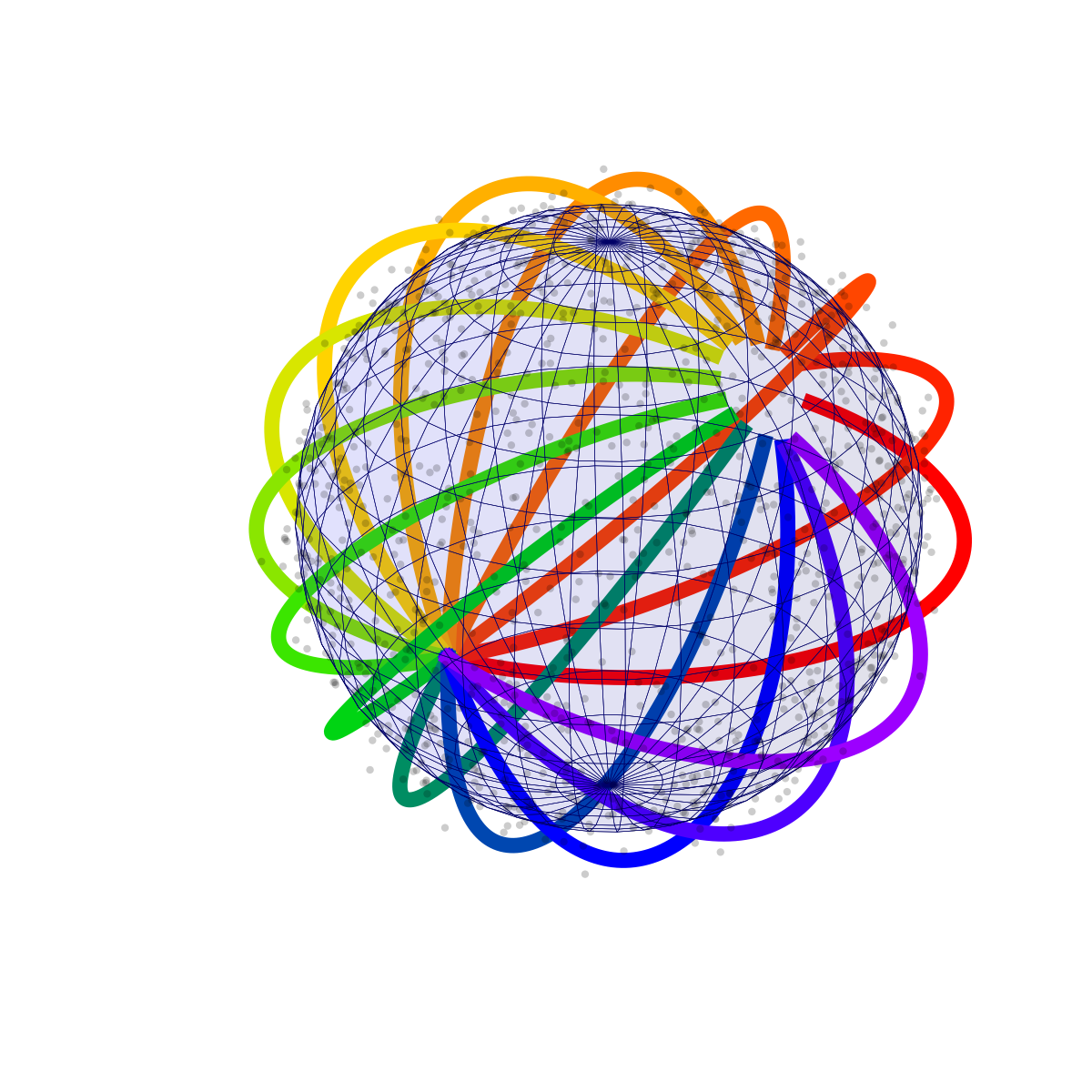}
    \caption{$h = 1$}
  \end{subfigure}%
  \begin{subfigure}[b]{0.32\textwidth}
    \centering
    \includegraphics[width=.85\linewidth]{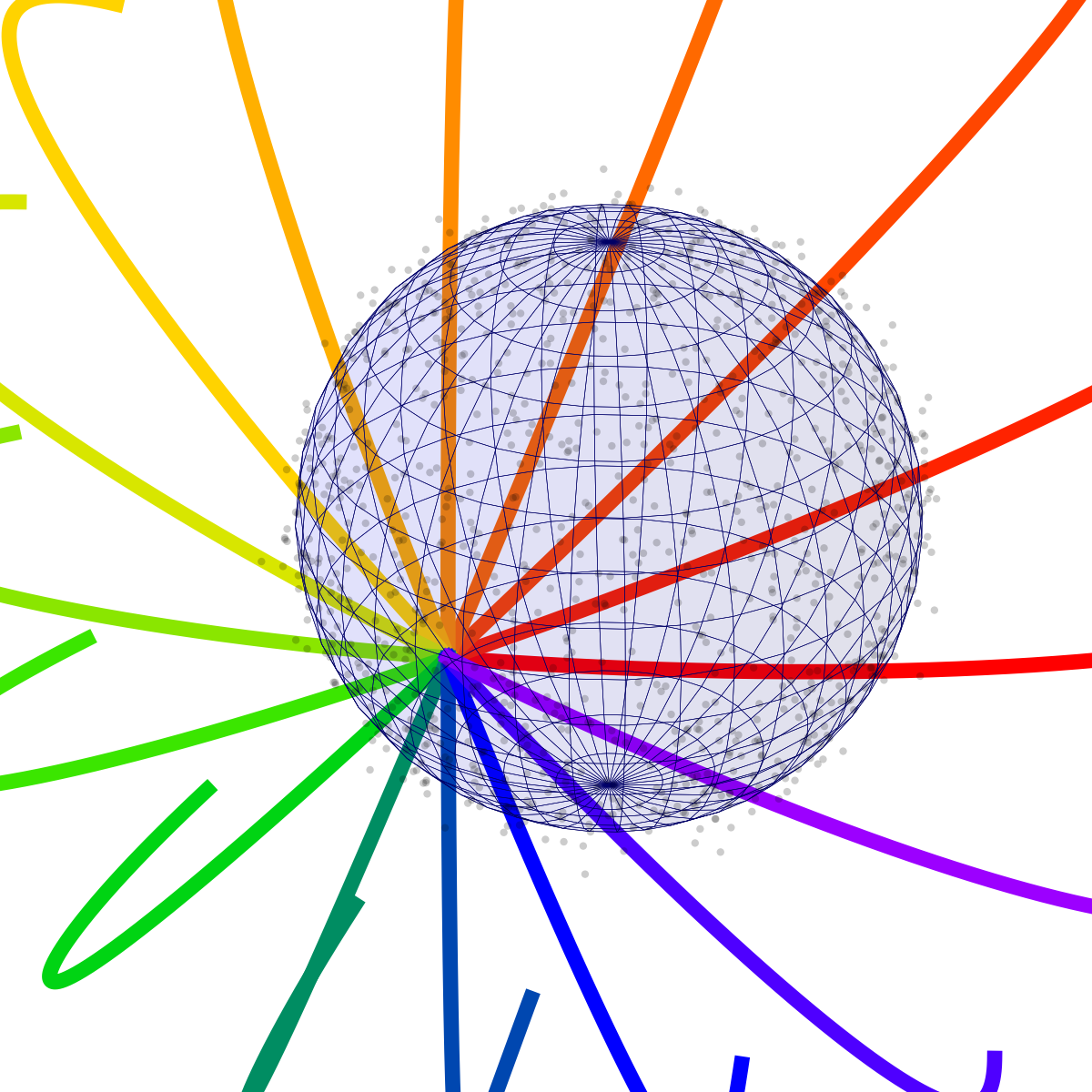}
    \caption{$h = 2$}
  \end{subfigure}% 
  \caption{Principal submanifold results for noise level $\sigma = 0.05$ and sample size $n = 1\,000$. In comparison to the simulation with $n = 10\,000$, the results are more bandwidth dependent but intermediate bandwidths $h = 0.2$ and $h = 0.5$ still yield excellent results.}
  \label{fig:bandwidth_1000}
\end{figure}

\begin{figure}[ht!]
  \centering
  \begin{subfigure}[b]{0.32\textwidth}
    \centering
    \includegraphics[width=.85\linewidth]{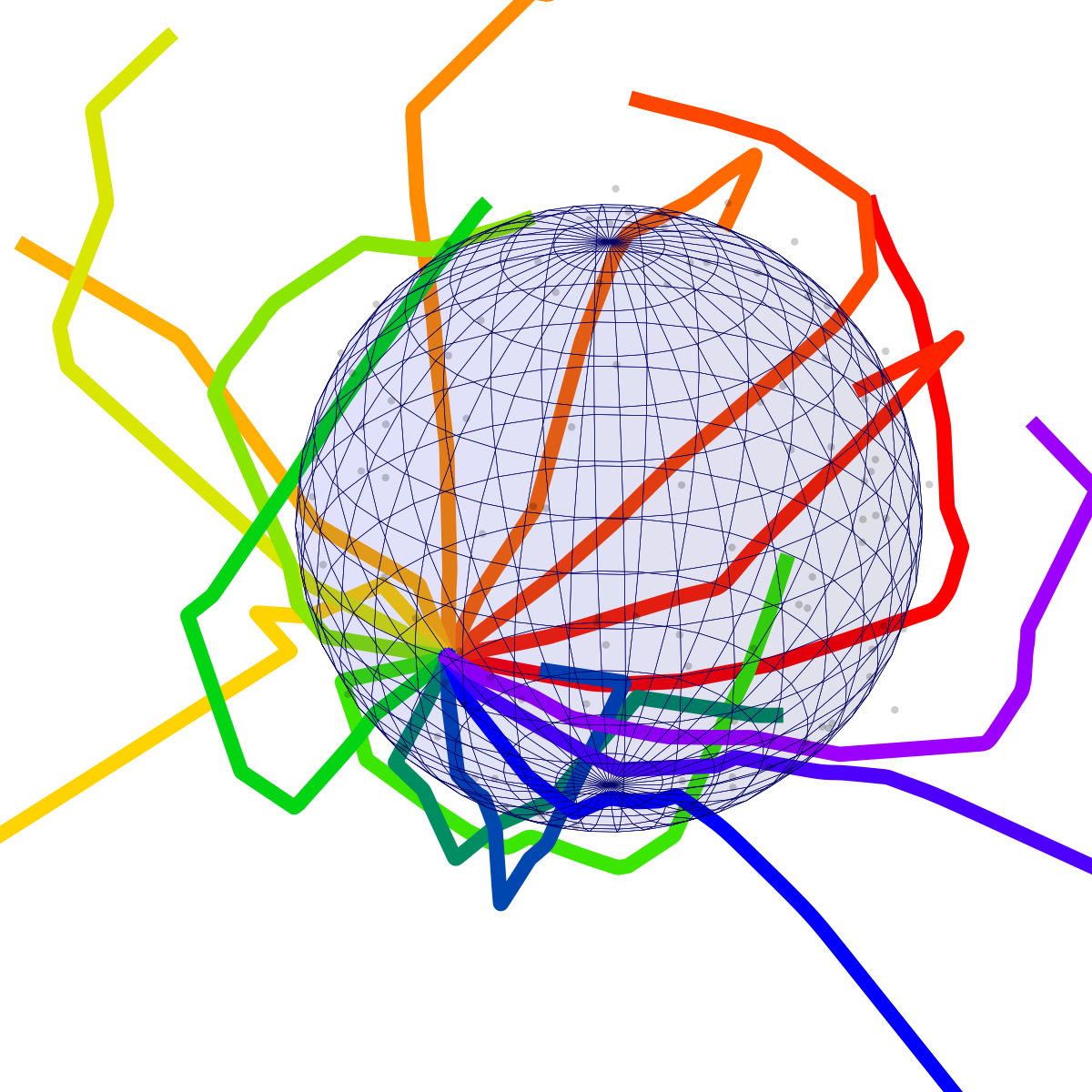}
    \caption{$h = 0.05$}
  \end{subfigure}%
  %\hspace{0.15 in}
  \begin{subfigure}[b]{0.32\textwidth}
    \centering
    \includegraphics[width=.85\linewidth]{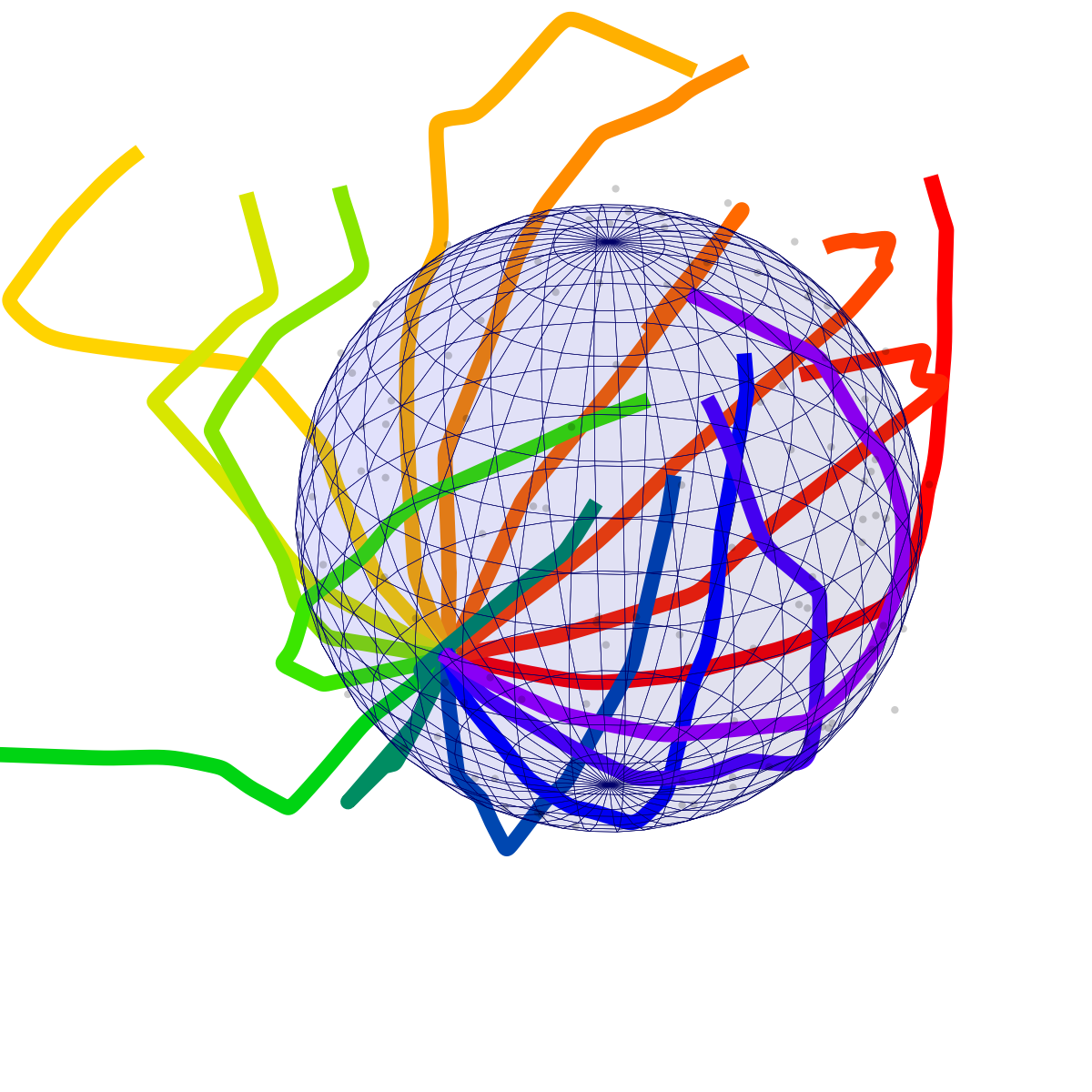}
    \caption{$h = 0.1$}
  \end{subfigure}%
  \begin{subfigure}[b]{0.32\textwidth}
    \centering
    \includegraphics[width=.85\linewidth]{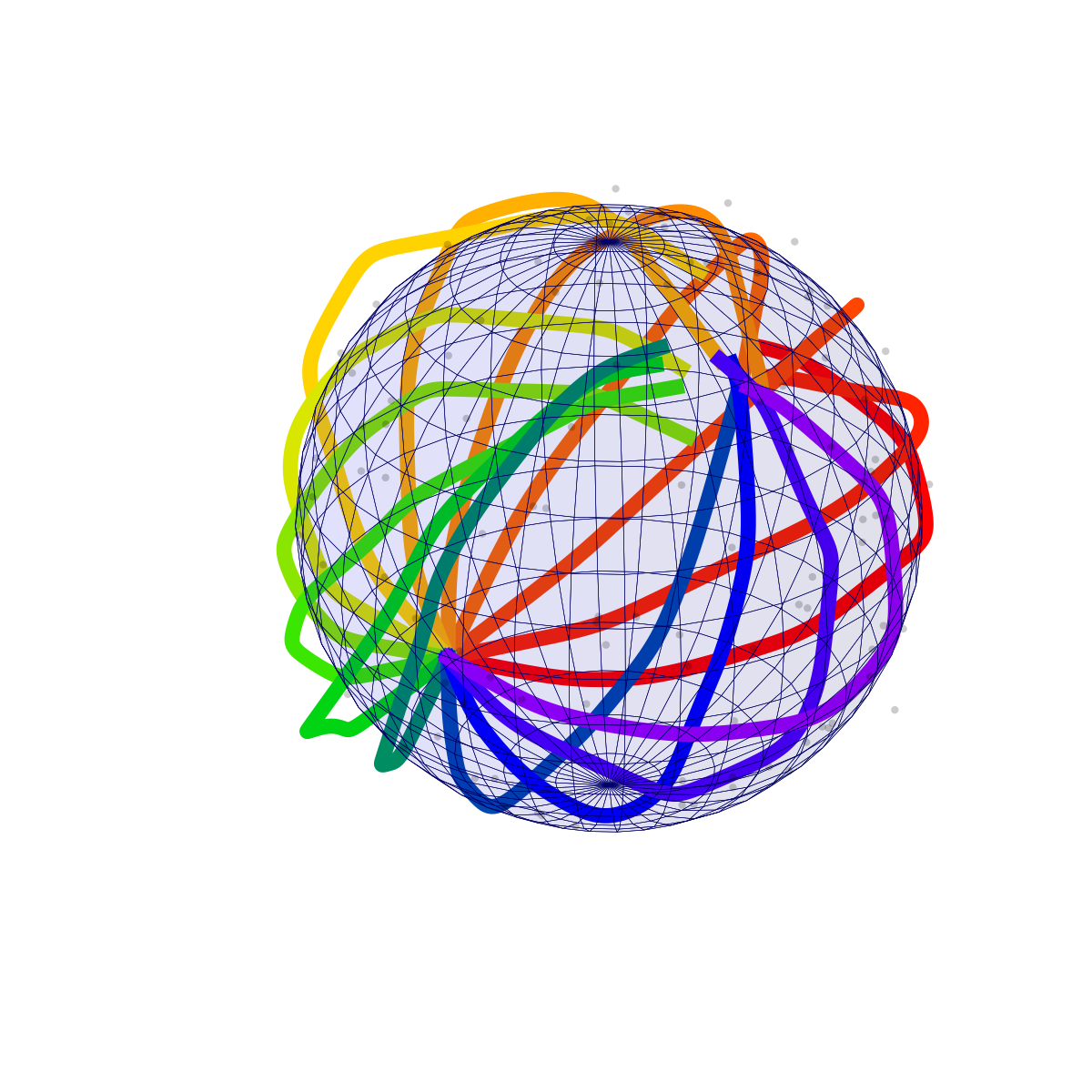}
    \caption{$h = 0.2$}
  \end{subfigure}\\
  \begin{subfigure}[b]{0.32\textwidth}
    \centering
    \includegraphics[width=.85\linewidth]{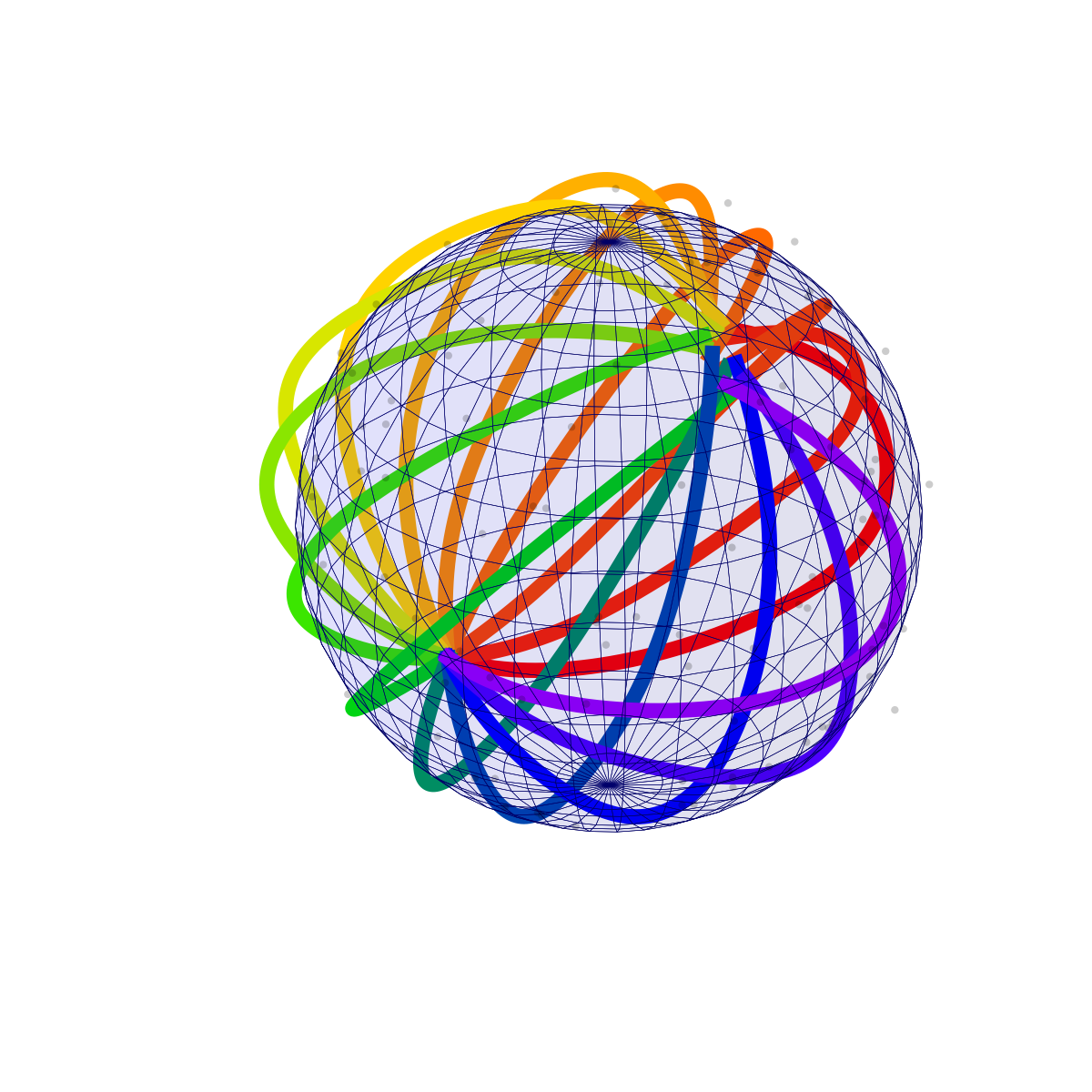}
    \caption{$h = 0.5$}
  \end{subfigure}%
  %\hspace{0.15 in}
  \begin{subfigure}[b]{0.32\textwidth}
    \centering
    \includegraphics[width=.85\linewidth]{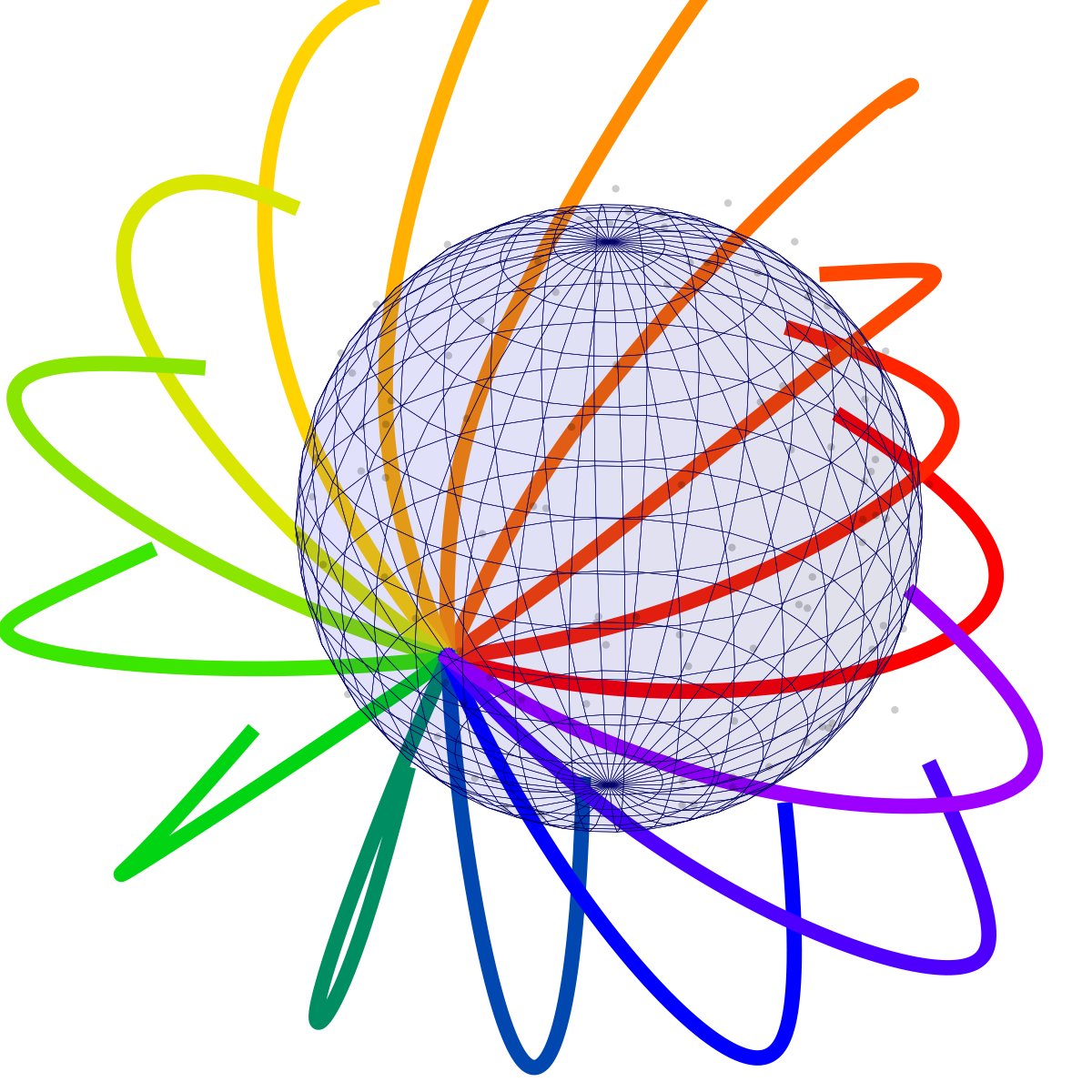}
    \caption{$h = 1$}
  \end{subfigure}%
  \begin{subfigure}[b]{0.32\textwidth}
    \centering
    \includegraphics[width=.85\linewidth]{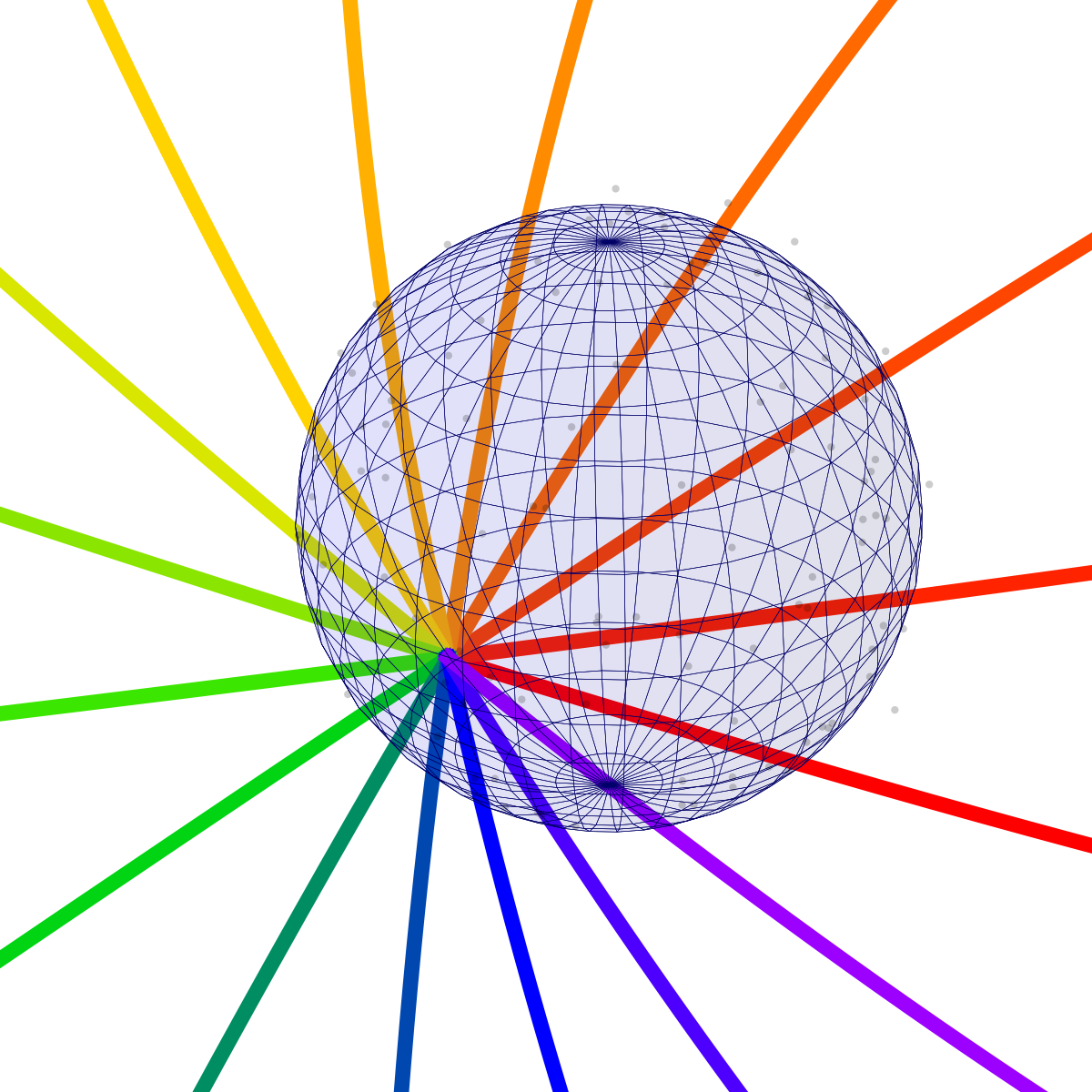}
    \caption{$h = 2$}
  \end{subfigure}% 
  \caption{Principal submanifold results for noise level $\sigma = 0.05$ and sample size $n = 100$. Even in these extremely sparse data sets, intermediate bandwidths $h = 0.2$ and $h = 0.5$ still yield quite good results and recover the spherical shape.}
  \label{fig:bandwidth_100}
\end{figure}

From Figures \ref{fig:bandwidth_10000}, \ref{fig:bandwidth_1000}, and \ref{fig:bandwidth_100}, one can clearly see that the principal submanifold algorithm requires a bandwidth which is more than a factor of $2$ larger than the noise level of the data. For lower bandwidth, the eigenvectors of local PCA become too variable and the two largest eigenvalues may not correspond to vectors whose span is close to tangential to the underlying sphere. For large bandwidths, which are less than a factor of $2$ below the diameter of the sphere, the whole data set contributes to the local PCA, leading to too weak dependence of eigenvectors on the neighborhood of a point and thus to too slow variation of the eigenspaces. As a result, the lines curve much less than the sphere does and therefore progressively move away from the data. These two effects are to be expected and constitute fundamental limitations of any local neighborhood approach.

By comparing Figures \ref{fig:bandwidth_10000}, \ref{fig:bandwidth_1000}, and \ref{fig:bandwidth_100}, one can see that the bandwidth dependence is exacerbated by reducing the sample size. However, even for $n=100$ the intermediate bandwidths $h = 0.2$ and $h = 0.5$ still achieve remarkably good results. This underscores the potential strength of principal submanifolds as a means of identifying low dimensional structure in data sets even of moderate size.

%MINST Data (added 2024)
\section{Principal variation on the MNIST data set}\label{app:12-digits-princ-var}
This section presents a detailed exploration of the principal sub-manifolds within the MNIST handwritten digit dataset, a comprehensive repository of handwritten digits widely utilized for the development and testing of various image processing algorithms. The dataset, accessible at \url{https://yann.lecun.com/exdb/mnist/}, encompasses a collection of 70,000 grayscale images, each of 28$\times$28 pixel resolution, depicting digits from ``0'' to ``9''. 

\begin{figure}[t]
  \centering
  \includegraphics[width=1\linewidth]{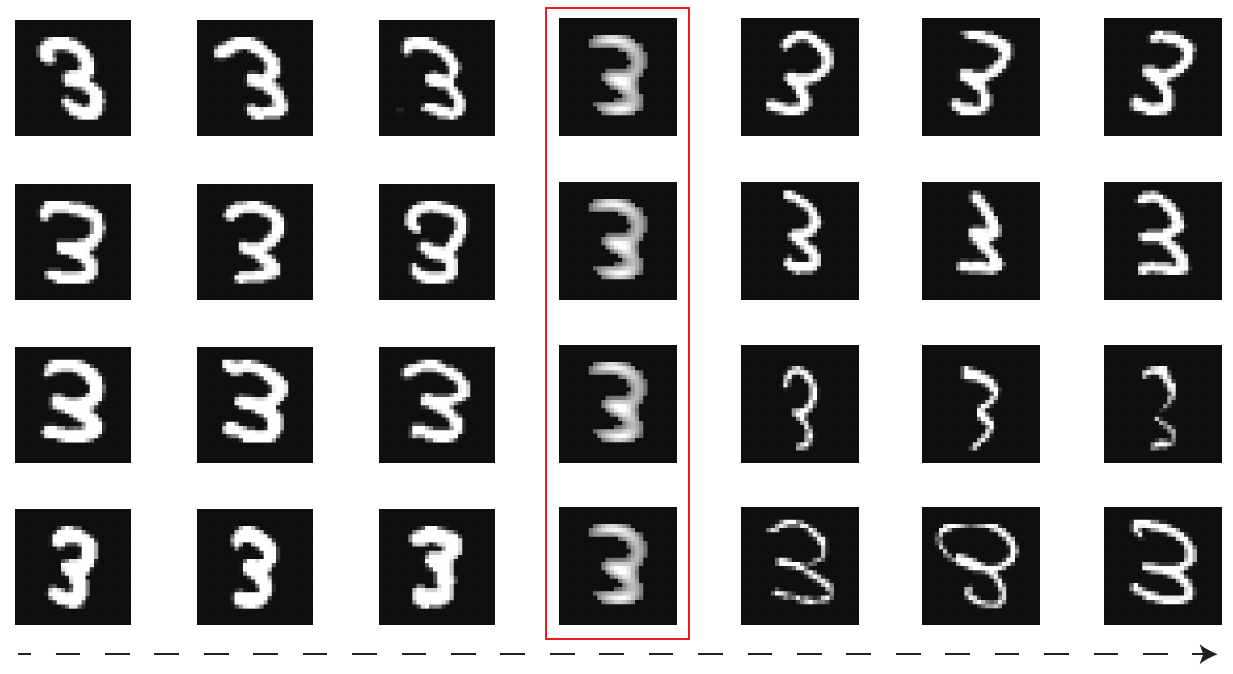}
  \caption{Principal sub-manifolds of the digit ``3" extracted from the MNIST dataset, originating from the dataset's mean. Each row illustrates the variation of the principal sub-manifolds across distinct principal directions, sequentially ordered from the first to the fourth principal direction, from the top row to the bottom. The digit displayed within a red frame at the center of each row denotes the mean.}
  \label{fig:PS_MNIST}
\end{figure}

In this analysis, we specifically focus on the digit ``3" to demonstrate the variability inherent in handwriting styles. This choice is motivated by the digit's capacity to exhibit a wide range of variations, such as differences in inclination angles, stroke thickness, opening sizes, and the curvature of junctions. Through the analysis of 7,141 instances of the digit ``3" from the MNIST dataset, we construct the principal sub-manifold centered around the digit's mean and examine it across four principal directions, see Figure \ref{fig:PS_MNIST}. This approach allows us to systematically describe the morphological changes of the digit ``3" across distinct dimensions of variation.

The first principal direction reveals a significant variation in the inclination of the digit ``3", transitioning from a leftward lean at the top to a rightward orientation. The second principal direction highlights a transformation in the corners of the ``3", ranging from smooth to pronouncedly sharp angles. In the third dimension, we observe the two semi-circular arcs of the ``3" evolving from fuller to more slender forms, indicating a variation in the digit's overall robustness. Finally, the fourth direction underscores changes in stroke thickness, illustrating a spectrum from markedly thick to fine lines. 

This multifaceted analysis not only underscores the diversity of handwriting styles captured within the MNIST dataset but also demonstrates the utility of principal sub-manifold analysis in uncovering the underlying patterns of variation in digital representations of handwritten digits. Such insights are invaluable for enhancing the accuracy and robustness of machine learning models in tasks related to handwriting recognition and image processing.

\section{Principal variation of leaf growth}\label{app:13-leaves}
\begin{figure}[ht!]
  \centering
  \begin{subfigure}[b]{0.4\textwidth}
    \includegraphics[width=\textwidth]{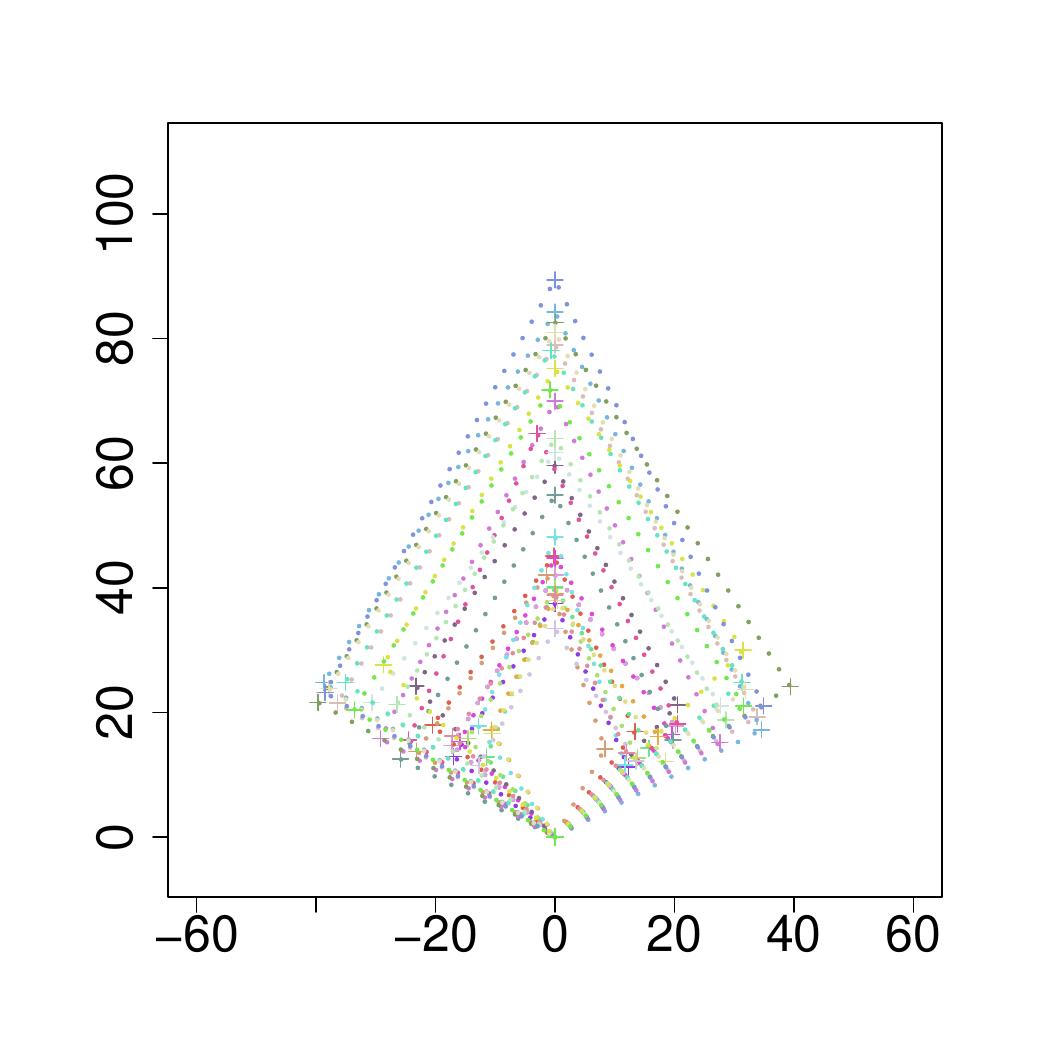}
    \caption{}
  \end{subfigure}
  \begin{subfigure}[b]{0.4\textwidth}
    \includegraphics[width=\textwidth]{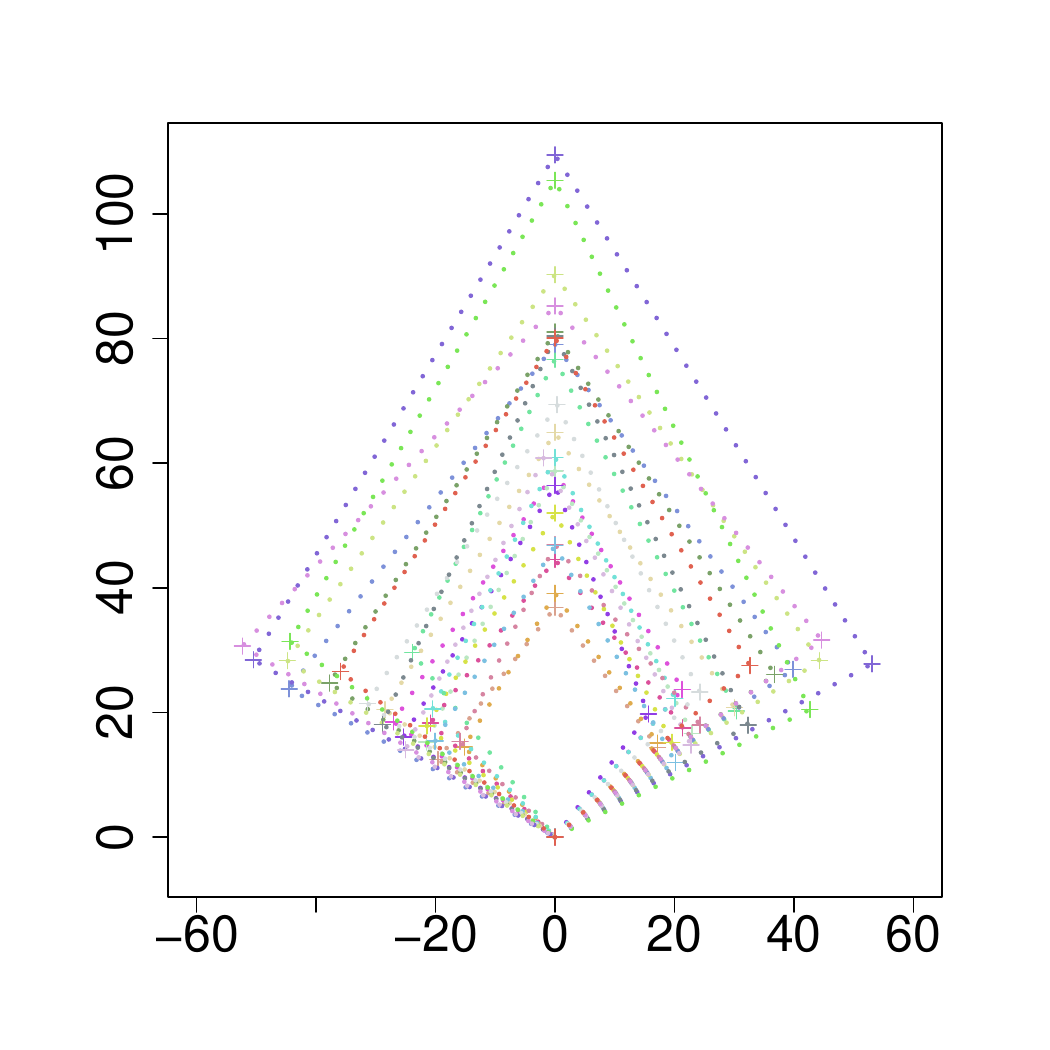}
    \caption{}
  \end{subfigure}\\
  \begin{subfigure}[b]{0.4\textwidth}
    \includegraphics[width=\textwidth]{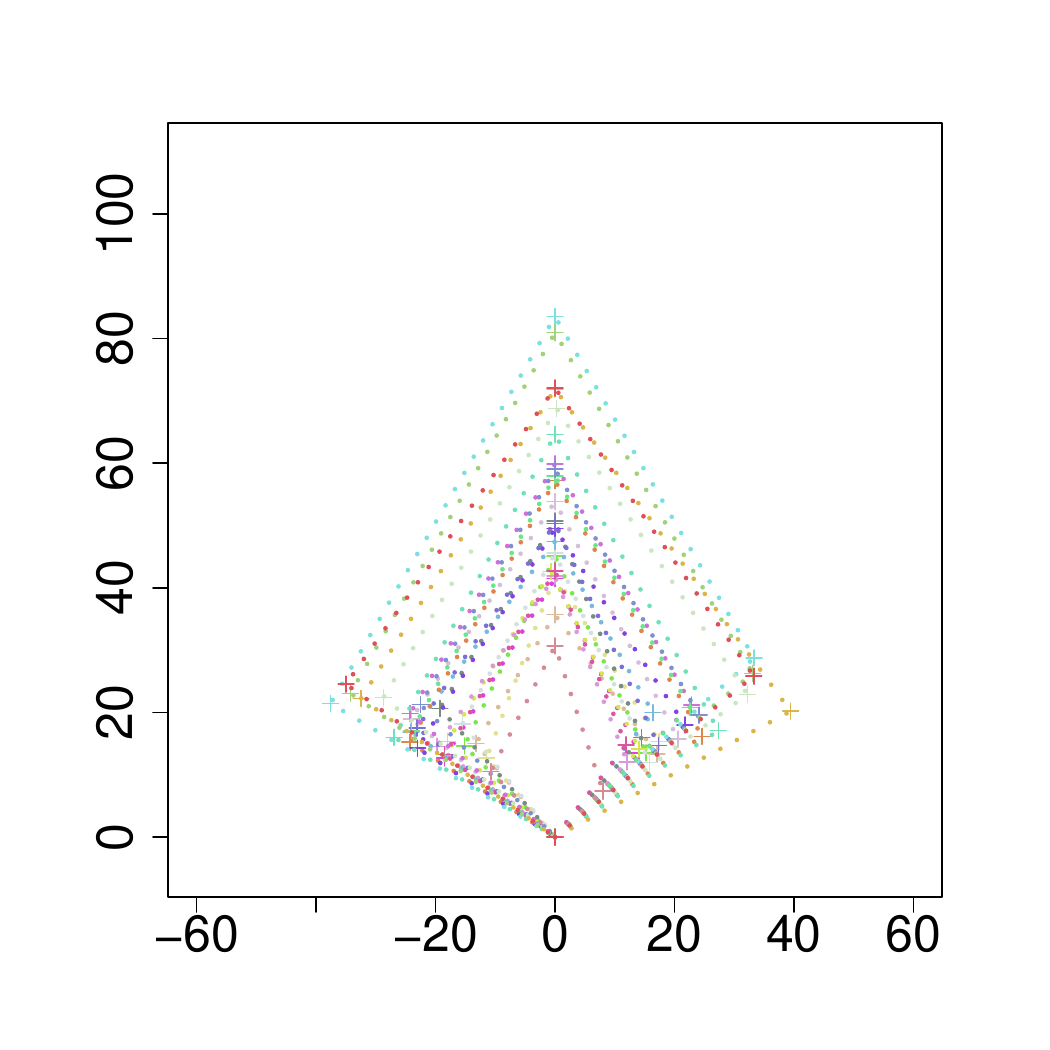}
    \caption{}
  \end{subfigure}
  \begin{subfigure}[b]{0.4\textwidth}
    \includegraphics[width=\textwidth]{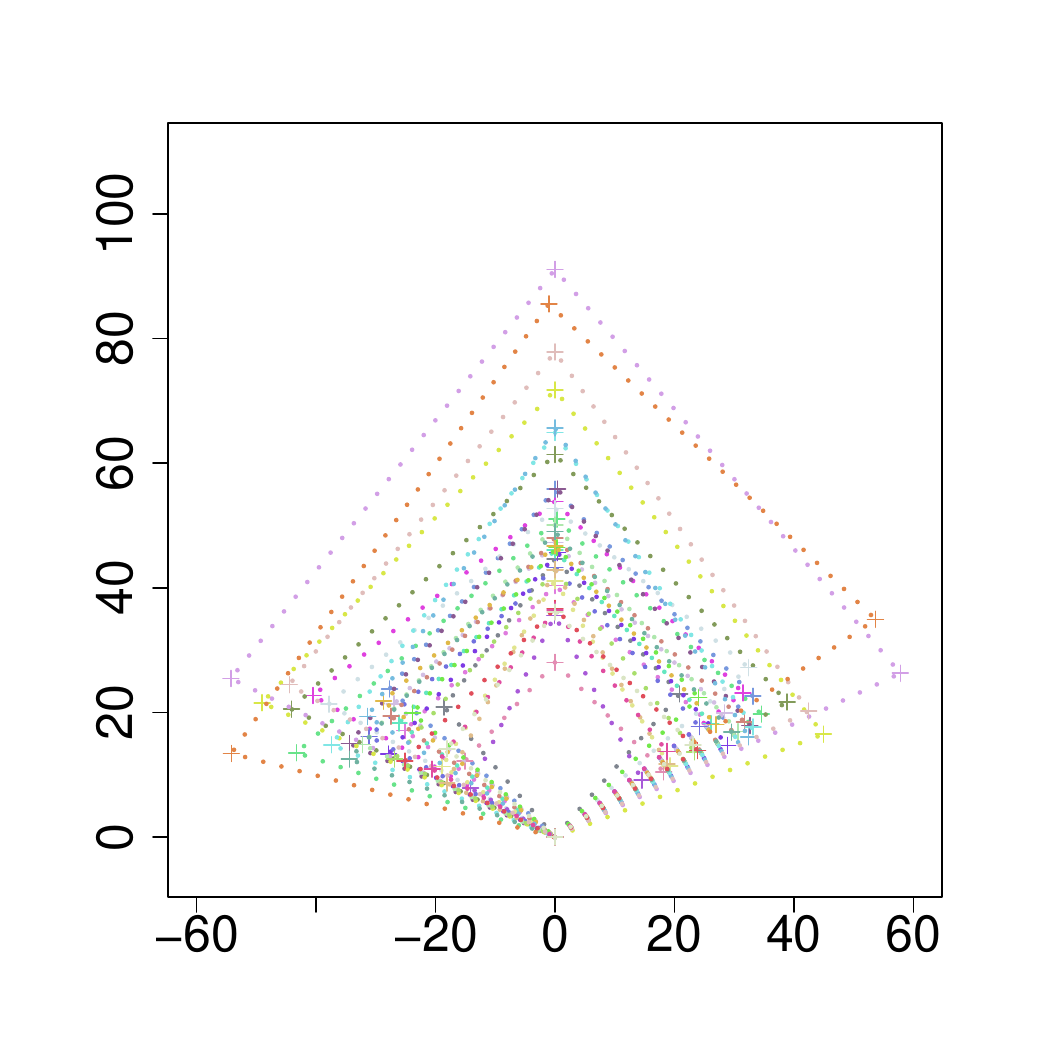}
    \caption{}
  \end{subfigure}
  \caption{Leaf growth over a growing period of Clone 1 (a), Clone 2 (b), Clone 3 (c), and a reference tree (d). (a) Four landmarks on the leaf of Clone 1 have been connected and represented by a polygon at each growing period (27 polygons totally); (b)-(d) provide the same information for Clone 2 (22 polygons), Clone 3 (24 polygons) and the reference tree (31 polygons).}
  \label{leaf-plot}
\end{figure}

We also considered a landmark data set consisting of leaf growth, collected from three Clones and a reference tree of young black Canadian poplars at an experimental
site at the University of G\"{o}ttingen (\url{http://stochastik.math.uni-goettingen.de/~huckeman/ishapes_1.0.1.tar.gz}). The landmark configurations of the leaves were collected from three Clones (`C1', `C2', `C3') and a reference tree (`r') collected at two different levels: breast height (Level 1) and the crown (Level 2).  They consist of the shapes of 27 leaves (nine from Level 1 and eighteen from Level 2) from Clone 1;  of 22 leaves (six from Level 1 and sixteen from Level 2) from Clone 2; and of 24 leaves (eighteen from Level 1 and seventeen from Level 2) from Clone 3 as well as of the shapes of 21 leaves (thirteen from Level 1 and eighteen from Level 2) from the reference tree, all of which have been recorded non-destructively over several days during a major portion of their growing period of approximately one month. There are four landmarks corresponding to quadrangular configuration at petiole, tip, and largest extensions orthogonal to the connecting line. Figure \ref{leaf-plot} represents the four landmarks extracted from the contour image of each leaf on a flat plane, the four landmarks contain, in particular, the information of length, width, vertical and horizontal asymmetry. 

Although it is known that the leaf growth of the genetically identical trees along a period of time reveals a non-Euclidean pattern \cite{huckemann2011}, the study only focused on the mean geodesic difference (therefore essentially a one-dimensional variation), which is used for the discriminant analysis across the trees. However, the shape change along different directions---especially the principal directions in shape space---has not been fully explored. We will investigate the shape variation using principal sub-manifold among three Clones and the reference tree.  As can be expected (see in Section 2.2 in the paper), each landmark configuration, represented by a polygon in Figure \ref{leaf-plot}, corresponds to a point in Kendall shape space. We focus on the non-geodesic shape variation primarily in vertical and horizontal direction of the leaf growth, the analysis of which requires a multi-dimensional scale treatment. %The goal is to study the shape variation among different directions in the shape space. 

\begin{figure}[ht!]
  \centering
  \begin{subfigure}[b]{0.24\textwidth}
    \includegraphics[width=\textwidth]{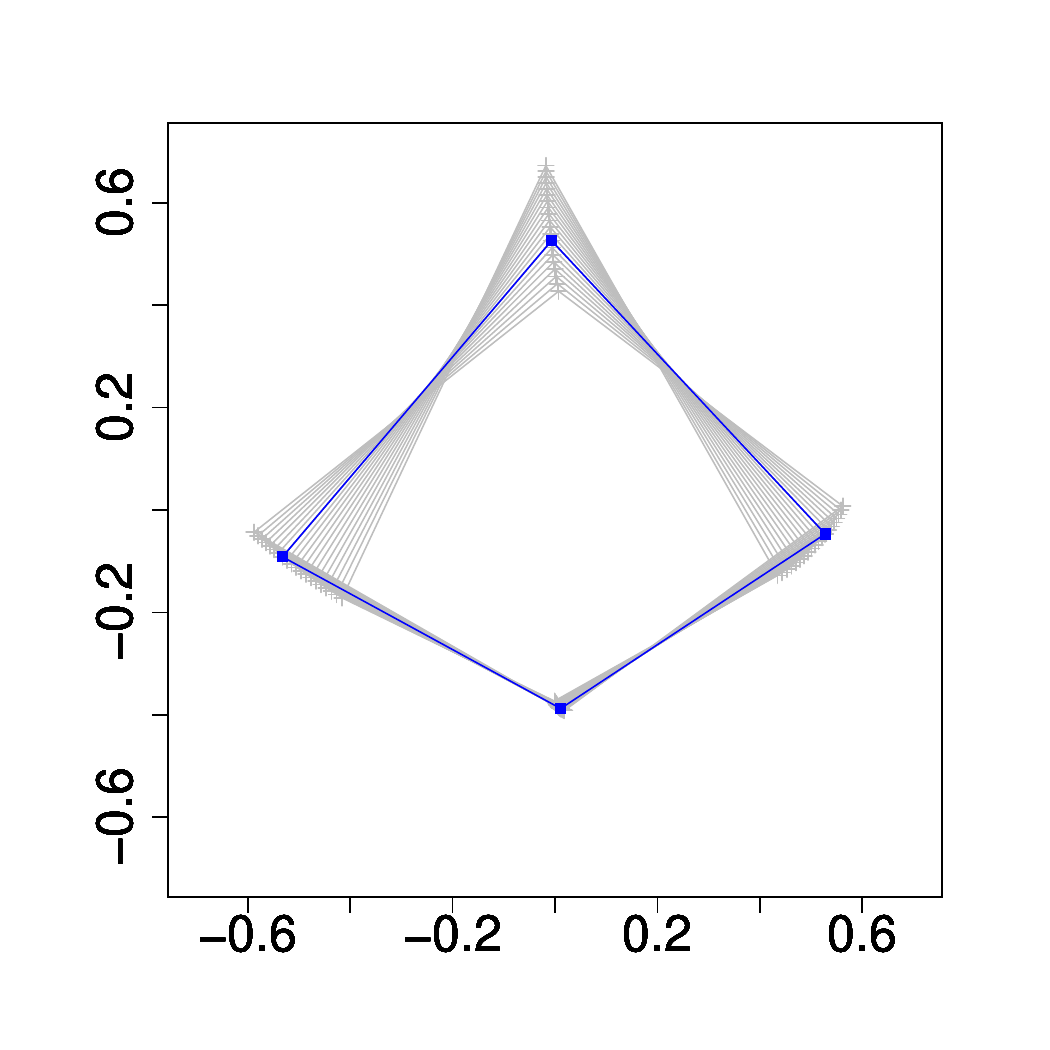}
    \caption{}
  \end{subfigure}
  \begin{subfigure}[b]{0.24\textwidth}
    \includegraphics[width=\textwidth]{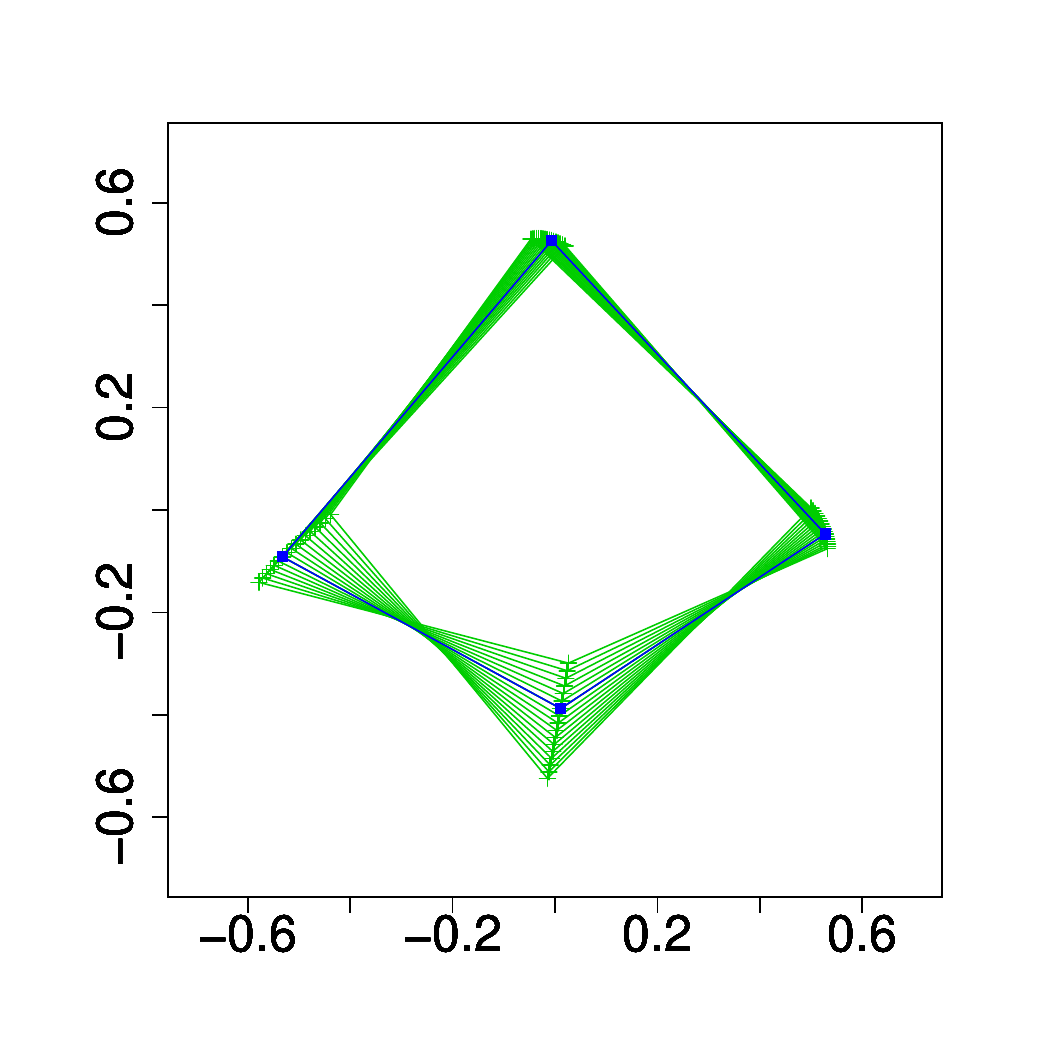}
    \caption{}
  \end{subfigure}
  \begin{subfigure}[b]{0.24\textwidth}
    \includegraphics[width=\textwidth]{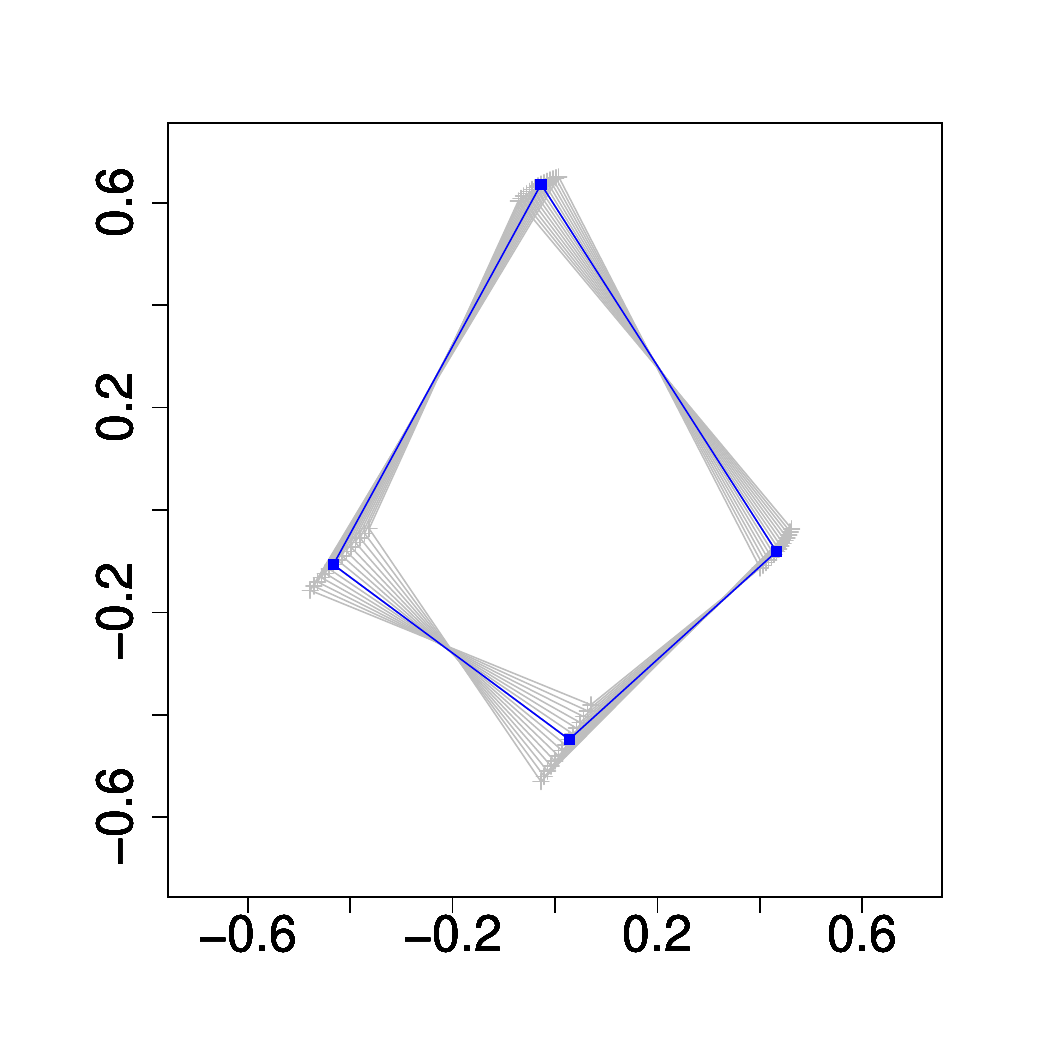}
    \caption{}
  \end{subfigure}
  \begin{subfigure}[b]{0.24\textwidth}
    \includegraphics[width=\textwidth]{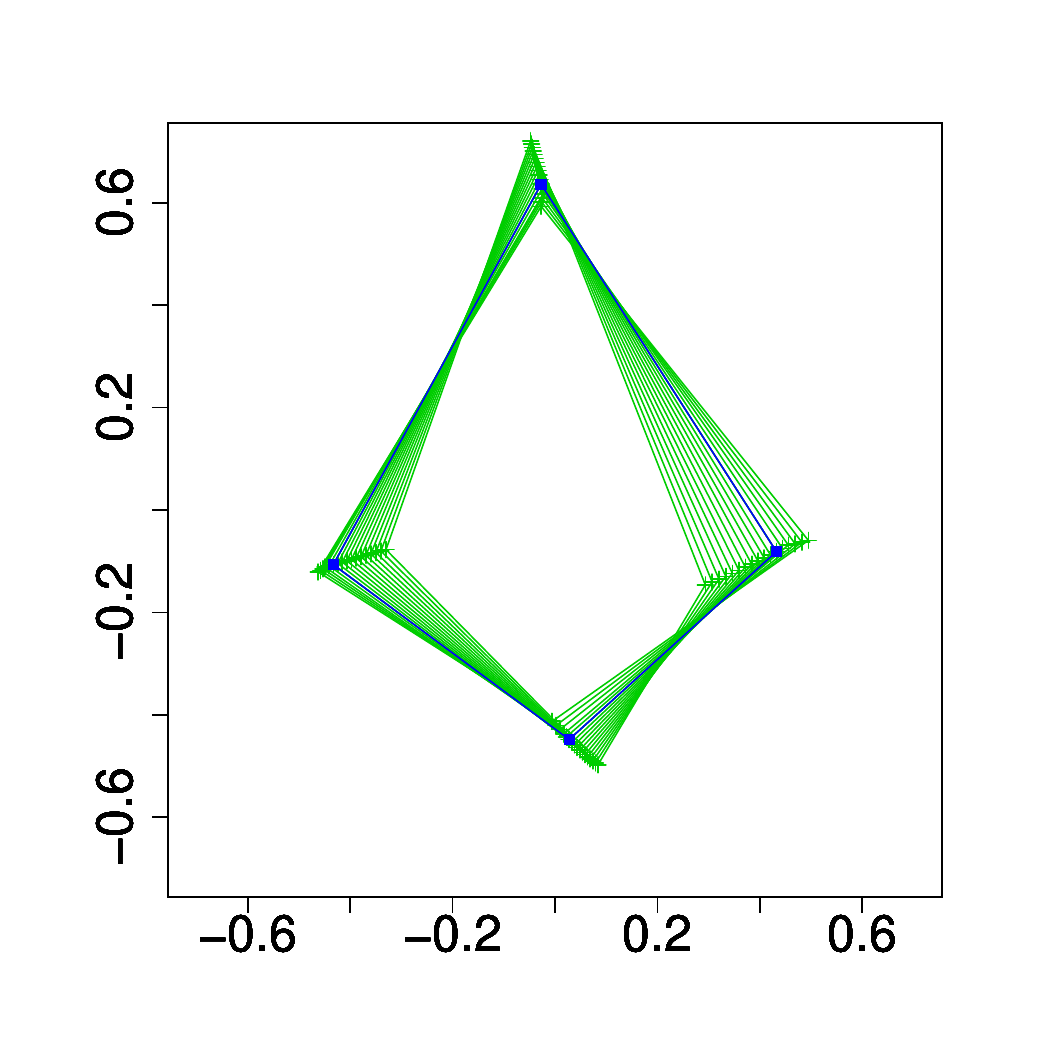}
    \caption{}
  \end{subfigure}\\
  \begin{subfigure}[b]{0.24\textwidth}
    \includegraphics[width=\textwidth]{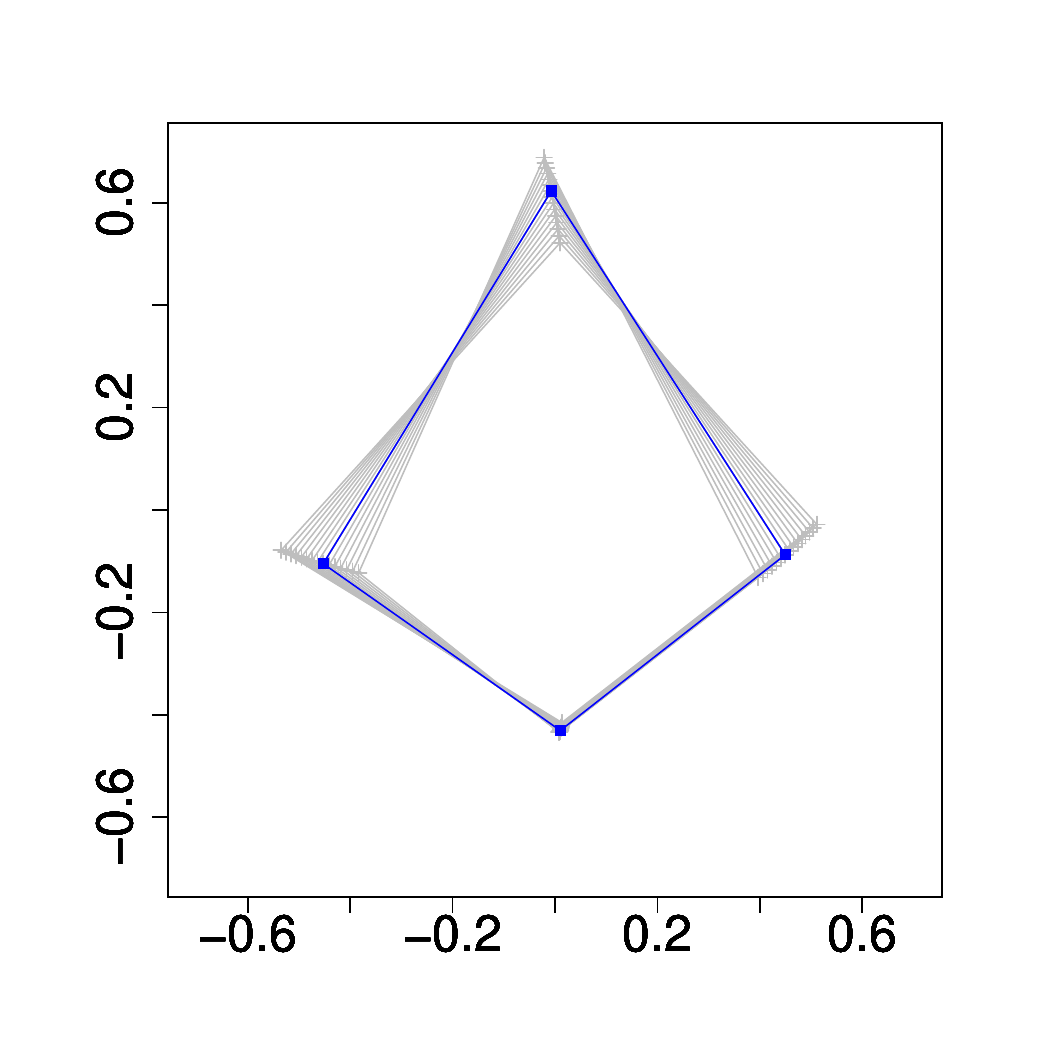}
    \caption{}
  \end{subfigure}
  \begin{subfigure}[b]{0.24\textwidth}
    \includegraphics[width=\textwidth]{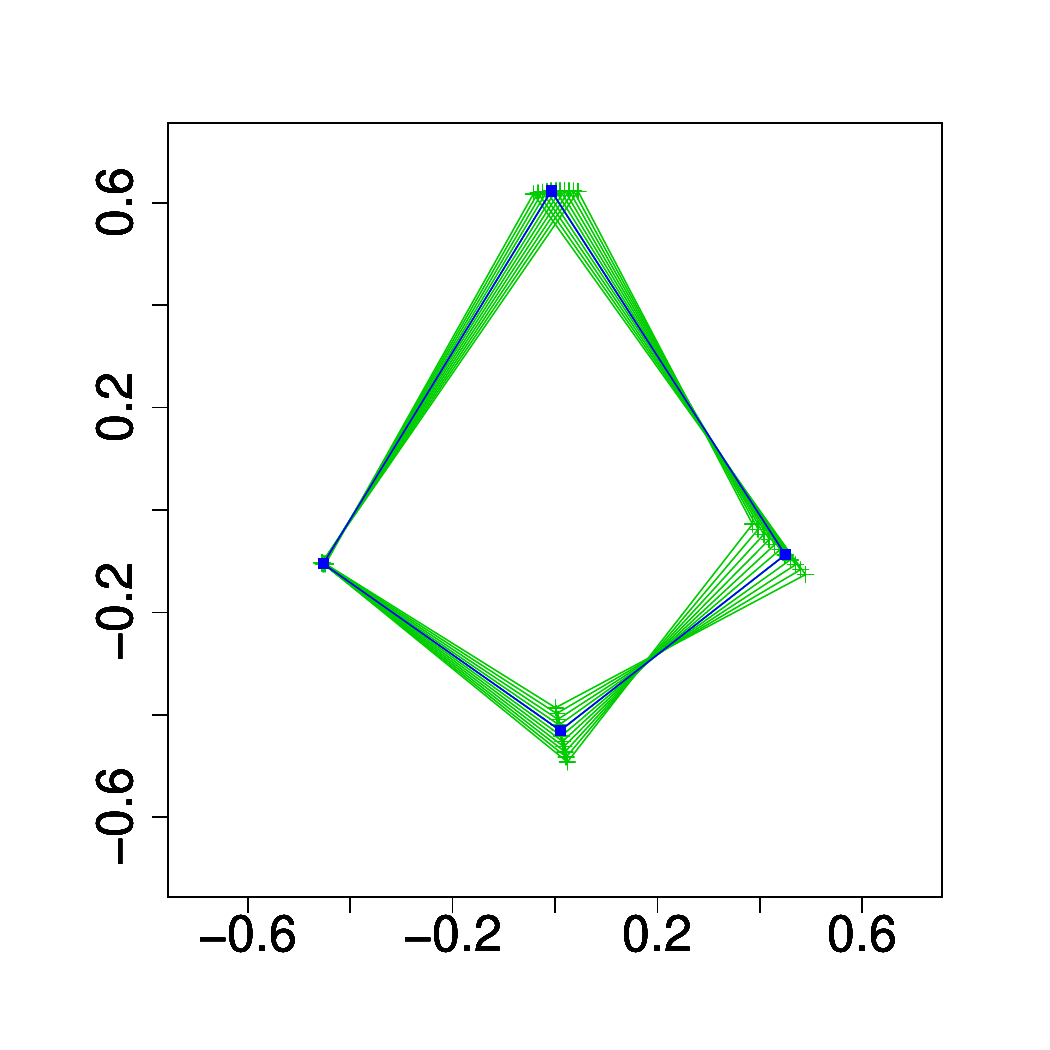}
    \caption{}
  \end{subfigure}
  \begin{subfigure}[b]{0.24\textwidth}
    \includegraphics[width=\textwidth]{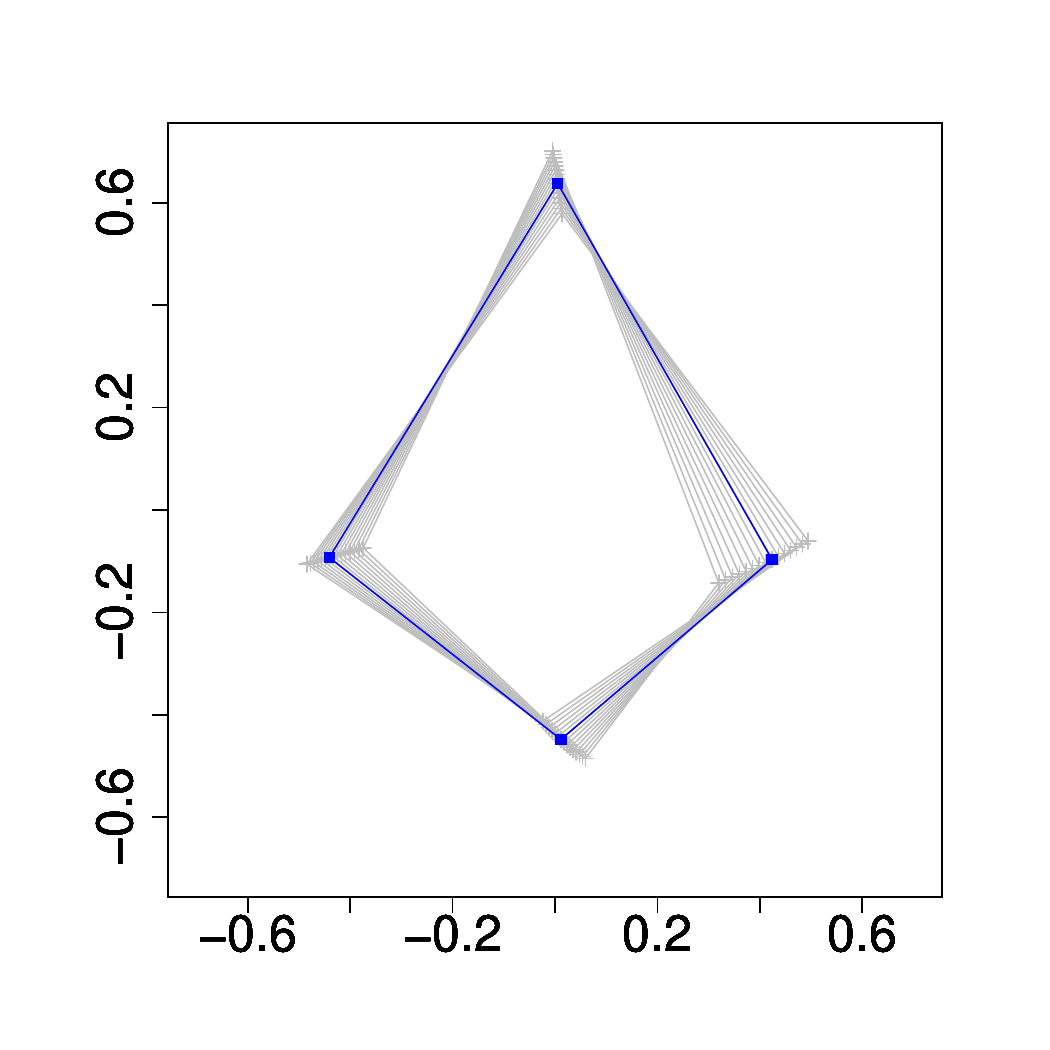}
    \caption{}
  \end{subfigure}
  \begin{subfigure}[b]{0.24\textwidth}
    \includegraphics[width=\textwidth]{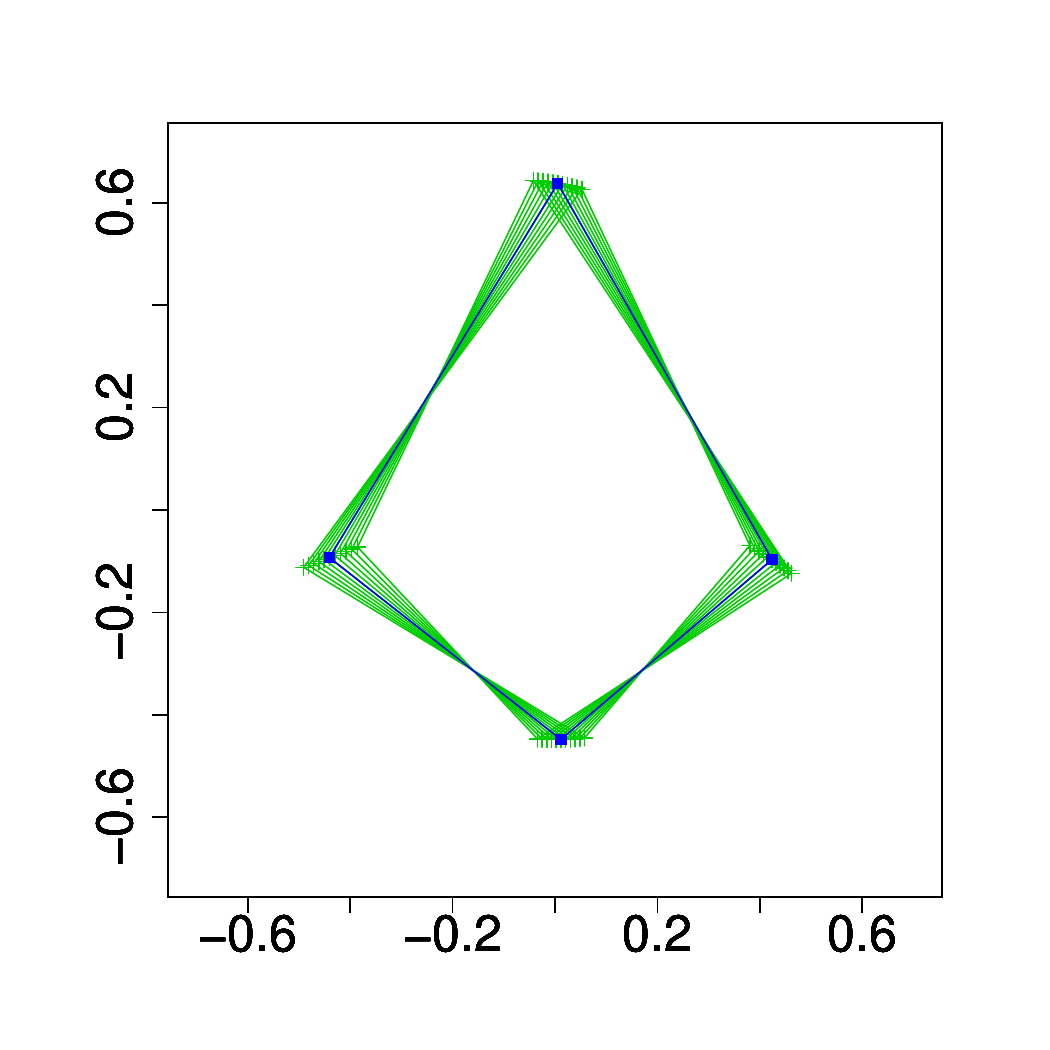}
    \caption{}
  \end{subfigure}
  \caption{Principal sub-manifolds of the leaf growth data. (a) First principal direction obtained from the combined leaves at breast height and the crown of the reference tree; (b) Second principal direction obtained from the combined leaves at breast height and the crown of the reference tree. (c)-(h) provide the same information for Clone 1, 2 and 3.}
  \label{combined}
\end{figure}
%\end{sidewaysfigure}
%\end{figure}

%\begin{sidewaysfigure}[htbp]
\begin{figure}[ht!]
  \centering
  \includegraphics[width=0.24\textwidth]{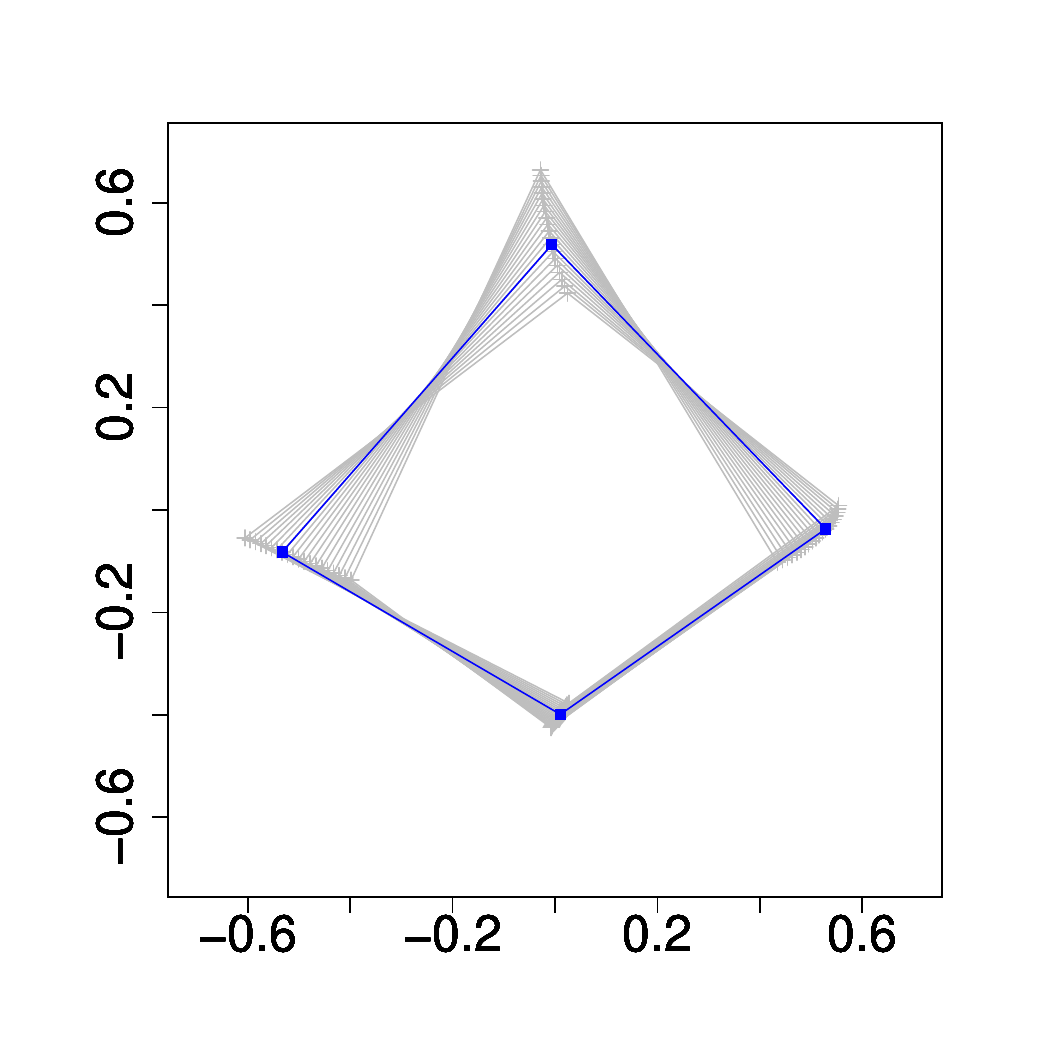}
  \includegraphics[width=0.24\textwidth]{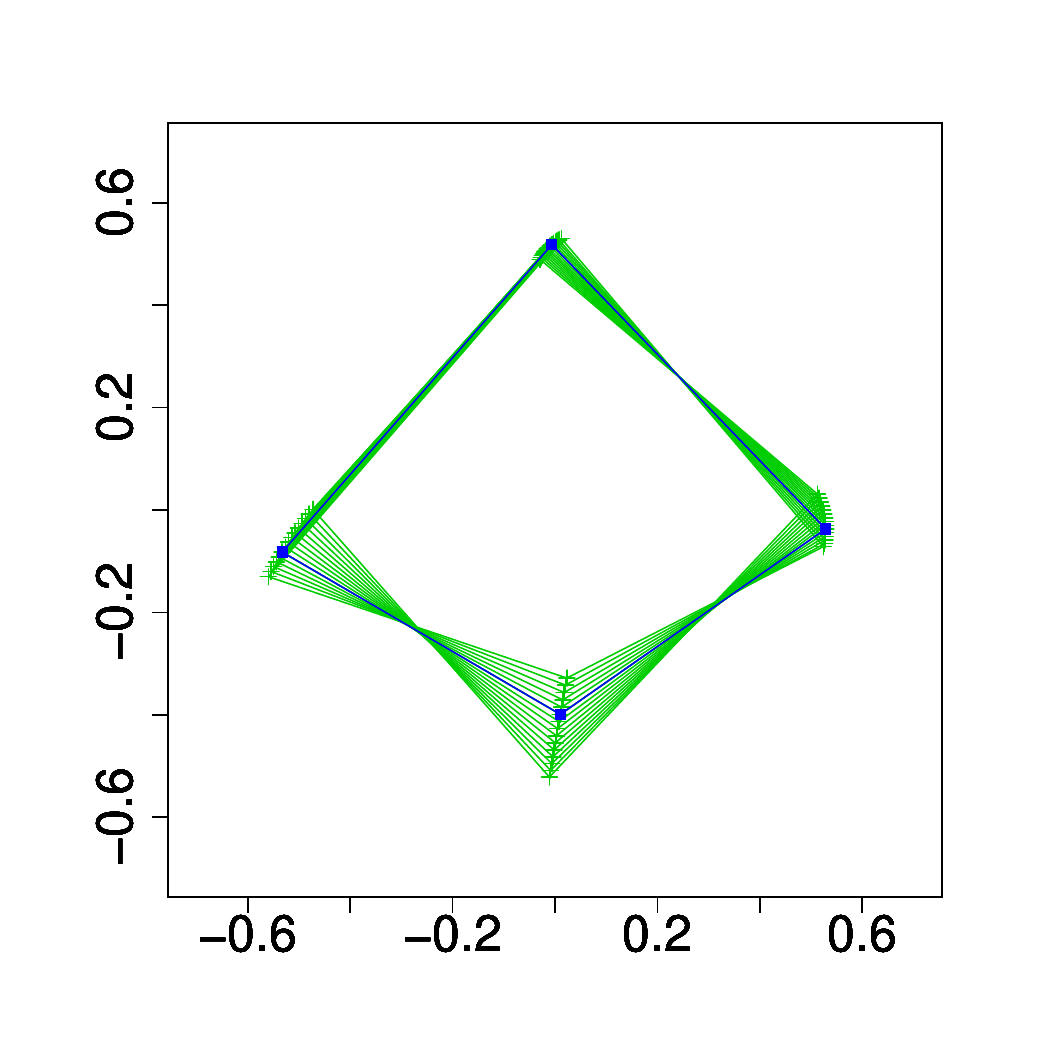}
  \includegraphics[width=0.24\textwidth]{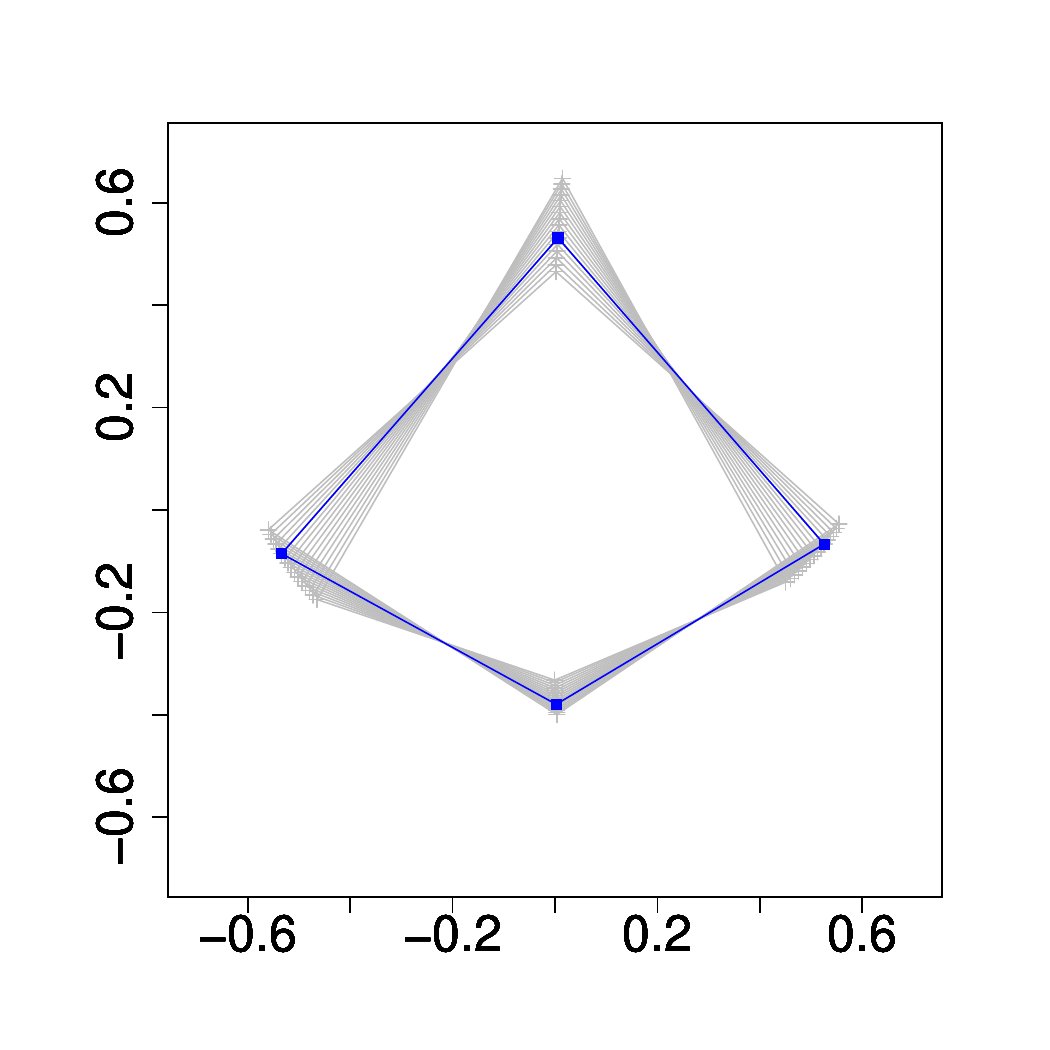}
  \includegraphics[width=0.24\textwidth]{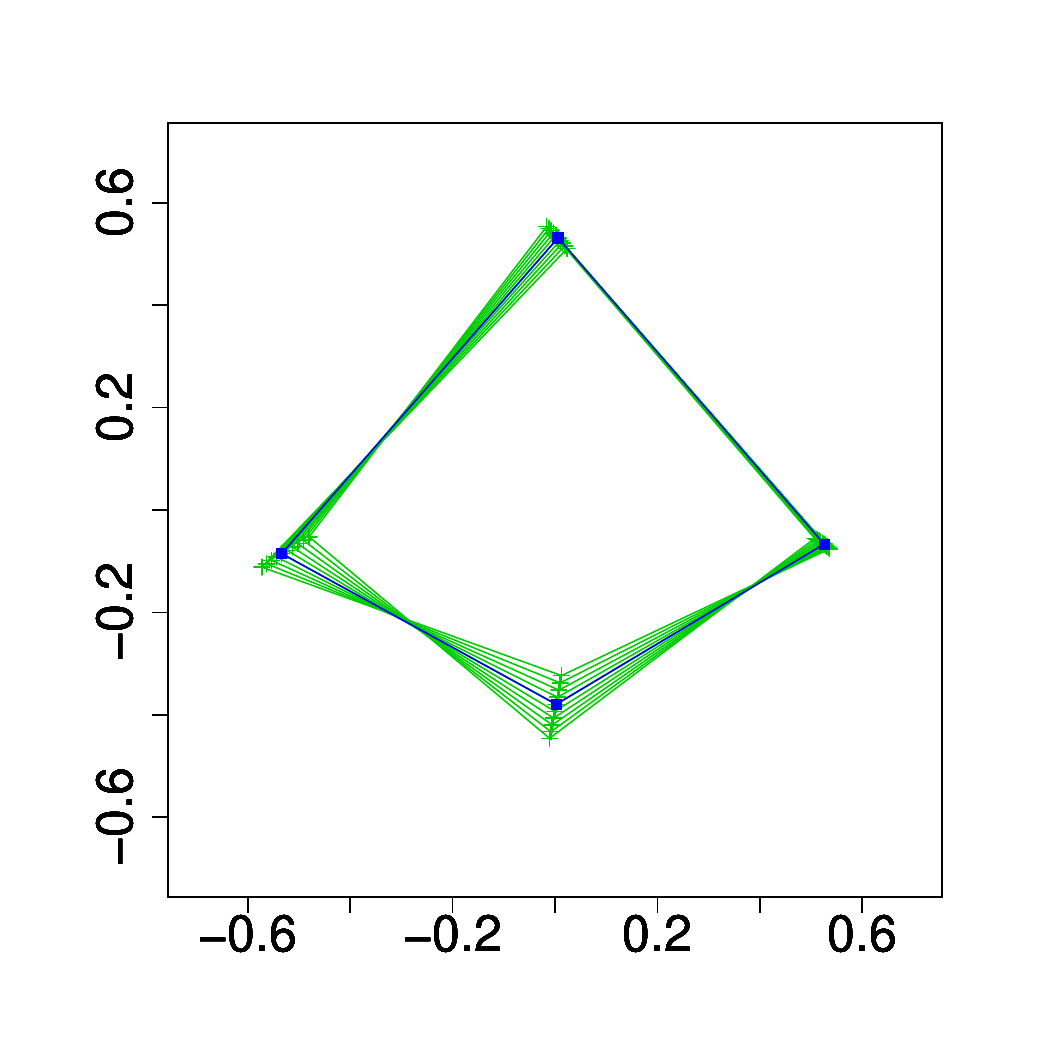}\\
  \includegraphics[width=0.24\textwidth]{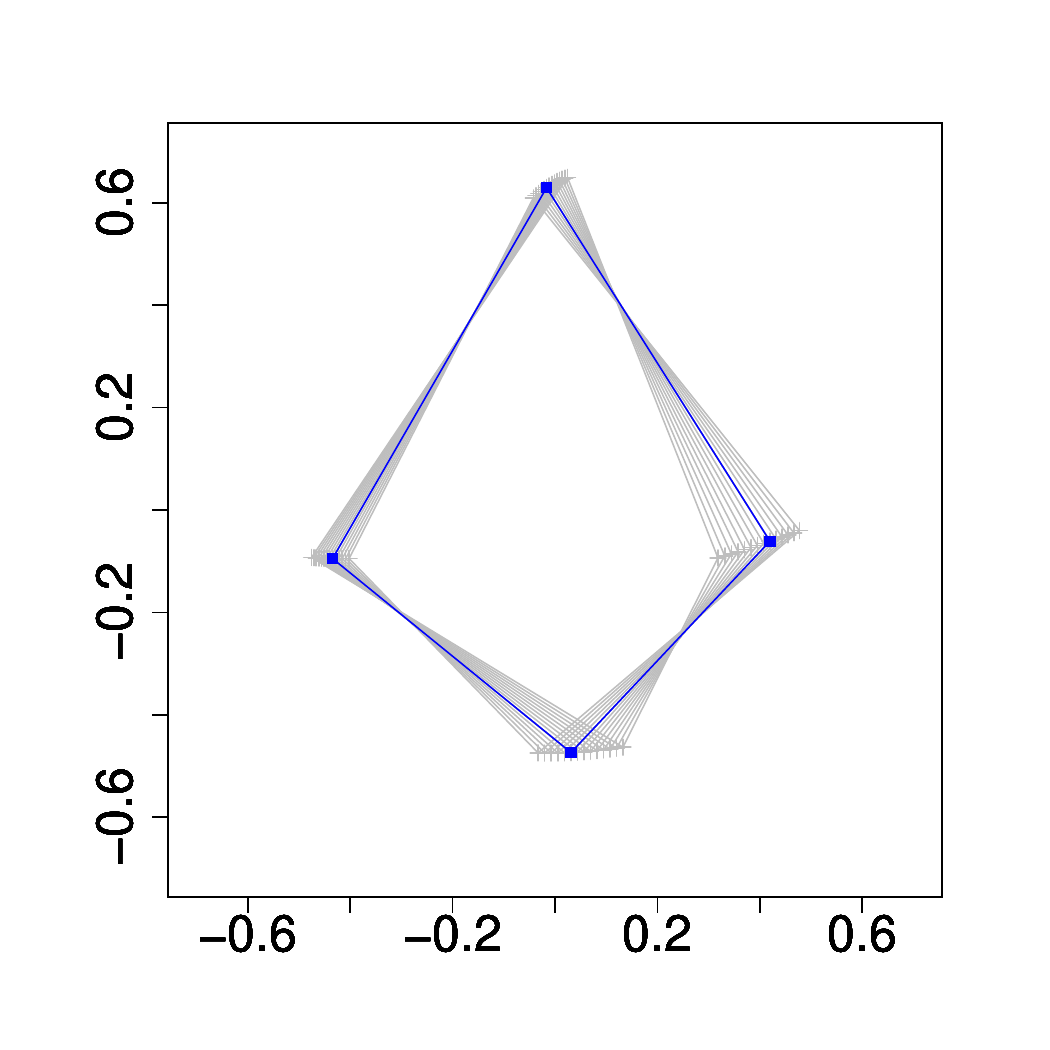}
  \includegraphics[width=0.24\textwidth]{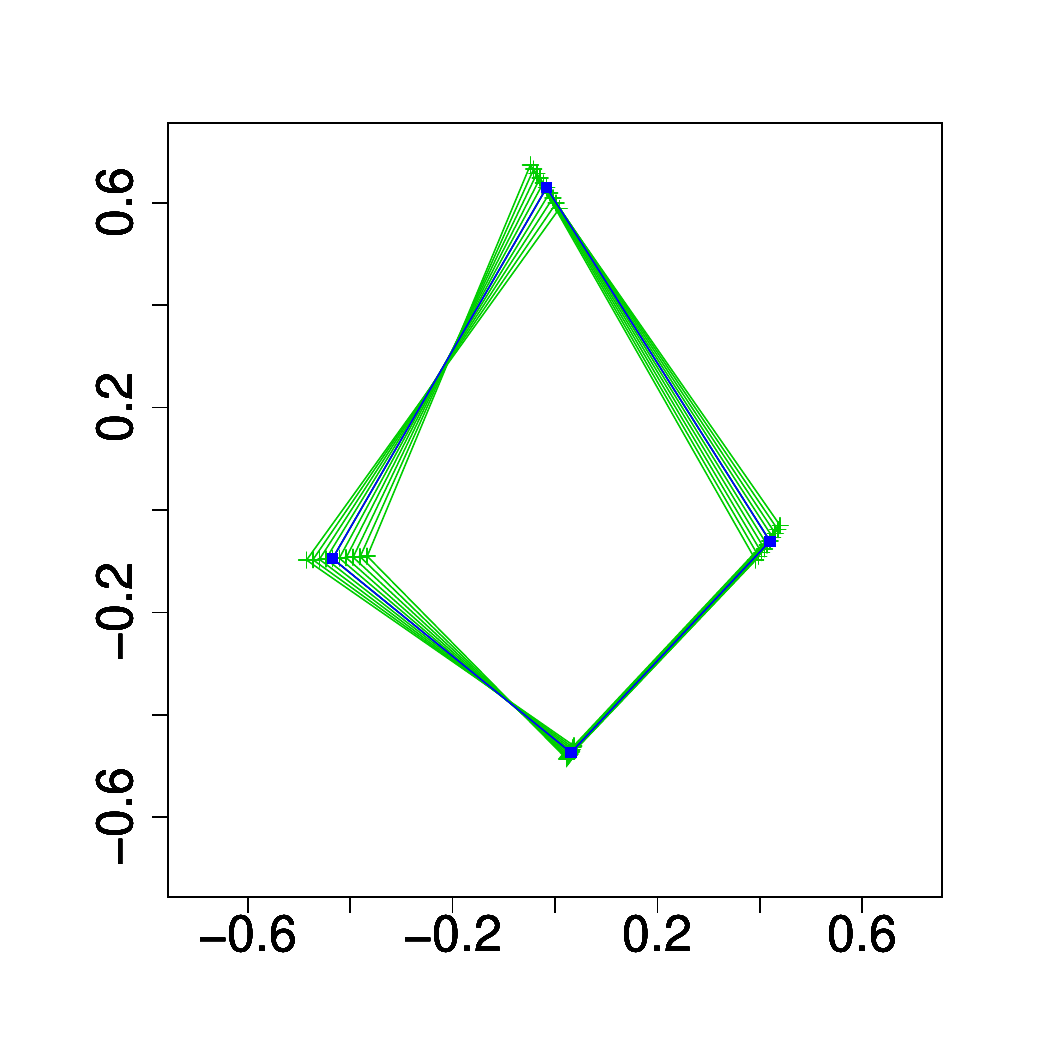}
  \includegraphics[width=0.24\textwidth]{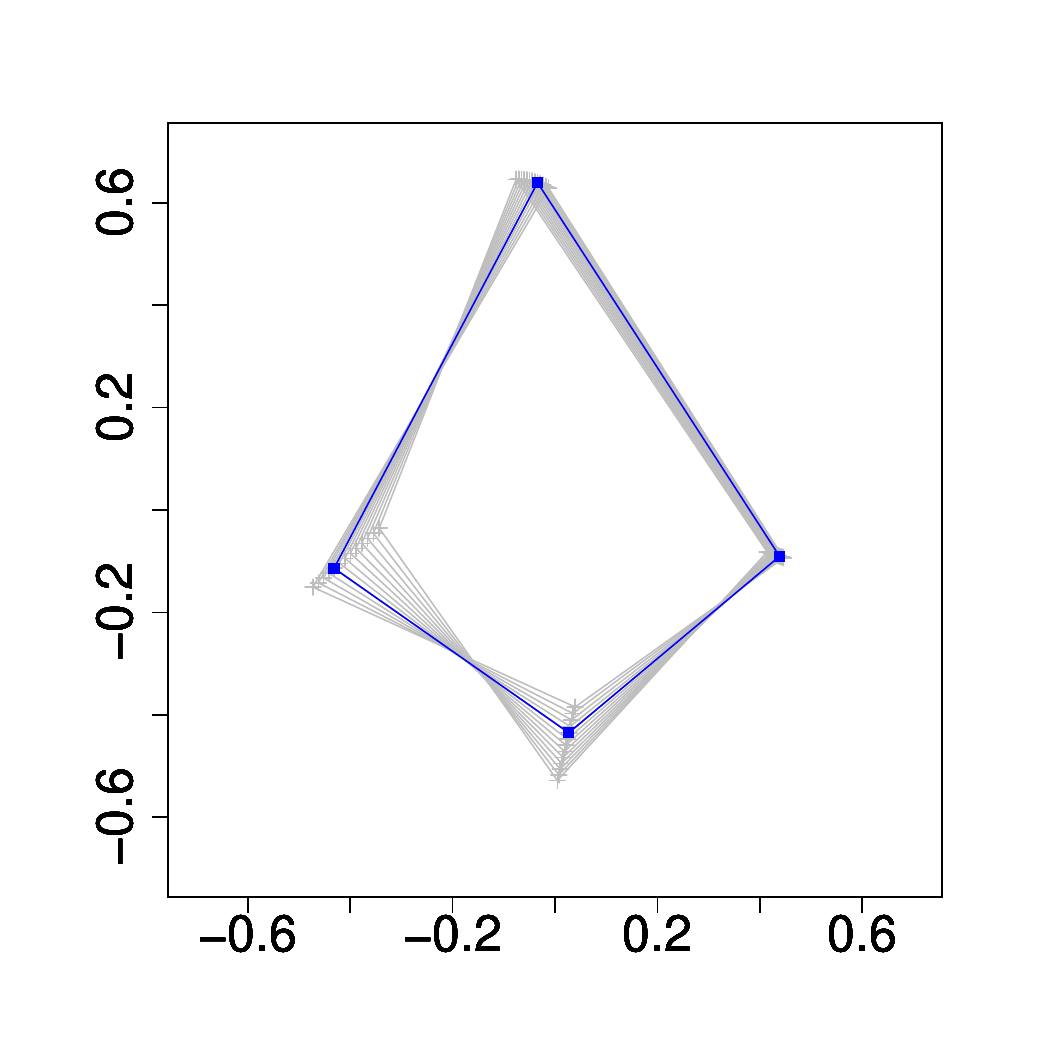}
  \includegraphics[width=0.24\textwidth]{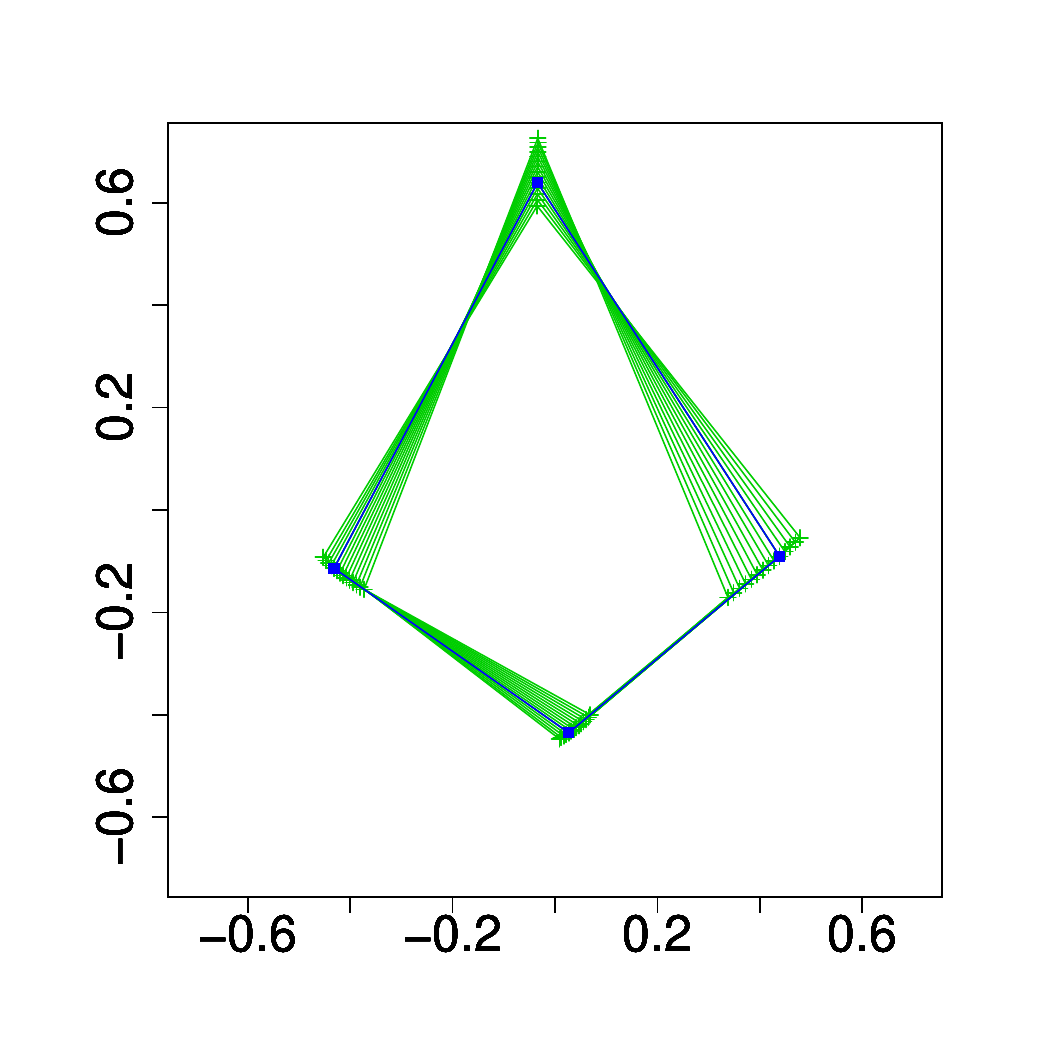}\\
  \includegraphics[width=0.24\textwidth]{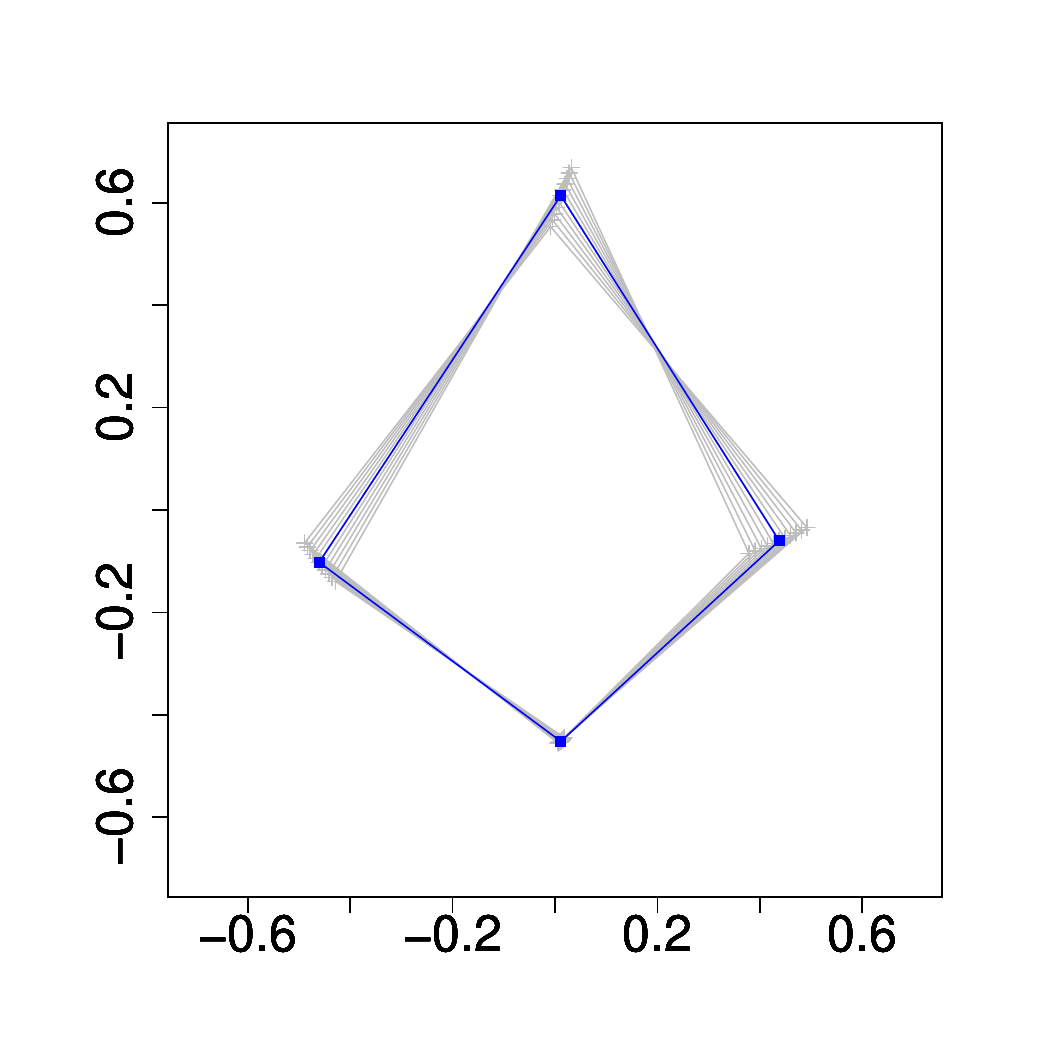}
  \includegraphics[width=0.24\textwidth]{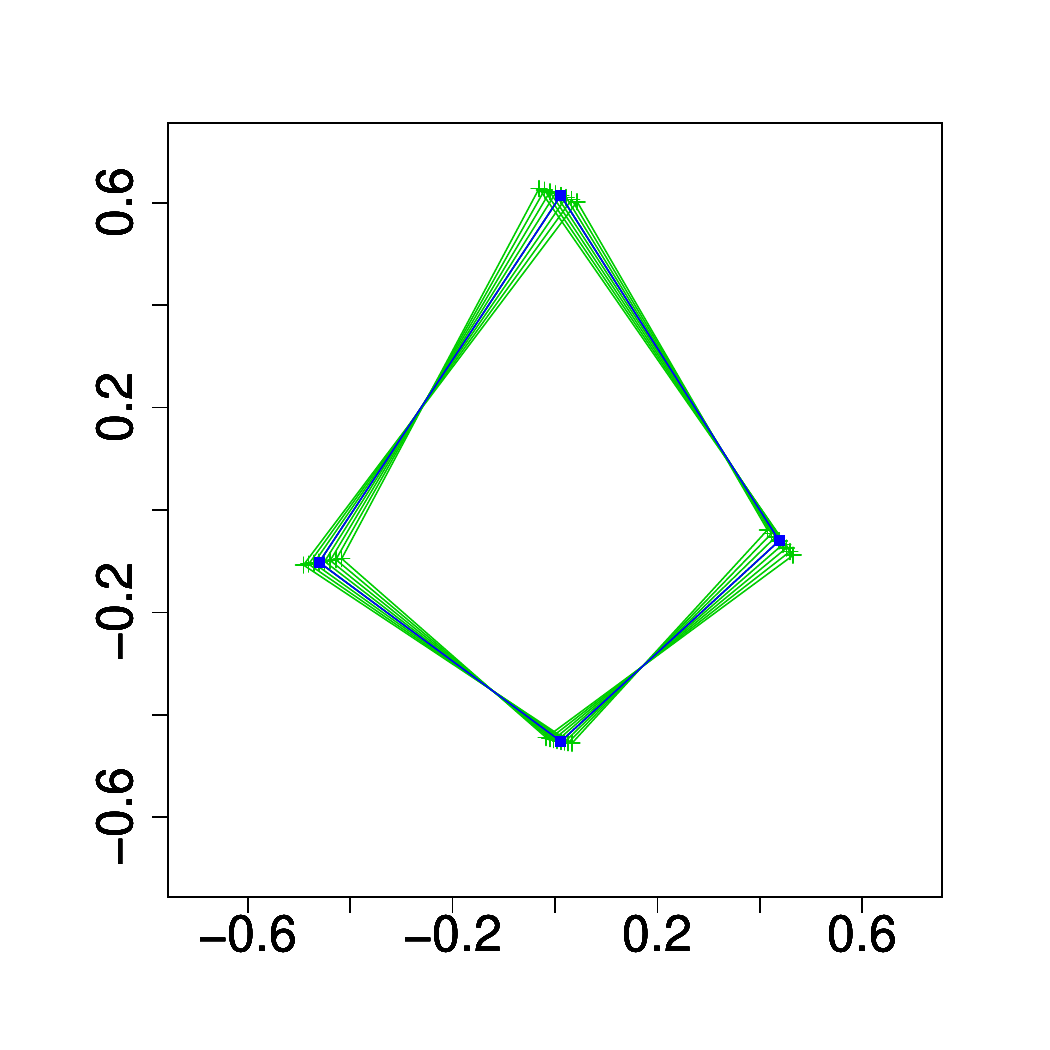}
  \includegraphics[width=0.24\textwidth]{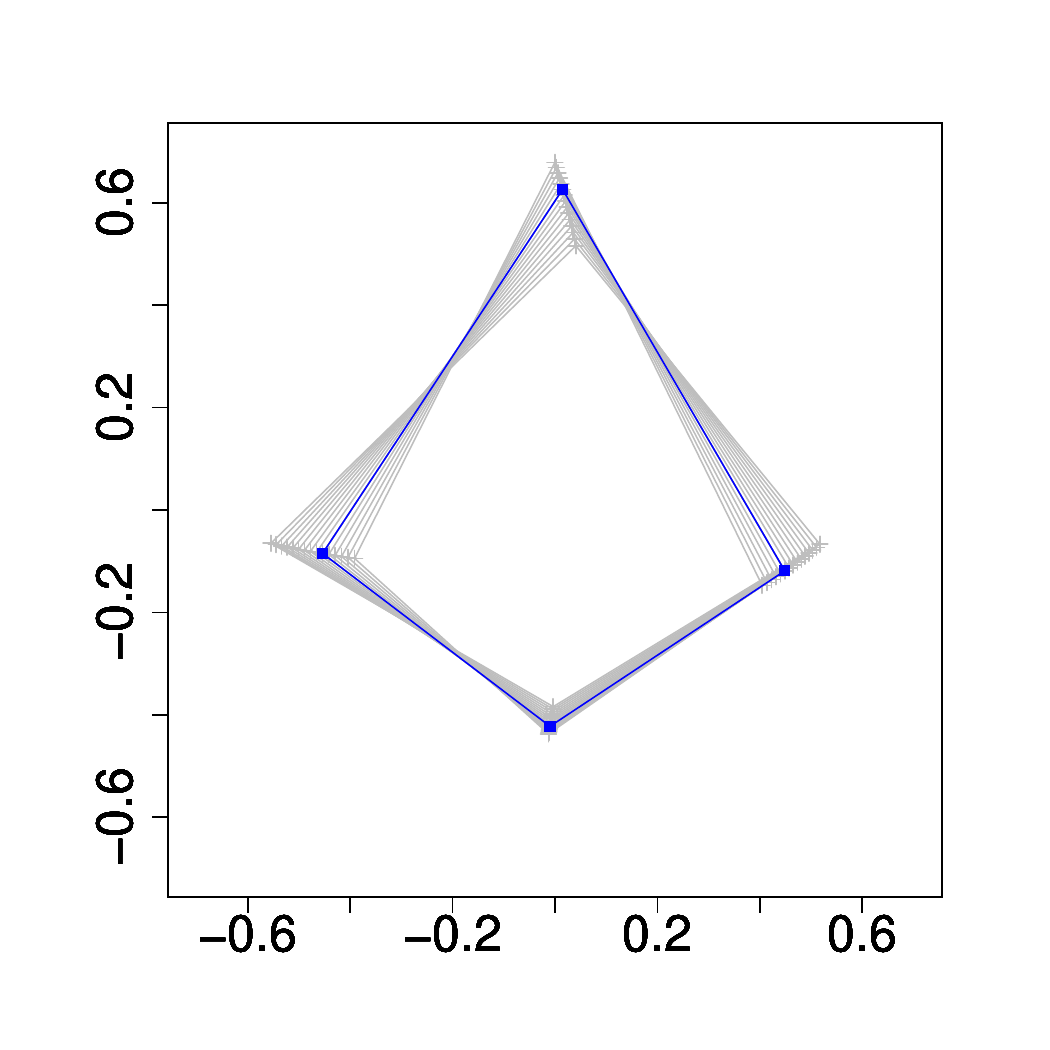}
  \includegraphics[width=0.24\textwidth]{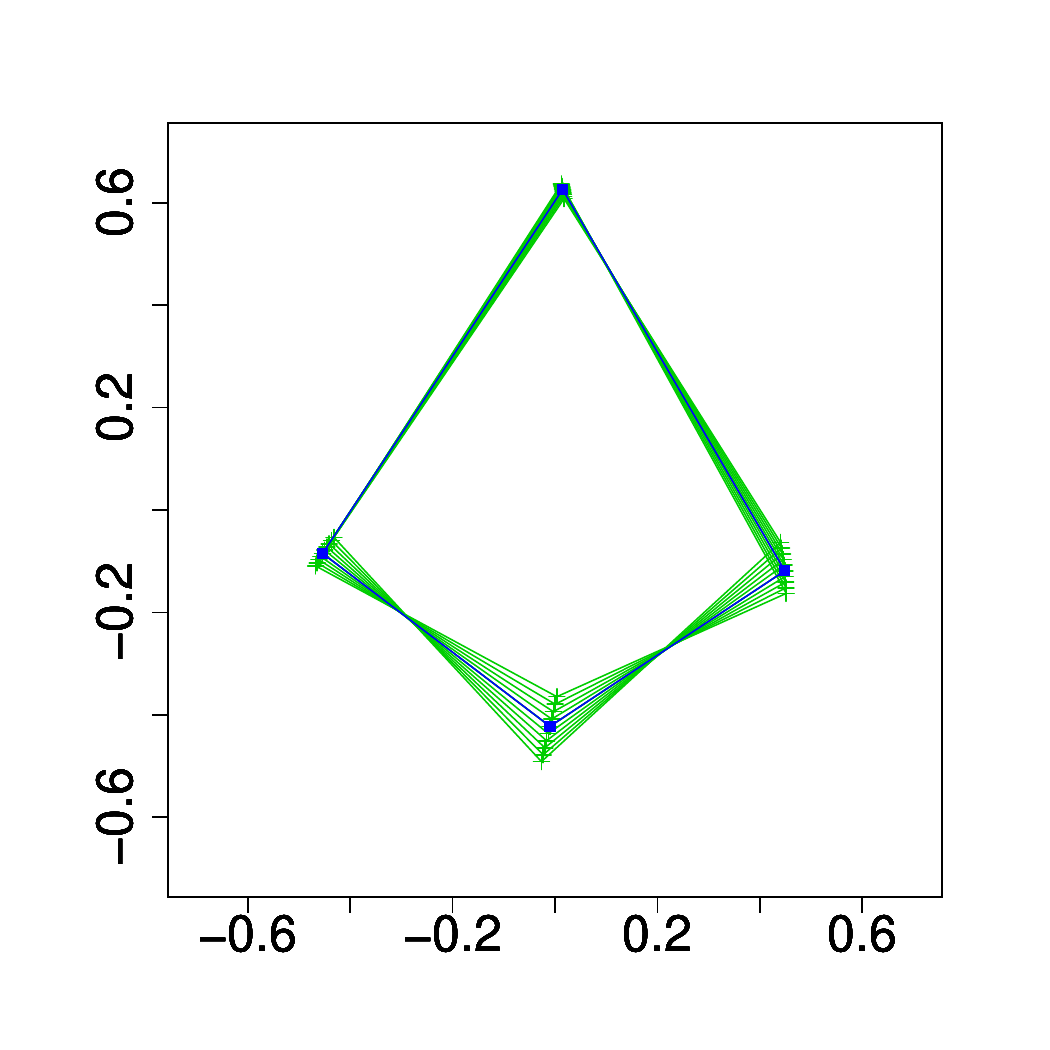}\\
  \includegraphics[width=0.24\textwidth]{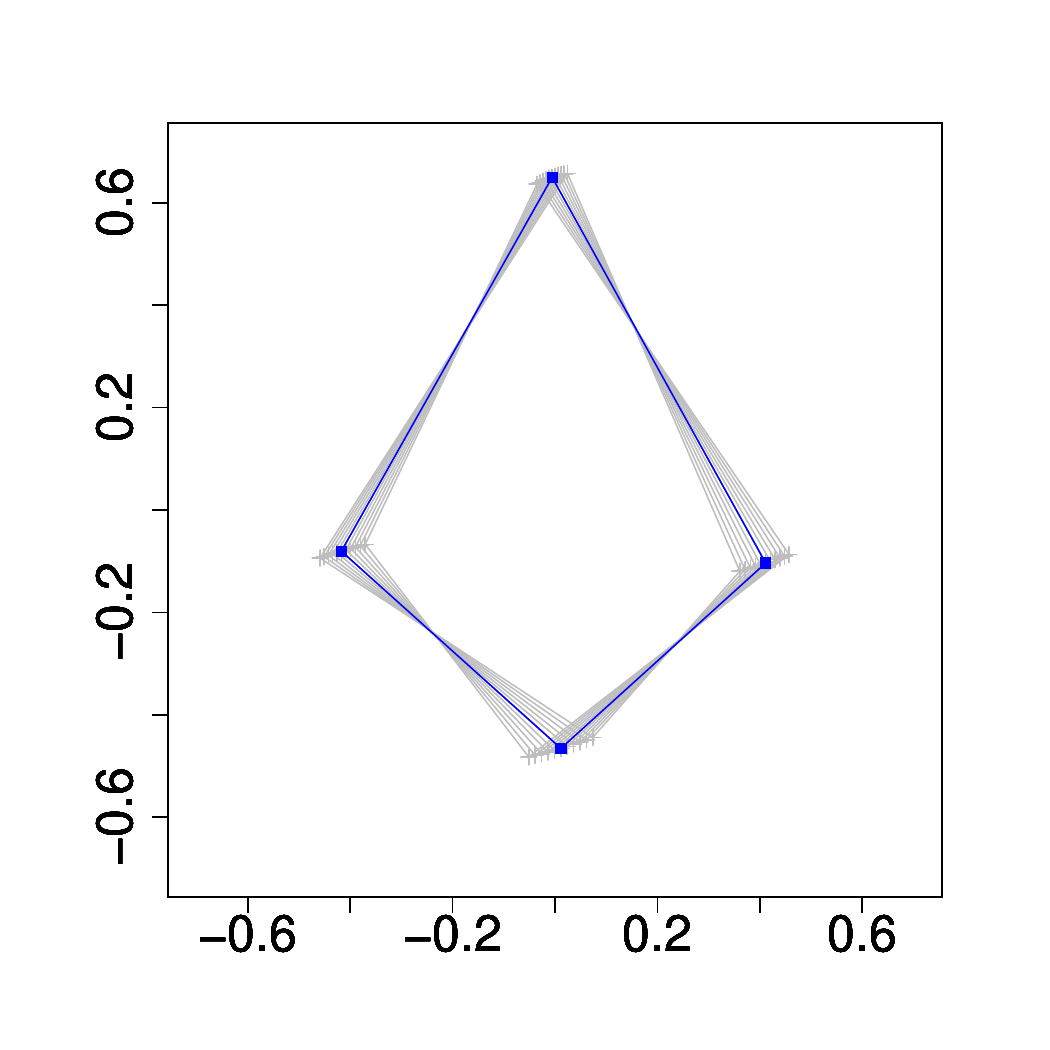}
  \includegraphics[width=0.24\textwidth]{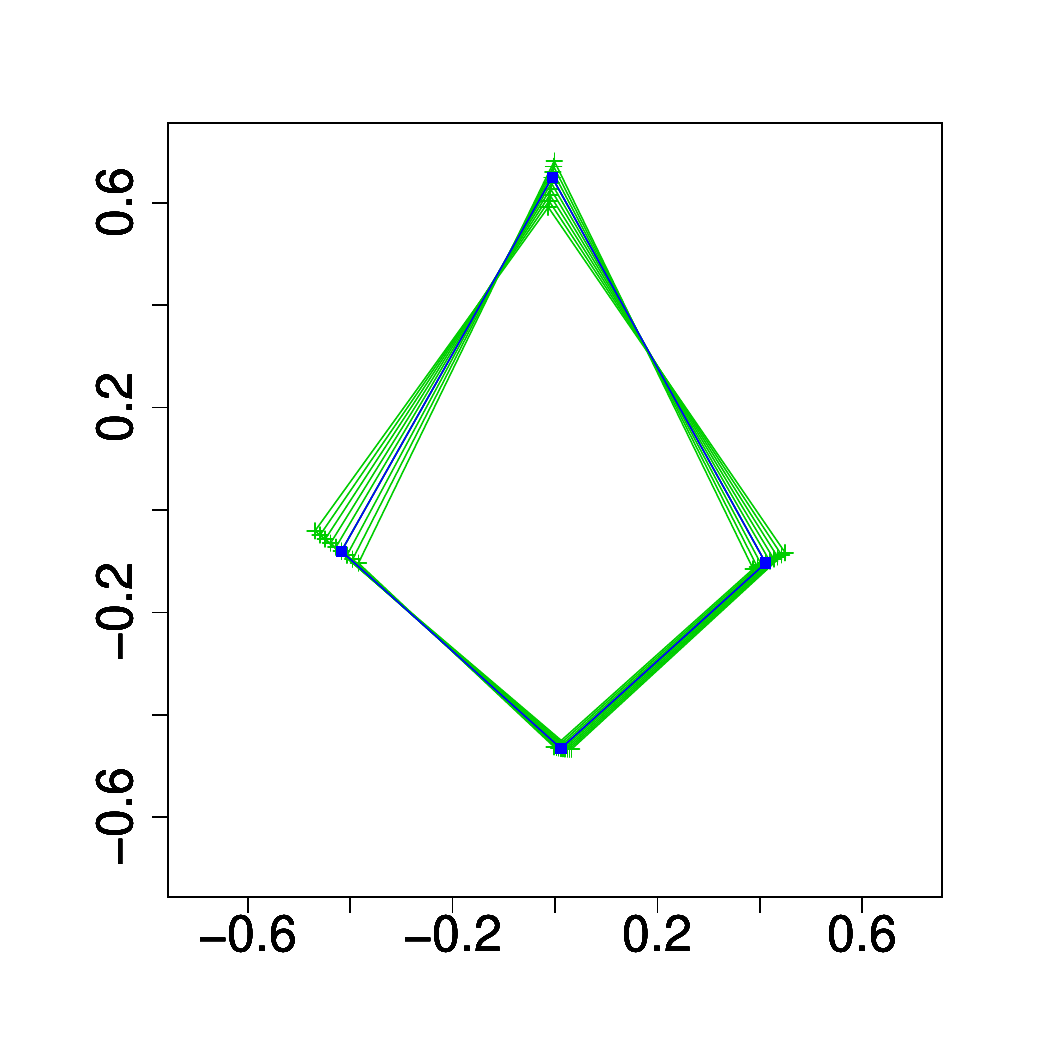}
  \includegraphics[width=0.24\textwidth]{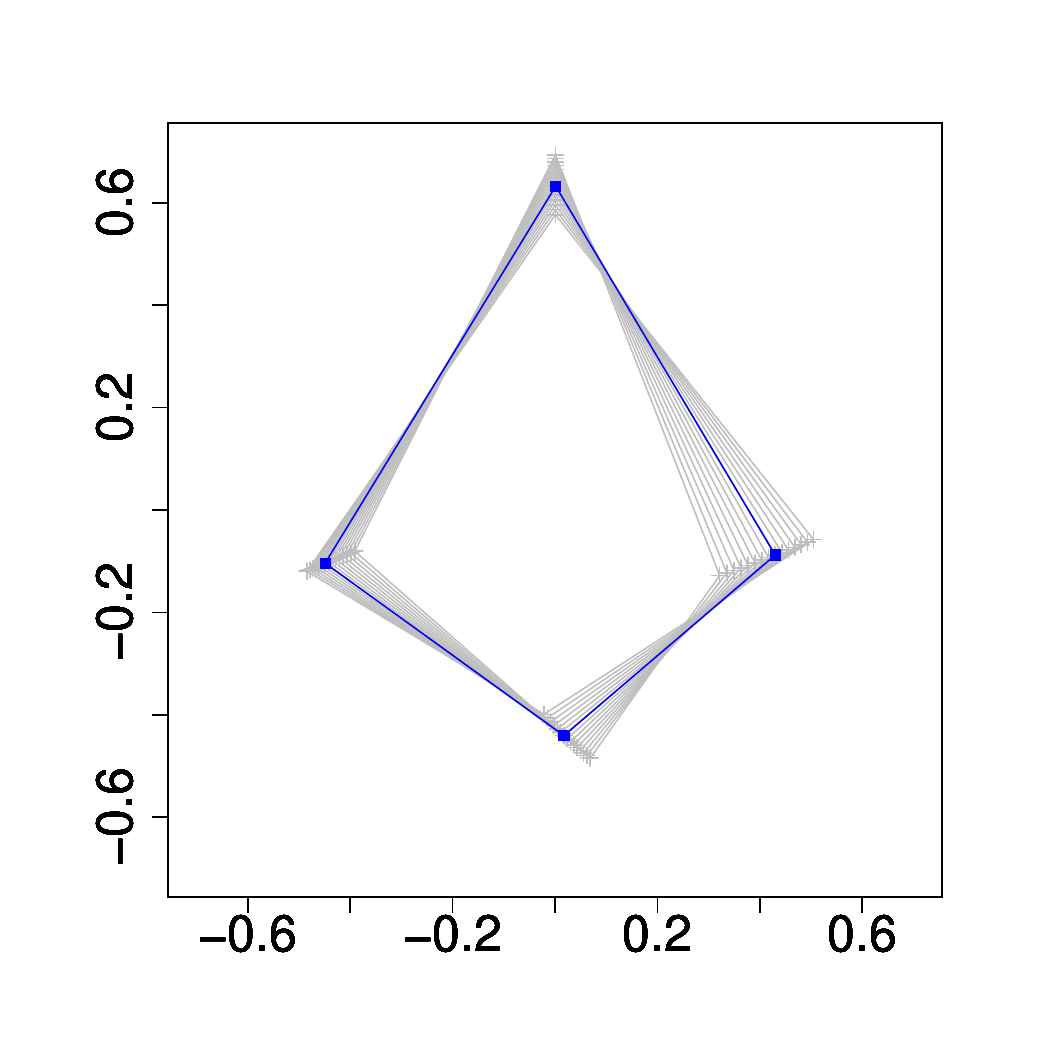}
  \includegraphics[width=0.24\textwidth]{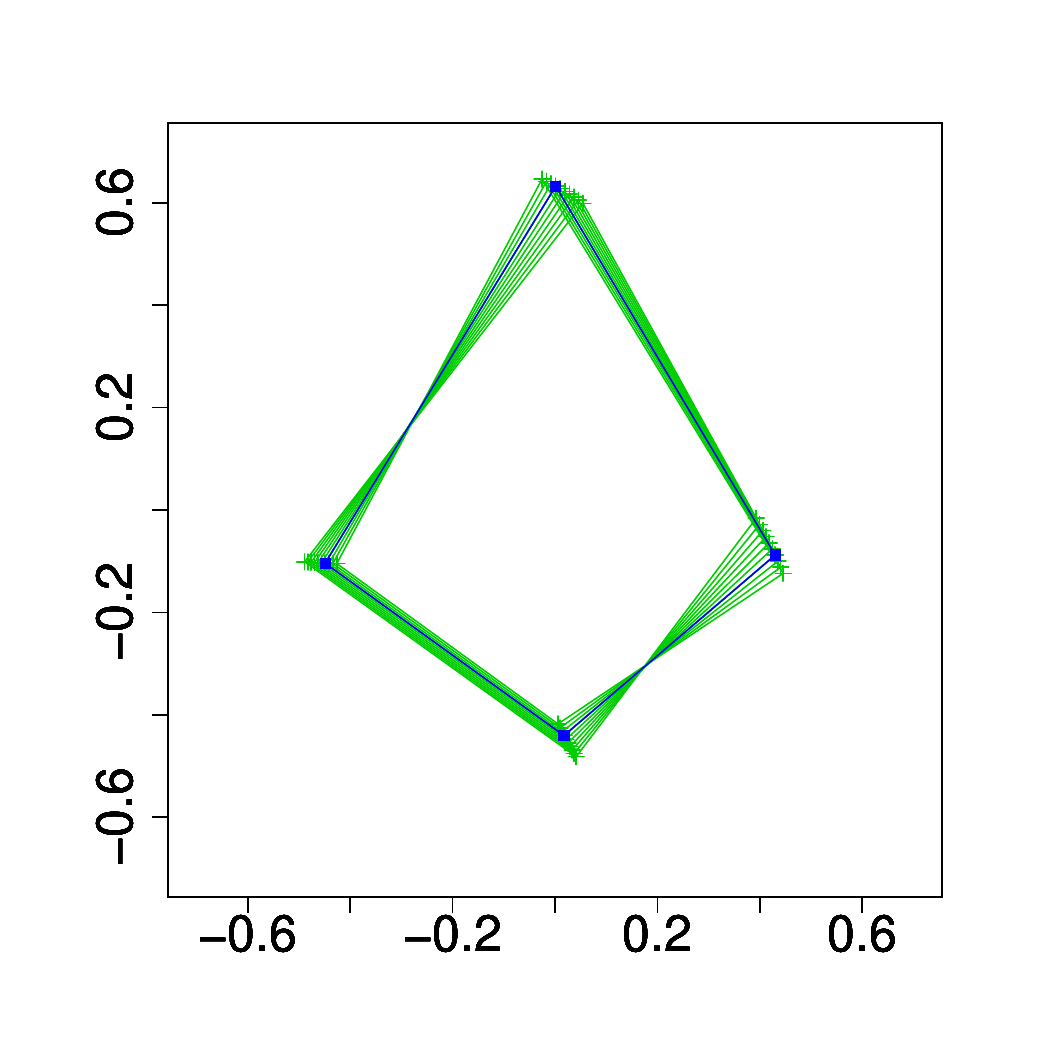}\\
  \caption{Principal sub-manifolds of the leaf growth data. Row 1 (reference tree): first principal direction at breast height; second principal direction at breast height; first principal direction at the crown; second principal direction at the crown. Row 2 - Row 4 provide the same information for Clone 1, 2 and 3.}
  \label{separate}
\end{figure}

As all the leaves are very young, we first combine the leaves from the breast height and crown for each tree. For each tree, a principal sub-manifold is found, where two principal directions are extracted from the fitted sub-manifold. The two principal directions are then transformed to the preshape space and all the landmarks recovered are superimposed. Results for all the three Clones and the reference tree are displayed in Figure \ref{combined}. The leaves of the reference tree exhibit two main kinds of variation: the first one tends to follow the horizontal direction with some effects along the vertical direction at tip. This can be well seen by the first principal direction in Figure \ref{combined}(a); the second one concentrates on petiole, which is displayed by the second principal direction in Figure \ref{combined}(b). The three Clones reveal different patterns of variation from the reference tree; between them,  each differs from the other. Clone 1 shows more variation at the petiole and the left extension in the first principal direction, while the second principal direction shows more variation at the right extension; the two principal directions of Clone 2 behave more similarly as that of the reference tree, with some other variation appearing in the second principal direction of Clone 2 at the right extension and the tip; unlike Clone 1 and 2, variation in both vertical and horizontal directions appear evenly in either the first or the second principal direction for Clone 3. The same analysis for the leaves at breast height and crown alone has also been performed separately with a similar outcome, as shown in Figure \ref{separate}, the result suggesting no different conclusion.   

\section{Introduction to landmark shape spaces}\label{app:14-landmark-shapes}% \label{spherical}
Here we introduce the notion of landmark shape spaces, which are used in one of the applications below. From the shape analysis point of view, landmark coordinates retain the geometry of a certain point configuration. The landmarks are observations, which are usually positions or correspondences on an object in an appropriate coordinate system. See, e.g., \cite{Dryden1998} for an accessible
overview for a rapid introduction. Consider a suitable ordered set of $k$ landmarks of an object, namely a $k$-ad (where $k \geq 2$), with each landmark lying in $\mathbb{R}^{d'}$. That is, 
\begin{align*}
  z=\left\{z^j \in \mathbb{R}^{d'}: 1 \leq j \leq k \right\},
\end{align*}
To compare the shapes of objects described by $k$ landmarks $z_j$, one can define the Kendall shape space $\Sigma_{d'}^k$ of configurations, which are invariant under translation, scaling, and rotation. This is achieved by transforming $k$-ads, $z=(z_1^T, \dots, z_k^T)^T$, to points on the unit sphere:
\begin{itemize}
  \item[] \textbf{Translation invariance}: $z^{*}=((z_1-\bar{z})^T, \dots, (z_k-\bar{z})^T)^T$, where $\bar{z}=\frac{1}{k} \sum_{j=1}^{k} z_j$
  \item[] \textbf{Scale invariance}: $z_{\mbox{{\tiny pre}}}=\frac{z^{*}}{\left\|z^{*}\right\|}$
  \item[] \textbf{Rotation invariance}: $\left[z\right]=Rz_{\mbox{{\tiny pre}}}$ for $R := \text{id}_k \otimes \widetilde{R}$ with the Kronecker product $\otimes$ and $\widetilde{R} \in SO(d')$. For $d'=2$ this reduces to $\left[z\right]=R(\theta)z_{\mbox{{\tiny pre}}}$ or $\left[z\right]=e^{i \theta}z_{\mbox{{\tiny pre}}}$ if $\mathbb{R}^2$ is identified with $\mathbb{C}$, where $-\pi < \theta \leq \pi$.
\end{itemize}

\begin{remark}
  Kendall shape space only leads to a manifold if $d'=2$, therefore we restrict to $d'=2$ here: the translation and scale invariant $z_{\mbox{\tiny pre}} \in S^{2k-3} \subset \mathbb{R}^{2k-2}$ is called the \emph{preshape} of $z$. Centering the data to achieve translation invariance reduces dimension by 2 and projecting to the unit sphere $S^{2k-3} =\left\{ v \in \mathbb{R}^{2k-2}: \left\| v \right\|=1 \right\}$ to achieve scale invariance reduces dimension by 1. Then, $\left[z \right]$ is the \emph{shape} of $z$ given by the \emph{orbit} of the preshape $z_{\mbox{\tiny{pre}}}$ under rotation. $\Sigma_2^k$ is a quotient space of $S^{2k-3}$ with dimension $2k-4$ of equivalence classes of $k$-ads.
\end{remark}

\FloatBarrier
\newpage

%% use bibfile 
\bibliographystyle{Chicago}      % Chicago style, author-year citations
\bibliography{Zhigang}   % name your BibTeX data base

%%%%%%%%%%%%%%%%%%%%%%%%%%%%%%%%%%%%%%%%%%%%%%%%%%%%%%%%%%%%%%%%%%%%%%%%%%%%%%%%%%%%%%%%%%%%%%%%%%%%%%%%%%%%%%%%%%%%%%%%%%%%
\vskip .65cm
\noindent
Zhigang Yao\\
Department of Statistics and Applied Probability  \\
National University of Singapore\\
Singapore 117546
\vskip 2pt
\noindent
E-mail: \texttt{zhigang.yao@nus.edu.sg}\\
\vskip 2pt

\noindent
Benjamin Eltzner\\
Department of Theoretical and Computational Biophysics \\
Max-Planck-Institute for Multidisciplinary Sciences \\
37077 Goettingen, Germany
\vskip 2pt
\noindent
E-mail: \texttt{benjamin.eltzner@mpinat.mpg.de}\\
\vskip 2pt

\noindent
Tung Pham\\
School of Mathematics and Statistics \\
University of Melbourne\\
Victoria 3010 Australia
\vskip 2pt
\noindent
E-mail: \texttt{pham.t@unimelb.edu.au}

% \vskip .3cm
%\centerline{(Received ???? 20??; accepted ???? 20??)}\par
\end{document}